%% file: main.tex
\documentclass[a4paper,11pt]{book}
\usepackage[utf8]{inputenc}
\usepackage[T1]{fontenc}
\usepackage{type1cm}
\usepackage{graphicx}
\usepackage{amsmath}
\usepackage{amssymb}
\graphicspath{{Images/}}
\usepackage[verbose=false,margin=25mm,includeheadfoot,bindingoffset=6mm]{geometry}
\usepackage{fancyhdr}
\usepackage{titlesec}
\usepackage{emptypage}
\usepackage{caption}
\usepackage{subcaption}
\usepackage{hyperref}
\usepackage{lettrine}
\usepackage{lipsum}
\usepackage{color}
\usepackage[normalem]{ulem}
\usepackage{setspace}
\usepackage{cprotect}
\usepackage{multicol}
\usepackage{tikz}
\usetikzlibrary{angles}
\usepackage{multirow}

\usepackage{charter}

\usepackage[backend=biber,style=ieee,citestyle=numeric-comp]{biblatex}
\usepackage[nohyperlinks]{acronym}

\titleformat{\chapter}[display]{\normalfont\huge\bfseries\raggedleft}{\chaptertitlename\ \thechapter}{20pt}{\Huge}
\titlespacing*{\chapter} {0pt}{0pt}{40pt}

\newcommand{\bxl}{b\times\ell}
\newcommand{\bl}{(b,\ell)}

\begin{document}

\pagestyle{empty}

\input{FirstPages/cover}

\newpage
\
\newpage
\input{FirstPages/citation}
\newpage
\
\newpage

\onehalfspacing
\chapter*{Acknowledgements}
\input{FirstPages/acknowledgements}

\thispagestyle{empty}
\singlespacing

\setcounter{page}{-1}
\tableofcontents

\onehalfspacing
\pagestyle{plain}

\chapter*{Abstract}
\addcontentsline{toc}{chapter}{Abstract}
\input{FirstPages/abstract}

\chapter*{R\'esum\'e}
\input{FirstPages/resume}

\chapter*{Acronyms}
\input{FirstPages/acronyms}

\chapter*{Introduction}
\addcontentsline{toc}{chapter}{Introduction}
\input{Chapters/introduction}

\chapter{Dark matter or Physics' greatest investigation}
\label{chap:DMintro}
\input{Chapters/chap1}

\chapter[Following dark matter's steps: Indirect detection]{Following dark matter's steps:\\ Indirect detection}
\label{chap:ID}
\input{Chapters/chap2}

\chapter{Constraining sub-GeV dark matter from diffuse \texorpdfstring{$X$}{X}-rays}
\label{chap:subGeV}
\input{Chapters/chap3}

\chapter{Improving the results with a realistic propagation setup}
\label{chap:prop}
\input{Chapters/chap4}

\chapter{Constraining primordial black holes from Galactic emissions}
\label{chap:PBH}
\input{Chapters/chap5}

\chapter*{Conclusion}
\addcontentsline{toc}{chapter}{Conclusion}
\label{conclusion}
\pagestyle{plain}
\input{Chapters/conclusion}

\appendix{

\chapter{Trigonometry for integrations over regions of interest}
\label{apx:trigo}
\pagestyle{fancy}
\input{Chapters/appendixA}

\chapter{Electron-positron energy-loss functions}
\label{apx:elosses}
\input{Chapters/appendixB}

}

\pagestyle{fancy}
\printbibliography[heading=bibintoc, title={Bibliography}]

\end{document}

%% file: FirstPages/cover.tex

\thispagestyle{empty}
\newgeometry{left=10mm,right=10mm,top=10mm,bottom=20mm,bindingoffset=15mm}

\begin{titlepage}
    \begin{center}
    	
	\vspace*{0.25cm}
	
        \includegraphics[width=0.28\textwidth]{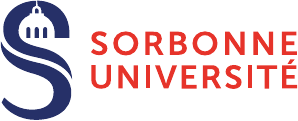}
        \hfill
        \includegraphics[width=0.28\textwidth]{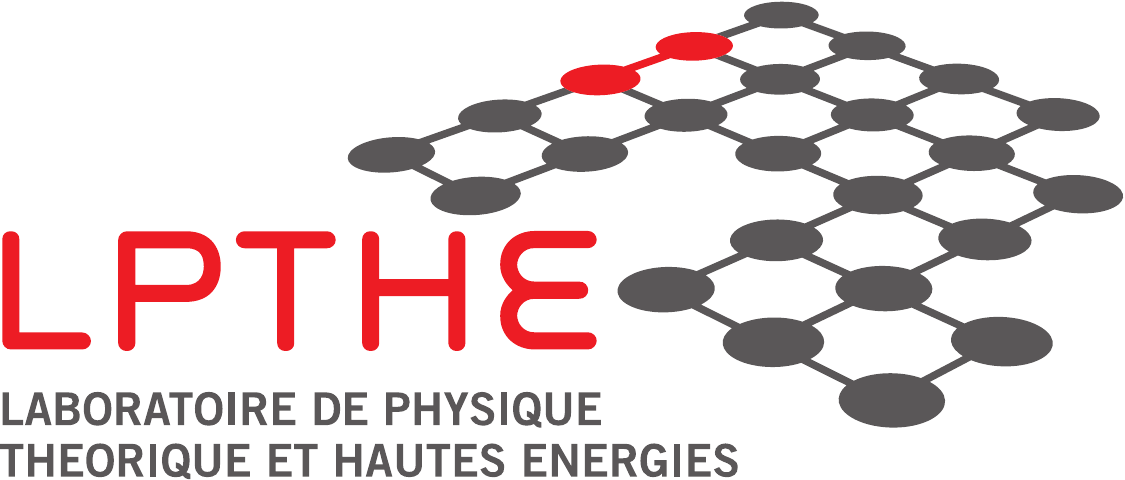}
        
        \vspace*{0.75cm}

        \LARGE
        \textbf{TH\`ESE DE DOCTORAT}\\
        \textbf{DE SORBONNE UNIVERSIT\'E}

        \vspace*{0.75cm}
        \Large
        \textbf{Sp\'ecialit\'e : Physique}

        \vspace*{0.25cm}
        \large
        \textbf{\'Ecole doctorale n\textsuperscript{o}564 : Physique en \^Ile-de-France}

        \vspace*{0.75cm}

        \normalsize
        \textbf{r\'ealis\'ee au}
        
        \vspace*{0.25cm}
        
        \Large
        \textbf{Laboratoire de Physique Th\'eorique et Hautes \'Energies}

        \vspace*{0.25cm}

        \normalsize
        \textbf{sous la direction de Marco CIRELLI}

        \vspace*{0.5cm}
        \textbf{pr\'esent\'ee par}

        \vspace*{0.25cm}
        \LARGE
        \textbf{Jordan KOECHLER}

        \vspace*{0.5cm}
        \normalsize
        \textbf{pour obtenir le grade de}

        \vspace*{0.5cm}
        \Large
        \textbf{DOCTEUR DE SORBONNE UNIVERSIT\'E}

        \vspace*{0.5cm}
        \textbf{Sujet de la th\`ese :}

        \vspace*{0.25cm}
        \LARGE
        \textbf{Ph\'enom\'enologie de la d\'etection indirecte \\ de mati\`ere noire}

        \vspace*{0.5cm}
        \Large
        \textbf{soutenue le 16 Septembre 2024}

        \vspace*{0.5cm}
        \large
        \textbf{devant le jury compos\'e de}

        \setlength{\tabcolsep}{12pt}
        \renewcommand{\arraystretch}{1}
        \vspace*{0.5cm}
        \begin{tabular}{l l}
             \Large \textbf{Marco CIRELLI}                          						& 	\Large \textbf{Directeur de th\`ese} \\
             \large Directeur de recherche (CNRS), Sorbonne Universit\'e			&	\\
             \Large \textbf{Malcolm FAIRBAIRN}								& 	\Large \textbf{Rapporteur} \\
             \large Professeur des universit\'es, King's College de Londres			&	\\
             \Large \textbf{Antoine LETESSIER-SELVON}						& 	\Large \textbf{Pr\'esident du jury} \\
             \large Directeur de recherche (CNRS), Sorbonne Universit\'e			&	\\
             \Large \textbf{Laura LOPEZ-HONOREZ}		       					& 	\Large \textbf{Rapporteuse} \\
             \large Chercheuse qualifi\'ee (FNRS), Universit\'e Libre de Bruxelles		&	\\
             \Large \textbf{Farvah Nazila MAHMOUDI}				      			& 	\Large \textbf{Examinatrice} \\
             \large Professeure des universit\'es, Universit\'e Claude Bernard Lyon 1	&	\\
             \Large \textbf{Gabrijela ZAHARIJAS}					     			& 	\Large \textbf{Examinatrice} \\
             \large Ma\^itresse de conf\'erences, Universit\'e de Nova Gorica			& 	\\
        \end{tabular}
        
    \end{center}
\end{titlepage}

\restoregeometry

%% file: FirstPages/citation.tex

\vspace*{5cm}

\hfill\begin{tabular}{l@{}}
    \textit{Lo duca e io per quel cammino ascoso}\\
    \textit{intrammo a ritornar nel chiaro mondo;}\\
    \textit{e sanza cura aver d’alcun riposo,}\\
    \\
    \noindent\textit{salimmo s\`u, el primo e io secondo,}\\
    \textit{tanto ch’i’ vidi de le cose belle}\\
    \textit{che porta ’l ciel, per un pertugio tondo.}\\
    \\
    \noindent\textit{E quindi uscimmo a riveder le stelle.} \\
    \\
    \hspace{5cm}
    Canto XXXIV\\
    \hspace{5cm}
   Inferno, Divina Commedia\\
    \hspace{5cm}
    Dante Alighieri
\end{tabular}

\vspace*{\fill}

%% file: FirstPages/acknowledgements.tex

Words will never be enough to express my gratitude to everyone who supported me throughout this journey, but I hope this section conveys my appreciation. 

First and foremost, I wish to thank my PhD advisor, Marco Cirelli, for making this entire adventure possible. Your kindness, expertise, sense of humor, and support are unparalleled, and it has been an honor to be your student -- although you always treated me as an equal, which made the experience incredibly fulfilling. You helped me develop autonomy and critical thinking, invaluable skills for becoming a well-rounded researcher. I am also deeply grateful for the many opportunities you provided me to travel and share our work.

Next, I would like to thank my collaborators, Shyam Balaji, Pedro De la Torre Luque, Nicolao Fornengo, Elena Pinetti, and Brandon M. Roach, for the trust and the knowledge you generously shared with me. My thanks also go to the referees of my manuscript, Malcolm Fairbairn and Laura Lopez-Honorez, for kindly agreeing to review my thesis, as well as to the other members of the jury, Antoine Letessier-Selvon, Nazila Mahmoudi, and Gabrijela Zaharijas, for their thoughtful questions and comments during the defence.

I would like to acknowledge the IAP for its hospitality, allowing me to work there every Thursday. Special thanks to Joe Silk for our (more or less) weekly discussions. I was always impressed, and admittedly a bit intimidated, by your boundless enthusiasm and wealth of knowledge, all of which were complemented by your humbleness and kindness.

I also wish to thank the LPTHE for providing such a great working environment. In particular, I am grateful to the administrative staff, especially Michela, Fran\c{c}oise,  Carole and Laurent for their invaluable help with the sometimes tedious administrative procedures.

I cannot forget to thank to my fellow PhD students and the entire LPTHE social circle, with whom I shared these memorable years. Pursuing a PhD can often be a solitary endeavor, and I was incredibly fortunate to have you all to share lunch breaks, coffee breaks, parties and evenings along the Seine. I will deeply miss those moments.

Special thanks to my close friends, my favorite `pigeons' -- Cervane, Hugo, Paul and Victoria -- for creating the perfect, most supportive environment where we could all vent about our PhD experiences. I am also grateful to Alexane and Camille, whom I have known for years and who supported me throughout this journey.

Last but certainly not least, I wish to thank my family for their constant encouragement, from when I was a child dreaming of piloting the Large Hadron Collider at CERN. Even though I ultimately chose a more theoretical path, I will always be deeply grateful to have you in my life. Thank you for your unwavering support.

%% file: FirstPages/abstract.tex

\lettrine[lines=3, nindent=1pt]{A}{mong} the open problems of modern physics, dark matter (DM) is one of the most fascinating. It explains several gravitational anomalies observed at different scales: the flatness of rotation curves of spiral galaxies, the dynamics of galaxy clusters, the distribution of large-scale structures in the Universe, and the anisotropies in the temperature of the cosmic microwave background. Precise measurements of the latter, possibly combined with other techniques, show that DM constitutes about a quarter of the Universe's energy budget. Although we have reliable observational evidence of DM's existence, its nature remains a mystery, as no observation has yet shown that DM can interact with ordinary matter other than gravitationally. Numerous hypotheses about its nature remain. DM could exist as elementary particles not included in the Standard Model of particle physics, or as macroscopic compact objects such as primordial black holes (PBH).

\medskip

To reveal the nature of DM, or to rule out hypotheses concerning it, several observational techniques are available. In this thesis, we focus on the method of indirect detection, which involves looking for signals of the annihilation or decay of DM in the form of charged cosmic rays, photons or neutrinos. Each product carries different types of information. Photons and neutrinos, being neutral particles, can propagate without being deflected by the surrounding magnetic fields, making it easier to trace their source of emission. Charged cosmic rays, on the other hand, may consist of antimatter, which is less likely produced by astrophysical processes and can therefore be detected with a low background.

\medskip

In this thesis, we study the emission of secondary photons by the interaction of DM products with the Galactic environment. Specifically, we consider the case in which DM is a particle with a mass below a GeV. The electrons and positrons produced could interact with ambient photons in the Galaxy, producing $X$-rays through inverse Compton scattering. The prediction of the spectrum of this radiation, compared with data from $X$-ray observatories, provides strong constraints on this type of DM. Similarly, we apply this same principle to the case of PBH evaporation in order to impose strong constraints on them.

%% file: FirstPages/resume.tex

\lettrine[lines=3, nindent=1pt]{P}{armi} les probl\`emes ouverts de la physique moderne, la mati\`ere noire (MN) est l'un des plus fascinants. Elle explique plusieurs anomalies gravitationnelles pr\'esentes dans des syst\`emes \`a diff\'erentes \'echelles : la platitude des courbes de rotation des galaxies spirales, la dynamique des amas de galaxies, la distribution des structures \`a grande \'echelle dans l'Univers et les anisotropies de la temp\'erature du fond diffus cosmologique. Des mesures pr\'ecises de ces derni\`eres, \'eventuellement combin\'ees \`a d'autres techniques, montrent que la MN constitue environ un quart de l'\'energie totale de l'Univers. Bien que nous ayons des preuves observationnelles fiables de l'existence de la MN, sa nature reste un myst\`ere, car \`a ce jour, aucune observation n'a pu montrer que la MN peut interagir avec la mati\`ere ordinaire autrement que gravitationnellement. De nombreuses hypoth\`eses sur sa nature subsistent. La MN pourrait exister sous la forme de particules \'el\'ementaires qui ne font pas partie du Mod\`ele Standard de la physique des particules, ou sous forme d'objets macroscopiques compacts tels que les trous noirs primordiaux (TNP).

\medskip

Pour r\'ev\'eler la nature de la MN, ou \`a d\'efaut, pour \'ecarter des hypoth\`eses la concernant, plusieurs techniques d'observation existent. Dans cette th\`ese, nous nous concentrons sur la m\'ethode de d\'etection indirecte, qui consiste \`a rechercher des signaux d'annihilation ou de d\'esint\'egration de la MN sous forme de rayons cosmiques charg\'es, de photons ou de neutrinos. Chaque produit porte diff\'erents types d'informations. Les photons et les neutrinos, \'etant des particules neutres, peuvent se propager sans \^etre d\'evi\'es par les champs magn\'etiques environnants, ce qui facilite la localisation de leur source d'\'emission. Les rayons cosmiques charg\'es, en revanche, peuvent \^etre constitu\'es d'antimati\`ere, qui est tr\`es peu produite par des processus astrophysiques et peut donc \^etre d\'etect\'ee avec un faible signal de fond.

\medskip

Dans cette th\`ese, nous \'etudions l'\'emission de photons secondaires par l'interaction des produits issus de la MN avec le milieu Galactique. En particulier, dans le cas o\`u la MN est une particule de masse inf\'erieure \`a un GeV, les \'electrons et positrons produits pourraient interagir avec des photons ambiants de la Galaxie, produisant des rayons $X$ par effet Compton inverse. La pr\'ediction du spectre de ce rayonnement, compar\'ee aux donn\'ees des observatoires \`a rayons $X$, permet d'obtenir de fortes contraintes sur ce type de MN. De m\^eme, nous appliquons ce m\^eme principe au cas de l'\'evaporation des TNP afin de leur imposer de fortes contraintes.

%% file: FirstPages/acronyms.tex

\vspace*{\fill}
\vspace*{-2cm}

\begin{multicols}{2}
    \begin{acronym}[MOND]
        \acro{LCDM}[$\Lambda$CDM]{Lambda cold dark matter}
        \acro{BH}{Black hole}
        \acro{BSM}{Beyond the Standard Model}
        \acro{CMB}{Cosmic microwave background}
        \acro{CR}{Cosmic ray}
        \acro{DD}{Direct detection}
        \acro{DM}{Dark matter}
        \acro{dSph}{Dwarf spheroidal galaxy}
        \acro{FSR}{Final state radiation}
        \acro{GC}{Galactic centre}
        \acro{GMF}{Galactic magnetic field}
        \acro{GP}{Galactic plane}
        \acro{GRXE}{Galactic ridge X-ray emission}
        \acro{ICS}{Inverse Compton scattering}
        \acro{ID}{Indirect detection}
        \acro{IR}{Infrared}
        \acro{ISM}{Interstellar medium}
        \acro{l.o.s.}{Line of sight}
        \acro{MW}{Milky Way}
        \acro{NFW}{Navarro-Frenk-White}
        \acro{Ps}{Positronium}
        \acro{PBH}{Primordial black hole}
        \acro{ROI}{Region of interest}
        \acro{SL}{Starlight}
        \acro{SM}{Standard Model}
        \acro{WIMP}{Weakly interacting massive particle}
    \end{acronym}
\end{multicols}

\vspace*{\fill}

%% file: Chapters/introduction.tex

\lettrine[lines=3, nindent=1pt]{T}{he} problem of dark matter (DM) remains to this day one of the main open problems in modern physics. So far, we only witness its existence through its gravitational impact in the Universe, such as in the rotation curves of spiral galaxies, the dynamics of galaxy clusters, the distribution of large-scale structures and the temperature anisotropies in the cosmic microwave background (CMB). We also know, thanks to precise measurements of the latter, that DM is not baryonic and constitutes almost 85\% of the total amount of matter contained in the Universe as well as around 25\% of its energy budget. Despite its imposing presence, we know very little about its properties and how it was produced in the early Universe. In turn, its nature remains a mystery and currently there exists a plethora of well-motivated candidates that can describe the observational properties of DM. Experimental methods such as collider searches, direct and indirect detection are crucial to help us understand the big picture, as their role is to test these candidates. All of these aspects -- observational evidence of the existence of DM, possible production scenarios and candidates, and experimental ways to test them -- are explored in Chapter~\ref{chap:DMintro}.

\medskip

In this thesis, we focus on the indirect detection (ID) of DM, which consists of probing signals from DM in the form of stable, charged cosmic rays (such as $e^\pm$ and light antinuclei) or radiation (photons and neutrinos). In particular, if DM is made of unknown particles, their annihilation or decay are expected to produce such signals, which can be hopefully detected in observatories and other experiments. Alternatively, if DM is constituted of macroscopic objects called primordial black holes (PBHs), they can produce similar signals through Hawking evaporation.

On the theoretical side, the goal is to predict what is the expected flux of these DM products, which essentially consists of i) computing the spectrum of these products at their production point and ii) dealing with their propagation through the (extra-)galactic medium. These steps depict the interdisciplinary nature of DM ID: the first one is entirely related to particle physics, which predicts what happens between the sole DM annihilation, decay or evaporation and the production of stable states, whereas the second one is related to astrophysics, that predicts what constitutes the interstellar medium through where the DM-produced particles propagate. After computing the expected flux of DM products at Earth, we can compare them with data in order to either provide a DM interpretation of an excess on top of the known astrophysical background, or otherwise set bounds on DM properties.

We can conduct this phenomenological approach on various targets such as celestial bodies, our own Galaxy, dwarf spheroidal galaxies and galaxy clusters. The bottom line is that an ideal target for DM ID should be sufficiently rich in DM, to be able to measure a strong signal from it, as well as sufficiently low in terms of astrophysical activity, as its modeling can be challenging and therefore be a source of large uncertainties. No target is perfect and compromise has to be made. More details on general DM ID phenomenology are provided in Chapter~\ref{chap:ID}.

\medskip

The phenomenology DM ID is very rich, in the sense that there is a plethora of possible DM candidates that each can provide many types of indirect signals. For example, in Chapter~\ref{chap:subGeV}, we consider that DM consists of sub-GeV particles, which is an increasingly attractive assumption, especially due to the lack of signals from paradigmatic GeV$-$TeV DM candidates in experiments. However, sub-GeV DM can also be challenging to probe indirectly: their annihilation or decay produce sub-GeV $e^\pm$ that are highly impacted by solar winds, decreasing drastically their flux at Earth. Moreover, there is currently no $\gamma$-ray observatory that can probe the $100$ keV $-$ $100$~MeV energy range with a good enough sensitivity, which can limit the detection of photons produced directly by the annihilation or decay of sub-GeV DM. A way to circumvent these issues is to study low-energy photons emitted through final state radiations, radiative decays of DM-produced unstable states but also during the propagation of DM-produced $e^\pm$, mainly through up-scattering ambient photons in the Galaxy via the inverse Compton effect. These photons can then be probed efficiently in $X$-ray observatories.

An advantage of sub-GeV DM is that there are only a few kinematically open DM annihilation and decays channels. We considered the following final states: electrons, muons and charged pions. After predicting semi-analytically the flux of $X$-rays for each channel, using a minimalistic propagation setup for DM-produced $e^\pm$, we perform an analysis over various diffuse $X$-rays datasets from {\sc NuStar}, {\sc Integral}, {\sc Suzaku} and {\sc Xmm-Newton}. We find that the latter provide us among the most stringent constraints on annihilating and decaying sub-GeV DM, due to i) the region where the data have been taken (close to the GC) and ii) how the data were cleaned from point sources. However, our bounds suffer from the uncertainties on the different astrophysical ingredients we used (DM profile, gas and ambient photon densities), whose impact we quantify.

\medskip

In Chapter~\ref{chap:prop}, we perform a similar analysis as in Chapter~\ref{chap:subGeV} but instead we study the impact on the bounds of adopting a realistic cosmic ray propagation setup. In particular, $e^\pm$ produced by lower mass DM can be reaccelerated through their interaction with the turbulent component of the Galactic magnetic field, and would in turn up-scatter ambient photons to $X$-ray energies, which was not the case without this effect. This significantly improves the limits on sub-GeV DM, especially for DM masses below $20$ MeV. For this study, we treat all propagation-related effects using the numerical code \verb|DRAGON2|, and perform computations of the secondary $X$-ray flux using \verb|HERMES|.

Unfortunately, the propagation parameters come with their uncertainties as well, and we show how they can impact the robustness of our $X$-ray limits. We also derive complementary bounds on sub-GeV using the local $e^\pm$ flux measured by {\sc Voyager 1} (with the same propagation setup). They appear to be comparable to other similar limits from the literature. In addition, they are weaker than the $X$-ray ones, although more robust, since they depend less on the choice of the DM profile. 

\medskip

PBHs can also constitute a credible DM candidate. These are black holes that are not produced by the collapse of a heavy star, but instead that could appear through the collapse of localised matter over densities in the early Universe. A way to probe these PBHs is to look for the products of their evaporation due to the Hawking mechanism, which has been proven to be efficient at excluding PBHs as a DM component for PBH masses below $5\times10^{17}$~g. The mass range between $\sim 5\times10^{17}$ and $10^{22}$ g is still unconstrained, therefore PBHs constituting the totality of DM remains a possibility as long as their mass is contained in this range. Refining the evaporation limits can therefore help us in probing the lower bound of this mass range, and if nothing is found, constrain it even further.

In Chapter~\ref{chap:PBH}, we investigate three complementary probes of PBH evaporation. Since PBHs with a mass above $7.5\times10^{14}$ g are expected to evaporate the same final states as for sub-GeV DM, we evaluate the expected PBH-produced $e^\pm$ flux at Earth as well as the secondary $X$-ray emissions in the same manner as in Chapter~\ref{chap:prop}. Compared to the case of decaying sub-GeV DM, the only thing that changes is the injection spectrum of $e^\pm$ which shows similarities with the one of a black body, convoluted with decay spectrum of evaporated muons and charged pions. The last probe we consider is the expected emission of $511$ keV photons due to both the annihilation of PBH-produced $e^+$ in the interstellar medium and the decay of para-positronium formed by PBH-produced $e^+$ and free $e^-$ in ionised interstellar gas. We then use longitude profiles of the $511$ keV line measured by {\sc Integral} in order to set a bound on PBHs.

In the end, we find that the $511$ keV limit is the strongest and most robust one among the three probes, and, in our fiducial scenario of astrophysical parameters, can even compete with other bounds in the literature. Moreover we discuss the impact of different spin and mass distributions on the limits we derive.

\medskip

The work presented in this thesis has led to three publications:
\begin{enumerate}
	\item[\cite{Cirelli:2023tnx}] M. Cirelli, N. Fornengo, \textbf{J. Koechler}, E. Pinetti and B. M. Roach, \textit{Putting all the X in one basket: Updated X-ray constraints on sub-GeV Dark Matter}, \href{http://dx.doi.org/10.1088/1475-7516/2023/07/026}{\textit{JCAP} \textbf{07} (2023) 026}, [\cprotect{\href{http://arxiv.org/abs/2303.08854}}{\verb|2303.08854|}],
	\item[\cite{DelaTorreLuque:2023olp}] P. De la Torre Luque, S. Balaji and \textbf{J. Koechler}, \textit{Importance of cosmic ray propagation on sub-GeV dark matter constraints}, \href{http://dx.doi.org/10.3847/1538-4357/ad41e0}{\textit{Astrophys. J.} \textbf{968} (2024) 1, 46}, [\cprotect{\href{http://arxiv.org/abs/2311.04979}}{\verb|2311.04979|}],
	\item[\cite{DelaTorreLuque:2024qms}] P. De la Torre Luque, \textbf{J. Koechler} and S. Balaji, \textit{Refining Galactic primordial black hole evaporation constraints}, \cprotect{\href{http://arxiv.org/abs/2406.11949}}{\verb|2406.11949|},
\end{enumerate}
which constitute more or less the main body text of Chapters~\ref{chap:subGeV}, \ref{chap:prop} and \ref{chap:PBH} respectively, with some introductory details and refinements.

\medskip

Finally, in the~\nameref{conclusion}, we summarise the main results, as well as present an outlook on some of the possible prospects on the research conducted in this thesis.

%% file: Chapters/chap1.tex

\pagestyle{fancy}

\lettrine[lines=3, nindent=1pt]{T}{hough} the title of this first chapter seems boastful at the very least, we cannot deny the fact that the dark matter (DM) problem is one of the most important and fascinating mystery of the Universe, which involves researchers from three of the main fields of Physics: astrophysics, particle physics and cosmology. First postulated by Fritz Zwicky in 1933 in his seminal work on the Coma cluster~\cite{Zwicky:1933gu}, this idea of an `invisible' type of matter --~that can explain the discrepancy between the cluster's mass computed with the virial theorem -- was however not taken seriously for over 40 years. Thanks to Vera Rubin and Kent Ford’s work on the measurement of the Andromeda galaxy rotation curve~\cite{Rubin:1970zza}, the DM hypothesis was then popularised. Combination of $X$-ray emission measurements and gravitational lensing on merging galaxy clusters also back up Zwicky's original claim. Finally, the precise measurement of the cosmic microwave background (CMB) by the {\sc Planck} experiment~\cite{Planck:2018vyg} leads to the same finding at cosmological scales. Although the existence of DM is a scientific consensus, its nature remains elusive. We have an idea of what could be some of the properties of DM, but many candidates can fit such a description. To this day, many physicists around the globe are committed to build and maintain sophisticated experiments and develop theoretical tools in order to find the culprit. In this chapter, we outline some well-known observational evidence of DM (Section \ref{sec:evidence}), then we enumerate the established properties of DM as well as fitting candidates (Section \ref{sec:DMpropcand}), to finally explain what are some of the current ways to probe them (Section \ref{sec:detecmethods}).

\newpage
\section{Evidences of the existence of dark matter}
\label{sec:evidence}

Since its original claim in 1933, we have accumulated observational evidence of DM at different scales. This section lists some of the most well-known ones: the discrepancies in rotation curves of spiral galaxies (Section~\ref{subsec:rotcurves}), in the mass of galaxy clusters and evidence from merging ones (Section~\ref{subsec:galclust}) and finally evidence from measurements of temperature anisotropies of the CMB (Section~\ref{subsec:CMB}).

\subsection{Discrepancies in rotation curves of spiral galaxies}
\label{subsec:rotcurves}

We begin by computing the circular velocity $v_c$ of stars populating a spiral galaxy. Let us first assume the mass distribution of the spiral galaxy to be spherical. For a given star of mass $m$ in uniform circular motion around the centre of the galaxy at a distance $r$, Newton's second law gives
\begin{equation}
    m\frac{v^2_c(r)}{r} = m\frac{GM(r)}{r^2} \implies v_c(r) = \sqrt{\frac{GM(r)}{r}}\;,
\end{equation}
where $M(r)=4\pi \int_0^r r'^2\rho(r') dr'$ is the mass enclosed within the radius $r$ and $G$ the gravitational constant (and $\rho(r)$ is the mass density of the galaxy at a radius $r$). In a typical spiral galaxy, all of visible matter is contained in a sphere of radius of $10-20$ kpc, meaning that for larger radii we retrieve the Keplerian behavior $v_c(r) \propto 1/\sqrt{r}$.

\begin{figure}[t]
    \centering
    \begin{subfigure}[c]{0.38\linewidth}
        \centering
        \includegraphics[width=\linewidth]{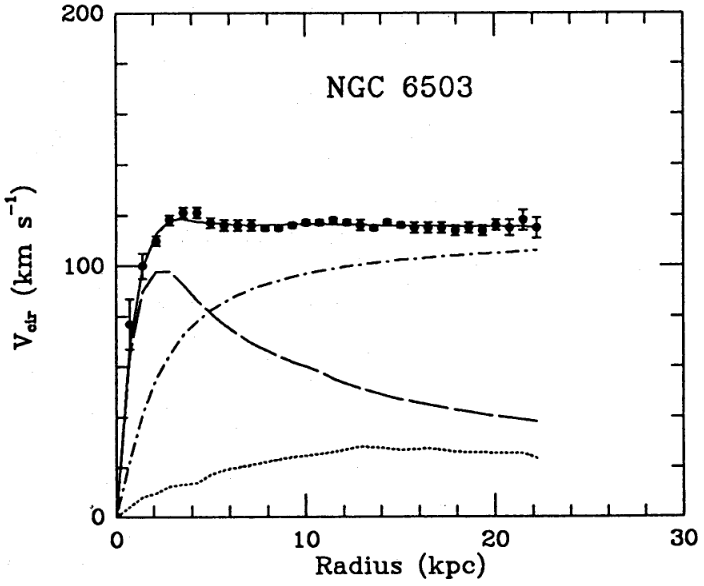}
    \end{subfigure}
    \hfill
    \begin{subfigure}[c]{0.6\linewidth}
        \centering
        \includegraphics[width=\linewidth]{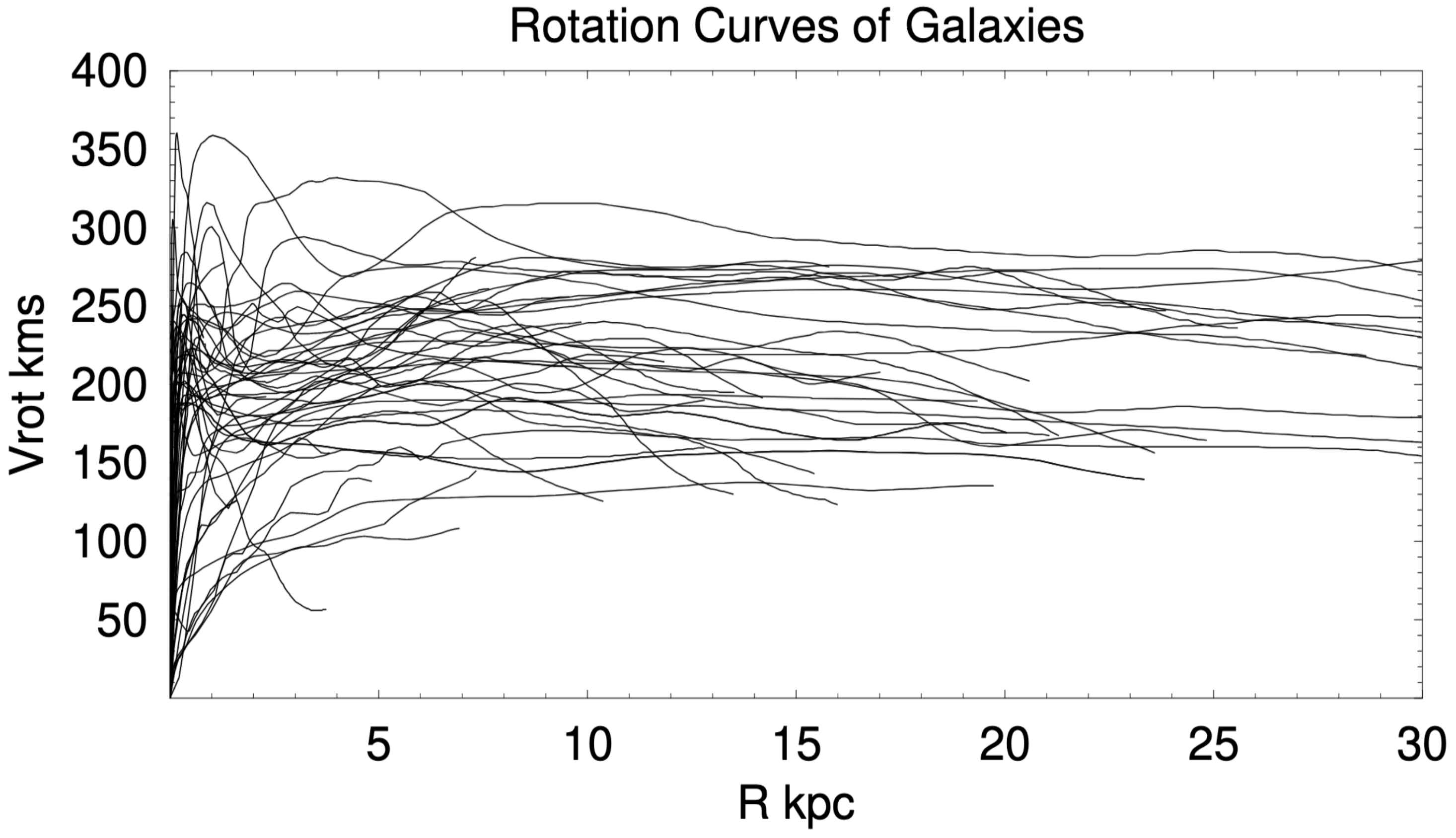}
    \end{subfigure}
    \caption{Left panel: Rotation curve of the galaxy NGC 6503~\cite{Begeman:1991iy}. The dashed line represents the contribution from visible components, the dotted one from the gas, the dot-dashed one from the DM halo and the solid one the sum. Right panel: Rotation curves of a sample of numerous spiral galaxies~\cite{1999ApJ...523..136S}.}
    \label{fig:rotcurve}
\end{figure}

In a given spiral galaxy, we can measure the circular velocity $v_c$ of visible matter at a distance $r$ from its centre by combining two types of observations: i) for lower $r$, measurements of emission lines in the optical range, ii) for higher $r$, of the $21$ cm line, which is the spectral line created by the spin-flip of the electron of an hydrogen atom. First highlighted by Vera Rubin and Kent Ford in 1970 in the Andromeda galaxy~\cite{Rubin:1970zza}, such observations show that $v_c$ is approximately constant at the outskirts of galaxies, and therefore we infer that galaxies are not entirely composed of visible matter: there exists a halo of matter which \emph{a priori} only interacts gravitationally and which extends beyond the galactic disk (typically $100-200$ kpc from the centre of the galaxy) and satisfies $M(r)\propto r$ ($\rho(r) \propto 1/r^2$) at larger $r$\footnote{This relation does not hold for $r\rightarrow\infty$ as $M(r)$ would diverge. We can introduce an effective cut-off radius $r_c$ from which $M(r\geq r_c)$ is constant.}. DM designates this exotic type of matter. The left panel of Figure~\ref{fig:rotcurve} illustrates the point, showing the rotation curve of the galaxy NGC 6503 as an example~\cite{Begeman:1991iy}, with the contribution of visible matter and DM, whereas the right panel shows the same behavior in a sample of spiral galaxies~\cite{1999ApJ...523..136S}.

Another explanation to this discrepancy was proposed by Mordehai Milgrom in 1983, and involves a modification of Newton's second law in the regime of low accelerations~\cite{Milgrom:1983ca}: $F = m \,a\, \mu(a/a_0)$ where $a_0$ defines the interface between the high and low-acceleration regimes, and $\mu(x)$ is an interpolation function that satisfies $\mu(x\gg1) = 1$ (to retrieve Newtonian dynamics) and $\mu(x\ll1) = x$. When $a\ll a_0$ we have $F = m a^2/a_0 = m (v_c^2/r)^2/a_0$, and when applied to the same system as above we obtain
\begin{equation}
    \frac{m}{a_0} \left(\frac{v_c^2(r)}{r}\right)^2 = m \frac{GM(r)}{r^2} \implies v_c(r) = (a_0 GM(r))^{1/4}\;,
\end{equation}
and therefore $v_c$ is a constant for $r$ larger than the spatial extension of all visible matter in the galaxy. This is the founding principle of \emph{Modified Newtonian Dynamics} (or MOND), where the main goal is to explain the discrepancy in rotation curves of spiral galaxies without introducing any exotic matter. While this solution is arguably elegant, it does not constitute a realistic candidate to explain the discrepancy at the scale of galaxy clusters or some properties of the CMB anisotropies.

\subsection{Evidence at the scale of galaxy clusters}
\label{subsec:galclust}

Actually, DM was first postulated at the scale of galaxy clusters. In 1933, Fritz Zwicky computed the mass of the Coma Cluster~\cite{Zwicky:1933gu} by measuring the velocity dispersion of galaxies inside it. This can be done by using the virial theorem, assuming the Coma cluster to be a stationary system. This theorem links the average kinetic energy $\langle T\rangle$ and potential energy $\langle U\rangle$ of the system through the relation $2\langle T\rangle + \langle U\rangle = 0$. On one hand, the kinetic energy is simply written as $\langle T\rangle = M\langle v^2 \rangle /2$ where $M$ is the mass of the cluster and $v$ the velocity of individual galaxies in the cluster. On the other hand, when assuming the distribution of galaxies in the cluster to be uniform in a sphere of radius $R$, we obtain $\langle U\rangle = -3GM^2/(5R)$. Therefore the velocity dispersion is 
\begin{equation}
    \label{eq:veldisp}
    \sigma_v = \sqrt{\langle v^2 \rangle} = \sqrt{\frac{3GM}{5R}} \sim 80\,\textrm{km/s}\;,
\end{equation}
since the Coma cluster has a radius of $R \simeq 0.3$ pc and contains around 800 galaxies with an individual mass of $10^9 M_\odot$. However the measured value of $\sigma_v$ lies around $1000$ km/s. Paired with the measurement of the mass-to-light ratio of $M/L\sim 500M_\odot/L_\odot$, it again suggests the presence of a dominant component of invisible matter in the Coma cluster. As the time goes by, other techniques have been developed to measure the mass of a galaxy cluster and are complementary to the one using kinematics and the virial theorem. The first one is by measuring the $X$-ray surface brightness of the target cluster. 

Galaxy clusters contain hot gas with temperature ranging from around $10$ to $100$ keV. The ionised electrons can produce $X$-rays through two processes: i) inverse Compton scattering (ICS) on CMB photons and ii) bremsstrahlung with ions in the gas. Measurements of the $X$-ray spectrum and intensity can infer the temperature $T_\textrm{gas}$ of the hot intracluster gas and its density $n_\textrm{gas}$, respectively. Then, to relate these measurements with the cluster mass $M$, we have to combine the hydrostatic equilibrium equation of the gas (assuming spherical symmetry)
\begin{equation}
\label{eq:hydroeq}
    \frac{\partial P_\textrm{gas}}{\partial r}=-\mu_\textrm{gas} n_\textrm{gas}\frac{\partial\Phi}{\partial r}\;,
\end{equation}
with the Poisson equation of the cluster's content
\begin{equation}
\label{eq:poisson}
    \vec{\nabla}^2\Phi = 4\pi G \rho_\textrm{tot}\implies \frac{\partial \Phi}{\partial r} = \frac{4\pi G}{r^2}\int_0^r dr'r'^2\rho_\textrm{tot}(r') = \frac{GM}{r^2}\;,
\end{equation}
where $P_\textrm{gas}$ is the gas pressure, $\mu_\textrm{gas}$ its mean atomic (or molecular) mass and $M$ the total mass of the cluster within a radius $r$. Combining Equations~\ref{eq:hydroeq} and \ref{eq:poisson} as well as adding the ideal gas law $P_\textrm{gas} = n_\textrm{gas} k_B T_\textrm{gas}$, we get
\begin{equation}
    k_B\frac{\partial \left(n_\textrm{gas}T_\textrm{gas}\right)}{\partial r} = -\mu_\textrm{gas}n_\textrm{gas}\frac{GM}{r^2}\;,
\end{equation}
and after some work we can write
\begin{align}
    M & = - \frac{k_Br^2}{G\mu_\textrm{gas}n_\textrm{gas}}\frac{\partial \left(n_\textrm{gas}T_\textrm{gas}\right)}{\partial r} \\
    & = - \frac{k_BT_\textrm{gas}r}{G\mu_\textrm{gas}}\left[\frac{r}{n_\textrm{gas}}\frac{\partial n_\textrm{gas}}{\partial r} + \frac{r}{T_\textrm{gas}}\frac{\partial T_\textrm{gas}}{\partial r}\right] \\
    & = -\frac{k_BT_\textrm{gas}r}{G\mu_\textrm{gas}}\left[\frac{\partial\ln n_\textrm{gas}}{\partial\ln r}+\frac{\partial\ln T_\textrm{gas}}{\partial\ln r}\right]\;.
\end{align}
Thanks to this equation we are able to link properties of the $X$-ray spectrum from a galaxy cluster to its total mass. First uses of this observational technique in 1971 limited the amount of hot intracluster gas to less than 2\% of what is required for gravitational binding~\cite{1971Natur.231..107M}.

\begin{figure}[t]
    \centering
    \begin{subfigure}[c]{0.48\linewidth}
        \centering
        \includegraphics[width=\linewidth]{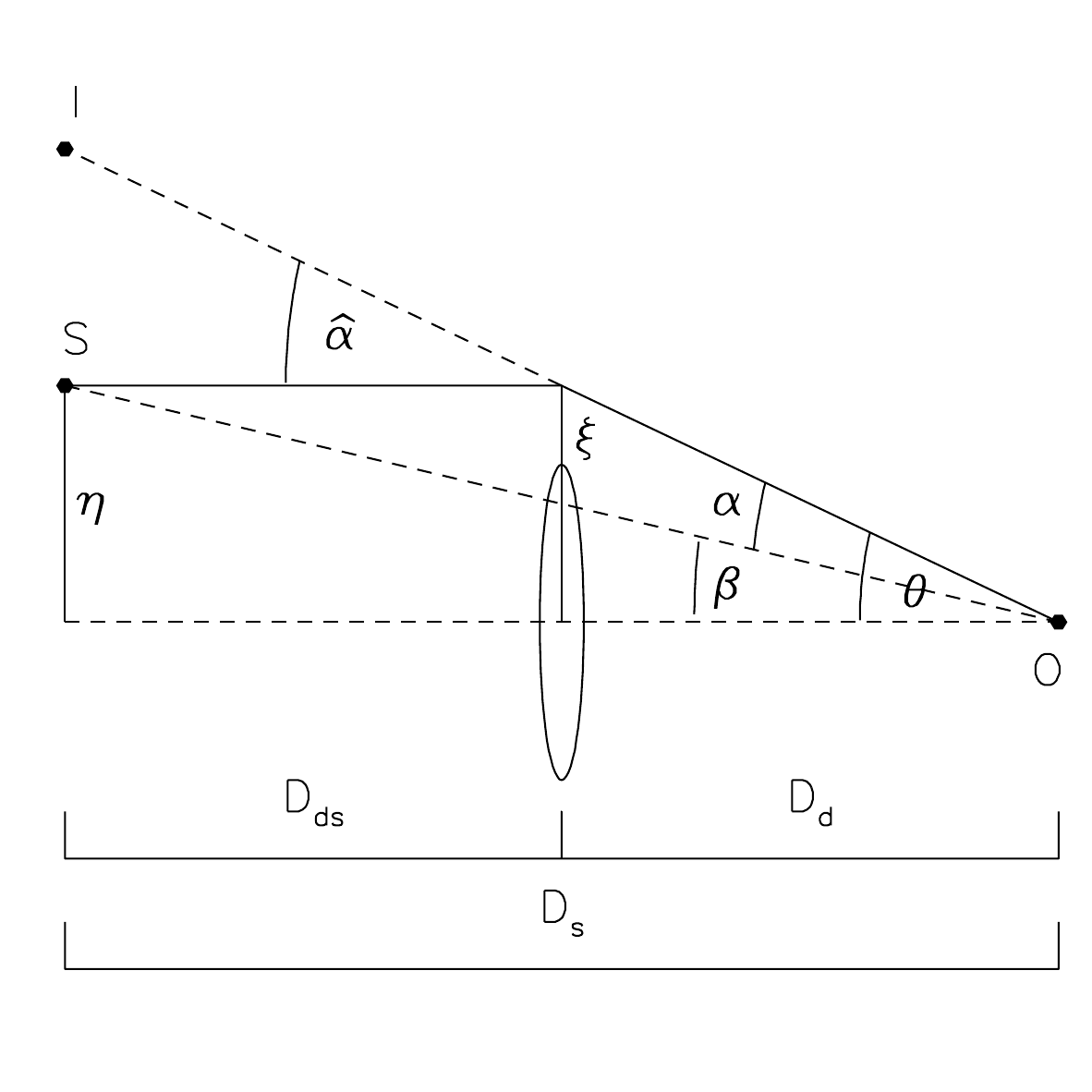}
    \end{subfigure}
    \hfill
    \begin{subfigure}[c]{0.48\linewidth}
        \centering
        \includegraphics[width=\linewidth]{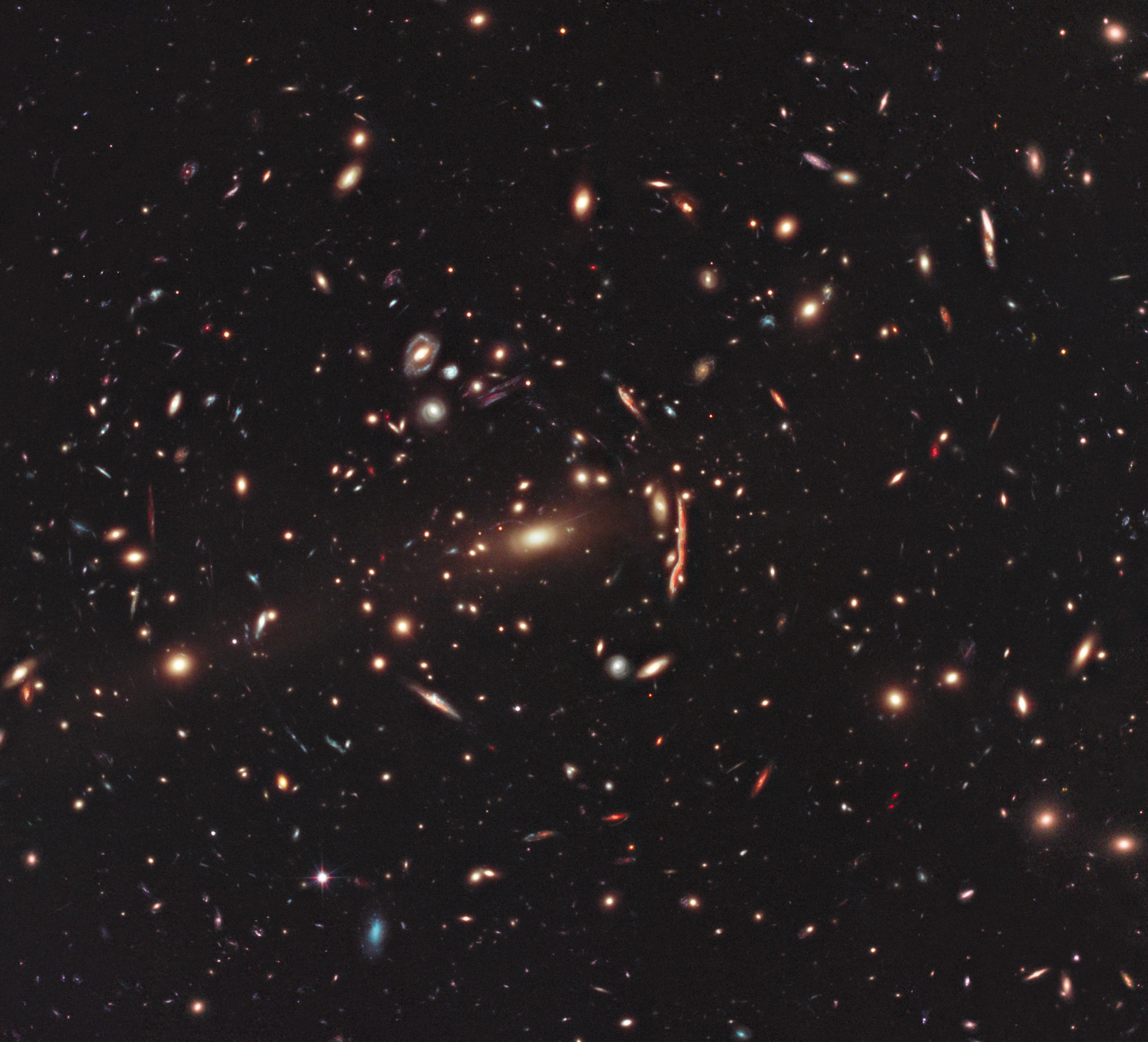}
    \end{subfigure}
    \caption{Left panel: Illustration of a gravitational lens system when the observer $O$, lens (oval shape) and source $S$ are not aligned, producing an image $I$~\cite{Narayan:1996ba}. Right panel: {\sc Hubble} image of the galaxy cluster MACS J1206~\cite{MACSJ1206}.}
    \label{fig:grav-lens}
\end{figure}

Another technique to infer the mass of a galaxy cluster is through weak gravitational lensing. Here the cluster acts as a `gravitational lens' which deflects the light coming from a source that is behind it from the perspective of the observer, resulting in perceived slight shape distortions of the source. An example of this effect is shown in the right panel of Figure~\ref{fig:grav-lens}, where the galaxy cluster MACS J1206 acts as the gravitational lens to some other background galaxies. 

Here we use the \emph{thin lens approximation}, meaning that we assume the distance between the source, lens and observer to be much larger than the size of the lens (\emph{i.e.}\ the source and lens are considered to be point-like). The left panel of Figure~\ref{fig:grav-lens} illustrates a gravitational lens system and defines all of the relevant angles and lengths we are about to use.

The modulus of the deflection angle of the light from the source due a lens of mass $M$ is written
\begin{equation}
    \label{eq:deflect}
    \hat{\alpha}(\xi) = \frac{4GM}{c^2\xi}
\end{equation}
where $\xi$ is the distance between the lens and the light-ray in the lens plane. Also, from Figure~\ref{fig:grav-lens} we can notice that the distance from the source $D_s$ and between the lens and the source $D_{ds}$ can be linked through
\begin{equation}
    \label{eq:geometry}
    \theta D_s = \beta D_s + \hat{\alpha}(\xi)D_{ds}\;,
\end{equation}
and finally by inserting Equation~\ref{eq:deflect} in Equation~\ref{eq:geometry} and realising that $\xi = \theta D_d$ from Figure~\ref{fig:grav-lens}, we get
\begin{equation}
    \label{eq:lenseq}
    \beta(\theta) = \theta - \frac{D_{ds}}{D_d D_s}\frac{4GM}{c^2\theta}\;.
\end{equation}
To ease up the notation, we introduce the \emph{Einstein radius} $\theta_E$
\begin{equation}
    \label{eq:redlenseq}
    \theta_E = \sqrt{\frac{4GM}{c^2}\frac{D_{ds}}{D_d D_s}} \implies \beta(\theta) = \theta - \frac{\theta_E^2}{\theta}\;.
\end{equation}
In the context of strong gravitational lensing, when the lens and source are perfectly aligned with the observer ($\beta = 0$), the image would be a ring of angular radius $\theta_E$, namely an \emph{Einstein ring}. Solving Equation~\ref{eq:redlenseq} provide us with two possible images
\begin{equation}
    \label{eq:lenseqsol}
    \theta_\pm = \frac{1}{2}\left(\beta \pm \sqrt{\beta^2+4\theta_E^2}\right)\;. 
\end{equation}
The surface brightness of the source and the image should be the same as dictated by Liouville's theorem. Therefore the total flux received by the image should vary by a factor $\mu$ named \emph{magnification} which is the ratio between the image and source solid angles. Assuming spherical symmetry, we have 
\begin{equation}
    \label{eq:magni}
    \mu = \frac{\theta}{\beta}\frac{d\theta}{d\beta}\;.
\end{equation}
The magnification of each of the two images are given by inserting Equation~\ref{eq:lenseqsol} in Equation~\ref{eq:magni}
\begin{equation}
    \mu_\pm = \left[1-\left(\frac{\theta_E}{\theta_\pm}\right)^2\right]^{-1} = \frac{u^2+2}{2u\sqrt{u^2+4}}\pm\frac{1}{2}\;,
\end{equation}
where $u = \beta/\theta_E$. When $\theta_\pm < \theta_E$, we get $\mu_\pm < 0$, meaning that the two images switched their positions from the observer's perspective. Therefore the total magnification should be
\begin{equation}
    \mu = \lvert\mu_+\rvert + \lvert\mu_-\rvert = \frac{u^2+2}{u\sqrt{u^2+4}}\;.
\end{equation}
Even though we considered a very simplified model, it still shows that weak gravitational lensing observables can help us in inferring the mass of a galaxy cluster.

\begin{figure}[t]
    \centering
    \begin{subfigure}[c]{0.48\linewidth}
        \centering
        \includegraphics[width=\linewidth]{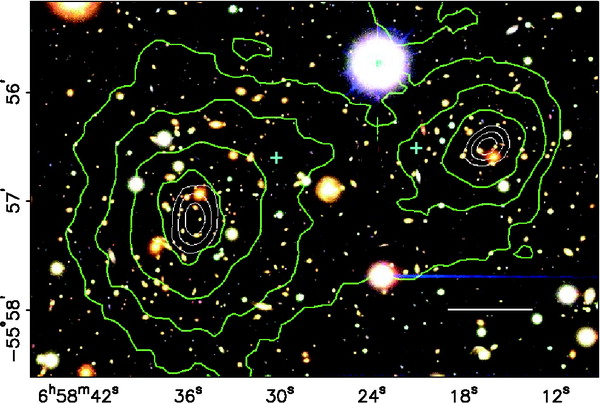}
    \end{subfigure}
    \hfill
    \begin{subfigure}[c]{0.48\linewidth}
        \centering
        \includegraphics[width=\linewidth]{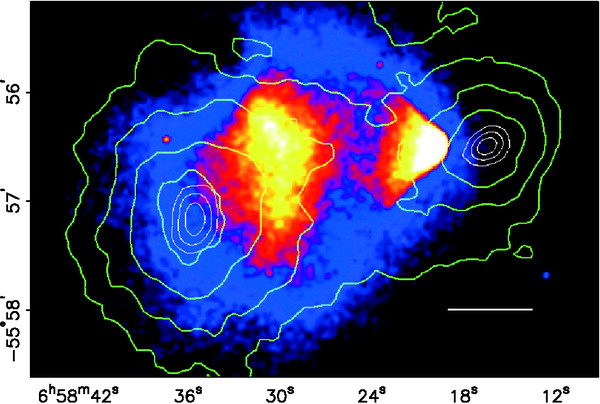}
    \end{subfigure}
    \caption{Observation of the Bullet Cluster by {\sc Magellan} in the optical range showing the position of individual galaxies (left panel) and by {\sc Chandra} in the $X$-ray range showing the intracluster gas heating from the galaxy cluster collision (right panel). The green contours show the mass distribution obtained from weak gravitational lensing. $x$-~and $y$-scales are respectively right ascension and declination \cite{Clowe:2006eq}.}
    \label{fig:bullet}
\end{figure}

A very strong evidence of the existence of DM in the Universe, that also shows how important it is to have complementarity in observational techniques, is the Bullet Cluster. This is a system of two colliding galaxy clusters $3.72$ billion light-years away from Earth that has been observed by the {\sc Magellan} space probe and the {\sc Hubble} Space Telescope in the optical range, by the {\sc Chandra} $X$-ray Observatory and where weak gravitational lensing measurements have been performed (see Figure~\ref{fig:bullet}). While optical observations show that individual galaxies were not much affected by the collision, the $X$-ray map shows that the two intracluster gas clouds have interacted electromagnetically with each other during the collision, therefore heated up and slowed down. But most importantly, weak gravitational lensing measurements show that the mass distribution of the merging cluster is not centred on the interacting gas clouds. Instead, it is observed that two massive halos have passed through each other during the collision. While the halos are collisionless almost in the same fashion as the individual galaxies, the mass of the halos is measured to be way greater than the total mass of the galaxies, meaning that the halos are mostly constituted of DM. Observations of dozens of other merging galaxy clusters~\cite{Harvey:2015hha} also back up this inference. This also constitutes a problem for the MOND interpretation of the missing mass problem, since MOND assumes that all matter is baryonic and therefore we would see the mass distribution of a merging cluster to be centred around the interacting intracluster gas clouds.

\subsection{The cosmic microwave background}
\label{subsec:CMB}

In the previous subsections, we showed that DM is a compelling solution to the missing mass problem at the scale of galaxies and galaxy clusters. However, these observations are not able to infer the total amount of DM in the whole Universe. For this, we turn to cosmological probes of DM, especially measurements of the temperatures anisotropies in the CMB. The CMB originated during the recombination epoch (378,000 years after the Big Bang). Before this epoch, the Universe was very hot and contained unbound electrons and protons, therefore it was opaque to radiation. As the Universe continued to expand, its temperature decreased, thus free electrons and protons could form the first hydrogen atoms. Radiation was then able to propagate through this new state of matter. The CMB is the oldest source of radiation we will ever be able to probe. It was actually discovered inadvertently by Arno Penzias and Robert Wilson in 1965~\cite{1965ApJ...142..419P}. 

\begin{figure}[t]
    \centering
    \begin{subfigure}[c]{0.49\linewidth}
        \centering
        \includegraphics[width=\linewidth]{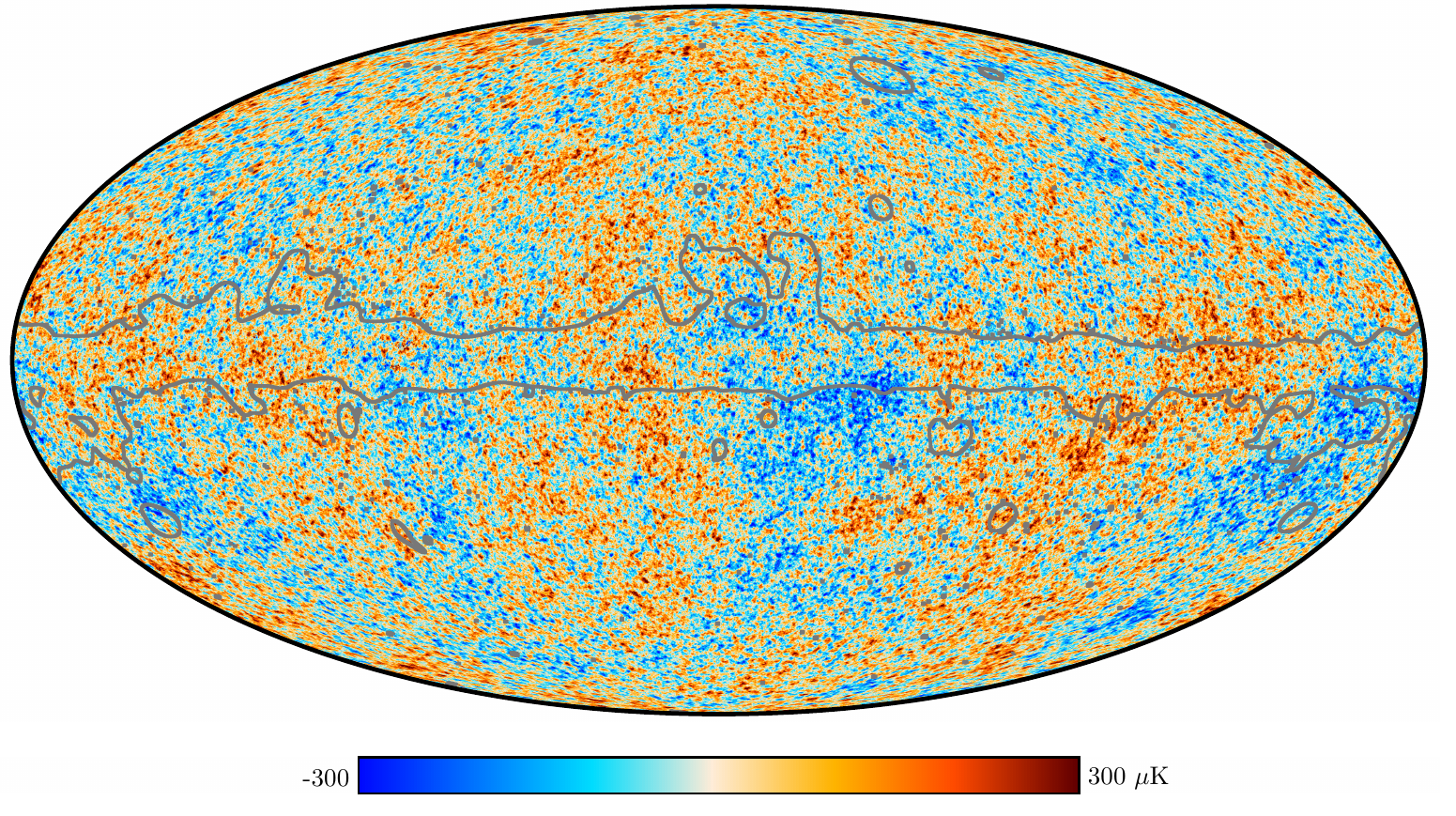}
    \end{subfigure}
    \hfill
    \begin{subfigure}[c]{0.49\linewidth}
        \centering
        \includegraphics[width=\linewidth]{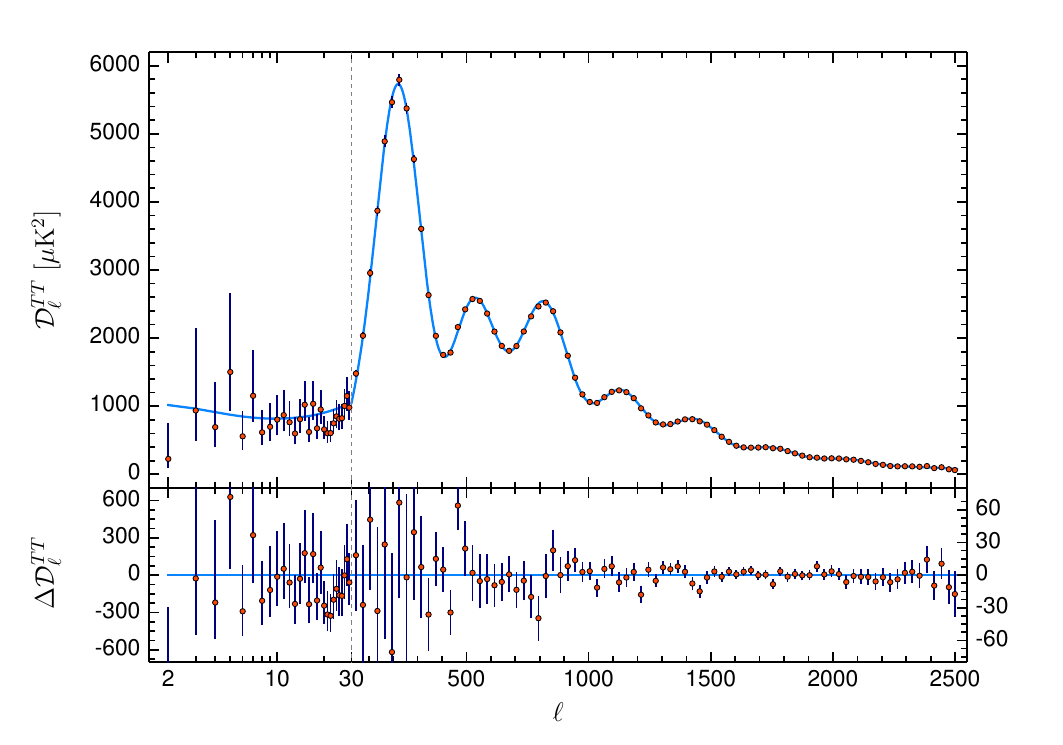}
    \end{subfigure}
    \caption{Left panel: Sky map of the CMB temperature anisotropies measured by {\sc Planck}~\cite{Planck:2018nkj}. Right panel: {\sc Planck} measurements of the variance of the CMB temperature anisotropies across a range of angular scales (red points) together with a 6-parameter fit to the flat $\Lambda$CDM model (blue line)~\cite{Planck:2018vyg}.}
    \label{fig:CMB}
\end{figure}

The properties of the CMB are essentially the ones of a black body of temperature $T_0 = 2.7255 \pm 0.0006$ K~\cite{Fixsen:2009ug} while showing some temperature anisotropies at the $10^{-5}$ level across a wide range of angular scales. The left panel of Figure~\ref{fig:CMB} shows a map of these anisotropies measured by {\sc Planck}~\cite{Planck:2018nkj}. To take them into account, the CMB temperature at a specific position in the sky $(\theta, \phi)$ can be written in the following expansion
\begin{equation}
    T(\theta, \phi) = T_0\;\sum_{l=0}^{+\infty}\sum_{m=-\ell}^{+\ell} a_{\ell m} Y_{\ell m}(\theta,\phi)\;,
\end{equation}
where $Y_{\ell m}(\theta,\phi)$ are the spherical harmonics. The observable would be the variance of the coefficients $a_{\ell m}$ across various angular scales, defined as
\begin{equation}
    C_\ell = \langle a_{\ell m} a_{\ell m}^* \rangle = \frac{1}{2\ell+1} \sum_{m=-\ell}^\ell \lvert a_{\ell m} \rvert^2\;.
\end{equation}
These measurements are then used to fit a specific cosmological model in order to infer a range of the current properties of our Universe, such as the abundance of its different components. Right panel of Figure~\ref{fig:CMB} shows measurements of $\ell(\ell+1)C_\ell / (2\pi)$ by {\sc Planck} as a function of the multipole numbers $\ell$~\cite{Planck:2018vyg}, in comparison with a 6-parameter fit to the flat $\Lambda$CDM, \emph{i.e.}\ the standard model of cosmology. 

Before further developing this topic it would be useful to do a little reminder on cosmology and the $\Lambda$CDM model. First measurements of radial velocities of galaxies by Vesto Slipher in 1917 have suggested that most of the galaxies are moving away from us~\cite{1917PAPhS..56..403S}. This statement was then backed up by Edwin Hubble's work in 1929~\cite{1929PNAS...15..168H}. To explain this phenomenon, Alexander Friedman and Georges Lema\^itre proposed respectively in 1922 and 1927 that the Universe is expanding~\cite{Friedman:1922kd,Lemaitre:1927zz}. The radial velocity $v$ is inferred using the Fizeau-Doppler effect: galaxies moving away from Earth emit radiation at a wavelength $\lambda_e$ that will undergo a redshift and arrive at Earth with a greater wavelength $\lambda_o$. We define the \emph{redshift} by
\begin{equation}
    z = \frac{\lambda_o}{\lambda_e} - 1 = \sqrt{\frac{1+\frac{v}{c}}{1-\frac{v}{c}}} -1\;.
\end{equation}
We observe that galaxies moving away from us follow the Hubble-Lema\^itre law, where their radial velocity is proportional to their distance from the Earth: $v = H_0\,d$ where $H_0$ is known as the \emph{Hubble constant}.

The founding ansatz of cosmology is the \emph{cosmological principle}: the Universe is homogeneous and isotropic at very large scales. The metric of an expanding Universe that follows this principle is called the Friedmann-Lema\^itre-Robertson-Walker (FLRW) metric, where the associated infinitesimal spacetime interval is written (in spherical coordinates $x^\mu = (t, r, \theta, \phi)$)
\begin{equation}
    ds^2 = dt^2 - a^2(t) \left[\frac{dr^2}{1-kr^2}+r^2\left(d\theta^2 + \sin^2\theta d\phi^2\right)\right]\;,
\end{equation}
where the $(+,-,-,-)$ metric signature is adopted. The scale factor $a(t)$ encodes the expansion of the Universe, whereas the curvature of the Universe is represented by $k$. The three special values of $k$ are: i) $k=1$ for a closed universe with a spherical geometry, ii) $k=0$ for a flat one, obeying Euclidean geometry and iii) $k=-1$ for an open one with an hyperbolic geometry. The redshift $z$ can now be expressed in term of the scale factor
\begin{equation}
     \frac{\lambda_o}{a_0} = \frac{\lambda_e}{a(t)} \implies z = \frac{a_0}{a(t)} - 1\;,
\end{equation}
where $x_0$ denotes the value of a physical quantity $x$ today. 

Now we have to link the expansion of the Universe with its contents. To do this, we need the Einstein field equations
\begin{equation}
    \label{eq:EFE}
    G_{\mu\nu} = 8\pi G T_{\mu\nu} + \Lambda g_{\mu\nu}
\end{equation}
where $G_{\mu\nu}$ is the \emph{Einstein tensor} which purely depends on the metric $g_{\mu\nu}$ (defined by $ds^2 = g_{\mu\nu}dx^\mu dx^\nu$), $\Lambda$ is the \emph{cosmological constant} and $T_{\mu\nu}$ is the \emph{stress-energy tensor} which encodes the physics of the components that fill the spacetime associated to the metric.

We assume the Universe to contain comoving perfect fluids $i$ of energy density $\rho_i$ and pressure $P_i$ with equation of state $P_i = w_i\rho_i$. The stress-energy tensor of such fluids is written
\begin{equation}
    \label{eq:SEtensor}
    T_{\mu\nu} = \sum_{i\in\{\textrm{fluids}\}} T^i_{\mu\nu} = \sum_{i\in\{\textrm{fluids}\}} \rho_i\left[(1+w_i) U^i_\mu U^i_\nu - w_i g_{\mu\nu}\right]\;,
\end{equation}
where $U_\mu \equiv dx_\mu/dt$ is the 4-velocity. For comoving fluids, we have $U_\mu = (1,0,0,0)$. Looking at the cosmological constant term in Equation~\ref{eq:EFE} and at Equation~\ref{eq:SEtensor}, we can see that the cosmological constant can be treated as a comoving perfect fluid of energy density $\rho_\Lambda = \Lambda/(8\pi G)$ and $w_\Lambda = -1$. If a positive cosmological constant dominates the Universe today, this can explain its expansion, since its pressure would be negative. Now by injecting the FLRW metric in the left-hand side of Equation~\ref{eq:EFE} and the stress-energy tensor in the right-hand-side, we obtain the two \emph{Friedmann equations}
\begin{equation}
    \label{eq:friedmann}
    \frac{\dot{a}^2+k}{a^2} = \frac{8\pi G}{3}\sum_{i\in\{\textrm{fluids} + \Lambda\}} \rho_i \quad \textrm{and} \quad \frac{\ddot{a}}{a}=-\frac{4\pi G}{3}\sum_{i\in\{\textrm{fluids} + \Lambda\}} \rho_i\left(1+3w_i\right)\;,
\end{equation}
where the sum over the fluids $i$ includes now the cosmological constant. If we consider only one kind of fluid, differentiating the first equation and injecting it into the second one gives a conservation equation
\begin{equation}
    \frac{\dot{\rho}}{\rho} = -3(1+w)\frac{\dot{a}}{a}\;,
\end{equation}
that allows us to write $\rho$ in terms of the redshift $z$
\begin{equation}
    \rho = \rho_0 \left(\frac{a}{a_0}\right)^{-3(1+w)} = \rho_0 (1+z)^{3(1+w)}\;.
\end{equation}
By introducing the \emph{Hubble parameter} $H = \dot{a}/a$ and the \emph{cosmological parameters}
\begin{equation}
    \label{eq:cosmoparam}
    \Omega_{i,0} = \rho_{i,0} / \rho_{c,0} \quad \textrm{and} \quad \Omega_{k,0} = -k/(a_0^2H_0^2)\;,
\end{equation}
where $\rho_{c,0} = 3H_0^2/(8\pi G)$, we can rewrite the first Friedmann equation
\begin{equation}
    \label{eq:reducfriedmann}
    \left(\frac{H}{H_0}\right)^2 = \sum_{i\in\{\textrm{fluids} + \Lambda + k\}} \Omega_{i,0}(1+z)^{3(1+w_i)}\;,
\end{equation}
where $w_k = -1/3$ is an effective definition since the curvature does not obey an equation of state. We can see from this equation that the sum of the cosmological parameters today ($z=0$) is normalized to 1.

Now let us consider the flat $\Lambda$CDM model. In this model, the Universe has no curvature ($k=0$) and contains baryonic matter and DM ($w_b = w_\textrm{DM} = 0$), radiation ($w_r = 1/3$) and the cosmological constant ($w_\Lambda = -1$). Therefore Equation~\ref{eq:reducfriedmann} is written in this case
\begin{equation}
    H = H_0 \sqrt{\Omega_{m,0} (1+z)^3 + \Omega_{r,0} (1+z)^4 + \Omega_{\Lambda,0}}\;,
\end{equation}
where $\Omega_{m,0} = \Omega_{b,0} + \Omega_{\textrm{DM},0}$. 

\begin{figure}[t]
    \centering
    \includegraphics[width=0.85\linewidth]{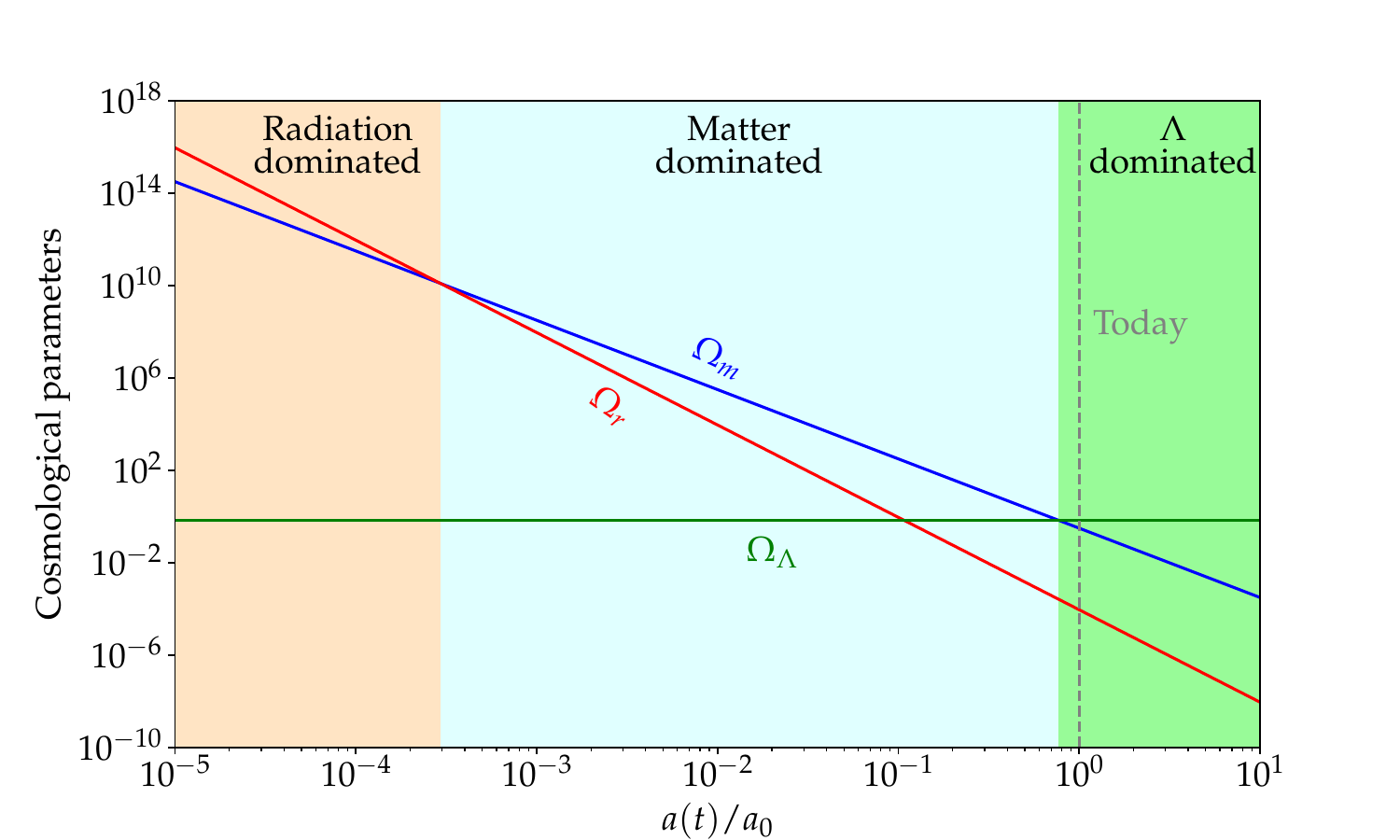}
    \caption{Evolution of the cosmological parameters in the $\Lambda$CDM model.}
    \label{fig:cosmoparam}
\end{figure}

By fitting the flat $\Lambda$CDM model to the CMB temperature anisotropies measured by {\sc Planck}, we obtain the following abundances for baryonic matter and DM today~\cite{Planck:2018vyg}
\begin{equation}
    \Omega_{b,0} = 0.0493 \pm 0.0006 \quad \textrm{and} \quad \Omega_{\textrm{DM},0} = 0.265 \pm 0.007\;,
\end{equation}
which shows that the Universe contains $5.375$ times more DM than baryonic matter. Also, {\sc Planck} measures a positive cosmological constant that dominates the Universe today, and a negligible radiation abundance:
\begin{equation}
    \Omega_{\Lambda,0} = 0.685 \pm 0.007 \quad \textrm{and} \quad \Omega_{r,0} = (5.38 \pm 0.15) \times 10^{-5}\;,
\end{equation}
The evolution of the cosmological parameters in the $\Lambda$CDM model is shown in Figure~\ref{fig:cosmoparam}. In this model, the Universe goes first through a radiation dominated era until $z \sim 3400$ to then a matter dominated era until $z \sim 0.3$. We now live in an Universe dominated by the cosmological constant and therefore in accelerating expansion, according to the second Friedmann equation ($\ddot{a}/a = \Lambda/3 >0$).

On a side note, the $\Lambda$CDM model provides an excellent fit to the measured CMB temperature anisotropies and allows us to infer a precise value for the Hubble constant: $H_0 = 67.4\pm 0.5$~km/s/Mpc~\cite{Planck:2018vyg}. However, this value is in tension with what we get from the Hubble-Lema\^itre law using cosmic distance ladder measurements, where $H_0 = 73\pm1$ km/s/Mpc~\cite{Riess:2021jrx}. This is known as the \emph{Hubble tension} and remains to this day one of the greatest problems of modern cosmology. From now on, we choose to use the {\sc Planck}/$\Lambda$CDM value of $H_0$.

So far, we have mentioned some of the evidences for the existence of DM, but we have yet to ask ourselves about its nature. In the following section, we describe the possible production mechanisms of DM as a particle, then mention what makes a good DM candidate to finally list some of the best known ones.

\section{Dark matter properties and candidates}
\label{sec:DMpropcand}

Although we are aware that the existence of DM is justified, its nature remains elusive. This section does not have the ambition to summarise every possible production mechanism for DM nor enumerate every DM candidates. We limit ourselves to some of the standard scenarios of particle DM production in the early Universe~\cite{Mambrini:2021}, while listing some of the best known candidates.

\subsection{Production in the early Universe}
\label{subsec:history}

Before recombination, the Universe was filled with diverse species of particles. Each species is characterised by their phase-space distribution $f(x^\mu, p^\mu)$ and may interact with other species to achieve thermal equilibrium with the primordial plasma of temperature $T$. The interaction rate per particle $\Gamma_i$ associated to a species $i$ is related to the average cross section of interaction times the relative velocity $\langle\sigma v\rangle_{ij}$ between $i$ and other species $j$ and the number density of targets per volume $n_j$
\begin{equation}
    \Gamma_i = \sum_j n_j \langle\sigma v\rangle_{ij}\;,
\end{equation}
where $j$ can also be $i$ to take into account self-interactions. As the Universe expands, the interaction rate $\Gamma_i$ decreases since $n_j \propto 1/V \sim a^{-3}$, where $V$ is the volume of the Universe. When $\Gamma_i \lesssim H$, the species $i$ is \emph{decoupled} from the primordial plasma, meaning that it no longer interacts with other species nor with itself, therefore its comoving density $n_i a^3$ remains constant\footnote{One can see that $\Gamma_i \lesssim H$ implies $\Gamma_i^{-1} \gtrsim H^{-1}$, where $\Gamma_i^{-1}$ is the mean free time of particles and $H^{-1}$ is approximately the age of the Universe at the time. This condition $\Gamma_i \lesssim H$ therefore imposes that particles are very unlikely to interact with one to another.}. Moreover it means that this species is no longer in thermal equilibrium with the plasma, hence has its own temperature $T_i$ different from the plasma one.

Recalling the cosmological principle, the phase-space distribution of the species of particles depends only on the magnitude of the momentum $p=\lvert \vec{p} \rvert$ of individual particles and time $x^0 = t$. We might as well use the temperature of the species $T_i$ as the evolutionary parameter instead of $t$, where $T_i = T$ if the species is coupled to the plasma or $T_i \neq T$ in the opposite case. We then can write some useful thermodynamical quantities associated to the different species, such as their number density per unit volume $n_i$, energy density per unit volume $\rho_i$, pressure $P_i$ and entropy density per unit volume $s_i$:
\begin{align}
    \label{eq:n}
    n_i(T_i) &= g_i\int \frac{d^3\vec{p}}{(2\pi)^3} f(p,T_i)\;, \\
    \label{eq:rho}
    \rho_i(T_i) &= g_i\int \frac{d^3\vec{p}}{(2\pi)^3} E(p)f(p,T_i)\;, \\
    \label{eq:P}
    P_i(T_i) &= g_i\int \frac{d^3\vec{p}}{(2\pi)^3} \frac{p^2}{3E(p)}f(p,T_i)\;, \\
    \label{eq:s}
    s_i(T_i) &= \frac{\rho(T_i) + P(T_i)}{T_i}\;,
\end{align}
where $E(p) = \sqrt{p^2+m_i^2}$ and $g_i$ is the number of internal degrees of freedom of the particles constituting the gas, for example the number of spin, helicity or colour states. The smallest phase-space volume a particle of the gas can occupy is $h^3$ according to quantum mechanics, hence the presence of the factor $(2\pi)^3$ in natural units.

Thermal equilibrium in the primordial plasma is realised through to the following process
\begin{equation}
    e^\pm + \gamma \longrightarrow e^\pm + \gamma\;,
\end{equation}
while the chemical equilibrium is achieved by the following processes
\begin{gather}
    e^+ + e^- \longleftrightarrow 2\gamma\;, \\
    e^+ + e^- \longleftrightarrow 3\gamma\;.
\end{gather}
The phase-space distribution of species in thermodynamical equilibrium with the primordial plasma is written
\begin{equation}
    \label{eq:feq}
    f_i^\textrm{eq}(E,T_i) = \frac{1}{e^{E/T_i}\pm 1}
\end{equation}
where $+$ ($-$) is for the fermions (bosons). We consider two extreme cases: i) the gas is constituted of relativistic particles ($T_i \gg m_i$), ii) of non-relativistic particles ($T_i \ll m_i$). Inserting Equation~\ref{eq:feq} in Equations~\ref{eq:n}, \ref{eq:rho}, \ref{eq:P} and \ref{eq:s} gives in the relativistic case
\begin{align}
    \label{eq:nrel}
    n^\textrm{eq}_i(T_i)_{T_i\gg m_i} &= g_i\frac{\zeta(3)}{\pi^2}T_i^3
    \begin{cases}
        3/4 & \textrm{(fermions)} \\
        1 & \textrm{(bosons)}
    \end{cases}\;, \\
    \rho^\textrm{eq}_i(T_i)_{T_i\gg m_i} &= g_i\frac{\pi^2}{30}T_i^4
        \begin{cases}
        7/8 & \textrm{(fermions)} \\
        1 & \textrm{(bosons)}
    \end{cases}\;, \\
    P^\textrm{eq}_i(T_i)_{T_i\gg m_i} &= \frac{1}{3}\rho^\textrm{eq}_i(T_i)_{T_i\gg m_i}\;, \\
    s^\textrm{eq}_i(T_i)_{T_i\gg m_i} &= \frac{4}{3}\frac{\rho^\textrm{eq}_i(T_i)_{T_i\gg m_i}}{T_i}\;,
\end{align}
(where $\zeta(3)\approx 1.202$) and in the non-relativistic case, identical for fermions and bosons
\begin{align}
    \label{eq:nnonrel}
    n^\textrm{eq}_i(T_i)_{T_i\ll m_i} &= g_i\left(\frac{m_iT_i}{2\pi}\right)^{3/2}e^{-m_i/T_i}\;, \\
    \rho^\textrm{eq}_i(T_i)_{T_i\ll m_i} &= m_i\,n^\textrm{eq}_i(T_i)_{T_i\ll m_i}\;, \\
    P^\textrm{eq}_i(T_i)_{T_i\ll m_i} &= 0\;, \\
    s^\textrm{eq}_i(T_i)_{T_i\ll m_i} &= \frac{\rho^\textrm{eq}_i(T_i)_{T_i\ll m_i}}{T_i}\;.
\end{align}
We can approximate the total energy density as a sum over the relativistic species only, since the energy density of non-relativistic species is Boltzmann suppressed:
\begin{equation}
    \rho(T) \simeq \sum_{i,\,\textrm{rel.}} \rho_i(T_i) = g_\rho(T)\frac{\pi^2}{30}T^4\;,
\end{equation}
where $g_\rho(T)$ is an effective weight function
\begin{equation}
    g_\rho(T) = \sum_{i=\{\textrm{rel.\,bosons}\}} g_i \left(\frac{T_i}{T}\right)^4 + \frac{7}{8}\sum_{i=\{\textrm{rel.\,fermions}\}} g_i \left(\frac{T_i}{T}\right)^4\;,
\end{equation}
and we can do the same reasoning for the total entropy density
\begin{equation}
    \label{eq:srel}
    s(T) \simeq \sum_{i,\,\textrm{rel.}} s_i(T_i) = g_s(T)\frac{2\pi^2}{45}T^3\;,
\end{equation}
where $g_s(T)$ is an other effective weight function
\begin{equation}
    g_s(T) = \sum_{i=\{\textrm{rel.\,bosons}\}} g_i \left(\frac{T_i}{T}\right)^3 + \frac{7}{8}\sum_{i=\{\textrm{rel.\,fermions}\}} g_i \left(\frac{T_i}{T}\right)^3\;.
\end{equation}
$g_\rho(T)$ and $g_s(T)$ essentially behave like a step function: when a species becomes non-relativistic as $T$ decreases, they are removed from the sum and the weight functions drop.

We consider now that DM particles can interact with Standard Model (SM) ones through 2-to-2 processes
\begin{align}
    \label{eq:DMprodann}
    \textrm{DM} + \overline{\textrm{DM}} &\longleftrightarrow \textrm{SM} + \overline{\textrm{SM}}\;, \\
    \label{eq:DMscat}
    \textrm{DM} + \textrm{SM} &\longleftrightarrow \textrm{DM} + \textrm{SM}\;,
\end{align}
without assuming any model that can predict such interactions. Equation~\ref{eq:DMprodann} explains the production and annihilation of DM, while Equation~\ref{eq:DMscat} is responsible for the thermodynamical equilibrium between DM and the plasma. The evolution of the number density of DM particles $n$ is encoded in the following Boltzmann equation (deriving this equation is not trivial, but the starting point is the Liouville's theorem)
\begin{equation}
    \label{eq:Boltz}
    \dot{n} +3Hn = \langle\sigma v\rangle \left(n_\textrm{eq}^2 - n^2\right)\;,
\end{equation}
where the second left-hand-side term takes into account the dilution of DM particles due to the expansion of the Universe, the first right-hand-side is related to the production of DM and the second right-hand-side its annihilation to SM particles. $\langle\sigma v\rangle$ is the averaged cross section of production/annihilation times relative velocity of DM particles. In general, $\langle\sigma v\rangle$ depends on the temperature of DM particles. We can rewrite Equation~\ref{eq:Boltz} by defining $Y=n/s$ (where $s$ is the entropy density of the plasma, not the DM one) which is related to the comoving abundance of DM. Knowing that $s \propto a^{-3} \implies \dot{s} = -3Hs$, we first obtain
\begin{equation}
    \dot{Y} = s\langle\sigma v\rangle \left(Y_\textrm{eq}^2 - Y^2\right)\;,
\end{equation}
where we can write $Y_\textrm{eq}$ from Equations~\ref{eq:nrel}, \ref{eq:nnonrel} and \ref{eq:srel}
\begin{align}
    Y_\textrm{eq}(T)_{T\gg m_\textrm{DM}} &= \frac{45\zeta(3)}{2\pi^4}\frac{g_\textrm{DM}}{g_s(T)}
    \begin{cases}
        3/4 & \textrm{(fermion)} \\
        1 & \textrm{(boson)}
    \end{cases}\;, \\
    Y_\textrm{eq}(T)_{T\ll m_\textrm{DM}} &= \frac{45}{4\sqrt{2}\pi^{7/2}}\frac{g_\textrm{DM}}{g_s(T)}\left(\frac{m_\textrm{DM}}{T}\right)^{3/2}e^{-m_\textrm{DM}/T}\;,
\end{align}
when DM is in thermodynamical equilibrium with the plasma. Finally by substituting $t$ and $T$ with $x = m_\textrm{DM}/T$ and inserting Equation~\ref{eq:srel} and the first Equation~\ref{eq:friedmann} we can obtain
\begin{equation}
    \label{eq:Boltzred}
    \frac{dY}{dx} = \frac{\lambda(x)}{x^2}\left(Y_\textrm{eq}^2 - Y^2\right)\;,
\end{equation}
where
\begin{equation}
    \lambda(x) = \sqrt{\frac{\pi}{45}}m_\textrm{DM}M_\textrm{Pl}\sqrt{g_\star(x)}\langle\sigma v\rangle\;,
\end{equation}
$M_\textrm{Pl} = 1/\sqrt{G}$ is the Planck mass and $g_\star(T)$ is another weight function
\begin{equation}
    g_\star(T) = \frac{g_s^2(T)}{g_\rho(T)}\left(1+\frac{1}{3}\frac{d\ln g_s(T)}{d\ln T}\right)^2 \sim \frac{g_s^2(T)}{g_\rho(T)}\;.
\end{equation}
Equation~\ref{eq:Boltzred} does not have an analytical solution, but we can identify different regimes. From now on, we assume that $\langle\sigma v\rangle$ does not depend on the temperature of DM, corresponding to what is known as \emph{$s$-wave} DM. $\lambda(x)$ then varies only with the weight function $g_\star(T)$, and it is therefore safe to assume that $\lambda(x)$ is constant as long as the number of relativistic species is conserved during a specific regime. These regimes are:

\begin{itemize}

    \item \textbf{Production:} In the very early Universe, the abundance of DM is zero. Therefore, we can study the behavior of Equation~\ref{eq:Boltzred} when $Y \ll Y_\textrm{eq}$. Since the primordial plasma is very hot at this stage, $x \ll 1$ hence $Y_\textrm{eq}$ is constant
    \begin{equation}
        \label{eq:DMprod}
        \frac{dY}{dx} \simeq \frac{\lambda}{x^2}Y_\textrm{eq}^2 \implies Y(x) \simeq \lambda Y_\textrm{eq}^2 \left(\frac{1}{x_i}-\frac{1}{x}\right)\;,
    \end{equation}
    where $x_i$ is the time when the production of DM particles begins, usually right after the inflationary phase of the Universe ($T\sim 10^{16}$ GeV). In this regime, $Y$ rapidly increases with $x$ until the thermodynamical equilibrium with the plasma is reached.

    \item \textbf{Equilibrium in the relativistic regime:} After the production, $Y$ reaches $Y_\textrm{eq}$ which is still a constant, hence $dY/dx = 0$ and therefore $Y(x) = Y_\textrm{eq}$.

    \item \textbf{Equilibrium in the non-relativistic regime:} When $x\gtrsim 1$, $Y_\textrm{eq}$ starts to be Boltzmann suppressed. Now $dY/dx \propto Y_\textrm{eq}^2 - Y^2 < 0$ and $Y$ decreases, following $Y_\textrm{eq}$ with a small latency.
    
\end{itemize}

Equation~\ref{eq:Boltzred} does not a good job showing what happens when DM particles decouple from the plasma. We can actually rewrite this equation, by noticing that (after some work)
\begin{equation}
    \label{eq:decoupl}
    \frac{\Gamma_\textrm{eq}}{H} = \frac{\langle\sigma v\rangle n_\textrm{eq}}{H} = \frac{\lambda(x)}{x}Y_\textrm{eq}\;,
\end{equation}
and therefore
\begin{equation}
    \label{eq:Boltzdec}
    \frac{dY}{dx} = \frac{\Gamma_\textrm{eq}}{H}\frac{1}{xY_\textrm{eq}}\left(Y_\textrm{eq}^2-Y^2\right)\;.
\end{equation}
When DM decoupled from the plasma, \emph{i.e.}\ $\Gamma_\textrm{eq} \ll H$, we see clearly from Equation~\ref{eq:Boltzdec} that $dY/dx \ll 1$ and therefore the DM abundance `freezes' from this point on (or at least decreases very slowly). Recalling Equation~\ref{eq:cosmoparam} we can write the expression of $\Omega_{\textrm{DM},0}$ knowing that DM is non-relativistic today
\begin{equation}
    \Omega_{\textrm{DM},0} = \frac{\rho_{\textrm{DM,0}}}{\rho_{c,0}} = \frac{m_\textrm{DM}n_0}{\rho_{c,0}} = \frac{ m_\textrm{DM}Y_0s_0}{\rho_{c,0}}\;.
\end{equation}
We can compute this quantity in the two following cases, both referred to as \emph{freeze-out} DM:

\begin{itemize}

    \item \textbf{Decoupling in relativistic equilibrium:} The DM abundance freezes while $Y(x_f) = Y_\textrm{eq}(x_f)$ and therefore we simply have $Y_0 \simeq Y_\textrm{eq}(x_f)$ where $x_f$ is the time of decoupling. In the end we obtain, when DM becomes non-relativistic
    \begin{equation}
        \label{eq:hotrelic}
        \Omega_{\textrm{DM},0} = \frac{8\zeta(3)m_\textrm{DM}T_0^3}{3H_0^2M_\textrm{Pl}^2}\frac{g_\textrm{DM}^\textrm{eff}g_s(x_0)}{g_s(x_f)} \sim 0.3\, g_\textrm{DM}^\textrm{eff} \left(\frac{10}{g_s(x_f)}\right)\left(\frac{m_\textrm{DM}}{6\,\textrm{eV}}\right)\;,
    \end{equation}
    where $g_\textrm{DM}^\textrm{eff} = g_\textrm{DM}$ ($3/4\,g_\textrm{DM}$) for bosonic (fermionic) DM. The value of $g_s(x_0)$ is
    \begin{equation}
        g_s(x_0) = 2 + \frac{7}{8}\times 2\times 3\times \frac{4}{11} = 3.91\;,
    \end{equation}
    where the two only remaining relativistic species are the photons and neutrinos. The latter have decoupled from the plasma and it can be shown that $T_{\nu} = (4/11)^{1/3}T_\gamma$. When DM freeze-out in the relativistic regime, we call it a \emph{hot relic}.

    \item \textbf{Decoupling in non-relativistic equilibrium:} Since we have $dY/dx \ll 1$, $Y$ decreases way slower than $Y_\textrm{eq}$. Not long after we have $Y\gg Y_\textrm{eq}$ and we can rewrite Equation~\ref{eq:Boltzred} to solve it by integrating it between the decoupling time $x_f$ and today $x_0$ while assuming a long time has passed ($x_0 \gg x_f$) and the DM abundance decreases substantially from that moment ($Y_0 \ll Y_f$):
    \begin{equation}
        \frac{dY}{dx} \simeq -\frac{\lambda}{x^2}Y^2 \implies Y_0 \simeq \frac{x_f}{\lambda}\;.
    \end{equation}
    We finally obtain
    \begin{equation}
        \label{eq:relic}
        \Omega_{\textrm{DM},0} = \frac{16\pi^{5/2}}{9\sqrt{5}}\frac{T_0^3}{H_0^2M_\textrm{Pl}^3}\frac{g_s(x_0)}{\sqrt{g_\star(x_f)}}\frac{x_f}{\langle\sigma v\rangle} \sim 0.3 \left(\frac{30}{g_\star(x_f)}\right)^{1/2}\left(\frac{x_f}{25}\right)\left(\frac{3\times 10^{-26}\,\textrm{cm}^3/\textrm{s}}{\langle\sigma v\rangle}\right)\;.
    \end{equation}
    The decoupling time can be obtained by setting $\Gamma_\textrm{eq}(x_f)/H(x_f)=1$ and by recalling Equation~\ref{eq:decoupl}
    \begin{equation}
    	\label{eq:decoupltime}
        x_f^{1/2}e^{-x_f} = \frac{4\pi^3}{3}\sqrt{\frac{2}{5}}\frac{\sqrt{g_\rho(x_f)}}{g_\textrm{DM}}\frac{1}{m_\textrm{DM}M_\textrm{Pl}\langle\sigma v \rangle}\;,
    \end{equation}
    giving $x_f\simeq 20-30$ for $m_\textrm{DM} \simeq 1-10^{4}$ GeV and $\langle\sigma v \rangle \simeq 3\times 10^{-26}$ cm$^3$/s. When DM freeze-out in the non-relativistic regime, we call it a \emph{cold relic}.
    
\end{itemize}

Actually, a nice coincidence appears when we consider that DM interacts with the SM through weak interactions, one of the three fundamental forces in the SM, mediated by the $W^\pm$ and $Z^0$ bosons. Such DM particles are named \emph{weakly interacting massive particles} (WIMPs). One can compute the WIMP annihilation cross section, when WIMPs annihilates into fermions through the exchange of a $Z^0$ boson and assuming the mass of the fermions to be negligible compared to the WIMP mass $m_\chi$ and $Z^0$ boson mass $m_Z = 91.19$ GeV
\begin{equation}
    \label{eq:wimpcs}
    \langle\sigma v\rangle = \frac{32G_F^2\cos^4\theta_W}{\pi}\frac{m_\chi^2m_Z^4}{\left(4m_\chi^2-m_Z^2\right)^2}\;,
\end{equation}
where $G_F = 1.166\times 10^{-5}$ GeV$^{-2}$ is the Fermi constant and $\cos\theta_W = 0.877$ denotes the ratio between the mass of the $W^\pm$ and the $Z^0$ bosons. By assuming that WIMPs are a cold relic, we can insert Equation~\ref{eq:wimpcs} in Equation~\ref{eq:relic} and use Equation~\ref{eq:decoupltime} to show that for WIMP masses between around $2$ GeV and $1$ TeV, their relic abundance is lower than the DM one measured by {\sc Planck}. This means that, in this mass range, WIMPs can constitute a fraction of DM, or even its totality at the boundaries of this range. This peculiar coincidence is referred to as the `WIMP miracle'. 

The last DM production process we investigate is an out-of-equilibrium one. Until now, we only considered $s$-wave DM as a first approximation. And actually, this is good approximation for the scenarios we considered: i) cold relics were initially decoupled from the non-relativistic equilibrium, therefore any temperature dependency in $\langle\sigma v\rangle$ would be suppressed, ii) recalling Equation~\ref{eq:hotrelic}, the abundance of a hot relic does not depend on $\langle\sigma v\rangle$. However, a more general, temperature dependent, description of DM production/annihilation is needed in order to investigate this next production process. In this context, we can generalise Equation~\ref{eq:wimpcs} where we consider a mediator $Y$ with a mass $m_Y$, coupled to the DM and SM sectors through the constant $\alpha$
\begin{equation}
    \langle\sigma v\rangle = \frac{\alpha^2s}{\left(s-m_Y^2\right)^2}\;,
\end{equation}
where $\sqrt{s} \simeq T + m_\textrm{DM}$ is the centre-of-mass energy. In the non-relativistic regime ($T \ll m_\textrm{DM}$), we retrieve $s$-wave DM. However, in the relativistic regime ($T \ll m_\textrm{DM}, m_Y$), we get $\langle\sigma v\rangle \sim \alpha^2/T^2 = \alpha^2x^2/m_\textrm{DM}^2$. Inserting this new expression in the left hand side of Equation~\ref{eq:DMprod} allows us to obtain
\begin{equation}
    Y(x) \simeq \lambda_0Y_\textrm{eq}^2x\;,
\end{equation}
assuming $x \gg x_i$ and where
\begin{equation}
    \lambda_0 = \alpha^2\sqrt{\frac{\pi}{45}}\frac{M_\textrm{Pl}}{m_\textrm{DM}}\sqrt{g_\star(x)}\;.
\end{equation}
In this scenario, DM production is much slower compared to the $s$-wave DM case. For very small couplings with the SM, DM may never reach thermodynamical equilibrium with the plasma and will stop being produced when the plasma becomes non-relativistic as its density becomes Boltzmann-suppressed. The DM abundance \emph{freezes-in} such that $Y_0 \simeq Y(x \simeq 1)$. For DM particles with $\alpha \sim 10^{-11}$, known as \emph{feebly interacting massive particles} (FIMPs), the observed DM relic density can be produced. This is known as the `FIMP miracle'~\cite{McDonald:2001vt, Hall:2009bx}.

\begin{figure}[t]
    \centering
    \includegraphics[width=0.85\linewidth]{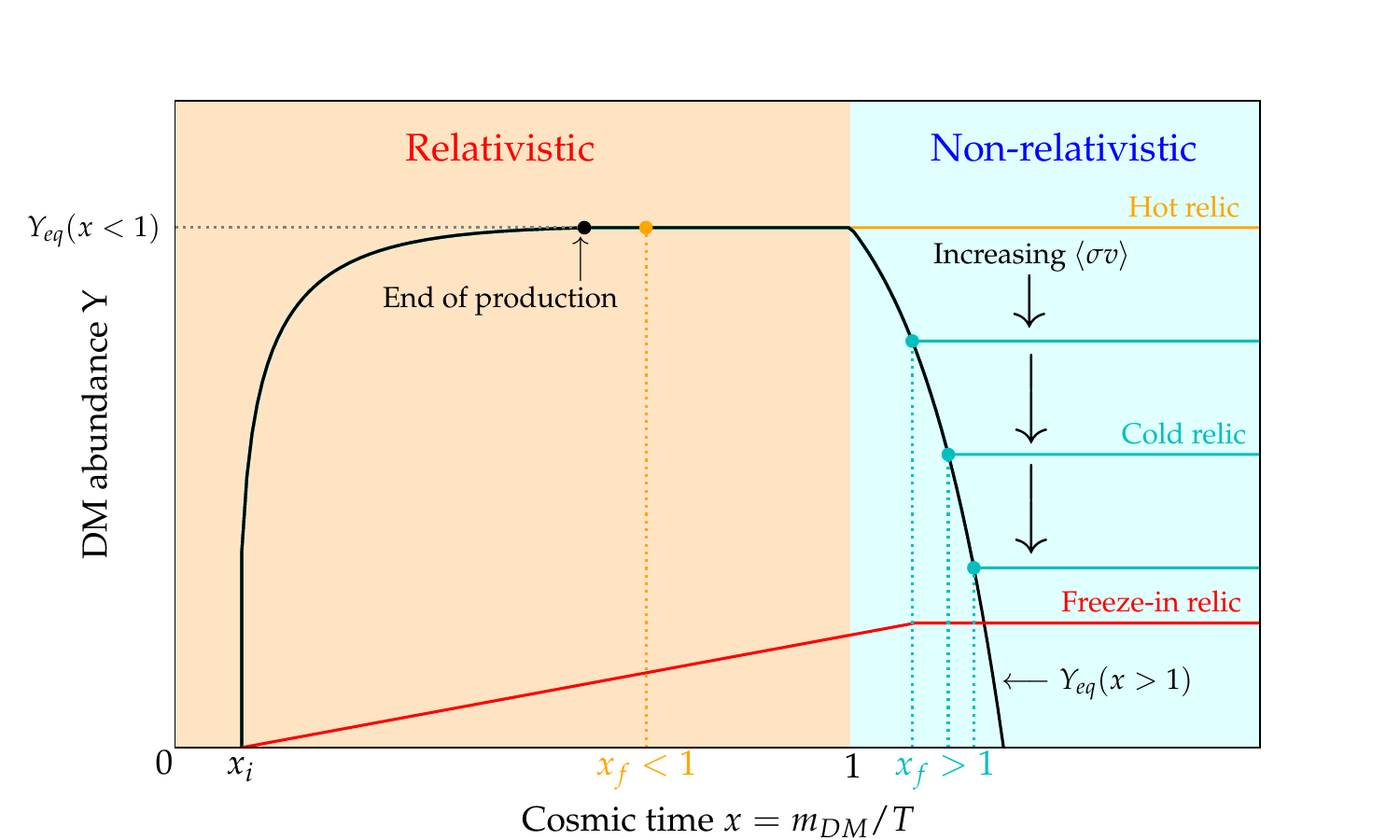}
    \caption{Evolution of the DM abundance across the history of the Universe. The DM relic abundance is set after decoupling from the primordial plasma during its thermodynamical equilibrium with the plasma (hot relic if DM particles are relativistic during decoupling, cold relic otherwise). If the production is slow enough so DM particles never reach equilibrium with the plasma, the relic abundance is set as the production stops, when the plasma becomes non-relativistic (freeze-in).}
    \label{fig:DMabundance}
\end{figure}

We described the most important particle DM production scenarios although many other DM production scenarios are possible \emph{e.g.}, cannibalism through $3\rightarrow 2$ annihilation processes~\cite{Carlson:1992fn,Pappadopulo:2016pkp}, asymmetric DM annihilation~\cite{Kaplan:2009ag} or decays of massive particles into DM. Figure~\ref{fig:DMabundance} summarises the evolution of DM abundance for the different aforementioned scenarios. We do not detail the production mechanisms of DM as macroscopic objects in this thesis, but we refer the reader to~\cite{Carr:2020gox} and references therein for more information instead.

\subsection{Properties of dark matter}
\label{subsec:DMprop}

Although we still do not know about the nature of DM, some of its properties are well constrained:

\begin{itemize}

    \item \textbf{Electric charge}: DM remains invisible even with the sensitivity of our current observational methods. Although DM is often assumed to be electrically neutral, some theoretical models allow DM to have a non-zero electric charge (known as \emph{millicharge} DM). The most stringent lower limits on the DM charge are $3.2\times 10^{-11}(m_\textrm{DM}/10\,\textrm{GeV})^{0.52}$ for $m_\textrm{DM} \gtrsim 10$ GeV and $1.0\times 10^{-9}(m_\textrm{DM}/10\,\textrm{GeV})^{0.25}$ for $m_\textrm{DM} \lesssim 10$ GeV, in units of the electron charge~\cite{Dunsky:2018mqs,cirelli2024dark}.

    \item \textbf{Self-interactions}: Observations of colliding galaxy clusters show that DM should have limited self-interactions (recalling what we stated in Section~\ref{subsec:galclust}). Constraints on the DM self-interaction cross section yield $\sigma_\textrm{DM-DM}/m_\textrm{DM} < 8.4 \times 10^{-25}$ cm$^2$/GeV~\cite{Harvey:2015hha}.

    \item \textbf{Lifetime}: The presence of DM in our Universe today indicates that DM has to be stable over cosmological timescales. Its decay time should therefore be large compared to the age of the Universe. Current constraints on the DM lifetime lead to $\tau_\textrm{DM} \gtrsim 10^{28}$ s for $m_\textrm{DM} \simeq 1$ GeV~\cite{DelaTorreLuque:2023olp}.
    
\end{itemize}

There exist some limits on the mass of DM, although the current allowed mass range for DM currently spans up to 90 orders of magnitude. There are two lower limits on the mass DM, depending on whether DM is a fermion or a boson. In the first case, the Pauli exclusion principle imposes that the phase-space distribution of fermions is bounded from above (see Equation~\ref{eq:feq})
\begin{equation}
    f(E) = \frac{1}{e^{E/T}+1} \leq \frac{1}{2}\;,
\end{equation}
therefore the number density of fermionic DM in a galaxy is bounded as well (recalling Equation~\ref{eq:n})
\begin{equation}
    n = g\int \frac{d^3\vec{p}}{(2\pi)^3}f(p) \lesssim \frac{1}{4\pi^2}\frac{p_\textrm{max}^3}{3}\;,
\end{equation}
where $g\sim \mathcal{O}(1)$ and assuming spherical symmetry. Here $p_\textrm{max} = m_\textrm{DM}v_\textrm{esc} = m_\textrm{DM}\sqrt{2GM/R}$ designates the maximal momentum to which the fermion is still gravitationally bounded to a galaxy of mass $M$ and radius $R$. Recalling Equation~\ref{eq:veldisp} obtained from the virial theorem, we have 
$v_\textrm{esc} = \sigma_v\sqrt{10/3}$ (where $\sigma_v$ is a better observable than $v_\textrm{esc}$). Finally, considering that DM contributes to a dominant part of the total mass of the galaxy, we get
\begin{equation}
    M \simeq \frac{4}{3}\pi R^3 n m_\textrm{DM} \lesssim \frac{1000}{81\pi} R^3 \sigma_v^3 m_\textrm{DM}^4 \implies m_\textrm{DM} \gtrsim 10 \,\textrm{eV} \left(\frac{20\,\textrm{kpc}}{R}\right)^{1/2}\left(\frac{300\,\textrm{km/s}}{\sigma_v}\right)^{1/4}\;.
\end{equation}
This is known as the \emph{Tremaine-Gunn limit}~\cite{Tremaine:1979we}. For bosonic DM, there is a lower bound on ultra-light particles. It is known that some of the dwarf spheroidal galaxies (dSphs) in our Local group have a greater mass-to-light ratio than the Milky Way (MW) (\emph{e.g.}, the Draco Dwarf has $M/L \simeq 75 M_\odot/L_\odot$ (see Table~\ref{tab:dSphs}) whereas the MW has $M/L \simeq 20 M_\odot/L_\odot$), indicating that they are rich in DM. In order for ultra-light DM to be present in a dSph, the associated de Broglie wavelength has to be lower than the typical diameter of dSphs (which is around 1 kpc):
\begin{equation}
    \lambda_\textrm{dB} = \frac{2\pi}{m_\textrm{DM} v} \simeq 1 \,\textrm{kpc} \left(\frac{10^{-21}\,\textrm{eV}}{m}\right)\left(\frac{10\,\textrm{km/s}}{v}\right) \lesssim 1\,\textrm{kpc}\;,
\end{equation}
where $v$ is the velocity of DM as a wave. The typical velocity dispersion in dSphs is $v \simeq 10$~km/s and therefore sets the lower bound on bosonic DM to $m_\textrm{DM} \gtrsim 10^{-21}$ eV. In a recent work, people have precisely computed this limit and have set it to $m_\textrm{DM} > 2.2\times 10^{-21}$ eV~\cite{Zimmermann:2024xvd}.

\begin{figure}[t]
    \centering
    \includegraphics[width=\linewidth]{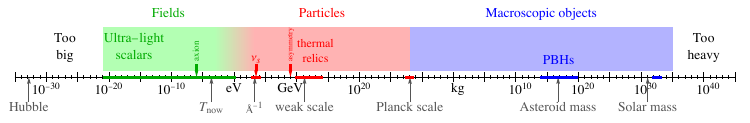}
    \caption{Allowed mass range for DM, spanning from around $10^{-21}$ eV to $10^5 M_\odot$. The allowed mass ranges for the following DM candidates are also represented: axions, sterile neutrinos ($\nu_s$), asymmetric DM, WIMPs and PBHs. Taken from~\cite{cirelli2024dark}.}
    \label{fig:DMmassrange}
\end{figure}

The presence of DM in dSphs also allows us to set an upper bound on the DM mass, which should be significantly lower than the mass of its host. Assuming that a dSph should be composed of at least $\mathcal{O}(100)$ DM `objects' to hold its structure, therefore we obtain $m_\textrm{DM} \lesssim 10^5 M_\odot$. The full range of DM masses allowed by the aforementioned limits is shown in Figure~\ref{fig:DMmassrange}. 

\subsection{Dark matter candidates}
\label{subsec:candidates}

As we are trying to picture what is the nature of DM, one of the first questions that would come to the mind of a non-expert is `Could DM be a particle of the SM?' and it would not be a foolish question. For a long time, SM neutrinos were a serious DM candidate as they fit all of the properties mentioned in Section~\ref{subsec:DMprop} and are a hot relic of the Universe. However the emergence of numerical simulations have discredited this hypothesis, as they show that SM neutrinos cannot coalesce due to the fact that they are highly relativistic, and therefore they cannot explain the observed formation of small-scale structures (such as galaxies)~\cite{1983ApJ...274L...1W}. As no other SM particle would fit the description of DM, the latter could either be a particle beyond the Standard Model (BSM), or could exist in the form of macroscopic, compact objects. In this section, we list some of the most often considered DM candidates as well as show the allowed mass range of some of them in Figure~\ref{fig:DMmassrange}, alongside the allowed mass range of DM without assuming any particular model. 

\begin{itemize}

    \item \textbf{Weakly interacting massive particles:} We already briefly mentioned this candidate and detailed its production mechanism in the end of Section~\ref{subsec:history}. The appeal of this paradigm comes from the coincidence that DM particles that can interact weakly with SM particles and with masses in the GeV--TeV range can set the observed relic abundance today through the freeze-out mechanism, recalling Equation~\ref{eq:relic}. Theoretically such particles can arise in \emph{supersymmetric} models. Supersymmetry is a renowned class of BSM models, known to be the most natural way to extend the SM in order to solve the Higgs hierarchy problem, by adding a bosonic \emph{superpartner} for each fermion and vice-versa. This hierarchy problem can be formulated by the question `Why is the mass of the Higgs boson ($\simeq 125$ GeV) so small compared to the Planck mass?' and cannot be answered by the SM alone. Even though WIMPs are an extremely well motivated DM candidate, dedicated experiments still have not found any evidence for the existence of these particles, putting this paradigm in peril.
    
    \item \textbf{Axions and axion-like particles:} It is known that strong interactions are expected to break the $CP$-symmetry, since the Lagrangian of quantum chromodynamics (QCD) contains the following $CP$-violating term
    \begin{equation}
        \label{eq:CPviolation}
        \mathcal{L}_\textrm{QCD} \supset \bar{\Theta}\,\frac{\alpha_s}{8\pi}G_a^{\mu\nu}\tilde{G}_{\mu\nu}^a\;,
    \end{equation}
    where $\bar{\Theta} \neq 0$ is a phase, $G_a^{\mu\nu}$ $(a = 1,...,8)$ are the eight gluon fields while $\tilde{G}_{\mu\nu}^a$ are their dual and $\alpha_s$ is the coupling of strong interactions. This $CP$-violating term should give rise to an electric dipole moment in the neutron $d_N \simeq 10^{-16}\ \bar{\Theta}\ e$ cm. However this dipole moment has not been observed yet, and the current lower bound is $\lvert d_N \rvert < 1.8\times 10^{-26}\ e$ cm~\cite{Abel:2020pzs}, imposing $\lvert \bar{\Theta} \rvert \lesssim 10^{-10}$. The fact that $\bar{\Theta}$ is so small without any apparent reason is known as the \emph{strong $CP$ problem}. A solution to this problem was proposed by Roberto Peccei and Helen Quinn in 1977~\cite{Peccei:1977ur, Peccei:1977hh}, which is to add a new global symmetry to the SM that can be spontaneously broken, effectively promoting $\bar{\Theta}$ to a field whose minimum is $0$. Frank Wilczek and Steven Weinberg then independently showed that such a symmetry breaking would give rise to a pseudo-Nambu-Goldstone boson: the \emph{axion}~\cite{Wilczek:1977pj, Weinberg:1977ma}. The $CP$-symmetry is restored by adding a term corresponding to the coupling of axions to gluons in the QCD Lagrangian
    \begin{equation}
        \mathcal{L}_\textrm{QCD} \supset \left(\bar{\Theta}-\frac{a}{f_a}\right)\frac{\alpha_s}{8\pi}G_a^{\mu\nu}\tilde{G}_{\mu\nu}^a\;,
    \end{equation}
    where $a$ is the axion field and $f_a$ the axion decay constant. The properties of the axion fit the ones mentioned in Section~\ref{subsec:DMprop}, making it a suitable DM candidate. The parameter space for the QCD axion is quite restricted, as it can be shown that all of the allowed couplings between the axion and SM particles are essentially a power-law of the axion mass. To open up the parameter space, other BSM models can predict particles with the same properties as the axion, while alleviating the dependency of the couplings to the particle mass. Such particles are called \emph{axion-like particles}. The most recent constraints on axions and axion-like particles can be found in~\cite{OHare:2021utq}.

    \item \textbf{Sterile neutrinos:} According to the SM, neutrinos are massless, only left-handed (or right-handed if they are anti-neutrinos) and can have three different flavors ($\nu_e, \nu_\mu, \nu_\tau$). However, experiments that probe neutrinos from the Sun, Earth's atmosphere, nuclear reactors and particle accelerators have all shown that neutrinos can undergo flavor oscillations, a phenomenon that occur only if the neutrinos are massive. The current lower bound on the sum of neutrino masses is $\sum m_\nu \gtrsim 0.06$ eV~\cite{ParticleDataGroup:2024cfk}. Moreover, the SM does not predict the existence of right-handed neutrinos (nor left-handed antineutrinos), whereas the other SM fermions exist in both helicities. To solve these problems, right-handed (or \emph{sterile}) neutrinos can be added to the SM in order to give a small mass to the left-handed neutrinos through the \emph{seesaw mechanism}. If sterile neutrinos do exist and are produced through the freeze-in mechanism, they can constitute a fitting DM candidate~\cite{Dodelson:1993je, Drewes:2016upu}. 

    \item \textbf{Primordial black holes:} Instead of being a particle or a field, there is also a possibility that DM could be in the form of compact objects, named \emph{primordial black holes} (PBHs). Instead of being formed during the collapse of a massive star, these black holes (BHs) are most likely formed from the collapse of overdense matter regions in the early Universe, but other scenarios are possible. The only open mass window in which PBHs can constitute all of the DM in the Universe is between $10^{17}$ and $10^{23}$ g~\cite{Carr:2020gox} (corresponding to $10^{-13}$ and $10^{-7}$ m in terms Schwarzschild radius $2GM/c^2$).

    \item \textbf{Sub-GeV DM:} This DM class has recently gained some attention from the community, due to the lack of experimental signals from WIMPs. There are some well motivated BSM models that can predict a DM candidate with a mass below one GeV, such as WIMPless DM~\cite{Feng:2008ya}, strongly interacting massive particles~\cite{Hochberg:2014dra} or asymmetric DM~\cite{Kaplan:2009ag}.
    
\end{itemize}

Up to this point we have described some well known DM candidates, however the whole landscape is large. In phenomenological studies of DM, it is usual to assume DM to be model-independent, \emph{i.e.}\ a field, particle or macroscopic object without presuming its production nor a model that predicts its existence. This allows to ease the readability and comparison between different phenomenological studies. In the following section, we describe the main experimental ways to probe particle DM.

\section{Bringing dark matter to light}
\label{sec:detecmethods}

Detection methods that consist of trying to figure out the nature of particle DM can be put in three categories: 
\begin{itemize}
    \item Collider and accelerator searches: producing DM through SM particle collisions,
    \item Direct detection (DD): probing nuclear recoils from DM particles scattering on atomic nuclei,
    \item Indirect detection (ID): looking at cosmic ray (CR) signals associated to DM annihilation or decay, or emissions from macroscopic DM.
\end{itemize}
DD and ID can also be used to probe axionic DM, in addition to specific methods such as light-shining-through-wall or microwave cavity experiments. Macroscopic DM searches solely rely on astrophysical observations and ID. In this section, we limit ourselves to describe collider, accelerator searches and DD, as ID will be deeply covered in Chapter~\ref{chap:ID}. Specific detection methods for axions are beyond the scope of this thesis.

In order to truly put a finger on the nature of DM, we have to rely on the complementary of the aforementioned detection methods. A possible DM signature appearing in experiments using one method should be confronted to a matching signal in experiments using other methods.

\subsection{Collider and accelerator searches}
\label{subsec:colliders}

A possible way to probe DM is to actually produce it in colliders or in particle accelerators. At the Large Hadron Collider operated by the CERN, proton beams are accelerated in opposite directions and then collide, producing a complex mix of various particles that travel to detectors, such as {\sc Atlas} or {\sc Cms}, that have been set up around the collision points. The idea is that DM particles can be produced from these collisions, and not interact with the detectors, leading to an `observation' of missing energy/momentum in the transverse plane of the collisions. The same principle can be applied in other types of accelerators, where a particle beam can create DM by hitting a fixed target or a beam dump.

These approaches are quite model-dependent, as we have to theorise what are the interactions between DM and SM particles that allow the latter to produce DM during collisions. DM production through an exchange of a $H$ or $Z$ boson are the simplest scenarios, but other theoretically motivated models can be considered \emph{e.g.}, supersymmetric models or hidden sectors. For now, no signal of missing energy has been observed and limits on the DM mass and couplings with the SM are set, depending on individual models.

The interested reader can find details on on-going efforts and future initiatives in trying to probe DM in colliders and accelerators in~\cite{Battaglieri:2017aum}.

\subsection{Direct detection}
\label{subsec:direct}

In 1985, Mark Goodman and Edward Witten were the first to discuss the possibility of detecting DM particles inhabiting the Galactic halo by probing their scattering off of nuclei in a detector~\cite{Goodman:1984dc}. The recoil spectrum $dR/dE_r$, as a function of the nuclear recoil energy $E_r$, is written
\begin{equation}
    \frac{dR}{dE_r}(E_r,t)=\frac{\rho_\odot M}{m_Nm_\textrm{DM}} \int_{v_\textrm{min}}^{v_\textrm{esc}} d^3\vec{v}\ v f(\vec{v}+\vec{v}_\oplus(t))\frac{d\sigma}{dE_r}(E_r,v)\;,
\end{equation}
where $M$ is the target mass of the detector, $m_N$ is the mass of the target nucleus, $\rho_\odot$ is the local DM energy density (which is around $0.4$ GeV/cm$^3$), $v = \lvert\vec{v}\rvert$ is the velocity of DM particles in the detector rest frame, $f(\vec{v}+\vec{v}_\oplus(t))$ their velocity distribution in Earth's rest frame, usually assumed to be roughly Maxwellian\footnote{$f(v) \propto e^{-v^2/(2\sigma^2)}$ where $\sigma \simeq 220 - 270$ km/s in the MW~\cite{Herzog-Arbeitman:2017fte}.}, $\vec{v}_\oplus$ is Earth's velocity in the Galactic rest frame and $d\sigma/dE_r$ is the differential cross section of DM particles scattering off of nuclei
\begin{equation}
    \frac{d\sigma}{dE_r}(E_r,v) = \frac{m_N}{2\mu^2v^2}\left(\sigma_\textrm{SI}F^2_\textrm{SI}(E_r) + \sigma_\textrm{SD}F^2_\textrm{SD}(E_r)\right)\;,
\end{equation}
where spin-independent (SI) and spin-dependent (SD) interactions are considered and $F^2(E_r)$ are the nuclear form factors. The minimal speed required for a DM particle to induce a nuclear recoil of energy $E_r$ is $v_\textrm{min} = \sqrt{E_rm_N/(2\mu^2)}$ with $\mu = m_Nm_\textrm{DM}/(m_N+m_\textrm{DM})$ being the reduced mass of the DM-nucleus system, and the maximal velocity of a DM particle is its escape velocity from the gravitational well of the MW $v_\textrm{esc}$. Essentially the recoil spectrum depends on the properties of the target (DD experiments are using a large variety of targets, such as Ge, liquid Xe, liquid Ar, NaI, etc.), the DM mass $m_\textrm{DM}$ and the DM-nucleon spin-(in)dependent cross sections $\sigma_\textrm{SD}$ ($\sigma_\textrm{SI}$).

Due to the motion of the Earth in the Galactic rest frame, which is a superposition of the Earth's orbit around the Sun and the Sun's motion around the Galactic centre (GC), in principle we should observe an annual modulation of the recoil spectrum in DD experiments. On June 2, Earth's velocity in the Galactic rest frame is maximal (the Earth faces a `DM wind' due to the motion of the Sun), and the recoil spectrum is also at its maximum. On the contrary, on December 2, the Earth faces a slightly weaker DM wind hence the recoil spectrum is at its minimum. Since the velocity of the Sun around the GC (around $250$ km/s) is way greater than the velocity of the Earth around the Sun (around $30$ km/s), the amplitude of expected annual modulation in the recoil spectrum should be around $5\%$.

\begin{figure}[t]
    \centering
    \includegraphics[width=0.85\linewidth]{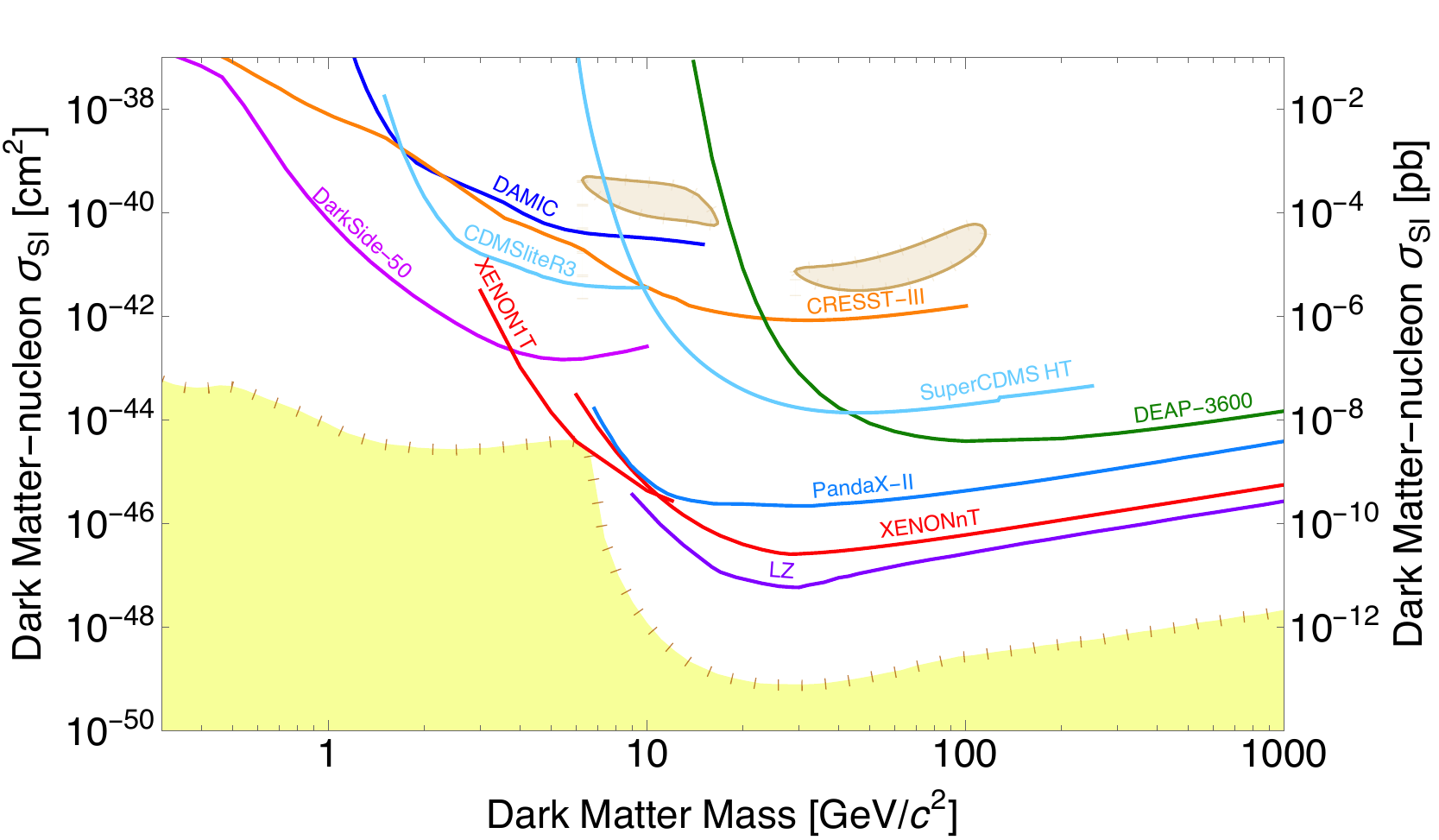}
    \caption{DD upper limits on the spin-independent DM-nucleon cross section as a function of DM mass from various experiments represented as solid lines. The regions of interest reported by {\sc Dama/Libra} are also represented in brown as well as the neutrino fog for a Xe target in yellow. Created using the Dark Matter Limit Plotter by the {\sc SuperCdms} collaboration~\cite{DMLimitPlotter}.}
    \label{fig:DDlimits}
\end{figure}

In 2008, the {\sc Dama/Libra} experiment reported such an annular modulation and constrained the properties of DM to be inside the brown regions in Figure~\ref{fig:DDlimits}~\cite{Savage:2008er}. However, current DD experiments do not observe DM-nucleus scatterings in these regions and exclude them by orders of magnitude in $\sigma_\textrm{SI}$ and $\sigma_\textrm{SD}$. This contradiction remains an open problem, even though most of the community dismisses the {\sc Dama/Libra} claim. In Figure~\ref{fig:DDlimits}, we show the current upper limits on $\sigma_\textrm{SI}$ as a function of $m_\textrm{DM}$ from some of the most important DD experiments: {\sc Cresst}~\cite{CRESST:2019jnq}, {\sc Deap}~\cite{DEAP:2019yzn}, {\sc SuperCdms}~\cite{SuperCDMS:2018gro}, {\sc CdmsSlite}~\cite{SuperCDMS:2017mbc}, {\sc PandaX-II}~\cite{PandaX-II:2020oim}, {\sc Damic}~\cite{DAMIC:2020cut}, {\sc Lux-Zeplin} ({\sc Lz})~\cite{LZ:2022lsv}, {\sc DarkSide}~\cite{DarkSide:2022dhx}, {\sc Xenon$n$T}~\cite{XENON:2023cxc} and {\sc Xenon1T}~\cite{XENON:2020gfr}.

In order for a non-relativistic DM particle to induce an observable nuclear recoil in the detector, its mass should be around or above the one of a typical target nucleus, \emph{i.e.}\ $\mathcal{O}(10\,\textrm{GeV})$. This is a current limitation on DD experiments, as shown by the weakening of the limits on $\sigma_\textrm{SI}$ for $m_\textrm{DM} \lesssim 10$ GeV in most of the experiments. A way to probe lighter DM is to either look at electronic recoils or the excitation/ionisation of a recoiled atom, known as the \emph{Migdal effect}~\cite{migdal1939ionizatsiya,migdal1978qualitative}. The {\sc DarkSide} limit in Figure~\ref{fig:DDlimits} includes this effect and shows an improvement of the sensitivity for light DM compared to the other experiments.

Finally, the sensitivity of DD experiments are getting closer to the \emph{neutrino fog}~\cite{Ruppin:2014bra, OHare:2021utq}, which is the unavoidable neutrino background coming from the coherent elastic scattering of neutrinos from the Sun, Earth's atmosphere and diffuse supernovae background on the nuclei in the detector. This would confirm that $\sigma_\textrm{SI}$ and $\sigma_\textrm{SD}$ for DM would be lower than the ones for neutrinos. This neutrino fog is shown in yellow in Figure~\ref{fig:DDlimits}.

%% file: Chapters/chap2.tex

\begin{figure}[t]
    \centering
    \includegraphics[width=\linewidth]{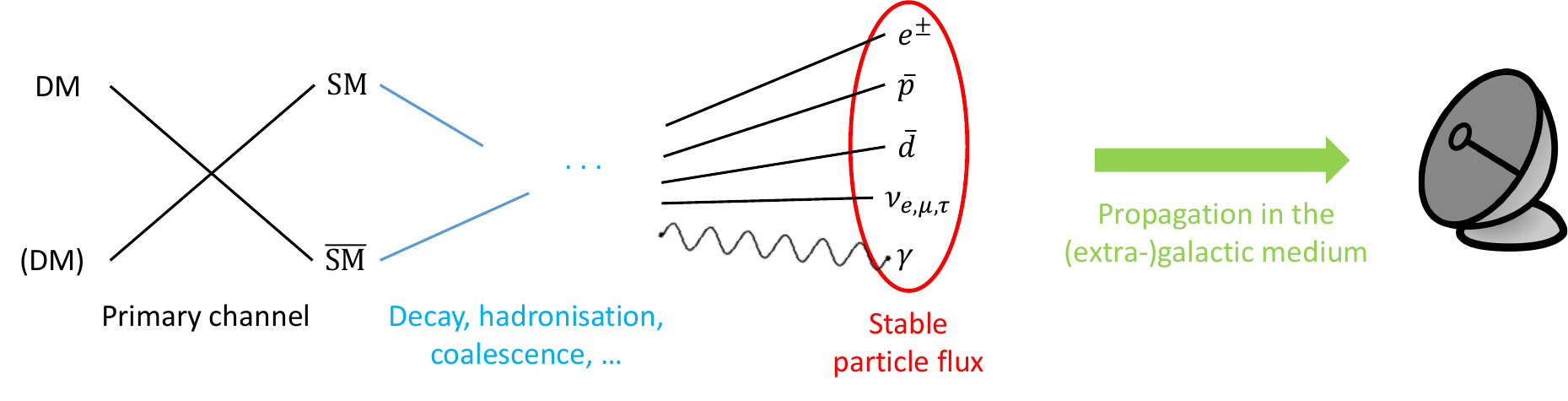}
    \caption{Illustration of the principle behind DM ID, from DM annihilation or decay to the production of stable particles that propagate until they reach our detectors.}
    \label{fig:DM ID}
\end{figure}

\lettrine[lines=3, nindent=1pt]{I}{n} the previous chapter, we elaborated on some of the main techniques that enable us to probe DM, including direct detection, collider, and accelerator searches. However, we conveniently avoided the subject of DM ID until now. The goal of the following chapter is to provide insight into what this type of detection entails, as it is the main topic of this thesis. The aim of ID is to look for the products of particle DM annihilation or decay, as well as those from PBH evaporation. Such products can be in the form of stable, charged CRs, photons, and neutrinos that propagate from their production point through the (extra-)galactic medium until they reach our detectors. Figure~\ref{fig:DM ID} illustrates the principle of this detection method. Each product carries different types of information, making the ability to detect multiple types of products highly beneficial. On one hand, photons and neutrinos are not deflected by surrounding magnetic fields (since they are neutral) and therefore allow us to precisely trace their source of emission. On the other hand, charged CRs can include antiparticles that can be detected with a low astrophysical background, since antiparticles are not easily produced by astrophysical processes. In this chapter, we investigate some of the most promising targets for DM ID (Section~\ref{sec:DMwhere}), then we expand on the production of these stable particles in the DM halo of the MW and their propagation (Section~\ref{sec:particlesDM}), and finally, provide a brief state-of-the-art review of the experiments useful for this detection method method (Section~\ref{sec:exp}).

\newpage

\section{Where should we look?}
\label{sec:DMwhere}

In this section we explore some of the most promising targets for DM ID. Essentially, the ideal target should meet two criteria: i) it has to be DM-rich to increase the likelihood of detecting a strong signal from it and ii) it has to be quite poor in astrophysical sources, as disentangling an indirect DM signal from the astrophysical background can be challenging. However, there is no perfect target and the ones we investigate in this section will always be subject to a compromise between the two criteria. We start by looking at the case of celestial bodies (Section~\ref{subsec:bodies}) and our own Galaxy (Section~\ref{subsec:MW}), to zoom out to the Local Group and consider the satellite dSphs of the MW (Section~\ref{subsec:dSph}). Finally we investigate targets outside of the Local Group, such as galaxy clusters (Section~\ref{subsec:galclustID}).

\subsection{Celestial bodies}
\label{subsec:bodies}

There is a possibility that DM particles from the MW halo could be captured by celestial bodies, such as stars and planets. This process can happen when DM particles successively scatter with the nuclei of a celestial body, in turn loosing kinetic energy and eventually sinking into its centre. Over time, they accumulate and essentially create a DM over-density in the centre of the body, that increases the production of indirect signals. 

In particular, this idea has been explored since the end of the 1980s, especially in the case of DM capture in the Sun~\cite{Gould:1987ir,Gould:1987ju,Gould:1987ww}. In this scenario, the only DM-produced particles that could escape the Sun are the neutrinos, and therefore detecting an excess of neutrinos from the Sun on top of the solar neutrino background (mainly from nuclear fusion) can be identified as a signal from DM. Typically the DM-produced neutrinos can have energies in the GeV--TeV range, whereas around the MeV for neutrinos produced by the Sun. Another credible possibility consists of DM annihilating into mediators particles that escape the Sun, that can then decay into any SM particle~\cite{Batell:2009zp,Schuster:2009au}.

The formalism for computing signals from DM captured in celestial objects is thoroughly described in~\cite{Leane:2023woh}, accompanied by a numerical framework named \verb|Asteria|~\cite{rebecca_k_leane_2023_8150110}.

\subsection{The Milky Way}
\label{subsec:MW}

We previously mentioned in Section~\ref{subsec:rotcurves} that galaxies are embedded in a DM halo that extends beyond their disk, which explains the flatness of their rotation curve at high distances from their centre. Our own Galaxy is therefore an excellent target for ID, since we are literally bathing in its DM halo. A crucial ingredient to compute indirect signals from DM in the MW is the distribution of its energy density $\rho_\textrm{DM}(\vec{x})$.

It is believed that galaxies form through the gravitational collapse of giant clouds of gas and DM. Baryonic matter can dissipate its energy through interactions with itself and in order to conserve its angular momentum during its cooling, the initially spherically symmetric gas cloud will rearrange itself into a disk. On the contrary, DM has limited interactions with itself and baryonic matter, thus cannot dissipate energy during the collapse: the distribution of DM remains approximately spherically symmetric. Therefore, when describing the DM distribution in a galaxy, it is usual to adopt a profile that is radial, \emph{i.e.}\ only dependent on the Galactocentric distance $r$.

The DM distribution in our Galaxy is still not known, as it is very challenging to reconstruct it while being inside the halo. Therefore, we have to assume a certain parametrisation of the profile when we are trying to predict indirect signals of DM in the MW. This would thus weaken the robustness of the prediction, especially when we are looking at a region of interest (ROI) close to the GC, as their behavior at $r\rightarrow 0$ can significantly differ from one profile to another, as we will see below. The most popular parametrisations of the DM profile are the following:
\begin{itemize}
    \item The \textbf{Navarro-Frenk-White} (NFW) profile~\cite{Navarro:1995iw}, which came from fitting the DM distribution of halos arising in seminal DM-only $N$-body simulations, remains to this day the standard benchmark for the DM profile in the MW. Its functional form is 
    \begin{equation}
    	\label{eq:NFW}
        \rho_\textrm{NFW}(r) = \rho_s \left(\frac{r_s}{r}\right) \left(1+\frac{r}{r_s}\right)^{-2}\;,
    \end{equation}
    showing that the DM density behaves like $r^{-1}$ near the GC and therefore diverges when $r\rightarrow 0$. To control how steep the DM density increases near the GC, the NFW can be generalised by introducing a free parameter $\gamma$ 
    \begin{equation}
    	\label{eq:gNFW}
        \rho_\textrm{gNFW}(r) = \rho_s \left(\frac{r_s}{r}\right)^\gamma \left(1+\frac{r}{r_s}\right)^{\gamma-3}\;,
    \end{equation}
    where the standard NFW profile is retrieved when $\gamma=1$. In this chapter and the following ones, the case $\gamma=1.26$ will be referred to as the contracted NFW profile (cNFW), which provides a better fit to the $\gamma$-ray excess in the GC~\cite{Daylan:2014rsa}. We further develop on this topic later in this section.

    \item The \textbf{Einasto} profile~\cite{1965TrAlm...5...87E,Graham:2005xx} provides a better fit to higher resolution $N$-body simulations. Contrary to the NFW profile, this profile does not diverge when $r\rightarrow 0$. Its functional form is written
    \begin{equation}
        \rho_\textrm{Ein}(r) = \rho_s \exp\left\{-\frac{2}{\alpha_\textrm{Ein}}\left[\left(\frac{r}{r_s}\right)^{\alpha_\textrm{Ein}}-1\right]\right\}\;,
    \end{equation}
    where $\alpha_\textrm{Ein} = 0.17$ is an averaged value from fits to different $N$-body simulations.

    \item The \textbf{Isothermal} profile~\cite{1980ApJS...44...73B} is obtained analytically by assuming DM particles are collisionless and therefore follow a Boltzmann velocity distribution. When doing the computation, one gets $\rho(r) \propto r^{-2}$ which is divergent for $r\rightarrow 0$. To alleviate this issue, a cutoff radius $r_s$ can be introduced, for which $\rho(r\lesssim r_s)$ is constant. The expression of this profile is therefore
    \begin{equation}
       \label{eq:Isothermal}
        \rho_\textrm{Iso}(r) = \frac{\rho_s}{1+(r/r_s)^2}\;.
    \end{equation}
    This profile was actually very popular before $N$-body simulations were available due to its theoretical motivation.

    \item The \textbf{Burkert} profile~\cite{Burkert:1995yz} initially arose from the measurement of rotation curves in a survey of dwarf galaxies, but then has proven to be a good fit to the ones of regular galaxies as well. This profile is parametrised as follows
    \begin{equation}
    	\label{eq:Burkert}
        \rho_\textrm{Bur}(r) = \frac{\rho_s}{(1+r/r_s)(1+(r/r_s)^2)}\;,
    \end{equation}
    which is also constant for $r\lesssim r_s$.
    
\end{itemize}

\begin{figure}[t]
    \begin{minipage}[c]{0.49\linewidth}
        \centering
        \includegraphics[width=\linewidth]{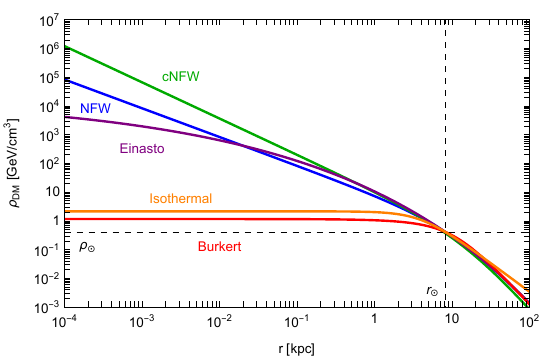}
    \end{minipage}
    \hfill
    \begin{minipage}[c]{0.49\linewidth}
        \centering
        \begin{tabular}{|c|c|c|}
            \hline
            DM halo  & $r_s$ [kpc] & $\rho_s$ [GeV/cm$^3$]  \\
            \hline
            NFW & $14.46$ & $0.566$ \\
            cNFW & $18.67$ & $0.272$ \\
            Einasto & $13.62$ & $0.154$ \\
            Isothermal & $4.38$ & $1.828$ \\
            Burkert & $10.62$ & $1.144$ \\
            \hline
        \end{tabular}
    \end{minipage}
    \caption{Left panel: Plot of the possible Galactic DM profiles. Right panel: Values of $r_s$ and $\rho_s$ adopted to compute the profiles shown in the left panel, and determined using the procedure explained in the text.}
    \label{fig:DMprofiles}
\end{figure}

In the case of the MW, one way to parametrise $\rho_s$ and $r_s$, that appear in the expression of the DM profiles, is by following two prescriptions~\cite{Cirelli:2009dv}:
\begin{enumerate}
	\item We assume the DM density at the position of the Sun ($r_\odot = 8.277$ kpc~\cite{GRAVITY:2021xju}) to be $\rho_\odot = \rho(r_\odot) = 0.4$ GeV/cm$^3$. Actually, the value of quantity is still debated in the literature, and can range between $0.2$ and $0.8$ GeV/cm$^3$~\cite{cirelli2024dark}. 
	\item We assume the DM mass contained in the MW within a radius $r_{200} \equiv 200$ pc to be $M_{200}^\textrm{DM} \equiv 4\pi \int_0^{r_{200}} r^2\rho(r)dr = 10^{12}M_\odot$. This value is an average of many results from the literature, using different techniques~\cite{Sakamoto:2002zr,SDSS:2008nmx,Przybilla:2010gd,2016MNRAS.463.2623H,McMillan:2016jtx,2019ApJ...873..118W,Callingham:2018vcf,Cautun:2019eaf}.
\end{enumerate}
These prescriptions allow us to fix $\rho_s$ and $r_s$ for all of the profiles except the Isothermal one. Indeed, the latter cannot fulfill both of the prescriptions at the same time, due to its slow steepness at large $r$. Instead we fix $r_s$ at $4.38$ kpc~\cite{Cirelli:2010xx} and apply the first prescription to obtain the value of $\rho_s$. The values of $r_s$ and $\rho_s$ of all the aforementioned profiles are tabulated in the right panel of Figure~\ref{fig:DMprofiles}, while showing a plot of the profiles in the left panel. 

We can divide them into two categories. On one hand, there are profiles, such as NFW and Einasto, where the DM density grows when approaching the GC, which is a feature that arises in DM-only $N$-body simulations. These are known as \emph{cusped} profiles. On the other hand, in DM profiles based on observations or on purely theoretical aspects, like the Isothermal or Burkert ones, the DM density remains constant when getting closer to the GC. These are known as \emph{cored} profiles. This clear discrepancy between the results from $N$-body simulations and observations/theory which is referred to as the \emph{cusp-core problem}~\cite{deBlok:2009sp}, although more recent $N$-body simulations have shown that including baryons has the effect of flattening the cusp we normally get in DM-only simulations, partially resolving the problem~\cite{Mollitor:2014ara}.

Now that we know more or less how DM is distributed in our Galaxy, we can start to think of which targets we should point our detectors to in order to probe indirect signals from DM. The most sensible answer would be to point them straight at the GC, since this is where the DM density is the highest. Actually, in recent years an excess of $\gamma$-rays has been reported in a region extending to a dozen of degrees around the GC~\cite{Hooper:2010mq} that could be attributed to DM. A more detailed analysis~\cite{Daylan:2014rsa} claimed that this excess is best fit by DM particles with $m_\textrm{DM}\simeq 45$ GeV annihilating into $b\bar{b}$ with a cross section of $\langle\sigma v\rangle \simeq 3 \times 10^{-26}$ cm$^3$/s, and distributed along a cNFW profile (with $\gamma = 1.26$). However, the GC is also extremely dense in astrophysical sources that our current detectors are not able to resolve. Therefore, the attempts to explain the GC excess depend greatly on our assumptions on the astrophysical background in this region. The other most credible explanation of the excess involves a high population of unresolved astrophysical sources, such as millisecond pulsars~\cite{Bartels:2015aea,Lee:2015fea}.

When exploring higher Galactic latitudes, both of the astrophysical background and DM density are reduced. This means that the modeling of astrophysical emissions becomes less important, but so does the expected DM indirect signal and therefore the likelihood of detecting one with our current observational strategies. There is no ideal ROI for DM ID in our Galaxy, we would always have to face a compromise.

\subsection{Dwarf spheroidal galaxies}
\label{subsec:dSph}

\begin{table}[t]
    \centering
    \begin{tabular}{|c|c|c|c|c|c|}
        \hline
        & \textbf{Name}     & \textbf{Position} & \textbf{Distance} & \textbf{Half-flux radius} & \textbf{Mass-to-light ratio} \\
        & (constellation)   & $(b,\ell)$           & $D$ [kpc]         & $r_h$ [pc]                & $M/L$ [$M_\odot/L_\odot$] \\
        \hline
        & Carina            & $(-22.2^\circ, 260.1^\circ)$ & $105\pm6$  & $250\pm39$ & $34$ \\
        & Draco             & $(+34.7^\circ, 86.4^\circ)$  & $76\pm6$   & $221\pm19$ & $75$ \\
        \parbox[t]{4mm}{\multirow{3}{*}{\rotatebox[origin=c]{90}{\textbf{Classical\;\;}}}}
        & Fornax            & $(-65.7^\circ, 237.1^\circ)$ & $147\pm12$ & $710\pm77$ & $6$ \\
        & Leo I             & $(+49.1^\circ, 226.0^\circ)$ & $254\pm15$ & $251\pm27$ & $4$ \\
        & Leo II            & $(+67.2^\circ, 220.2^\circ)$ & $233\pm14$ & $176\pm42$ & $12$ \\
        & Sculptor          & $(-83.2^\circ, 287.5^\circ)$ & $86\pm6$   & $283\pm45$ & $12$ \\
        & Sextans           & $(+42.3^\circ, 243.5^\circ)$ & $86\pm4$   & $695\pm44$ & $120$ \\
        & Ursa Minor        & $(+44.8^\circ, 105.0^\circ)$ & $76\pm3$   & $181\pm27$ & $67$ \\
        \hline
        \parbox[t]{4mm}{\multirow{3}{*}{\rotatebox[origin=c]{90}{\textbf{Ultrafaint\;\;\;}}}}
        & Bo\"otes II         & $(+68.9^\circ, 353.7^\circ)$ & $42\pm1$   & $51\pm17$  & $6300$ \\
        & Canes Venatici II & $(+82.7^\circ, 113.6^\circ)$ & $160\pm4$  & $74\pm14$  & $230$  \\
        & Segue II          & $(-38.1^\circ, 149.4^\circ)$ & $35\pm2$   & $35\pm3$   & $1500$ \\
        & Ursa Major II     & $(-37.4^\circ, 152.5^\circ)$ & $32\pm4$   & $149\pm21$ & $1900$ \\
        & Willman I         & $(-56.8^\circ, 158.6^\circ)$ & $38\pm7$   & $25\pm6$   & $530$ \\
        \hline
    \end{tabular}
    \caption{Properties of the classical and some of the ultrafaint satellite dSphs of the MW~\cite{McConnachie:2012vd}: position in the sky in terms of the Galactic latitude $b$ and longitude $\ell$, heliocentric distance $D$, half-flux radius $r_h$ and mass-to-light ratio $M/L$ within $r_h$.}
    \label{tab:dSphs}
\end{table}

Our Galaxy is surrounded by a number of satellite dSphs that are promising targets for DM ID. First, the number of stars contained in dSphs can range between a few dozens to several thousands of stars and, in addition to that, they are scarce in interstellar gas. These observations imply that the astrophysical background coming from them is low. Second, measurements of their rotation curve and mass-to-light ratio indicate they are dominated by DM. Moreover they are positioned in relatively high Galactic latitudes, therefore their faint emission is not covered by any foreground emission from the Galactic plane (GP). In Table~\ref{tab:dSphs} we list some of the measured properties of a few dSphs around our Galaxy. \emph{Classical} dSphs refer to the ones that were discovered before 1990, and therefore are the most luminous ones. The other category, \emph{ultrafaint} dSphs, is straightforward. The latter can have a way higher mass-to-light ratio compared to the classical ones, \emph{e.g.}\ the Bo\"otes II dSph reaches several thousands of $M_\odot/L_\odot$.

The star scarcity of dSphs can be a double edge-sword. On one hand it implies that astrophysical emissions from a dSph are limited, therefore increasing the likelihood of detecting an indirect signal from DM. On the other hand, stars are the only visible tracers of the gravitational potential in dSphs, and their scarcity makes the reconstruction of the DM profile challenging. 

\subsection{Galaxy clusters}
\label{subsec:galclustID}

\begin{table}[t]
    \centering
    \begin{tabular}{|c|c|c|}
        \hline
        \textbf{Name}   & \textbf{Position}            & \textbf{Distance} \\
                        & $(b,\ell)$                      & $D$ [Mpc] \\
        \hline
        Bullet          & $(-21.2^\circ, 266.0^\circ)$ & $1300$ \\
        Coma            & $(+88.0^\circ, 58.1^\circ)$  & $100$ \\
        Fornax          & $(-53.6^\circ, 236.7^\circ)$ & $20$  \\
        Hercules        & $(+44.5^\circ, 31.6^\circ)$  & $170$ \\
        Virgo           & $(+74.4^\circ, 283.8^\circ)$ & $24$ \\
        \hline
    \end{tabular}
    \caption{Properties of a few of the most well-known galaxy clusters~\cite{NEP}: position in the sky in Galactic coordinates $(b,\ell)$ and their distance $D$ from the MW.}
    \label{tab:galclust}
\end{table}

As mentioned in Section~\ref{subsec:galclust}, observations of galaxy clusters were historically the first to reveal the existence of DM in the Universe. They are too DM dominated and can be positioned at high Galactic latitudes, making them good targets for searching indirect signals of DM. However, their distance from us can range between $10$ and $1000$ Mpc and therefore their emission is very contaminated by foreground sources. Being the largest gravitationally bound systems in the Universe, constituted of galaxies with billions of stars each and containing diffuse intracluster gas, the astrophysical emissions from a galaxy cluster can be large and also difficult to disentangle with a potential signal from DM. We list the position and their distance from the MW of the most well-known galaxy clusters in Table~\ref{tab:galclust}.

\section{Predicting indirect signals from dark matter in the Milky~Way}
\label{sec:particlesDM}

The goal of this section is to provide an overview on the theoretical framework useful to predict indirect signals coming from DM in the Galactic halo. We choose to solely focus on this target, since the research conducted during this thesis (summarised in Chapters~\ref{chap:subGeV}, \ref{chap:prop} and \ref{chap:PBH}) has only involved DM searches in the MW. We begin with the production of stable particles (Section~\ref{subsec:particleprod}), then investigate the propagation of DM-produced charged CRs (Section~\ref{subsec:CCRs}), photons and neutrinos (Section~\ref{subsec:gammaneutrinos}) in the interstellar medium (ISM) until possibly reaching our detectors.

\subsection{Particle production}
\label{subsec:particleprod}

In principle, DM particles can annihilate or decay into any particle-antiparticle pair, either from the SM or BSM, called \emph{final states}, as long as it is kinematically possible\footnote{In the case where DM is non-relativistic, we can assume that DM particles annihilate (decay) at rest. To insure the conservation of energy, $m_\textrm{DM}$ must be greater or equal than (half of) the mass of the final state particle.}. After that, the final states can undergo decays, hadronisation, coalescence and emit soft radiations. In the end, only stable particles such as $e^\pm$, $\overset{\scriptscriptstyle(-)}{p}$, $\overset{\scriptscriptstyle(-)}{d}$, $\gamma$ and $\overset{\scriptscriptstyle(-)}{\nu}_{e,\mu,\tau}$ survive and propagate through the ISM until possibly reaching our detectors. We usually consider that the annihilation or decay of DM particles and the production of stable ones are occurring at the same point, since the length scale at which the latter propagate is way larger than the one at which particle physics processes happen. The general expression of the energy spectrum \emph{at production} $dN_i/dK_i$ (where $K$ is the kinetic energy) of stable particles $i$ is the sum of individual energy spectra at production over all annihilation or decay channels, weighted by their branching ratio
\begin{equation}
    \label{eq:sumBR}
    \frac{dN_i}{dK_i} = 
    \left\{
    \begin{array}{ll}
	\displaystyle
        \sum_\textrm{FS} \frac{\langle\sigma v\rangle_\textrm{FS}}{\langle\sigma v\rangle_\textrm{tot}} \frac{dN_i^\textrm{FS}}{dK_i} = \sum_\textrm{FS} \textrm{BR}_\textrm{FS} \frac{dN_i^\textrm{FS}}{dK_i} & \textrm{(annihilation)} \\
    \\
    \displaystyle
        \sum_\textrm{FS} \frac{\Gamma_\textrm{FS}}{\Gamma_\textrm{tot}} \frac{dN_i^\textrm{FS}}{dK_i} = \sum_\textrm{FS} \textrm{BR}_\textrm{FS} \frac{dN_i^\textrm{FS}}{dK_i} & \textrm{(decay)}
    \end{array}
    \right.\;,
\end{equation}
where the branching ratios BR$_\textrm{FS}$ are expressed in terms of the annihilation cross section $\langle\sigma v\rangle_\textrm{FS}$ or decay rate $\Gamma_\textrm{FS}$ of DM into the final states FS and their sum over all final states $\langle\sigma v\rangle_\textrm{tot}$ and $\Gamma_\textrm{tot}$. 

For a given BSM model, one can write the expression of the branching ratios, annihilation cross sections and decay rates in terms of the couplings between DM and SM particles. However, from now on, we adopt a model-independent approach where DM annihilation and decay channels are treated independently, the final states are only SM particles, and $\langle\sigma v\rangle$ and $\Gamma$ are free parameters. This approach is often used in phenomenological studies of DM ID, as it is unbiased and allows for a common ground where results can be easily compared from one study to another.

\begin{figure}[t]
    \centering
    \begin{subfigure}[c]{0.325\linewidth}
        \centering
        \includegraphics[width=\linewidth]{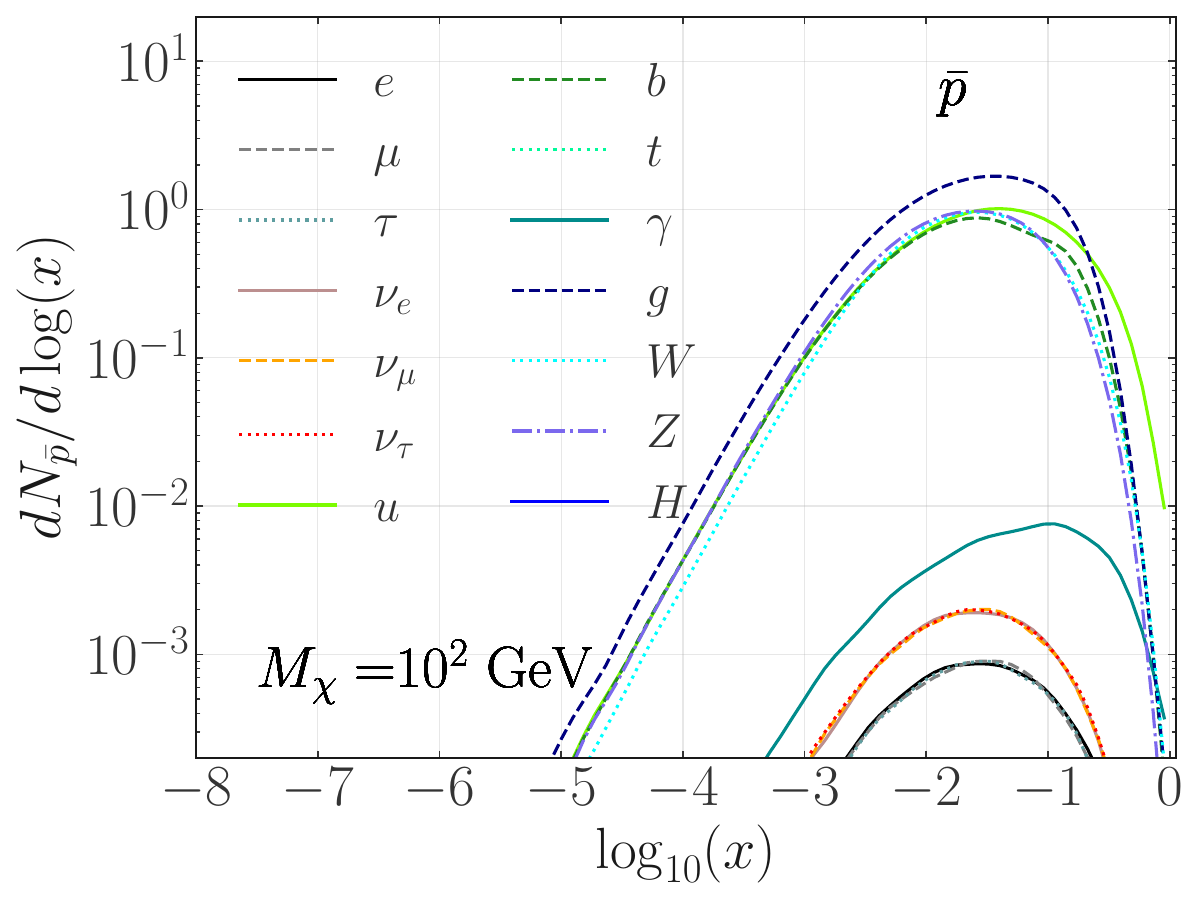}
    \end{subfigure}
    \hfill
    \begin{subfigure}[c]{0.325\linewidth}
        \centering
        \includegraphics[width=\linewidth]{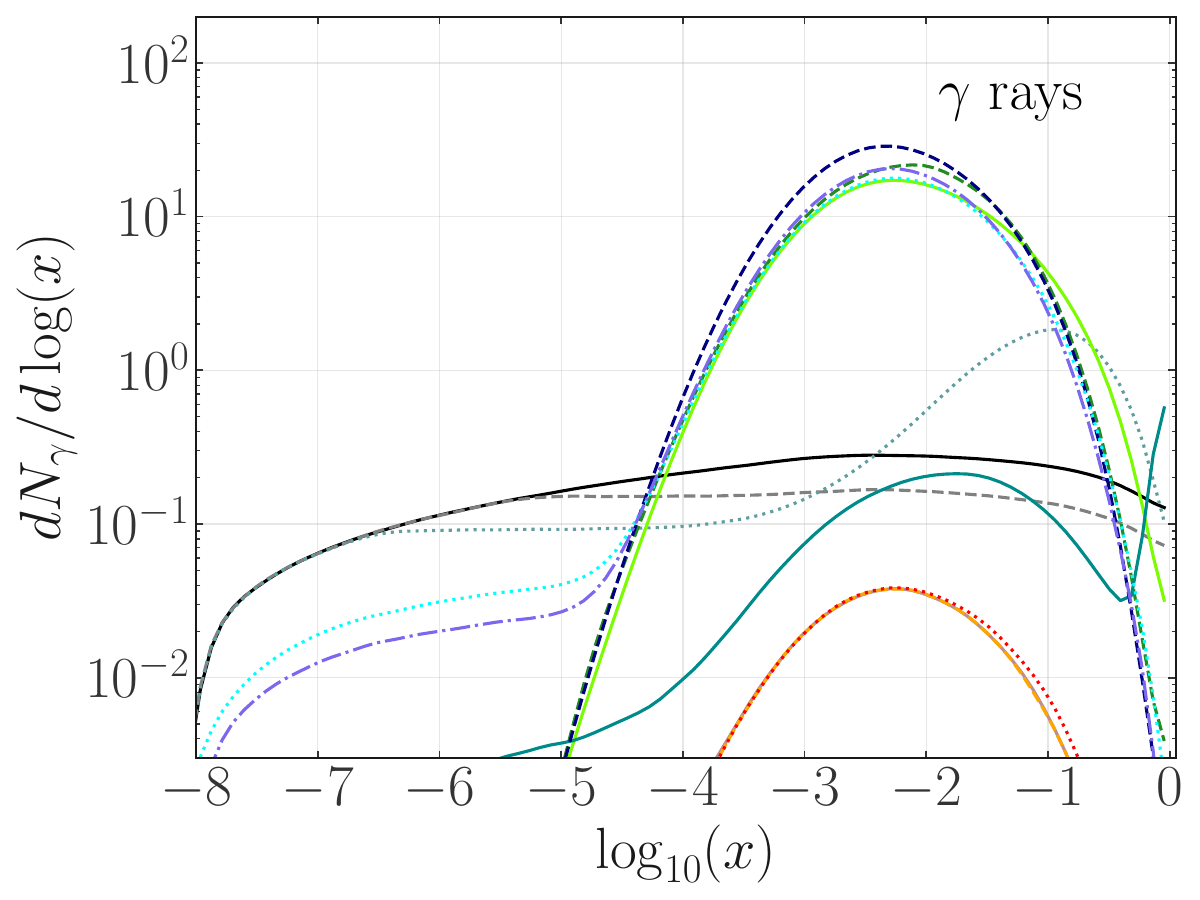}
    \end{subfigure}
    \hfill
    \begin{subfigure}[c]{0.325\linewidth}
        \centering
        \includegraphics[width=\linewidth]{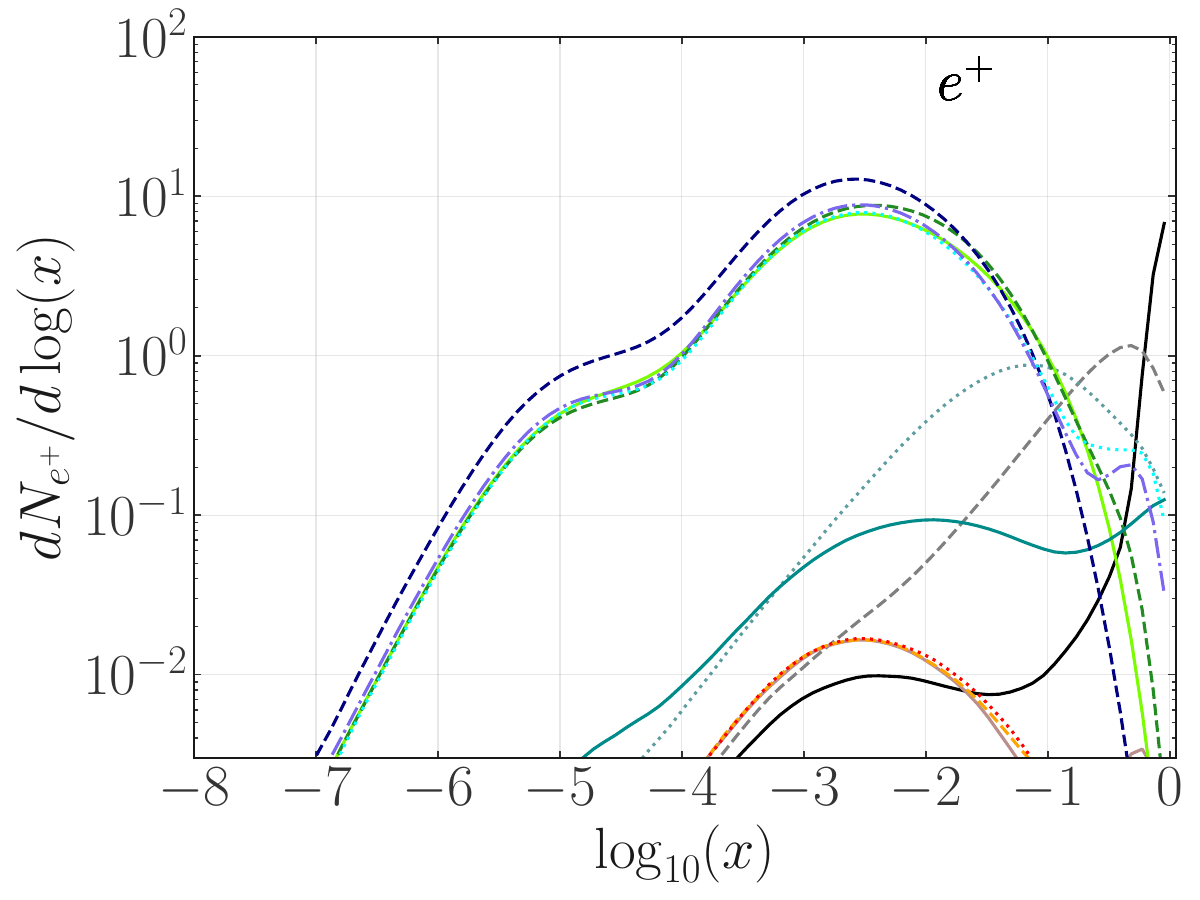}
    \end{subfigure}
    \hfill
    \begin{subfigure}[c]{0.325\linewidth}
        \centering
        \includegraphics[width=\linewidth]{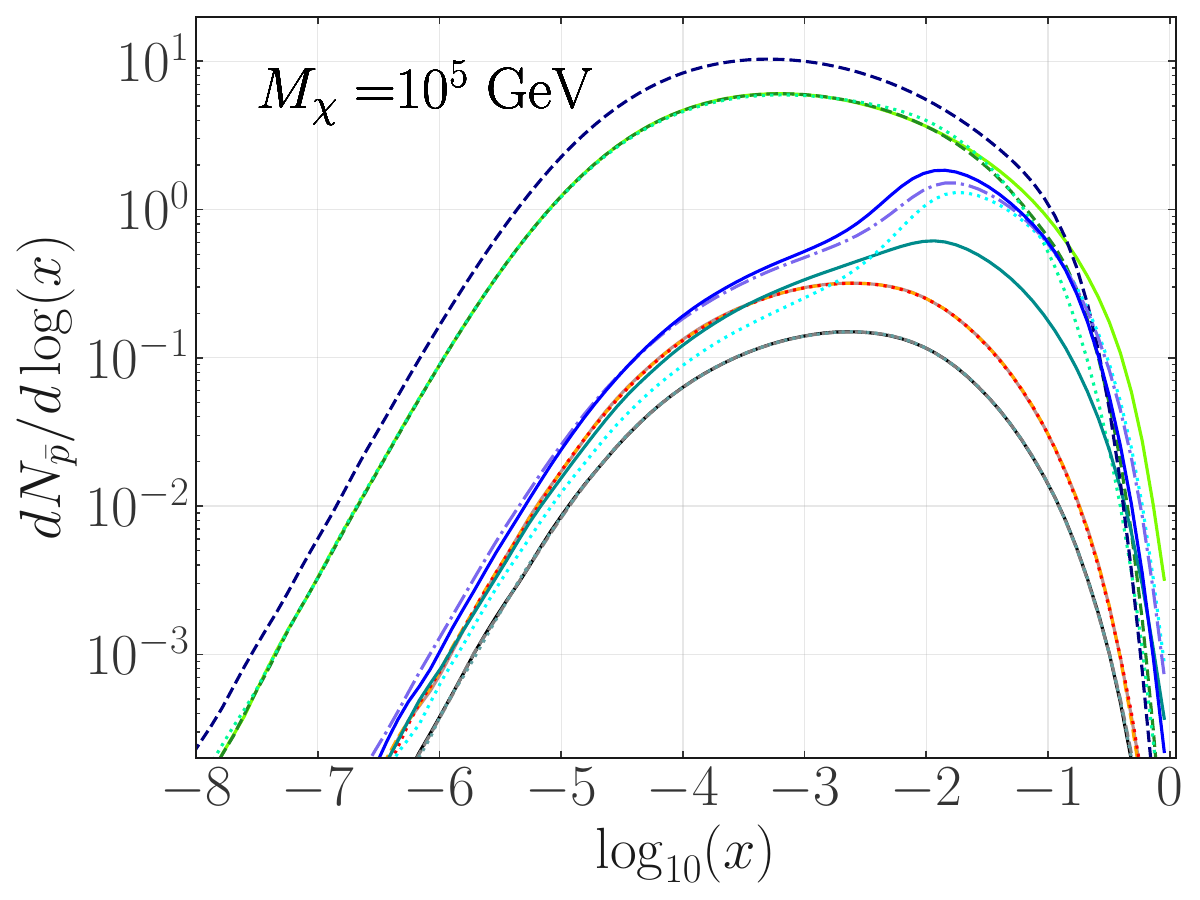}
    \end{subfigure}
    \hfill
    \begin{subfigure}[c]{0.325\linewidth}
        \centering
        \includegraphics[width=\linewidth]{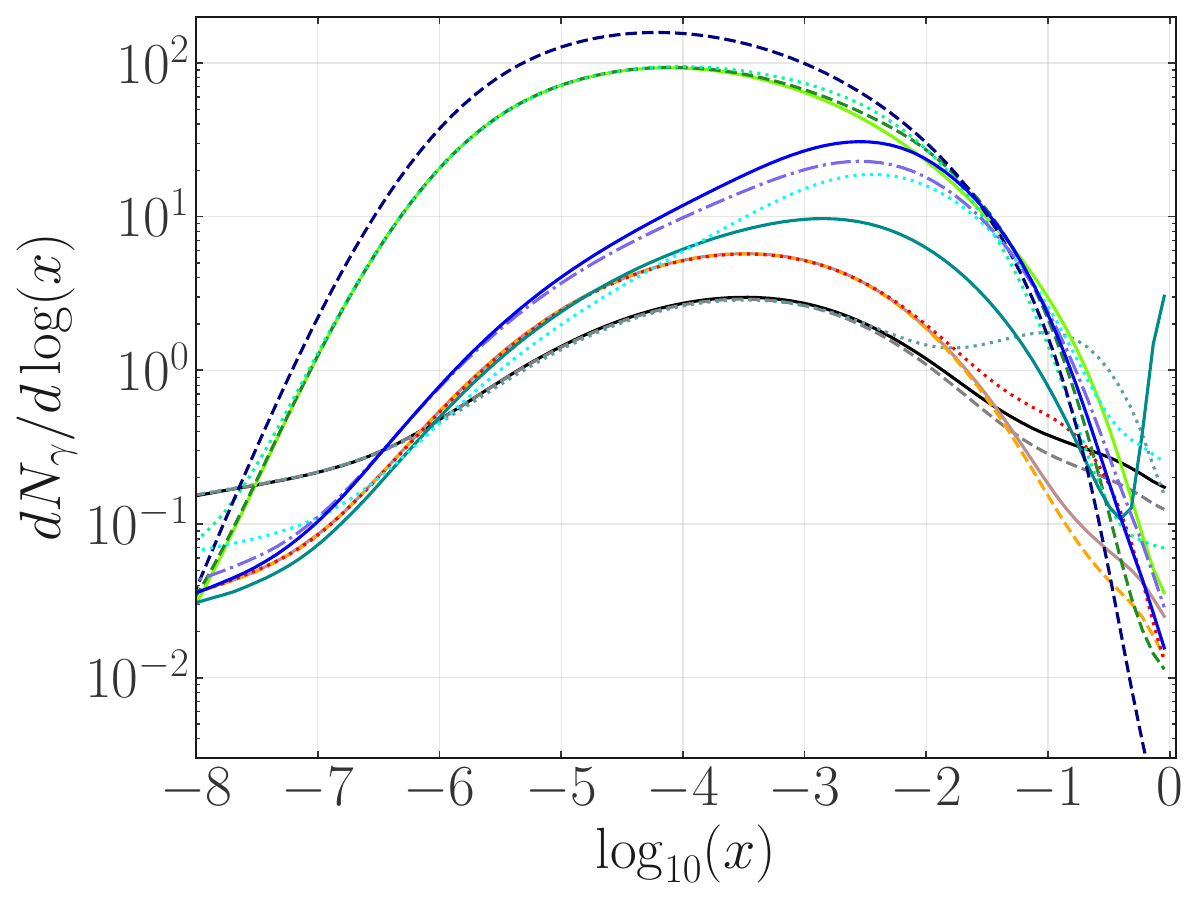}
    \end{subfigure}
    \hfill
    \begin{subfigure}[c]{0.325\linewidth}
        \centering
        \includegraphics[width=\linewidth]{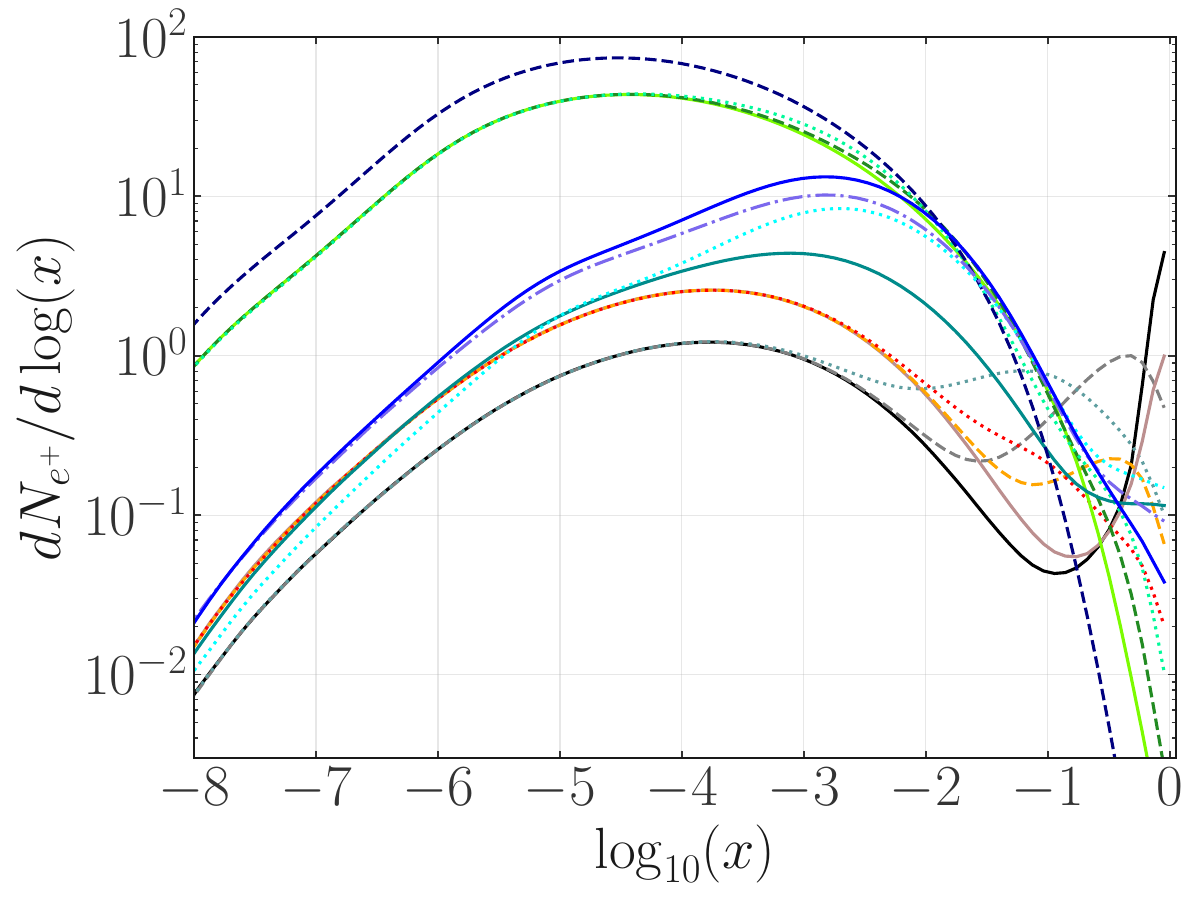}
    \end{subfigure}
    \cprotect\caption{Spectra at production of $\bar{p}$ (left panels), $\gamma$ (middle panels) and $e^+$ (right panels) for DM particles with a mass of $100$ GeV (top panels) or $100$ TeV (bottom panels) annihilating in the following final states: $e^+e^-$, $\mu^+\mu^-$, $\tau^+\tau^-$, $\nu_{e,\mu,\tau}\bar{\nu}_{e,\mu,\tau}$, $u\bar{u}$, $b\bar{b}$, $t\bar{t}$, $\gamma\gamma$, $gg$, $W^+W^-$, $ZZ$ and $HH$. Taken from~\cite{Arina:2023eic}.}
    \label{fig:fluxatprod}
\end{figure}

The spectrum at production of stable particles $dN_i/dK_i$ for a specific channel can be computed using Monte-Carlo simulations, \emph{e.g.}\ \verb|PYTHIA|~\cite{Bierlich:2022pfr} or \verb|HERWIG|~\cite{Bahr:2008pv,Bellm:2019zci}, however they require a good understanding of their tunable parameters. Instead, other numerical codes can provide these spectra, such as \verb|PPPC4DMID|~\cite{Cirelli:2010xx}, \verb|Hazma|~\cite{Coogan:2019qpu,Coogan:2022cdd}, \verb|HDMSpectra|~\cite{Bauer:2020jay} or \verb|CosmiXs|~\cite{Arina:2023eic}, that are based on the interpolation of results from well-tuned Monte-Carlo simulations and/or analytical computations. To illustrate, we show in Figure~\ref{fig:fluxatprod} the spectra at production of $\bar{p}$, $\gamma$ and $e^+$ computed using \verb|CosmiXs|, for various DM annihilation channels and for $m_\textrm{DM} = 100$ GeV and $100$ TeV, in terms of the energy fraction $x = K/m_\textrm{DM}$.

\begin{figure}[t]
    \centering
    \begin{subfigure}[c]{0.49\linewidth}
        \centering
        \includegraphics[width=\linewidth,trim= 0.5cm 0 2.5cm 1.5cm]{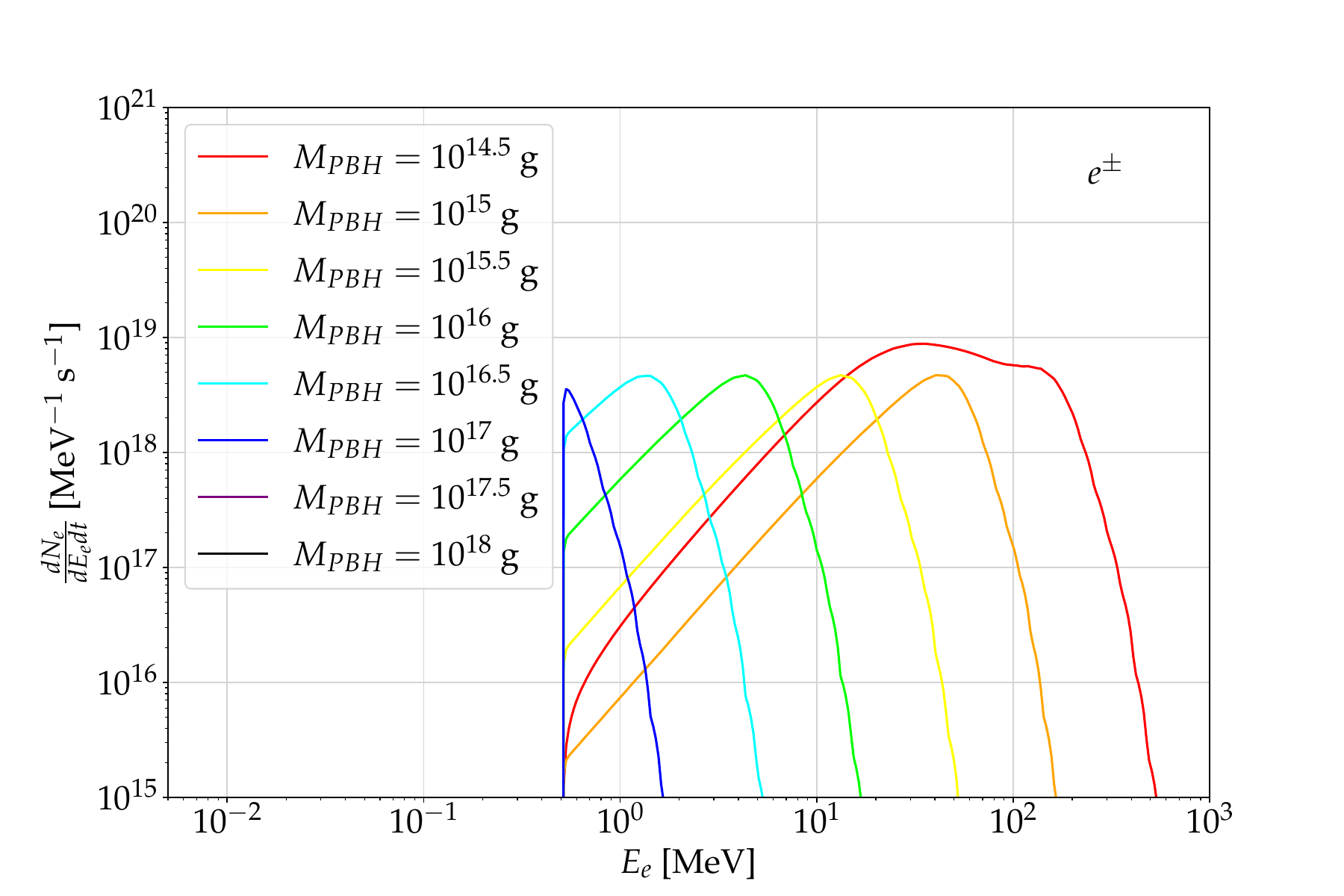}
    \end{subfigure}
    \hfill
    \begin{subfigure}[c]{0.49\linewidth}
        \centering
        \includegraphics[width=\linewidth,trim= 0.5cm 0 2.5cm 1.5cm]{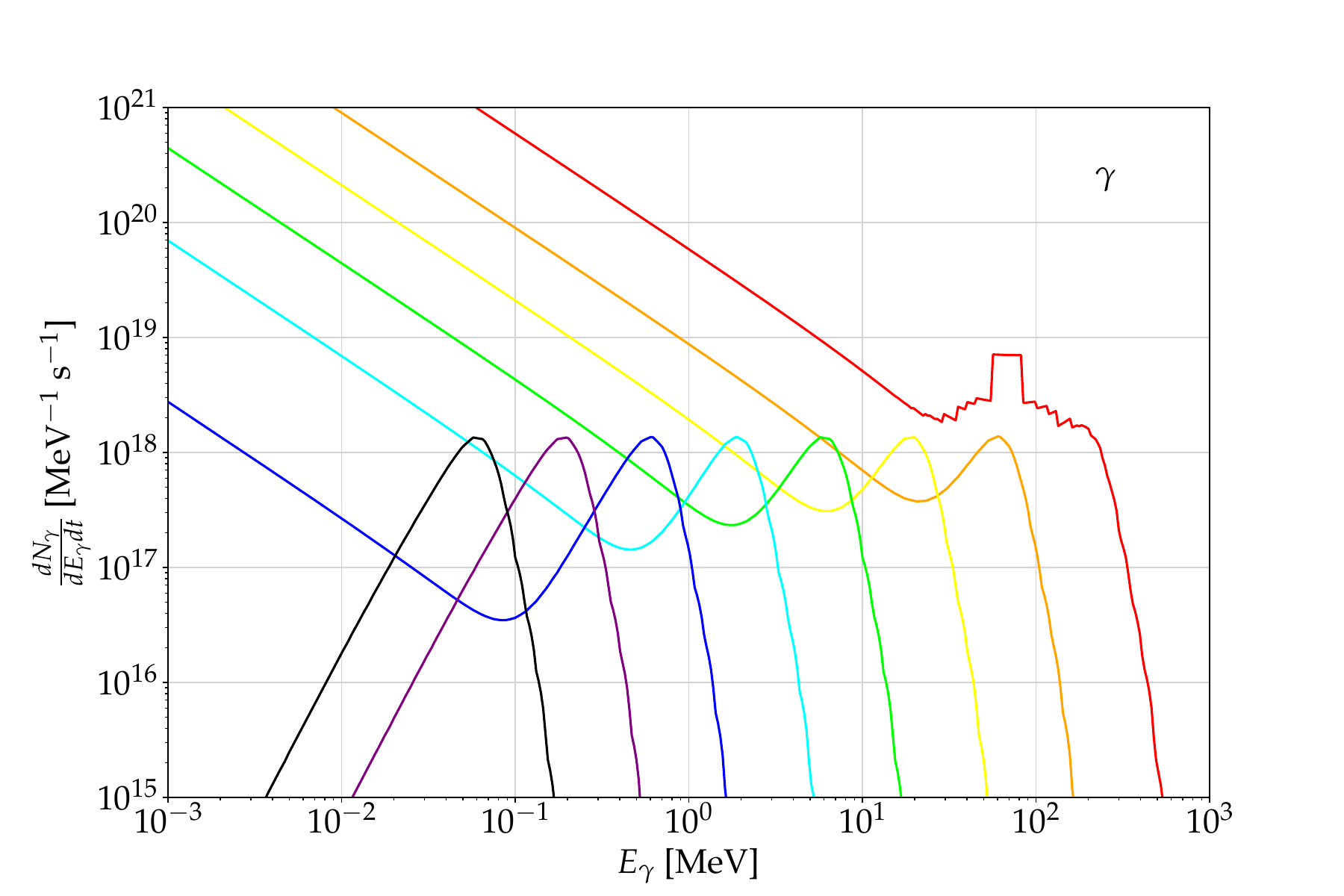}
    \end{subfigure}
    \cprotect\caption{Emission spectra of $e^+$ (left panel) and $\gamma$ (right panel) from the evaporation of a single PBH, for masses varying between $10^{14.5}$ g and $10^{18}$ g. Computed using \verb|BlackHawk+Hazma|.}
    \label{fig:PBHevap}
\end{figure}

In the case where DM is composed of PBHs, we can also expect an indirect signal from them. In 1975, Stephen Hawking has predicted that BHs can emit particles thanks to quantum effects at their event horizon~\cite{Hawking:1975vcx}. This process is known as BH \emph{evaporation}. For a Schwarzschild (electrically neutral and non-spinning) PBH, the emission rate of particles $i$ per unit time and energy through its evaporation is given by
\begin{equation}
    \label{eq:evap}
    \frac{d^2N_i}{dtdK_i} = \frac{1}{2\pi} \sum_\text{d.o.f.} \frac{\Gamma_i(K_i,M)}{e^{K_i/T}\pm 1} \quad \textrm{(PBH evaporation)}\;,
\end{equation}
where $+$ ($-$) is taken if $i$ is a fermion (boson), $M$ is the PBH mass, $T = 1/(8\pi G M)$ its temperature, and the sum is performed over the degrees of freedom (d.o.f.)\ of the emitted particles (spin, color, helicity and charge multiplicities). The deviation of this spectrum from a black body one is characterised by the grey body factor $\Gamma_i$ that can be computed semi-analytically or using numerical codes \emph{e.g.}, \verb|BlackHawk|~\cite{Arbey:2019mbc,Arbey:2021mbl}. Secondary processes (hadronisation, decay, ...) can then happen, in the same manner as in the case of annihilating or decaying DM particles. In Figure~\ref{fig:PBHevap} we show the emission spectra of $e^+$ and $\gamma$ from PBHs with different masses, computed using \verb|BlackHawk| combined with \verb|Hazma| to deal with the secondary processes. In this DM scenario the unknowns are the fraction of PBHs constituting the total amount of DM in the Universe $f_\textrm{PBH}\equiv \rho_\textrm{PBH}/\rho_\textrm{DM}$ and the mass distribution of PBHs.

Now that we have the tools to compute the spectrum of stable particles produced on the spot of particle DM annihilation or decay and PBH evaporation, we can go into the physics of their propagation in the MW.

\subsection{Charged cosmic-rays}
\label{subsec:CCRs}

As we previously mentioned, DM-produced charged CRs are a promising indirect signal as they can be antimatter particles that are detectable over a sufficiently low background, since antimatter particles are scarcely produced in astrophysical processes. In order to compute the flux of DM-produced, stable charged CRs at Earth, we need to take into account their transport in the ISM. To do so, we have to solve the \emph{diffusion-convection-loss equation} in order to obtain the number of particles of species $i$ per energy and volume unit $f_i(K_i,\vec{x})$ at a position $\vec{x}$ in the Galaxy~\cite{Ginzburg:1969,Evoli:2016xgn}
\begin{equation}
    \label{eq:transporteq}
    -\vec{\nabla}\cdot\left(D_i\vec{\nabla}f_i + \vec{v}_cf_i\right)-\frac{\partial}{\partial K_i}\left[\dot{K}_i f_i - K_i^2 D_{pp}\frac{\partial}{\partial K_i}\left(\frac{f_i}{K_i^2}\right)-\frac{K_i}{3}\left(\vec{\nabla}\cdot\vec{v}_c \right)f_i\right] = Q_i - c\beta_if_i \sum_j n_j\sigma_{ij}\;,
\end{equation}
where different processes are included:
\begin{itemize}
    \item \textbf{Spatial diffusion:} As the charged CRs propagate, they interact with the Galactic magnetic fields (GMFs). The turbulent component of GMFs induces a random walk of the CRs, well-described by a spatial diffusion term in Equation~\ref{eq:transporteq} that involves a diffusion tensor $D_i(K_i,\vec{x})$.
    \item \textbf{Momentum space diffusion:} The interactions between charged CRs and the turbulent component of the GMFs can also provoke the diffusion of CRs in their momentum space, possibly causing their stochastic reacceleration. This effect is described by the term involving the momentum diffusion coefficient $D_{pp}(K_i,\vec{x})$ in Equation~\ref{eq:transporteq}.
    \item \textbf{Energy losses:} CRs can interact with various components of the ISM and lose some of their energy in the process. $e^\pm$ propagating through GMFs can lose energy through synchrotron radiation, they can up-scatter ambient photons (via ICS), emit bremsstrahlung when passing by nuclei in the ISM gas, which is mostly comprised of hydrogen -- that can be atomic (HI), ionised (HII) and molecular (H$_2$) -- as well as neutral He. Hadronic CRs ($\overset{\scriptscriptstyle(-)}{p}, \overset{\scriptscriptstyle(-)}{d}$) can lose their energy through the production of pions by colliding with nuclei in the ISM. All charged CRs can also lose their energy by ionising the gas in the ISM and through Coulomb scattering on free $e^-$ in ionised gas. In Equation~\ref{eq:transporteq}, energy losses are represented by the term in $\dot{K}$. They are encoded in the energy loss function $b_\textrm{loss}(K_i,\vec{x})\equiv -\dot{K}$ that can be computed analytically~\cite{Schlickeiser:2002pg,Buch:2015iya,Evoli:2016xgn} and given in Appendix~\ref{apx:elosses} (only for $e^\pm$).
    \item \textbf{Convection:} Stellar winds and supernovae induce a flow of particles averagely directed outwards the Galaxy, called the \emph{Galactic wind}. This wind, modeled by the velocity $\vec{v}_c(\vec{x})$ in Equation~\ref{eq:transporteq}, also contributes to the transport of charged CRs by deflecting them.
    \item \textbf{Interactions with the ISM:} $p$ and $d$ CRs can collide with nuclei present in the ISM gas, essentially shattering them, in turn producing more $p$ and $d$ CRs. This process is known as \emph{spallation}. On the other hand, $\bar{p}$ and $\bar{d}$ CRs can annihilate with their matter counterpart present in the ISM gas. Both of these interactions are encoded in the second right-hand-side term of Equation~\ref{eq:transporteq}, where $\beta_i$ is the CR velocity in units of $c$, $n_j(\vec{x})$ is the ISM gas density, $\sigma_{ij}$ is the spallation (annihilation) cross section of $p$ and $d$ ($\bar{p}$ and $\bar{d}$) with the ISM gas, and $j \in$ \{HI, HII, H$_2$, He\}.
\end{itemize}

Finally, the number of DM-produced CRs injected at a position $\vec{x}$ in the Galaxy per unit time, volume and energy is encoded in the source term $Q_i(K_i,\vec{x})$ of Equation~\ref{eq:transporteq}. For particle DM, this term is written
\begin{equation}
    \label{eq:source}
    Q_i (K_i,\vec{x}) = 
    \left\{
\begin{array}{ll}
	\displaystyle  
        \xi\langle \sigma v\rangle \left(\frac{\rho_\textrm{DM}(\vec{x})}{m_\textrm{DM}}\right)^2\frac{dN_i^{\textrm{ann}}}{dK_i} &\text{(annihilation)}\\
	\\
        \displaystyle
        \quad \Gamma \ \; \left(\frac{\rho_\textrm{DM}(\vec{x})}{m_\textrm{DM}}\right)\; \frac{dN_i^{\textrm{dec}}}{dK_i} &\text{(decay)}
    \end{array}
    \right.\;,
\end{equation}
where $\xi = 1/2$ if DM is its own antiparticle and $1/4$ otherwise, $dN_i/dK_i$ is the injection spectrum of particle $i$ per annihilation or decay of DM-produced CRs that can be computed by the tools described in Section~\ref{subsec:particleprod}. In the case where DM is in part constituted of PBHs we have
\begin{equation}
    \label{eq:sourcePBH}
    Q_i(K_i,\vec{x}) = f_\textrm{PBH}\rho_\textrm{DM}(\vec{x}) \int \frac{dM}{M}\frac{dN_\textrm{PBH}}{dM}\frac{d^2N_i}{dtdK_i} \quad \textrm{(PBH evaporation)}\;,
\end{equation}
where $d^2N_i/dtdK_i$ is the emission rate of particles $i$ through the evaporation of a single PBH, as described in Section~\ref{subsec:particleprod}, and $dN_\textrm{PBH}/dM$ represents the mass distribution of PBHs in the MW. The case of PBH evaporation is essentially similar to the one of decaying DM particles, except that the decay rate $\Gamma$ is contained in the emission rate $d^2N_i/dtdK_i$. This is especially true when we consider the PBH mass distribution to be monochromatic ($dN_\textrm{PBH}/dM = \delta(M-M_\textrm{PBH})$) where
\begin{equation}
    Q_i(K_i,\vec{x}) = f_\textrm{PBH}\left(\frac{\rho_\textrm{DM}(\vec{x})}{M_\textrm{PBH}}\right) \frac{d^2N_i}{dtdK_i} \quad \textrm{(monochromatic)}\;.
\end{equation}

Equation~\ref{eq:transporteq} has to be solved within a certain boundary of space. The different processes that affect the propagation of charged CRs involve ingredients that have a cylindrical symmetry in a first approximation, such as the GMFs, ISM gas and ambient photon densities. It is therefore customary to solve Equation~\ref{eq:transporteq} within the boundaries of a cylinder of radius $R_\textrm{max}$ and height $L$ (commonly known as the \emph{halo height}) that represents the boundaries of the Galaxy~\cite{Taillet:2002ub,Strong:1998pw}. In other words, $R_\textrm{max}$ and $L$ are implicit components of the propagation equation.

Although an analytical solution to the full Equation~\ref{eq:transporteq} does not exist, one can make some assumptions to simplify and solve it. Alternatively, it can be solved numerically as long as the different ingredients involved are parametrised properly. As such, many studies consider the spatial diffusion tensor $D$ to be a rigidity-dependent\footnote{The rigidity is defined as the ratio between the CR momentum and charge: $R \equiv p/\lvert q \rvert$.} coefficient, and can be parametrised as the following double broken power-law~\cite{Genolini:2021doh}
\begin{equation}
    \label{eq:diffcoef}
    D(R) = \beta^\eta D_0 \left(\frac{R}{1\,\textrm{GV}}\right)^\delta \left[1+\left(\frac{R_l}{R}\right)^{(\delta-\delta_l)/s_l}\right]^{s_l}\left[1+\left(\frac{R}{R_h}\right)^{(\delta-\delta_h)/s_h}\right]^{-s_h}\;,
\end{equation}
where $D_0$ is a normalisation factor, $\eta$ is the velocity index, $R_{l(h)}$ indicate the rigidity at which the spectral breaks occur in the diffusion coefficient, $\delta_{l(h)}$ are the spectral indices at the (high-) low-rigidity regimes, $\delta$ at the mid-rigidity regime, and $s_{l(h)}$ controls the smoothness around the breaks. The momentum space diffusion coefficient $D_{pp}$ can also be parametrised (assuming the turbulence in the GMFs is characterised by Alfv\'en waves~\cite{Berezinsky:1990qxi,1994ApJ...431..705S})
\begin{equation}
    \label{eq:momdiffcoef}
    D_{pp}(p) = \frac{4}{3\delta(4-\delta^2)(4-\delta)}\frac{p^2v_A^2}{D(p)}\;,
\end{equation}
where $v_A$ is the Alfv\'en velocity. The value of the propagation parameters involved in Equations~\ref{eq:transporteq}, \ref{eq:diffcoef} and \ref{eq:momdiffcoef} (alongside $R_\textrm{max}$ and $L$) can have a sizeable impact on the CR flux predictions, but they can be constrained by fitting the latter on local fluxes and flux-ratios of different CR species (such as $e^\pm$, $p/\bar{p}$, but also lithium, beryllium, carbon and boron nuclei) measured by the various experiments listed in Section~\ref{subsec:expCCRs}. When it comes to the ISM gas, GMFs and ambient photons maps, a plethora of models is available, all more or less based on observations. Once all of the aforementioned ingredients are carefully set, Equation~\ref{eq:transporteq} can then be solved using numerical codes such as \verb|GALPROP|~\cite{2011CoPhC.182.1156V,galpropweb}, \verb|DRAGON2|~\cite{Evoli:2016xgn,Evoli:2017vim} and \verb|USINE|~\cite{Maurin:2018rmm} or semi-analytically. In the end, we obtain the differential flux of propagated DM-produced CRs of species $i$ at any position $\vec{x}$ in the Galaxy, which is written
\begin{equation}
    \label{eq:CRflux}
    \frac{d\Phi_i}{dK_i}(K_i,\vec{x}) \equiv \frac{c\beta_i}{4\pi}f_i(K_i,\vec{x})\;,
\end{equation}
and, when evaluated at Earth's position, can then be directly used to either explain excesses in CR data, allowing us to pinpoint the value of the free parameters we previously mentioned (such as $\langle\sigma v\rangle$, $m_\textrm{DM}$, ...), or otherwise set constraints on them.

\subsection{Photons and neutrinos}
\label{subsec:gammaneutrinos}

In this section, we go through the computation of the flux of DM-produced photons and neutrinos. These particles can be directly emitted by DM annihilation, decay or PBH evaporation, which are referred to as \emph{prompt emissions}. In comparison to charged CRs, photons and neutrinos are not deviated by the GMFs and do not interact much with the ISM, therefore their propagation is not described by Equation~\ref{eq:transporteq}. This also means that their source of emission can be easily identified. On the other hand, DM-produced charged CRs can interact with the ISM, producing (more diffuse) \emph{secondary emissions} of photons. In this section, we specifically investigate the case of secondary photons emitted by DM-produced $e^\pm$ through ICS, synchrotron and bremsstrahlung.

\subsubsection{Prompt emissions of photons and neutrinos}

\begin{figure}[t]
    \centering
    \begin{tikzpicture}[scale=1, every node/.style={transform shape}]
        \draw [line width=0.5mm] (0,2.75) ellipse (5 and 1);
        \draw [line width=0.5mm] (-5,0) arc (180:360:5 and 1);
        \draw [line width=0.5mm,dashed] (-5,0) arc (180:360:5 and -1);
        \draw [line width=0.5mm] (-5,-2.75) arc (180:360:5 and 1);
        \draw [line width=0.5mm,dashed] (-5,-2.75) arc (180:360:5 and -1);
        \draw [line width=0.5mm] (-5,-2.75) -- (-5,2.75);
        \draw [line width=0.5mm] (5,-2.75) -- (5,2.75);
        \draw [line width=0.5mm] (5,-2.75) -- (5,2.75);
        
        \draw [line width=0.5mm] (-5,0) -- (5,0);

        \draw [line width=0.5mm, black!40!green] (-4,0) -- (2.5,1.75) node[midway,above,xshift=10mm,yshift=3mm] {\large $s$};
        \draw [line width=0.5mm, blue] (2.5,1.75) -- (2.5, 0.75) node[midway,right] {\large $z$};
        \draw [line width=0.5mm] (-4,0) -- (2.5, 0.75);
        \draw [line width=0.5mm, orange] (0,0) -- (2.5,1.75) node[midway,above right,yshift=1.5mm] {\large $r$};
        \draw [line width=0.5mm, red] (0,0) -- (2.5, 0.75) node[midway,below right,xshift=0.5cm,yshift=0.25cm] {\large $R$};

        \node {} (2.5, 0.75) coordinate (TZ);
        \filldraw [red] (2.5,1.75) circle (3pt) coordinate (T);
        \filldraw [red] (0,0) circle (3pt) node[below] {\large GC} coordinate (C);
        \filldraw [red] (-4,0) circle (3pt) node[below] {\large Sun} coordinate (S);

        \pic [draw, -> , line width=0.5mm, angle radius=3.5cm, orange] {angle=C--S--T} node[midway,xshift=-0.3cm,yshift=0.7cm, orange] {\large $\theta$};
        \pic [draw, ->, line width=0.5mm, angle radius=2.4cm, blue] {angle=TZ--S--T} node[midway,xshift=-1.3cm,yshift=0.55cm, blue] {\large $b$};
        \pic [draw, ->, line width=0.5mm, angle radius=2.9cm, red] {angle=C--S--TZ} node[midway,xshift=-0.8cm,yshift=0.2cm, red] {\large $\ell$};
        
    \end{tikzpicture}
    \caption{Illustration of the coordinate systems describing the position of a point in the MW.}
    \label{fig:coords}
\end{figure}
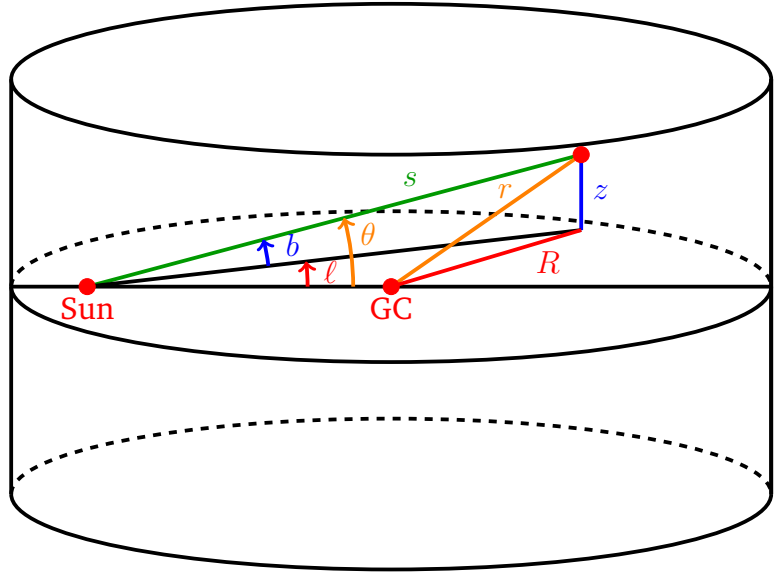

The differential flux of prompt photons and neutrinos from particle DM annihilation and decay in the Galaxy, from an infinitesimal solid angle $d\Omega$ centred on the line of sight~(l.o.s.), is given by
\begin{equation}
    \label{eq:fluxphDM}
    \frac{d\Phi_{\gamma,\nu}}{dE_{\gamma,\nu}d\Omega}=\frac{1}{4\pi}
    \left\{
    \begin{array}{ll}
    	\displaystyle  
        \frac{\xi\langle \sigma v\rangle}{m_\textrm{DM}^2} \frac{dN_{\gamma,\nu}^{\textrm{ann}}}{dE_{\gamma,\nu}} J(\theta) &\text{(annihilation)}\\
    	\\
    	\displaystyle  
        \frac{\Gamma}{m_\textrm{DM}} \ \; \frac{dN_{\gamma,\nu}^{\textrm{dec}}}{dE_{\gamma,\nu}} D(\theta) &\text{(decay)}\\
        \end{array}
    \right.\;,
\end{equation}
where $J(\theta)$ and $D(\theta)$ are the so-called \emph{$J$-factor} and \emph{$D$-factor}, defined as the integral of the DM profile along the l.o.s.
\begin{gather}
    J(\theta) = \int_\textrm{l.o.s.} ds\,\rho_\textrm{DM}^2(r(s,\theta))\;, \\
    D(\theta) = \int_\textrm{l.o.s.} ds\,\rho_\textrm{DM}(r(s,\theta))\;,
\end{gather}
where $s$ is the coordinate running along the l.o.s., linked to the Galactocentric distance $r$ and the angle $\theta$ between the l.o.s.\ and the axis between the GC and Earth by the relation $r = \sqrt{r_\odot^2+s^2-2sr_\odot\cos\theta}$. $\theta$ is also related to the Galactic latitude $b$ and longitude $\ell$ of the target by $\cos\theta = \cos b\cos\ell$. Figure~\ref{fig:coords} illustrates the relation between the different Galactic coordinate systems, which also includes the cylindrical one ($R$, $z$).

For PBHs we can write the differential flux of evaporated photons and neutrinos, again showing some resemblance with respect to the case of decaying DM particles by involving the $D$-factor.
\begin{equation}
    \label{eq:fluxphPBH}
    \frac{d\Phi_{\gamma,\nu}}{dE_{\gamma,\nu}d\Omega}=\frac{1}{4\pi}f_\textrm{PBH}D(\theta)\int \frac{dM}{M}\frac{dN_\textrm{PBH}}{dM}\frac{d^2N_{\gamma,\nu}}{dtdE_{\gamma,\nu}} \quad \textrm{(PBH evaporation)}\;.
\end{equation}

In order to obtain the flux of photons and neutrinos coming from a specific ROI, we then integrate the differential fluxes of Equations~\ref{eq:fluxphDM} and \ref{eq:fluxphPBH} over the relevant solid angle $\Delta\Omega$
\begin{equation}
    \label{eq:intROI}
    \frac{d\Phi_{\gamma,\nu}}{dE_{\gamma,\nu}} = \int_{\Delta\Omega}d\Omega\,\frac{d\Phi_{\gamma,\nu}}{dE_{\gamma,\nu}d\Omega}\;.
\end{equation}
This integration is described for different types of ROIs in Appendix~\ref{apx:trigo}.

\subsubsection{Secondary emissions of photons}

DM-produced $e^\pm$ can interact with the ISM and emit photons through three processes: ICS on ambient photons, synchrotron emission with the GMFs and bremsstrahlung with the ISM gas. In the following, we consider that DM-produced $e^\pm$ are relativistic, which is justified for $m_\textrm{DM} \gtrsim 5$ MeV\footnote{From Appendix~\ref{apx:elosses}, we can show that a $5$ MeV $e^\pm$ loses $\sim1.5$ MeV by crossing the whole GP horizontally.}, and therefore their kinetic energy constitutes their total energy ($K_e \approx E_e$). Essentially, the differential flux of secondary photons produced in these processes is given by
\begin{equation}
    \label{eq:secphflux}
    \frac{d\Phi_\gamma^\textrm{sec}}{dE_\gamma d\Omega} = \frac{1}{E_\gamma} \int_\textrm{l.o.s.} ds\, \frac{j_\textrm{sec}(E_\gamma, \vec{x})}{4\pi}\;,
\end{equation}
where $j_\textrm{sec}(E_\gamma, \vec{x})$ is the emissivity of secondary photons at a specific position $\vec{x}$ in the Galaxy from $\textrm{sec}=\{\textrm{ICS}, \textrm{syn}, \textrm{brems}\}$, that is expressed as a convolution of the density $f_e(E_e,\vec{x})$ of DM-produced $e^\pm$, solution to Equation~\ref{eq:transporteq}, with the radiating power $\mathcal{P}_\textrm{sec}(E_\gamma,E_e,\vec{x})$
\begin{equation}
    \label{eq:emiss}
    j_\textrm{sec}(E_\gamma, \vec{x}) = 2 \int_{m_e}^{m_\textrm{DM}(/2)}dE_e\,\mathcal{P}_\textrm{sec}(E_\gamma,E_e,\vec{x})f_e(E_e,\vec{x})\;,
\end{equation}
where the upper integration bound is the maximum energy of the DM-produced $e^\pm$: $m_\textrm{DM}$ for annihilating DM and $m_\textrm{DM}/2$ for decaying DM. The factor $2$ takes into account the sum the individual contribution of $e^+$ and $e^-$. In the following, we write the expression of the radiating power for the three secondary processes. Their derivation is cumbersome and out of the scope of this thesis, but the reader may find further information in the undermentioned references. 

The ICS radiating power, in the approximation where the up-scattered photon energy $E_\gamma$ is negligible compared to the initial DM-produced $e^\pm$ energy $E_e$ (also known as the \emph{Thomson limit}), is written~\cite{Cirelli:2009vg}
\begin{equation}
    \label{eq:PICS}
    \mathcal{P}_\textrm{ICS}(E_\gamma,E_e,\vec{x})=\frac{3\sigma_\textrm{T}}{4\gamma_e^2}E_\gamma \int_0^1 dy\,n_\gamma\left(E_\gamma^0(y),\vec{x}\right)\left(2\log y-2y+y^{-1}+1\right)\;,
\end{equation}
where $\sigma_\textrm{T} \equiv 8\pi r_e^2/3 = 6.65\times10^{-25}$ cm$^2$ is the Thomson scattering cross section (where $r_e = \alpha/m_e$ is the $e^\pm$ classical radius), $\gamma_e=E_e/m_e$ is the $e^\pm$ Lorentz factor, $n_\gamma(E_\gamma^0,\vec{x})$ is the number density of ambient photons with energy $E_\gamma^0$ at a position $\vec{x}$ in the Galaxy, and $y=E_\gamma/(4\gamma_e^2E_\gamma^0)$. Ambient photons are constituted of three different sources: optical and ultraviolet starlight (SL), infrared (IR) from the scattering of SL on Galactic dust, and the CMB. The energy range of these ambient photons can vary between $0.1$ meV to $10$ eV. The maximum energy at which these photons are up-scattered by a given $e^\pm$ is $E_\gamma^\textrm{max} = 4\gamma_e^2E_\gamma^0$ in the Thomson limit $E_e \gg E_\gamma^\textrm{max}$. Therefore it can be shown that this approximation is valid when $m_\textrm{DM} \sim E_e \lesssim 1$ TeV, in the case of ICS on Galactic ambient photons.

The radiating power of synchrotron emission for relativistic DM-produced $e^\pm$ depends on the strength of the GMF $B(\vec{x})$ at a given position $\vec{x}$, and is given by~\cite{1988ApJ...334L...5G}
\begin{equation}
    \mathcal{P}_\textrm{syn}(E_\gamma,E_e,\vec{x})=2\sqrt{3}\,\frac{e^3B}{m_ec^2}y^2\left[K_{4/3}(y)K_{1/3}(y)-\frac{3}{5}y\left(K_{4/3}^2(y)-K_{1/3}^2(y)\right)\right]\;,
\end{equation}
where $K_n$ is the modified Bessel function of the second kind of order $n$, $y = E_\gamma/E_\gamma^c$ and $E_\gamma^c$ is the critical photon energy at which $e^\pm$ emits the most through synchrotron (where $\mathcal{P}_\textrm{syn}$ reaches its maximum) $E_\gamma^c = 3eB\gamma_e^2/(2\pi m_ec)$.

Lastly, we write the radiating power of the bremsstrahlung of $e^\pm$ with the ISM gas~\cite{Buch:2015iya}
\begin{equation}
    \mathcal{P}_\textrm{brems}(E_\gamma,E_e,\vec{x})=cE_\gamma \sum_i n_i(\vec{x}) \frac{d\sigma_i}{dE_\gamma}(E_\gamma,E_e)\;,
\end{equation}
where $n_i$ is the ISM gas density of the species $i \in$ \{HI, HII, H$_2$, He\} at a position $\vec{x}$ in the MW, and $d\sigma_i/dE_\gamma$ are the differential bremsstrahlung cross sections expressed in Appendix~\ref{apx:elosses}.

Once the different radiating powers and the DM-produced $e^\pm$ density are known, we can compute the differential flux of secondary photons by plugging everything in Equations~\ref{eq:emiss} and then \ref{eq:secphflux}. Finally we can integrate this flux over a given ROI, in the same manner as in Equation~\ref{eq:intROI}.

One important thing we did not mention is the absorption of energetic $\gamma$-rays. Indeed, the ISM is opaque to $\gamma$-rays with energies higher than $\sim 10^{5}$ GeV, due to $e^\pm$ pair production and photon-photon scattering on ambient photons, but also pair production on baryonic matter. On a side note, these same effects are responsible for the absorption of energetic $\gamma$-rays in the extragalactic medium as well, leading to an attenuation of the extragalactic $\gamma$-ray flux, characterised by an optical depth that depends on the redshift of emission and the $\gamma$-ray energy. We refer the reader to~\cite{Cirelli:2010xx} for more details on this topic.

Now that we have an idea of what an indirect signal from DM might look like, we need to compare the prediction with observations from experiments. In the next section, we give an overview of the available experiments allowing us to detect a possible indirect signal from DM or, if not, help us set constraints on DM.

\section{Experiments}
\label{sec:exp}

A serious advantage of indirect searches is that they capitalise on the fact that DM could produce astrophysical-like signals, meaning we can use the plethora of already available experiments in order to conduct these searches in addition to dedicated experiments. In this section, we describe some of the experimental techniques used to detect charged CRs (Section~\ref{subsec:expCCRs}), photons (Section~\ref{subsec:expphotons}) and neutrinos (Section~\ref{subsec:expneutrinos}), as well as list some of the most known old, current and future experiments.

\subsection{Charged cosmic-rays}
\label{subsec:expCCRs}

\begin{figure}[t]
    \centering
    \begin{subfigure}[c]{0.535\linewidth}
        \centering
        \includegraphics[width=\linewidth]{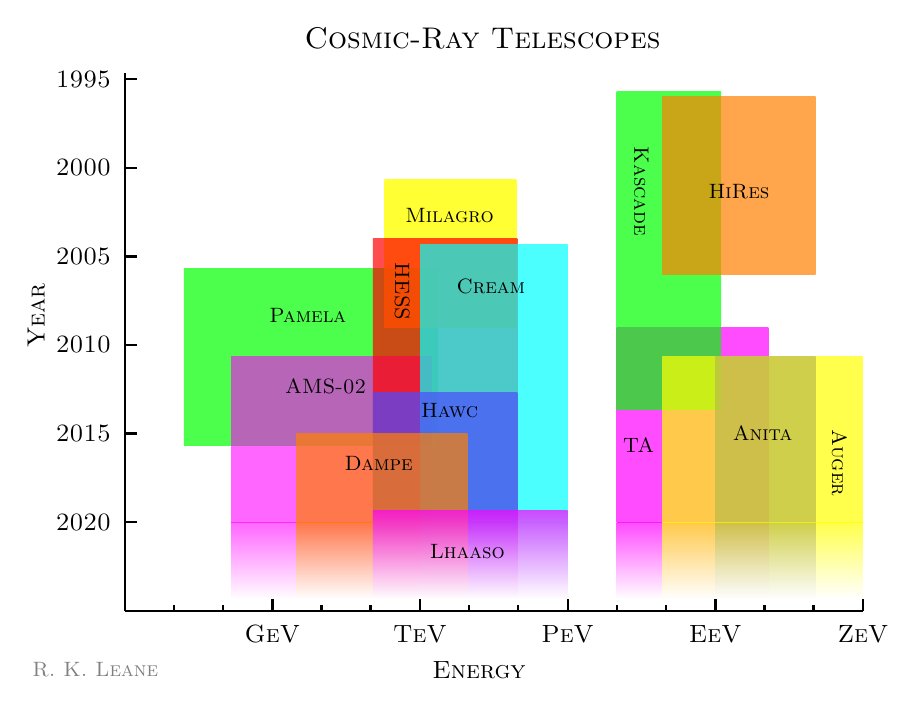}
    \end{subfigure}
    \hfill
    \begin{subfigure}[c]{0.455\linewidth}
        \centering
        \includegraphics[width=\linewidth]{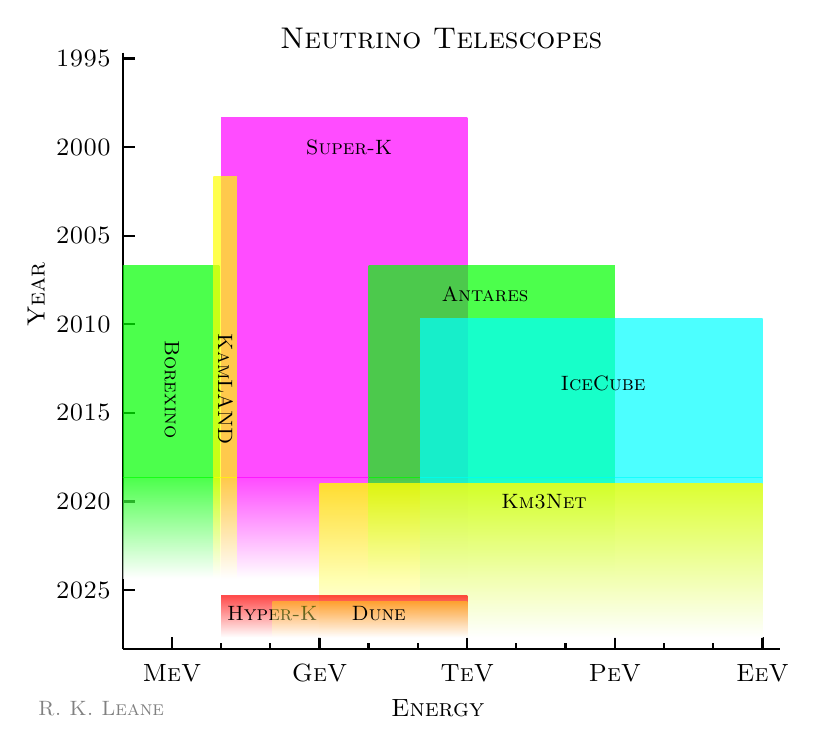}
    \end{subfigure}
    \caption{Representation of the energy range ($x$-axis) and operation dates ($y$-axis) of a selection of charged CRs (left panel) and neutrino (right panel) experiments. Taken from \cite{Leane:2020liq}.}
    \label{fig:expCRs}
\end{figure}

Charged CRs can be detected using three main techniques:
\begin{itemize}
	\item When a charged CRs enters Earth's atmosphere, their successive interactions with elements provoke a particle cascade called an \emph{air shower}. One technique consists of measuring the Cherenkov radiation emitted by shower particles when then travel faster than light in large tanks filled with water. Some of the experiments that use this \emph{water Cherenkov} technique are {\sc Hawc}, {\sc Lhaaso} or the Pierre {\sc Auger} Observatory, although some of them may also use other complementary detection techniques.
	\item Cherenkov radiation can also be emitted in the atmosphere during the particle cascade, and can be probed by so-called \emph{imaging atmospheric Cherenkov telescopes} such as {\sc Hess}, {\sc Magic} or {\sc Veritas}.
	\item Satellite and balloon-borne experiments are also an option. They can collect charged CRs directly and even be equipped with magnetic spectrometers to discriminate their charge and figure out their energy. Such experiments are {\sc Pamela}, {\sc Ams-02} or {\sc Dampe}.
\end{itemize}
We show the energy range and operation dates of a selection of charged CRs experiments in Figure~\ref{fig:expCRs}.

In 2008, the {\sc Pamela} experiment have reported an excess in the fraction $e^+/(e^++e^-)$ between $10$ GeV and $100$ GeV~\cite{PAMELA:2008gwm}, later confirmed by the {\sc Ams} experiment in 2014 and extended up to $300$ GeV~\cite{AMS:2013fma,AMS:2014bun}. This of course encouraged the community to provide a DM explanation of this excess and the most credible candidate was annihilations of leptophilic\footnote{DM has to be leptophilic to produce $e^\pm$ but also to avoid the production of $\bar{p}$, in order to match the absence of a $\bar{p}$ excess.} DM with a mass in the TeV range. However, it was difficult to satisfy both the excess in the $e^+$ fraction and the $e^+$ flux at the same time with a single candidate. Moreover, as the time passed, other ID searches involving measurements of the CMB~\cite{Slatyer:2015jla,Slatyer:2015kla} and the isotropic $\gamma$-ray background~\cite{Fermi-LAT:2015qzw} have then excluded this possibility. Other explanations involving a mismodeling of astrophysical sources seem to be the most probable ones~\cite{Serpico:2011wg}. 

Fortunately, exciting prospects are about to arrive, such as the launch of the {\sc Gaps} balloon in 2024 which will collect $\bar{d}$ and $\overline{\textrm{He}}$ in an effort to search for DM.

\subsection{Photons}
\label{subsec:expphotons}

\begin{figure}[t]
    	\centering
        	\includegraphics[width=\linewidth]{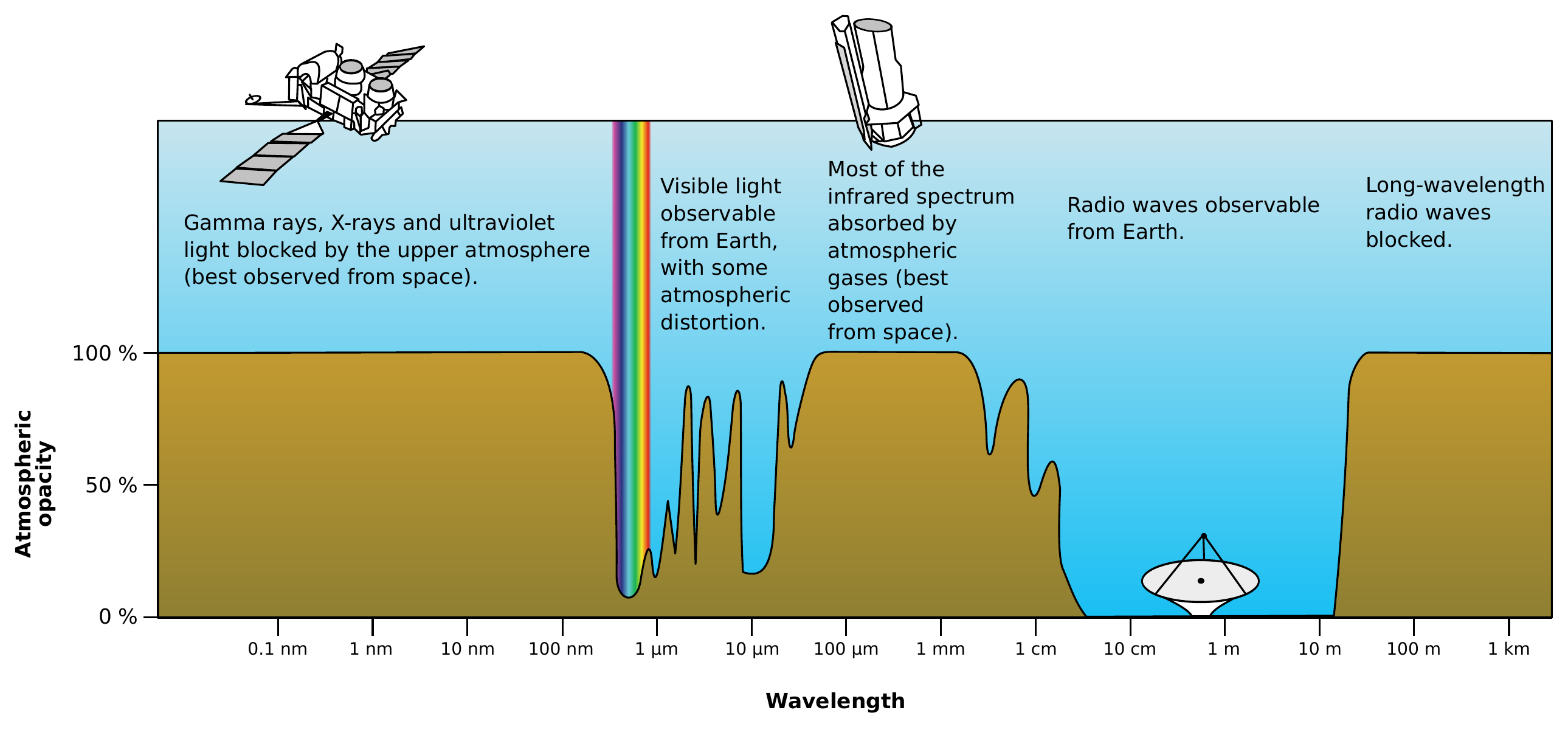}
	\caption{Illustration of the atmospheric opacity over the whole electromagnetic spectrum~\cite{AtmOpacity}.}
	\label{fig:AtmOpacity}
\end{figure}

Photons are probably the most investigated signals from space. Over the course of history, starting by using only our eyes, we have developed more and more precise instruments to have a better look at our sky. Nowadays we have a plethora of observatories that, in sum, can cover almost the whole electromagnetic spectrum. As illustrated by Figure~\ref{fig:AtmOpacity}, observatories use different experimental techniques, as the Earth's atmosphere blocks some parts of the electromagnetic spectrum. To probe these energy ranges, we use a satellite or balloon-borne experiments. Otherwise, ground experiments are more advantageous, since they can be as large as possible, without having the constraints of fitting in a spacecraft.

\subsubsection{Radio waves}

\begin{figure}[t]
    \centering
    \begin{subfigure}[c]{0.46\linewidth}
        \centering
        \includegraphics[width=\linewidth]{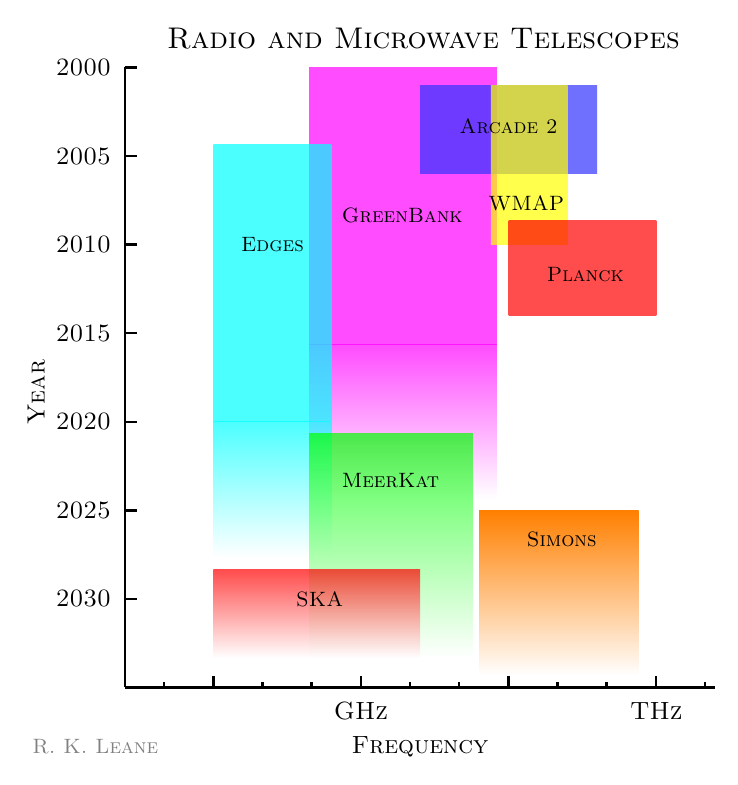}
    \end{subfigure}
    \hfill
    \begin{subfigure}[c]{0.53\linewidth}
        \centering
        \includegraphics[width=\linewidth]{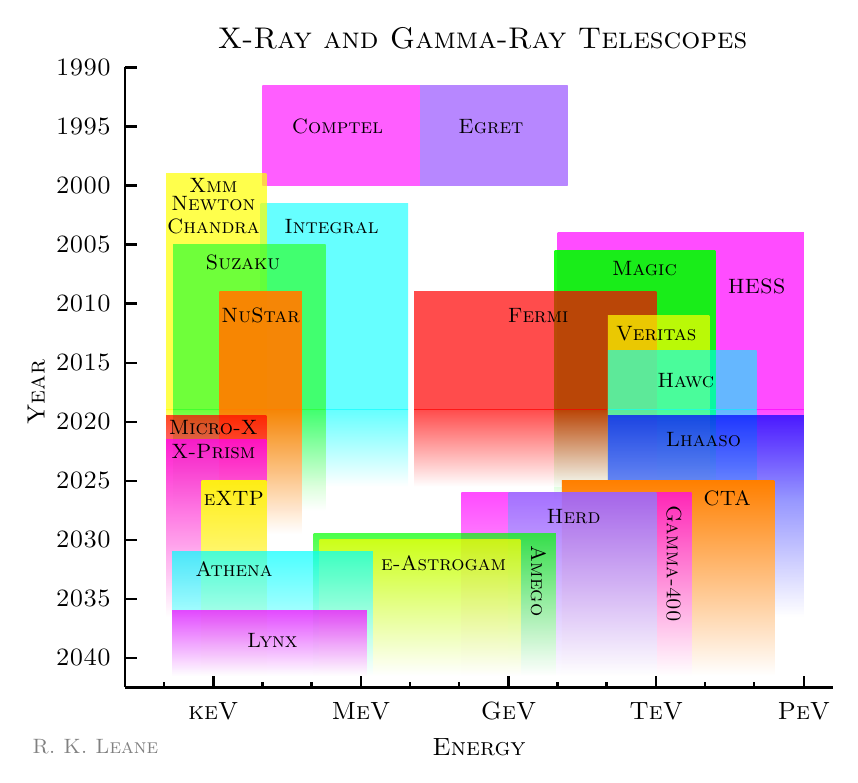}
    \end{subfigure}
    \caption{Representation of the energy range ($x$-axis) and operation dates ($y$-axis) of a selection of radio, microwave (left panel), $X$- and $\gamma$-ray (right panel) telescopes and observatories. Taken from \cite{Leane:2020liq}.}
    \label{fig:expphoton}
\end{figure}

The atmosphere is transparent to a large range of radio waves, allowing ground telescopes such as {\sc MeerKat} to probe them. Balloon-borne experiments, \emph{e.g.}\ {\sc Arcade-2}, are also an option, especially for wavelengths above $10$ m which are blocked by the atmosphere. We show the energy range and operation dates of a selection of radio (and microwave) experiments in the left panel of Figure~\ref{fig:expphoton}. 

Radio waves can originate from diverse astrophysical sources, the main ones being active galactic nuclei, the HI gas in the ISM emitting 21-cm lines, or synchrotron radiation. The latter is especially interesting for DM indirect searches, as DM-produced $e^\pm$ with energies around one GeV can produce synchrotron radio emissions due to their interaction with the GMFs. In particular, in 2009, the {\sc Arcade-2} have reported an excess 5 to 6 times above the prediction from astrophysical sources between $22$ MHz and $10$ GHz~\cite{Fixsen:2009xn} that could be explained by this scenario, for DM with a mass of $m_\textrm{DM} \simeq 10 - 25$ GeV annihilating into $\mu^+\mu^-$. Although this excess is now in tension with measurements from the {\sc Edges} telescope, the upcoming {\sc Ska} observatory will hopefully help us to figure out the origin of this excess.

\subsubsection{\texorpdfstring{$X$}{X}- and  \texorpdfstring{$\gamma$}{γ}-rays}

$X$- and soft $\gamma$-rays (with energies below $\simeq 50$ GeV) are blocked by our atmosphere, but more energetic $\gamma$-rays can produce an air shower, in the same manner as charged CRs. Therefore, the detection methods of hard $\gamma$-rays are similar to the ones for charged CRs: water/atmospheric Cherenkov detectors, satellite and balloon-borne experiments. To detect $X$- and soft $\gamma$-rays we exclusively use space based observatories. In the right panel of Figure~\ref{fig:expphoton} we show the energy range and operation dates of a selection of $X$- and $\gamma$-ray telescopes.

\begin{figure}[t]
    	\centering
        	\includegraphics[width=0.85\linewidth]{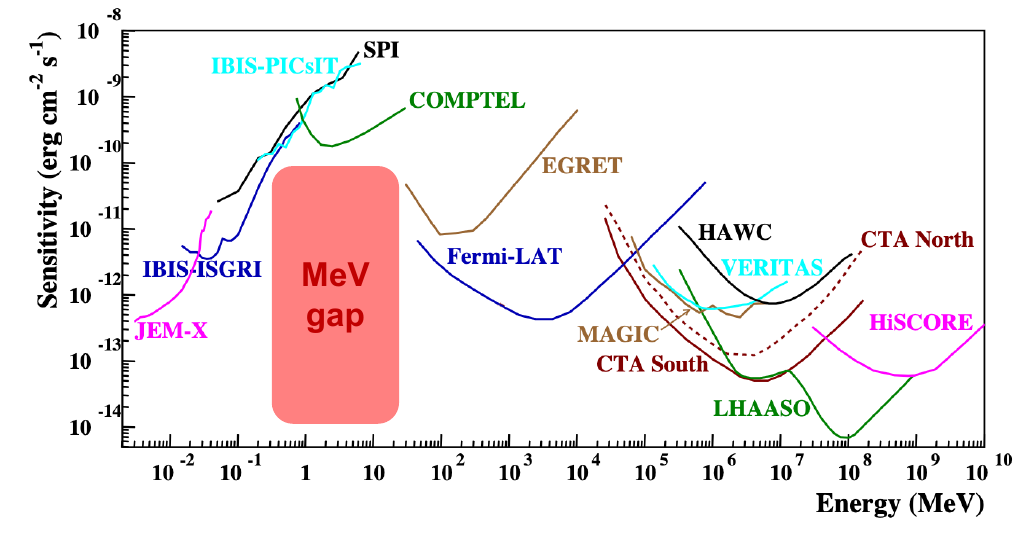}
	\caption{Sensitivities of a selection of $X$- and $\gamma$-ray observatories, illustrating the MeV gap. Adapted from~\cite{e-ASTROGAM:2016bph}.}
	\label{fig:MeVgap}
\end{figure}

Currently, space based observatories of $X$- and $\gamma$-rays work using three detection methods that are efficient at different energy ranges:
\begin{itemize}
	\item \textbf{Photoelectric effect:} Energetic enough photons can eject electrons in a material inside the detector, that in turn will get tracked and get its energy measured in a calorimeter. This process, useful for detecting photons up to $\sim 10$ MeV, is used in $X$-ray and soft $\gamma$-ray observatories such as {\sc NuStar}, {\sc Xmm-Newton} or {\sc Integral}.
	\item \textbf{Compton scattering:} By measuring the recoil of an electron from an incoming photon, and the energy of the scattered photon, we can deduce the total energy of the incident photon and its direction. This process was used to detect $\sim 1-30$ MeV photons in the {\sc Comptel} observatory.
	\item \textbf{Pair production:} Incoming photons produce a $e^+e^-$ pair that gets tracked in the detector and their energy measured in a calorimeter. This process is used in hard $\gamma$-ray experiments such as {\sc Fermi}, that can detect photons above $\sim 20$ MeV and up to $500$~GeV.
\end{itemize}
At the first sight, we would think that the full range of $X$- and harder $\gamma$-rays is equally covered thanks to the overlapping of the detection techniques. However the sensitivity of observatories using the photoelectric effect worsen at higher energies, and the {\sc Comptel} observatory, the only one using the Compton scattering technique, was built in the 1990s and therefore was not as sensitive as current observatories. This results in the so-called \emph{MeV gap} shown in Figure~\ref{fig:MeVgap}: there are no current observatories that are able to probe photons between $\sim 100$ keV and $100$ MeV with a satisfying enough sensitivity. This can be an issue when trying to probe indirect photon signals from sub-GeV DM, as we will discuss in Chapter~\ref{chap:subGeV}. Fortunately, proposed experiments such as {\sc e-Astrogam} and {\sc Amego} plan to fill this gap in the next decade.

Three main excesses have been reported in the $X$- and $\gamma$-ray energy range, the first one being the $\gamma$-ray excess around the GC we discussed thoroughly in Section~\ref{subsec:MW}. {\sc Integral} has confirmed another one but at lower energies and in the form of a spectral line at $511$~keV~\cite{Knodlseder:2005yq}. This line peaks exactly at the value of the electron mass, indicating us that this very likely originates from annihilations of an excess of $e^+e^-$ in the GC. At the moment, a DM interpretation is possible, the first proposition being DM lighter than $3$ MeV annihilating in $e^+e^-$~\cite{Beacom:2005qv} which is still not excluded by other probes to this day. Another line at $3.5$ keV have been initially reported in galaxy clusters surveys from {\sc Xmm-Newton} and {\sc Chandra}~\cite{Bulbul:2014sua,Boyarsky:2014jta}. However, this line is still questioned by the community. For example, the short-lived {\sc Hitomi} telescope\footnote{Launched on February 1st 2016, the mission has ended about a month later, on March 26th, due to multiple incidents that have led to the destruction of the telescope.} did not see the line, and recent reanalyses of the initial galaxy clusters surveys lead to actual disappearance of the line~\cite{Dessert:2023fen}. Hopefully, the {\sc Xrism} telescope, the recently launched successor of {\sc Hitomi}, might provide us with a definitive answer.

\subsection{Neutrinos}
\label{subsec:expneutrinos}

Finally, neutrinos can be detected in large underground water or ice Cherenkov detectors. When neutrinos interact with the material in the detector or its surroundings, secondary particles are produced (most of the time muons) which can emit Cherenkov radiations when they go through the material, and be detected. The energy and direction of incident neutrinos can then be reconstructed. We show the energy range and operation dates of a selection of neutrino experiments in the right panel of Figure~\ref{fig:expCRs}. For now, no neutrino excess compared to the astrophysical background has been observed, and constraints on DM-neutrino interactions can be set. For instance, {\sc IceCube} has observed a $290$ TeV neutrino event that coincides with a $\gamma$-ray burst from the same direction, caused by a flaring blazar~\cite{IceCube:2018dnn}. The DM-neutrino scattering cross section can be then constrained by the fact that this neutrino has crossed a large amount of DM: the possible DM spike around the BH at the blazar's centre, as well as the extragalactic and Galactic DM~\cite{Cline:2022qld}.

%% file: Chapters/chap3.tex

\lettrine[lines=3, nindent=1pt]{T}{he} possibility that DM consists of particles lighter than a few GeV has received significant attention recently, in part due to the lack of convincing signals from the paradigmatic WIMPs in current experiments~\cite{Schumann:2019eaa,Cirelli:2015gux,Gaskins:2016cha,Hooper:2018kfv,Buchmueller:2017qhf,Kahlhoefer:2017dnp}. Sub-GeV DM, as we call it in Section~\ref{subsec:candidates}, appears as a solid, theoretically well motivated alternative~\cite{Knapen:2017xzo,Boehm:2002yz,Boehm:2003hm,Fayet:2007ua,Boehm:2003bt,Ahn:2005ck,Boehm:2006mi,Ema:2020fit,Hochberg:2014dra,Boddy:2014yra,Hochberg:2014kqa,Choi:2017zww,Berlin:2018tvf,DAgnolo:2015ujb,Falkowski:2011xh,Lin:2011gj,Hooper:2007tu,Bertuzzo:2017lwt,Darme:2017glc,Katz:2020ywn}. However, detection of this particular DM candidate is more challenging than WIMPs~\cite{Cirelli:2020bpc}. In DM ID, recalling a discussion in Section~\ref{subsec:expphotons}, the two main obstacles that prevent us to have high-sensitivity data in the interval where the signals from sub-GeV DM particle annihilation or decay are expected are i) the solar activity, which screens DM-produced $e^\pm$ and ii) the MeV gap in $X$- and soft $\gamma$-ray experiments. A novel technique introduced in~\cite{Cirelli:2020bpc} allows to circumvent both of these problems. The idea is to focus on secondary emissions from DM, and in particular on the ICS of DM-produced $e^\pm$ on Galactic ambient light, which produces hard $X$-rays with typical keV energies.  As a result, one can leverage on the abundant data in $X$-ray keV observations, rather than the scarce MeV experiments, in order to test sub-GeV DM. In \cite{Cirelli:2020bpc} it was shown that the method is powerful: using data from a large region of the inner Galaxy observed by the {\sc Integral/Spi} spectrometer, the authors were able to obtain stringent constraints on annihilating DM, \emph{i.e.}\ on $\langle\sigma v\rangle$, in the $m_\textrm{DM}$ range between $1$ MeV and $5$ GeV, even with a conservative propagation setup for DM-produced $e^\pm$. This chapter follows up on that work, embarking in a systematic analysis of the available $X$-rays datasets in order to assess their full constraining power on sub-GeV DM, along the lines of the strategy described above. In addition, we consider both the case of annihilating and decaying DM.  We start by recalling the formalism and the relevant quantities necessary for computing prompt and secondary $X$-ray emissions from sub-GeV DM annihilations and decays in Section~\ref{sec:XrayDM}, in Section~\ref{sec:data} we detail the datasets that we use and our analysis, in Section~\ref{sec:results1} the main results, and in Section~\ref{sec:comparison} we compare them to related studies.

\newpage

\section{\texorpdfstring{$X$}{X}-rays from annihilating and decaying sub-GeV dark matter}
\label{sec:XrayDM}

We start by introducing the formalism for $X$-ray production from DM annihilations and decays. Since we focus on sub-GeV DM, only a few annihilation or decay channels in SM final states are kinematically open. Among them, we consider the three following ones~\footnote{DM annihilations and decays into other final states such as $\pi^0\pi^0$, $\gamma\gamma$ and $\nu\bar{\nu}$ can also be kinematically open, however they do not produce $X$-rays.}
\begin{align}
	{\rm DM \, (DM)} &\longrightarrow e^+e^-,  \label{eq:ee}\\
	{\rm DM \, (DM)} &\longrightarrow \mu^+\mu^-, \label{eq:mumu}\\
	{\rm DM \, (DM)} &\longrightarrow \pi^+\pi^-, \label{eq:pipi}
\end{align}
whenever $\sqrt{s} > m_i$, with $i=\{e,\mu,\pi\}$ and where $\sqrt{s} = m_\textrm{DM}$ ($m_\textrm{DM}/2$) for annihilating (decaying) DM. We study each channel independently although, recalling the discussion at the beginning of Section~\ref{subsec:particleprod}, DM could annihilate or decay in a combination of final states, in this case also including other light hadronic or mesonic resonances. A more thorough model-dependent study can be done by computing photon energy spectra using numerical codes that are valid for the considered DM mass range, \emph{e.g.}\ \verb|Hazma|, and applying them to our study.

Given a fixed channel, the total flux of photons is given by the sum of prompt and secondary emissions.  The prompt emission of photons consists of final state radiation (FSR) during the DM annihilation or decay, and of radiative decays (Rad) which occur whenever $\mu^\pm$ or $\pi^\pm$ undergo a decay with an extra photon ($\mu\to e\nu_e\nu_\mu\gamma$ and $\pi\to l\nu_l\gamma$ with $l=e,\mu$). On the other hand, secondary emissions originate from ICS of energetic DM-produced $e^\pm$ on ambient Galactic light. In Section~\ref{subsec:gammaneutrinos} we mentioned that synchrotron radiation and bremsstrahlung are also a source of secondary emissions. However, i) in the case of sub-GeV DM annihilations and decays, synchrotron radiations do not land on the energy range of $X$- and soft $\gamma$-rays and ii) in most of the datasets we use in the study, the GP, \emph{i.e.}\ the region where the ISM gas density is high and therefore bremsstrahlung is more important, was masked and therefore we do not include them in our prediction.

\subsection{Prompt emissions}
\label{subsec:prompt}

The spectrum of FSR photons is written, in the case of DM annihilating or decaying in leptons~\cite{Bystritskiy:2005ib}
\begin{equation}
	\label{eq:FSRll}
	\frac{dN_{\textrm{FSR} \gamma}^{l^+l^-}}{dE_\gamma} = \frac{\alpha}{\pi\beta(3-\beta^2)\,\sqrt{s}} 
	\left[\mathcal{A}(\nu) \, \ln \frac{1+R(\nu)}{1-R(\nu)}-2\, \mathcal{B}(\nu)\,  R(\nu)
	\right] \,,
\end{equation}
where $\alpha \simeq 1/137$ is the fine-structure constant, and
\begin{equation}
	\mathcal{A}(\nu) = \frac{\left(1+\beta^{2}\right)\left(3-\beta^{2}\right)}{\nu}-2\left(3-\beta^{2}\right)+2 \nu\,,
\end{equation}
\begin{equation}
	\mathcal{B}(\nu) = \frac{3-\beta^{2}}{\nu}(1-\nu)+\nu\,,
\end{equation}
where $\nu = E_\gamma/\sqrt{s}$,  $\beta^2 = 1-4\mu^2$ with $\mu=m_l/(2\sqrt{s})$ and $R(\nu) = \sqrt{1-4\mu^2/(1-\nu)}$.  And for DM annihilation and decay in $\pi^+\pi^-$ we have~\cite{Bystritskiy:2005ib}
\begin{equation}
	\label{eq:FSRpipi}
	\frac{dN_{\textrm{FSR} \gamma}^{\pi^+\pi^-}}{dE_\gamma} = 
	\frac{2 \alpha}{\pi\beta\sqrt{s}}\left[\left(\frac{\nu}{\beta^{2}}-\frac{1-\nu}{\nu}\right) R(\nu)+\left(\frac{1+\beta^{2}}{2 \nu}-1\right) \ln \frac{1+R(\nu)}{1-R(\nu)}\right] \,,
\end{equation}
with the same definitions as above with $m_l \rightarrow m_\pi$.

The spectrum of photons emitted from the radiative decay of $\mu^\pm$ in their rest frame is written~\cite{Kuno:1999jp}
\begin{equation}
	\left.\frac{dN_{\textrm{Rad}\gamma}^{\mu}}{dE_\gamma} \right|_{E_\mu = m_\mu} = \frac{\alpha(1-x)}{36 \pi E_\gamma}\left[12\left(3-2 x(1-x)^{2}\right) \log \left(\frac{1-x}{r}\right) +x(1-x)(46-55 x)-102 \right]\,,
	\label{eq:muon_rad}
\end{equation}
where $x = 2E_\gamma/m_\mu$, $r = (m_e/m_\mu)^2$ and the maximal photon energy is $E^\textrm{max}_\gamma = m_\mu (1-r)/2 \simeq 52.8$ MeV. For muons in flight, this spectrum is boosted to the frame where the muon has energy $E_\mu = \sqrt{s}$. This is done by doing the following computation
\begin{equation}
	\label{eq:boost}
	\frac{dN}{dE} = \frac{1}{2\beta\gamma} \; \int_{E'_-}^{E'_+} \frac{1}{p'}\;\frac{dN}{dE'}\;,
\end{equation}
where $\gamma = 1/\sqrt{1-\beta^2} = E_A/m_A$ is the Lorentz boost factor, $dN/dE'$ corresponds to the spectra of photons at the rest frame of the parent particle and $E'_\pm = \gamma (E\pm \beta\,p)$.  $E_A$ and $m_A$ refer respectively to the energy and mass of the parent particle $A$, therefore in the case of radiative decays of DM-produced $\mu^\pm$, we have $\gamma = \sqrt{s}/m_\mu$. Finally, a multiplicity factor of $2$ needs to be applied in Equation~\ref{eq:muon_rad}, since a pair of muons is produced for each DM annihilation or decay. 

On the other hand, the spectrum of photons emitted from the radiative decay of $\pi^\pm$ is written~\cite{Bryman:1982et}
\begin{equation}
	\left.\frac{dN_{\textrm{Rad}\gamma}^{\pi}}{dE_\gamma}\right|_{E_\pi = m_\pi} = \frac{\alpha\,(f(x)+g(x))}{24\, \pi \,m_{\pi} \,f_{\pi}^{2}\,(r-1)^{2}\,(x-1)^{2} \,r \,x}\;,
\end{equation}
where $x = 2E_\gamma/m_\pi$, $r = (m_\ell/m_\pi)^2$, $f_\pi = 92.2$ MeV is the pion decay constant and
\begin{eqnarray}
	f(x)  &=& (r+x-1)\left[m_{\pi}^{2}\, x^{4}\left(F_{A}^{2}+F_{V}^{2}\right)\left(r^{2}-r x+r-2(x-1)^{2}\right)\right. \nonumber\\
	&& -12 \,\sqrt{2} \,f_{\pi}\, m_{\pi} \,r(x-1) x^{2}\left(F_{A}(r-2 x+1)+x F_{V}\right) \nonumber\\
	&& \left.-24 \,f_{\pi}^{2}\, r(x-1)\left(4 r(x-1)+(x-2)^{2}\right)\right] \,,\\
	g(x) &=& 12\, \sqrt{2} \, f_\pi \, r(x-1)^2 \log\left(\frac{r}{1-x}\right) [m_{\pi}\, x^{2} (F_{A}\, (x-2r) - x \, F_{V}) \nonumber \\
	&&+\sqrt{2}\, f_{\pi} (2 r^{2}-2 r x-x^{2}+2 x-2)] \,,\nonumber
\end{eqnarray}
where $F_{A}=0.0119$ and $F_{V}\left(q^{2}\right)=F_{V}(0)\left(1+a q^{2}\right)$ (with $F_{V}(0)=0.0254$, $a=0.10$, $q^2 = (1-x)$) are respectively the axial and vectorial form factors~\cite{Coogan:2019qpu, ParticleDataGroup:2024cfk}.

When $\pi^\pm$ decays into $\mu^\pm$, the radiative decay of the latter is again a source of low-energy photons. The total radiative $\pi^\pm$ spectrum in their rest frame can be expressed by the following sum
\begin{equation}
	\left.\frac{dN_{\textrm{RadTot}\gamma}^\pi}{dE_\gamma}\right|_{E_\pi = m_\pi} =
	\sum_{\ell=e, \mu} \textrm{BR}\left(\pi \rightarrow \ell \nu_{\ell}\right) \,
	\left.\frac{dN_{\textrm{Rad}\gamma}^{\pi}}{dE_\gamma}\right|_{E_\pi = m_\pi}
	+\textrm{BR}(\pi \rightarrow \mu \nu_{\mu}) 
	\left.\frac{dN_{\textrm{Rad}\gamma}^{\mu}}{dE_\gamma}\right|_{E_\mu = E_\star}\;,
	\label{eq:pion_radtot}
\end{equation}
where $E_\star=(m_\pi^{2}+m_{\mu}^{2})/(2 m_\pi)$ is the muon energy in the pion rest frame.
For pions in flight, the spectrum above is boosted to the frame where $E_\pi = \sqrt{s}$ using Equation~\ref{eq:boost} with $\gamma = \sqrt{s}/m_\pi$. A multiplicity factor of $2$ is again applied, since each DM annihilation or decay produces a pair of charged pions.

The differential flux of the prompt emissions $d\Phi_{\textrm{prompt}\gamma}/dE_\gamma d\Omega$ is then computed using Equation~\ref{eq:fluxphDM}, by integrating the emissions along the l.o.s.\ in a given direction $\theta$ defined in Figure~\ref{fig:coords}.

\subsection{Secondary emissions}
\label{subsec:secondary}

To compute the differential flux of secondary emissions from ICS, we use the procedure described in Section~\ref{subsec:gammaneutrinos}. In particular, i) we compute the injection spectrum of DM-produced $e^\pm$ to insert it in the source term of Equation~\ref{eq:transporteq}, described in Equation~\ref{eq:source}, ii) we solve Equation~\ref{eq:transporteq} for the density of DM-produced $e^\pm$ in a specific propagation setup, iii) we evaluate the ambient photon density from the CMB, SL and IR components, iv) we insert everything in Equations~\ref{eq:PICS}, \ref{eq:emiss} and \ref{eq:secphflux}. 

For the first step, we need to compute the $e^\pm$ spectrum for each DM annihilation or decay channel. For $e^+e^-$ final states, the spectrum is simply a monochromatic line ($dN_e/dE_e = \delta(E_e - \sqrt{s})$). For $\mu^+\mu^-$, their decay into $e^\pm$ follows the Michel spectrum~\cite{Michel:1950}, expressed in the rest frame of the muon as follows
\begin{equation}
	\frac{dN^{\mu\rightarrow e\nu\bar{\nu}}_e}{d E_{e}}=\frac{4 \sqrt{\xi^{2}-4 \varrho^{2}}}{m_{\mu}}\left[\xi(3-2 \xi) + \varrho^{2}(3 \xi-4)\right]\;,
	\label{eq:mudecay}
\end{equation}
where $\varrho=m_e/m_\mu$, $\xi = 2E_e/m_\mu$. For DM annihilating or decaying into $\mu^+\mu^-$, this spectrum has to be boosted to the DM rest frame using Equation~\ref{eq:boost}. And for $\pi^+\pi^-$, we need to boost the Michel spectrum twice in order to follow the decay chain $\pi \rightarrow \mu \rightarrow e$. Here we consider that the direct decay of pions into $e\nu_e$ has a negligible branching ratio (BR = $1.23\times10^{-4}$) compared to the one into $\mu\nu_\mu$ (BR = $0.999877$) and therefore is not included in the prediction. For the DM density profile, we choose the NFW one, defined in Equation~\ref{eq:NFW}, as our fiducial case, although in Section~\ref{sec:results1} we discuss the impact on our results of choosing different DM profiles.

For the second step, we decide to adopt a minimalistic propagation model, where only the energy losses are relevant. This is partially justified by the SLIM propagation model~\cite{Genolini:2021doh}, which nullifies convection and momentum space diffusion and still can provide good fits to CR data. Only spatial diffusion and energy losses remain, and we here decide to neglect the first over the second, to be as conservative as possible. However, we still consider that DM-produced $e^\pm$ are confined in the Galaxy by the GMFs within the boundary $(R_\textrm{max}, L) = (20, 4)$~kpc. This prescription is applied when integrating the ICS emissivity along the l.o.s.\ as in Equation~\ref{eq:secphflux}, in order to obtain the differential flux. The maximal value of the l.o.s.\ coordinate $s$ in this setup is computed in Appendix~\ref{apx:trigo}. In Chapter~\ref{chap:prop}, we discuss the impact of setting a realistic propagation setup on the results derived in this chapter. With the minimalistic propagation setup we adopt in this chapter, Equation~\ref{eq:transporteq} can be solved analytically and give
\begin{equation}
	f_e(E_e,\vec{x}) = \frac1{b_\textrm{tot}(E_e,\vec{x})}\int_{E_e}^{\sqrt{s}}dE'_e\ Q_e(E'_e, \vec{x})\;,
\end{equation}
where $b_{\rm tot}(E,\vec x) \equiv -\dot{E} = b_{\rm Coul+ion} + b_{\rm brem} +  b_{\rm syn} + b_{\rm ICS}$ is the previously defined energy loss function, which takes into account all the energy loss processes that the $e^\pm$ suffer in the local Galactic environment in which they are injected. Its expression is detailed in Appendix~\ref{apx:elosses}.

\begin{figure}[t]
	\centering
	\begin{subfigure}[c]{0.49\linewidth}
		\centering
		\includegraphics[width=\linewidth]{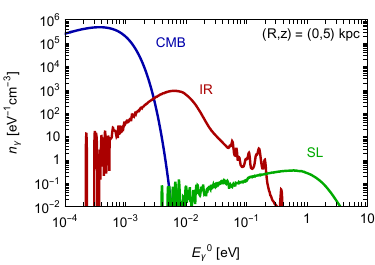}
	\end{subfigure}
	\hfill
	\begin{subfigure}[c]{0.49\linewidth}
		\centering
		\includegraphics[width=\linewidth]{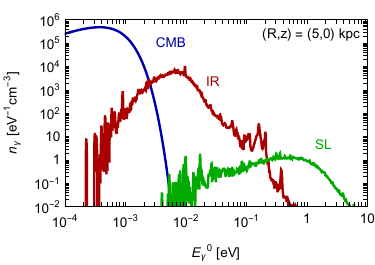}
	\end{subfigure}
	\caption{Spectra of the photon number density per unit energy of ambient photons above the GC (left panel) and in the GP (right panel), and separating the different components: SL (green), IR (red) and CMB (blue).}
	\label{fig:isrf}
\end{figure}

For the third step, the IR and SL components of the ambient photon bath are computed using maps extracted from \verb|GALPROP|~\cite{2011CoPhC.182.1156V}, in turn based on observations from the {\sc Cobe/Dirbe} telescope. The CMB component is computed analytically by assuming it to be an isotropic black body spectrum with a temperature $T=2.73$ K
\begin{equation}
	n_\textrm{CMB}(E_\gamma^0) = \frac{{E_\gamma^0}^2}{\pi^2\hbar^3c^3}\frac{1}{e^{E_\gamma^0/kT}-1}\;.
\end{equation}
Figure~\ref{fig:isrf} shows the spectrum of the number density per unit energy of all three components of the ambient photon bath at two positions in the MW.

\begin{figure}[p]
	\centering
	\begin{subfigure}[c]{0.49\linewidth}
		\centering
		\includegraphics[width=\linewidth]{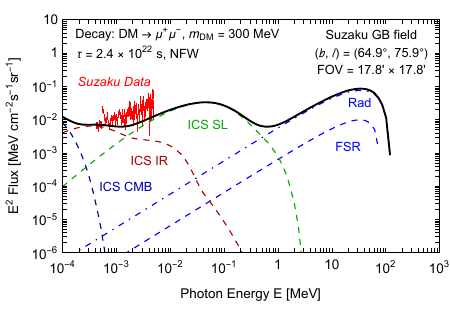}
	\end{subfigure}
	\hfill
	\begin{subfigure}[c]{0.49\linewidth}
		\centering
		\includegraphics[width=\linewidth]{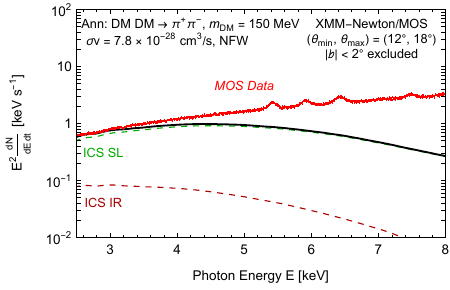}
	\end{subfigure}
	\hfill
	\begin{subfigure}[c]{0.49\linewidth}
		\centering
		\includegraphics[width=\linewidth]{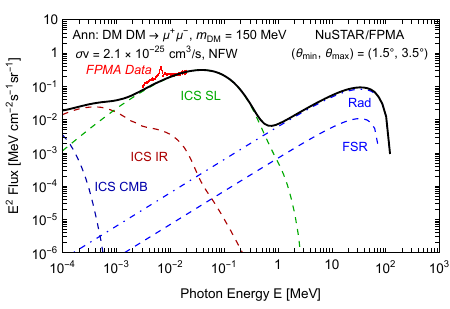}
	\end{subfigure}
	\hfill
	\begin{subfigure}[c]{0.49\linewidth}
		\centering
		\includegraphics[width=\linewidth]{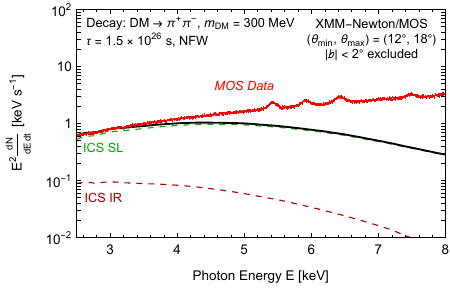}
	\end{subfigure}
	\hfill
	\begin{subfigure}[c]{0.49\linewidth}
		\centering
		\includegraphics[width=\linewidth]{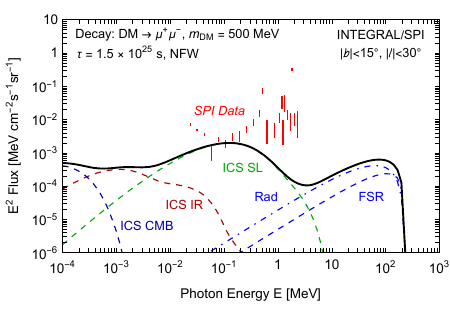}
	\end{subfigure}
	\hfill
	\begin{subfigure}[c]{0.49\linewidth}
		\centering
		\includegraphics[width=\linewidth]{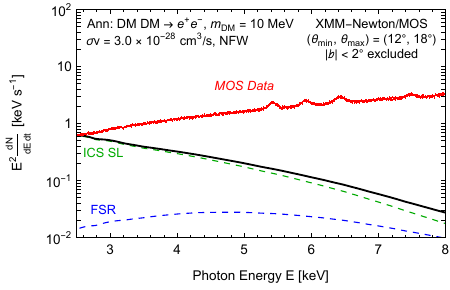}
	\end{subfigure}
	\caption{Illustration of some fluxes of hard $X$-rays from DM annihilation or decay, compared to the different datasets adopted in our analysis. In each panel we indicate the DM specifications (annihilation or decay channel, mass, annihilation cross section or decay rate, NFW profile) and the characteristics of the considered region of observation.}
	\label{fig:Xrayfluxes}
\end{figure}

Now we have all of the ingredients we need to make some flux predictions. The full photon flux from sub-GeV DM annihilation and decays is then obtained by integrating the sum of the computed prompt and secondary differential flux of photons over ROIs, following Equation~\ref{eq:intROI}. Figure~\ref{fig:Xrayfluxes} illustrates a few examples of the total flux, compared to the datasets that we considered in our analysis. Such datasets are discussed in the following section. 

\section{Datasets and analysis}
\label{sec:data}

\begin{figure}[t]
	\centering
	\includegraphics[width=0.80\linewidth]{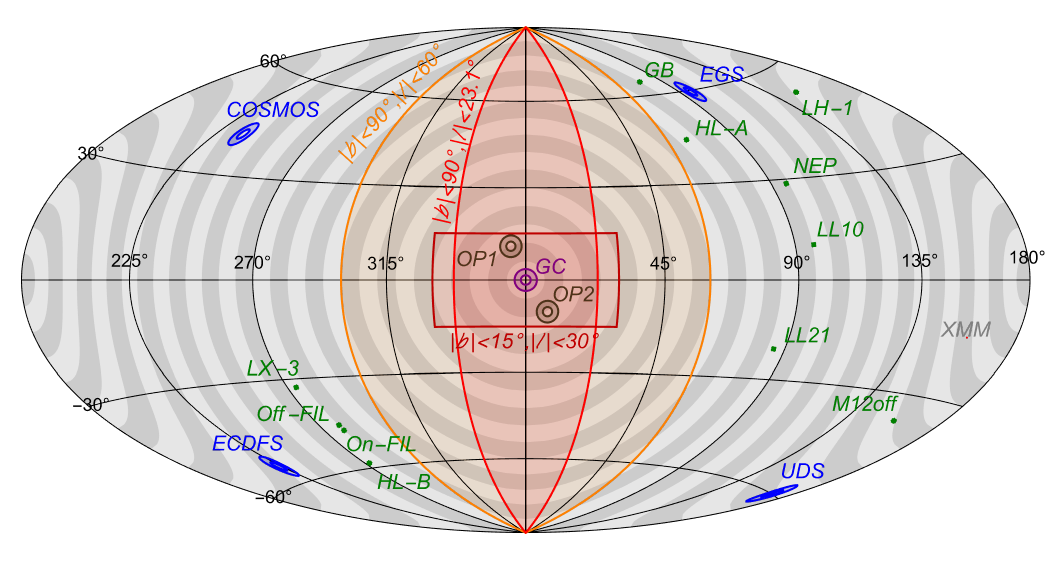}
	\caption{Chart of the Galaxy in Galactic coordinates $(b,\ell)$ with the location of the datasets we use. The three regions of observation relevant for the {\sc Integral} datasets are represented in orange and red. The four {\sc NuStar} blank-sky fields (COSMOS, EGS, ECDFS and UDS) are in blue, the {\sc NuStar} GC and off-plane (OP1 \& OP2) are in purple and dark brown, respectively. {\sc Xmm-Newton} rings are drawn in shades of grey and the eleven {\sc Suzaku} fields in green. All of the fields are to scale.} 
	\label{fig:galaxymap}
\end{figure}

In this study, we focus on the $X$-ray emission of the MW and exploit the datasets listed below. The locations of the respective ROIs on the Galactic sky are depicted for illustration in Figure~\ref{fig:galaxymap}.

\begin{itemize}

\item \underline{{\sc Integral}.} The data are reported in~\cite{Bouchet:2011fn}, which follows previous work in~\cite{Bouchet:2008rp,Bouchet:2005ys}. These datasets were previously used in~\cite{Cirelli:2020bpc}. 
The data were collected by the {\sc Spi} $X$-ray spectrometer onboard the {\sc Integral}  satellite, in the period 2003$-$2009, corresponding to a significant total exposure of about $100$ Ms, and cover a range in energy between $20$ keV and a few MeV. They are provided either in the form of a spectrum of the total diffuse flux in a rectangular region of observation centered around the GC ($|b|<15^\circ$, $|\ell|<30^\circ$, Figures 6 and 7 in~\cite{Bouchet:2011fn}) or in the form of an angular flux in latitude and longitude bins, in 5 energy bands ($27-49$ keV, $49-90$ keV, $100-200$ keV, $200-600$ keV and $600-1800$ keV) (Figures 4 and 5 in~\cite{Bouchet:2011fn}). 
As in~\cite{Cirelli:2020bpc}, we use the angular flux in latitude bins only, from which we cut out the GP. The longitude window is $|\ell|<23.1^\circ$ for the first four energy bands and $|\ell|<60^\circ$ for the fifth one.

\item \underline{{\sc NuStar} blank-sky fields.} These data are presented in~\cite{Krivonos:2020qvl}, which aims at measuring the cosmic $X$-ray background in the $3-20$ keV energy band. 
The data are collected from the {\sc NuStar} extragalactic survey program, which includes a number of fields with different sky coverage and exposure times, among which there are the COSMOS, EGS, ECDFS, UDS that we use. 
These are the same fields used in~\cite{Roach:2022lgo}, although in another context (namely, to probe sterile neutrino DM). 
The actual areas of observation have a complex shape: they consist of two partly overlapping `Pac-Man$^{\scriptscriptstyle \rm TM}$-like' regions located around the nominal pointing center of the field, with uneven coverage (see \emph{e.g.}, Figure 4 in~\cite{Perez:2016tcq}). We choose to approximate each of them as a square annulus of inner size $1.5^\circ$ and outer size $3.5^\circ$. This approximation is justified by the fact that the DM emissivity in those relatively small regions varies little, thus we can adopt a simpler geometrical area.
The nominal exposure is of about $7$ Ms.

\item \underline{{\sc NuStar} GC region.} The data are provided in~\cite{Mori:2015vba,Hong:2016qjq} and are the same than the ones used in~\cite{Perez:2016tcq} in another context (namely, to probe sterile neutrino DM). 
The shape of the areas of observation is the same as in the previous item: we just model it here as an annulus of inner radius $1.5^\circ$ and outer radius $3.5^\circ$.
We use the data provided in Figure 5 of~\cite{Perez:2016tcq}, restricting at $E_\gamma \le 20$ keV because the instrumental background becomes dominant for higher energies\footnote{We should note, however, that these spectra (even for $E_\gamma < 20$ keV) include a small contribution from internal detector background, which we do not model nor subtract. This implies that our bounds are derived from a nominal flux which is slightly larger that the true astrophysical emission: thus, the derived DM limits are conservative compared to the approach where the full background is modeled.}. Since this emission originated from regions close to the GC, it is subject to attenuation upon the dense ISM. However, using a HI column density of $10^{22}$ cm$^{-2}$~\cite{HI4PI} and the cross sections tabulated in~\cite{Wilms:2000ez}, we find that such attenuation is at most $\sim10\%$ at $E_\gamma = 3$ keV and quickly diminishes at higher energies, hence it is negligible for our purposes.  

\item \underline{{\sc NuStar} off-plane faint-sky observations.} The data are presented and used in \cite{Roach:2019ctw}. They correspond to the observation of two annuli, with shapes equivalent to those described for the previous datasets (which we model as in the `blank-sky' case), located about $10^\circ$ above and below the GP. The total exposure time amounts to about $100$ ks. 
The emission in these regions is understood to be essentially cosmic $X$-ray background only, since the Galactic component is estimated to be negligible. In particular, the Galactic ridge emission (GRXE)\footnote{The GRXE mostly comes from accreting compact objects, mainly white dwarfs. More specifically, it is believed to be produced in the accretion streams of magnetic cataclysmic variable stars, plus a $6.4$ keV Fe I line. The interested reader can find more information in~\cite{Roach:2019ctw,Krivonos:2020qvl}.} is expected to be small, since it falls off rapidly with increasing latitude, due to the lower stellar density. 
Hence, we use the same data as the {\sc NuStar} Blank-Sky fields, but with error bars scaled up by a factor $\sqrt{7 \, \textrm{Ms}/100 \, \textrm{ks}} = 8.4$ to account for the shorter exposure time. 
We stress that, given the weak constraining power that {\sc NuStar} turns out to provide (as we will discuss in the following section), these approximations are sufficient for our purposes. As a side remark, note that the {\sc NuStar} data we use were collected by the {\sc Fmpa} and {\sc Fmpb} detectors on board of the satellite. Because the photon spectra measured by the two detectors are similar, the computed constraints have only a negligible difference, thus we only show the results using the {\sc Fmpa} detector.

\item \underline{{\sc Xmm-Newton} whole-sky observations.} The data are used in~\cite{Dessert:2018qih,Foster:2021ngm} to search for decaying sterile neutrino DM. In particular, the data are provided in a very convenient form~\cite{xmm}, which we use extensively.
They correspond to the observation of the whole sky with the two cameras (called {\sc Mos} and {\sc Pn}) onboard the {\sc Xmm-Newton} satellite, over an extensive period of about 18 years, from the launch of the telescope (in late 1999) to September 2018.
After the removal of point sources, the data are combined into 30 concentric rings of width $6^\circ$ as measured in angular distance from the GC. A slice of $|b|\le 2^\circ$ is removed, \emph{i.e.}\ the GP is masked. 
The energy range initially covers $2$ eV to $20$ keV, however we restrict it as prescribed in~\cite{Foster:2021ngm} to avoid the dominant instrumental background. The final energy range is therefore $2.5$ to $8$ keV for {\sc Mos} and $2.5$ to $7$ keV for {\sc Pn}. Response matrices for both instruments are also provided.

\item \underline{{\sc Suzaku} high-latitude fields.} The data are provided in \cite{Yoshino:2009kv}, which focuses on measuring the soft diffuse $X$-ray emission from several small fields located at large Galactic longitudes ($65^\circ < \ell < 295^\circ$) and observed for a period of a few days each between 2006 and 2008, using the backside illuminated CCD (BI CCD) of the {\sc Xis} spectrometer on board of the {\sc Suzaku} satellite. We use the data\footnote{The data are shown in Figures 2 and 5 of \cite{Yoshino:2009kv} and we obtained in digital form from M.~Kazuhisa, private communication. The NEP field combines the data from NEP1 and NEP2. We could not obtain the data for the LH-2 field, which we therefore neglect.}
from the 11 fields denoted as: GB, HL-B, LH-1, Off-FIL, On-FIL, HL-A, M12off, LX-3, NEP, LL21 and LL10. We refer to Table 1 of \cite{Yoshino:2009kv} for the details of the regions (coordinates, exposures and the original references). We do not consider the R1 and R2 fields, which include bright point sources. From the data, the point sources and the $X$-ray emission induced by the solar wind proton flux have been carefully removed by the {\sc Suzaku} collaboration. The energy range is $0.4-5$ keV for all fields, and the typical exposures vary between $16$ and $60$ ks. The effective area of the experiment in the range of interest roughly equals $100$ to $300$ cm$^2$. However, we use the detailed published determination (see below).

\end{itemize}

In order to derive the constraints, we first compute the total photon flux from DM annihilation or decay, for each channel considered in Equations~\ref{eq:ee}, \ref{eq:mumu} and \ref{eq:pipi} and ROI. For the {\sc Integral/Spi} dataset, we compute the photon flux for each latitude bin and energy band. For the remaining datasets we compute the photon flux for each energy bin. Then we correct some of the predicted flux in order to take into account instrumental features:
\begin{itemize}
	\item For each ring of the {\sc Xmm-Newton} dataset, we convolve the photon energy spectrum with the instrumental response function as prescribed in~\cite{Kaastra:2016qwt}. Given a specific ring, where $\left(d\Phi_\gamma/dE_\gamma\right)_j=\left(dN_\gamma/dE_\gamma \, dA \, dt\right)_j$ is our predicted DM spectrum in the input energy bin  $j$, the discrete convolution with the instrument response is 
$\left(dN_\gamma/dE_\gamma \, dt\right)_i=\sum_jR_{ij}\left(dN_\gamma/dE_\gamma \, dA\, dt \right)_j$ 
in the output energy bin $i$, where $R_{ij}$ is the instrument response matrix.\footnote{Here by `input' and `output', we mean the predicted flux before and after the convolution with the instrumental response matrix, respectively.} The matrices are different for each ring and take into account the effective area of the instrument (in units of cm$^2$).
	\item For the {\sc Suzaku} dataset, we multiply the calculated photon energy spectrum by the {\sc Xis} effective area function as provided on the {\sc Nasa} archives~\cite{suzaku}. We use the function for the BI CCD.
\end{itemize}

We infer the constraints for each dataset separately via the test statistic
\begin{equation}
	\chi_>^2 = \sum_i \left(\frac{{\rm max}[\Phi_{\gamma,i}(p, m_\textrm{DM})-\phi_i,0]}{\sigma_i}\right)^2,
	\label{eq:chi2}
\end{equation}
where $p = \langle \sigma v \rangle$ or $\Gamma$, $\Phi_{\gamma,i}$ is the predicted photon flux from DM annihilation or decay\footnote{For {\sc Xmm-Newton} and {\sc Suzaku} the flux is actually replaced by the rate of photons per second per keV, the quantity provided by the experiment. For all the other experiments, we use the proper flux.} at the energy (or latitude for {\sc Integral}) bin $i$, $\phi_i$ is the observed flux and $\sigma_i$ its uncertainty. We then impose a $2\sigma$ bound on the parameter $p$ (for each value of $m_\textrm{DM}$) whenever we obtain $\chi_>^2=4$.
This procedure means, in particular, that we directly compare the DM prediction with the data, without including any $X$-ray astrophysical background. Including an astrophysical background would in most cases reduce the room for the DM flux and therefore strengthen the constraints. Our procedure thus allows us to derive conservative bounds. 
In the next section, we discuss the obtained constraints.

\section{Results and discussion}
\label{sec:results1}

\begin{figure}[t]
	\centering
	\begin{subfigure}[c]{0.49\linewidth}
		\centering
		\includegraphics[width=\linewidth]{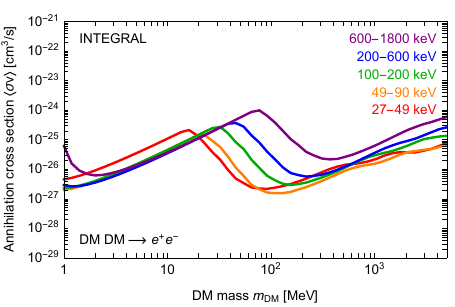}
	\end{subfigure}
	\hfill
	\begin{subfigure}[c]{0.49\linewidth}
		\centering
		\includegraphics[width=\linewidth]{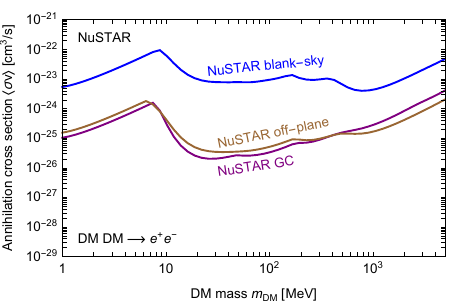}
	\end{subfigure}
	\hfill
	\begin{subfigure}[c]{0.49\linewidth}
		\centering
		\includegraphics[width=\linewidth]{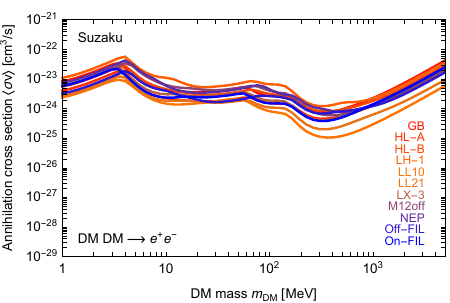}
	\end{subfigure}
	\hfill
	\begin{subfigure}[c]{0.49\linewidth}
		\centering
		\includegraphics[width=\linewidth]{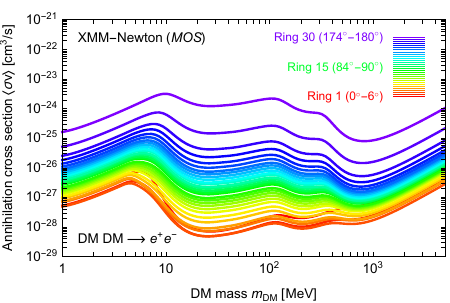}
	\end{subfigure}
	\caption{Conservative constraints on annihilating DM from the different portions of the datasets that we consider. Top left panel: constraints from the different energy bands of the {\sc Integral} dataset (different colours). Top right panel: constraints from the three different regions of observation that we use in the {\sc NuStar} dataset. Bottom left panel: constraints from the eleven different fields of the {\sc Suzaku} dataset (distinguished by the different colours as in the legend). Bottom right panel: constraints from the thirty rings of the {\sc Xmm-Newton} data (distinguished by the different colours as in the legend), for the {\sc Mos} camera for definiteness.}
	\label{fig:PartialResultsAnn}
\end{figure}

\begin{figure}[t]
	\centering
	\begin{subfigure}[c]{0.49\linewidth}
		\centering
		\includegraphics[width=\linewidth]{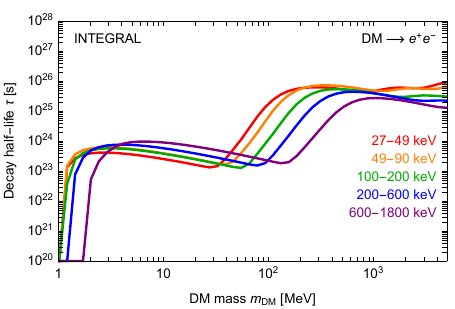}
	\end{subfigure}
	\hfill
	\begin{subfigure}[c]{0.49\linewidth}
		\centering
		\includegraphics[width=\linewidth]{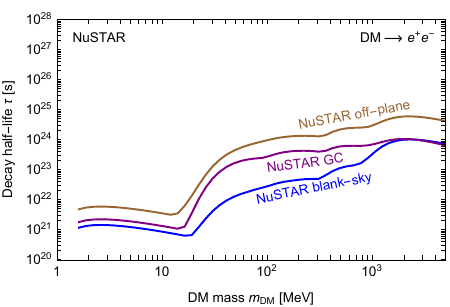}
	\end{subfigure}
	\hfill
	\begin{subfigure}[c]{0.49\linewidth}
		\centering
		\includegraphics[width=\linewidth]{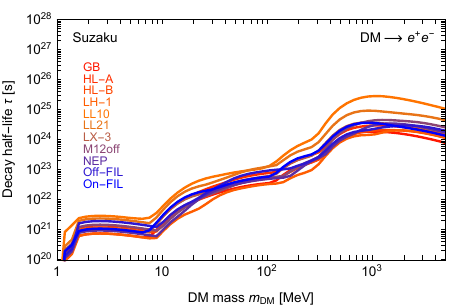}
	\end{subfigure}
	\hfill
	\begin{subfigure}[c]{0.49\linewidth}
		\centering
		\includegraphics[width=\linewidth]{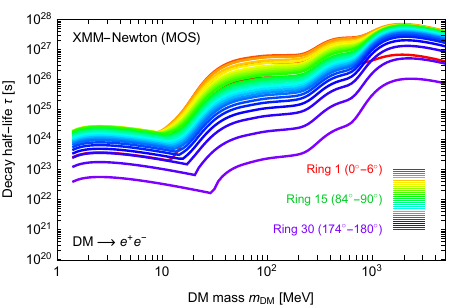}
	\end{subfigure}
	\caption{Same as in Figure~\ref{fig:PartialResultsAnn} but for decaying DM.}
	\label{fig:PartialResultsDec} 
\end{figure}

We start by presenting, in Figure~\ref{fig:PartialResultsAnn} for the annihilation case and in Figure~\ref{fig:PartialResultsDec} for the decay case, the conservative constraints obtained from each experiment for each portion of the dataset (either observation subfield or energy band). In each case the bounds are derived using the criterion in Equation~\ref{eq:chi2}. We focus here for definiteness on the $\textrm{DM (DM)} \to e^+e^-$ channel. 

In the top left panel we show the {\sc Integral} bounds imposed by each energy band separately (for the annihilation case, this figure reproduces the analogous one in~\cite{Cirelli:2020bpc}). The characteristic shape of the curves is motivated as follows: in the region of large DM masses a strong bound occurs because the ICS flux is constrained by the data points, as shown in the lower left panel of Figure~\ref{fig:Xrayfluxes}; the prompt emission is  instead responsible for the bound on small DM masses. In the intermediate mass range the bound is weaker because the data fall in the trough of the characteristic `double hump' shape of the prompt+ICS spectra. Note that the kink between large and small masses moves to larger DM masses for the higher energy bands and to lower masses for the lower energy bins. This is due to the fact that the DM spectrum shifts to the left with decreasing $m_\textrm{DM}$. Overall, given the configuration of the data points and the DM spectra, we find that the low-energy bands are more constraining for large masses while high-energy bins are more constraining for small masses. 

In the top right panel we show the bounds imposed by each {\sc NuStar} dataset separately. The shape of the constraints is analogous to that of {\sc Integral}, with the kink occurring at smaller masses ($m_\textrm{DM} \simeq 10$ MeV) since the {\sc NuStar} data cover lower energies. The limits from the GC region and the off-plane fields are more constraining, while those from the blank-sky fields are weaker. In absolute terms, the {\sc NuStar} results are weaker with respect to the {\sc Integral} ones for the following reasons. For the {\sc NuStar} blank-sky case, the fields are at very high latitudes, where the Galactic DM emission is small. For the {\sc NuStar} GC case, the main component of the measured flux is understood to be the GRXE~\cite{Roach:2019ctw}, and the DM flux has to compete with this sizeable foreground: for decaying DM, the DM flux is overwhelmed by the GRXE; for annihilating DM, the DM flux is boosted by the square of the large DM density in the central regions and hence better bounds occur. The off-plane case offers competitive limits overall because, as discussed above, the regions of observation are located enough far away from the plane that the GRXE has decreased and hence the DM contribution can emerge. 

In the bottom left panel we show the bounds imposed by each one of the 11 {\sc Suzaku} fields. Now the kink occurs at $m_\textrm{DM} \lesssim 10$ MeV because the {\sc Suzaku} data are even lower in energy compared to {\sc NuStar} and {\sc Integral}. The fields (green in Figure~\ref{fig:galaxymap}) are all positioned at high latitudes and large longitudes and offer comparable bounds, with LL10 and LL21 slightly more stringent than the other ones. 

Finally, in the bottom right panel we show the bounds imposed by {\sc Xmm-Newton} data considering each ring separately. We show for definiteness the data from the {\sc Mos} camera (those from the {\sc Pn} camera turn out to be very similar but slightly less stringent). Each line/colour in the plot corresponds to one $6^\circ$ degree ring as depicted in Figure~\ref{fig:galaxymap}. Not surprisingly, the inner rings (warmer colours in the figure) are more constraining because the DM density is higher in the inner Galaxy. However, due to the astrophysical foreground, the innermost ring does not provide the tightest constraints. The third ring from the GC ($12^\circ - 18^\circ$), and for some small mass intervals to the adjacent ones, provide the most constraining limits. Note that the spread of the limits is wider for annihilating DM compared to decaying DM, as expected because of the different dependence of the source with the DM density ($\rho_\textrm{DM}^2$ versus $\rho_\textrm{DM}$, respectively). 

\begin{figure}[t]
	\centering
	\begin{subfigure}[c]{0.49\linewidth}
		\centering
		\includegraphics[width=\linewidth]{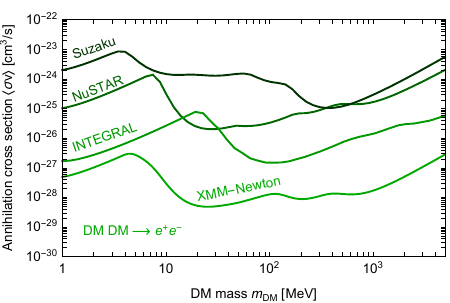}
	\end{subfigure}
	\hfill
	\begin{subfigure}[c]{0.49\linewidth}
		\centering
		\includegraphics[width=\linewidth]{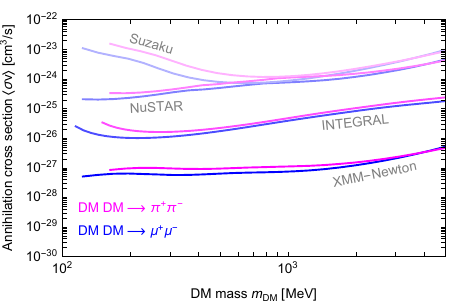}
	\end{subfigure}
	\caption{Summary of our conservative constraints on annihilating DM from each experiment and for all channels. The left panel refers to the $e^+e^-$ annihilation channel (green lines), while the right plot to the $\pi^+\pi^-$ (magenta) and $\mu^+\mu^-$ (blue) channels. From top (least constraining) to bottom (most constraining), the experiments are roughly ordered as {\sc Suzaku}, {\sc NuStar}, {\sc Integral} and {\sc Xmm-Newton}.}
	\label{fig:ResultsAnn} 
	\bigskip
	\begin{subfigure}[c]{0.49\linewidth}
		\centering
		\includegraphics[width=\linewidth]{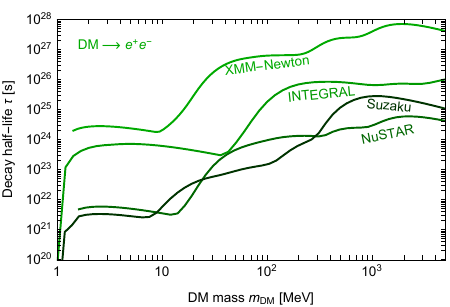}
	\end{subfigure}
	\hfill
	\begin{subfigure}[c]{0.49\linewidth}
		\centering
		\includegraphics[width=\linewidth]{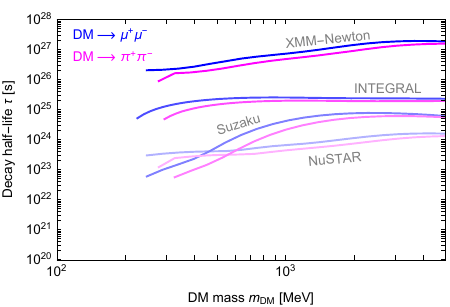}
	\end{subfigure}
	\caption{Summary of our conservative constraints on decaying DM from each experiment and for all channels. The ordering, now inverted as bottom (least constraining) to top (most constraining), is very similar to Figure~\ref{fig:ResultsAnn}.}
	\label{fig:ResultsDec} 
\end{figure}

In Figures~\ref{fig:ResultsAnn} and \ref{fig:ResultsDec} we show the combined bounds for each experiment. This means that we apply the statistical criterion in Equation~\ref{eq:chi2} to the whole dataset of each experiment: the {\sc Integral} bounds are obtained using all the data of the 5 energy bands and the {\sc NuStar}, {\sc Suzaku} and {\sc Xmm-Newton} ones using all the regions of observation. The left panels refer to the $\textrm{DM (DM)} \to e^+e^-$ channel while the right panels to the $\textrm{DM (DM)} \to \mu^+\mu^-$ and $\textrm{DM (DM)} \to \pi^+\pi^-$ channels. Along the entire mass range, the {\sc Xmm-Newton} bounds are the most stringent ones. 
\begin{figure}[t]
	\centering
	\includegraphics[width=0.8\linewidth]{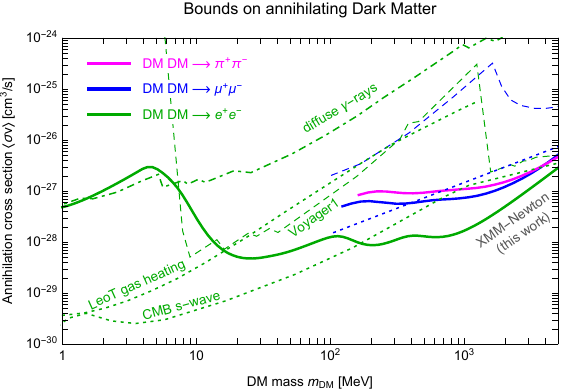} 
	\caption{Final combined results for annihilating DM from this work ({\sc Xmm-Newton}), compared with existing bounds. We report the bounds from Essig et al.~\cite{Essig:2013goa}, obtained using a compilation of $X$-ray and soft $\gamma$-ray data (dot-dashed green line marked `diffuse $\gamma$-rays'); the bounds from Boudaud et al.~\cite{Boudaud:2016mos} derived using data from {\sc Voyager 1} (dashed green and blue lines, corresponding to the $e^+e^-$ and $\mu^+\mu^-$ annihilation channels, respectively); the CMB bounds from Slatyer~\cite{Slatyer:2015jla} and Lopez-Honorez et al.~\cite{Lopez-Honorez:2013cua} (dotted green and blue lines, for the $e^+e^-$ and $\mu^+\mu^-$  channels); the bounds from gas heating in Leo T, obtained by Wadekar and Wang~\cite{Wadekar:2021qae} (also dotted, since the physics mechanism of energy injection is similar to the CMB one).}
	\label{fig:FinalResultsAnn} 
\end{figure}

\begin{figure}[t]
	\centering
	\includegraphics[width=0.8\linewidth]{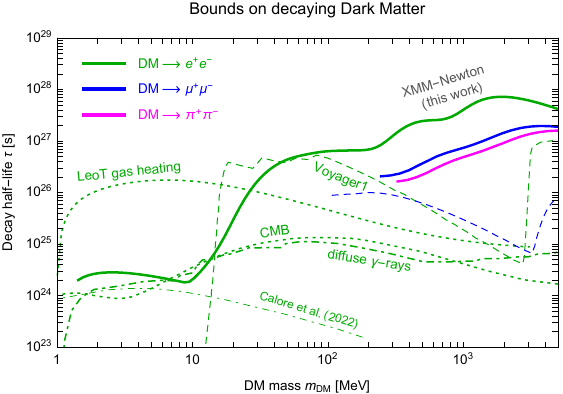} 
	\caption{Final combined results for decaying DM from this work ({\sc Xmm-Newton}), compared with existing bounds. The constraints and the references are the same as in Figure~\ref{fig:FinalResultsAnn}, except that the CMB ones are derived in Liu et al.~\cite{Liu:2016cnk}. In addition, we plot the subdominant constraints of Calore et al.~\cite{Calore:2022pks}.}
	\label{fig:FinalResultsDec} 
\end{figure}

Figures~\ref{fig:FinalResultsAnn} and \ref{fig:FinalResultsDec} represent our final results: we show only the most stringent constraints that we obtain (from {\sc Xmm-Newton}), for the three annihilation or decay channels. For the case of DM annihilating into $e^+e^-$, {\sc Xmm-Newton} imposes the bound $\langle \sigma v \rangle \lesssim 10^{-28}$~cm$^3$/s, over the wide range $m_\textrm{DM} \simeq 20\textrm{ MeV} - 1\textrm{ GeV}$. The bound loosens to $\langle \sigma v \rangle \lesssim 10^{-27}$~cm$^3$/s in the range $m_{\rm DM} \simeq 1 - 20$ MeV, the region where the dominant contribution of the ICS component is too low in energy to be constrained by the data. DM annihilating into $\mu^+\mu^-$ or $\pi^+\pi^-$ is constrained to $\langle \sigma v \rangle \lesssim 10^{-27}$ cm$^3$/s in the relevant mass interval. For the case of DM decaying into $e^+e^-$, {\sc Xmm-Newton} imposes the bound on the decay half-life $\tau \equiv 1/\Gamma \gtrsim 10^{27}$ s, over the range $m_\textrm{DM} \simeq 50\textrm{ MeV} - 1\textrm{ GeV}$. The limit approaches $\tau \sim 10^{28}$ s for $m_\textrm{DM} \sim$ few GeV.

\begin{figure}[p]
	\centering
	\begin{subfigure}[c]{0.49\linewidth}
		\centering
		\includegraphics[width=\linewidth]{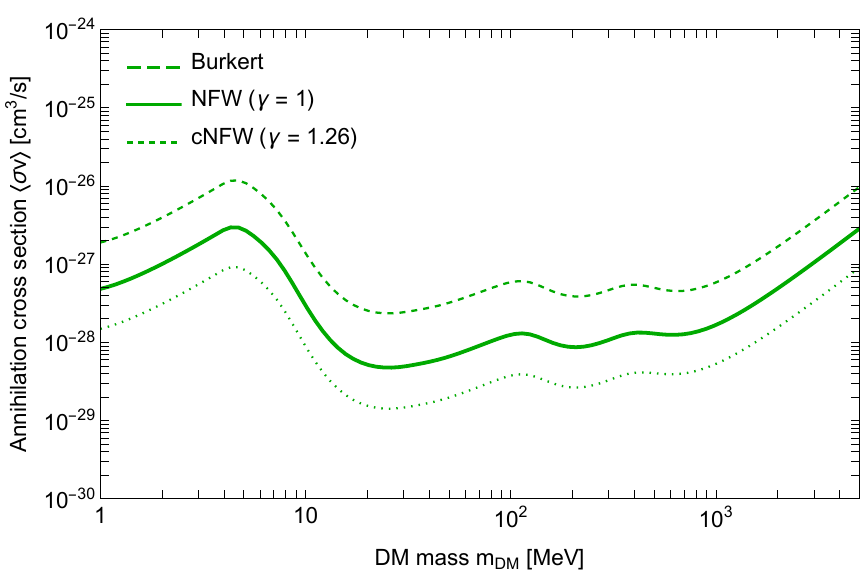}
	\end{subfigure}
	\hfill
	\begin{subfigure}[c]{0.49\linewidth}
		\centering
		\includegraphics[width=\linewidth]{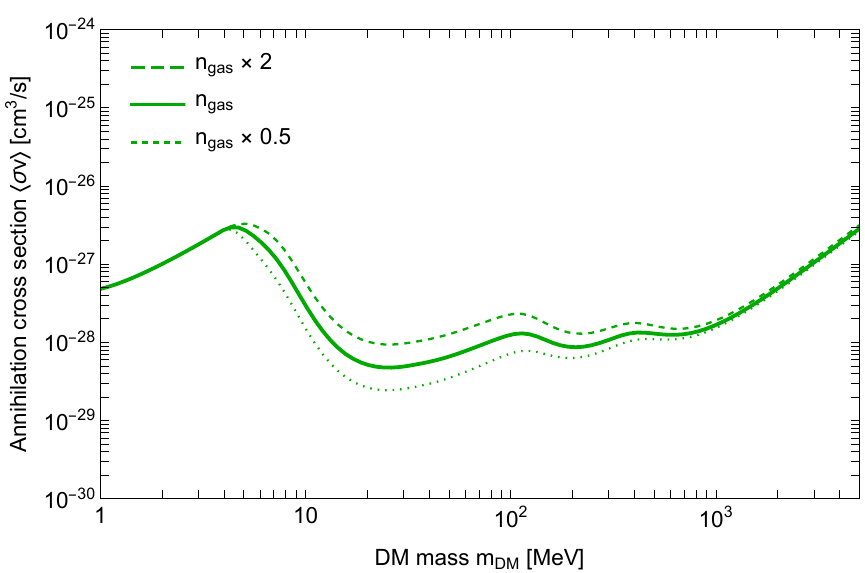}
	\end{subfigure}
	\hfill
	\begin{subfigure}[c]{0.49\linewidth}
		\centering
		\includegraphics[width=\linewidth]{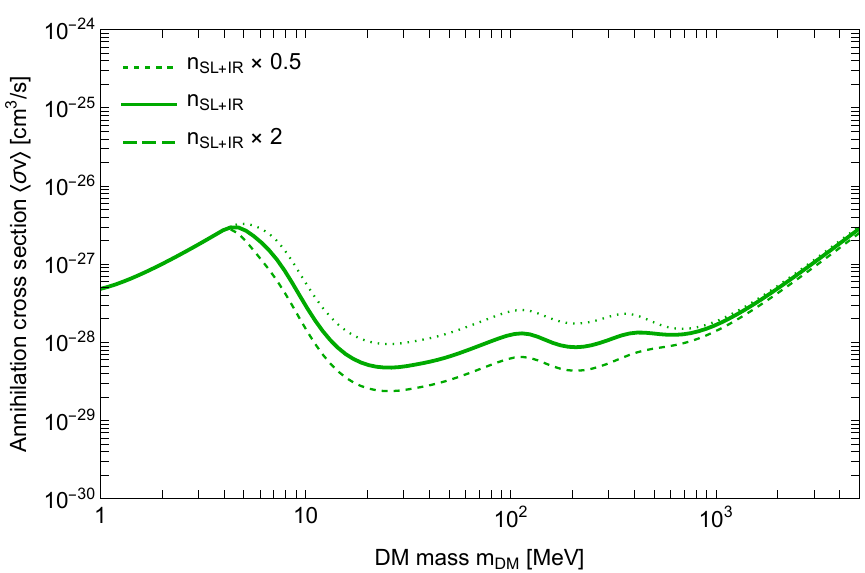}
	\end{subfigure}
	\hfill
	\begin{subfigure}[c]{0.49\linewidth}
		\centering
		\includegraphics[width=\linewidth]{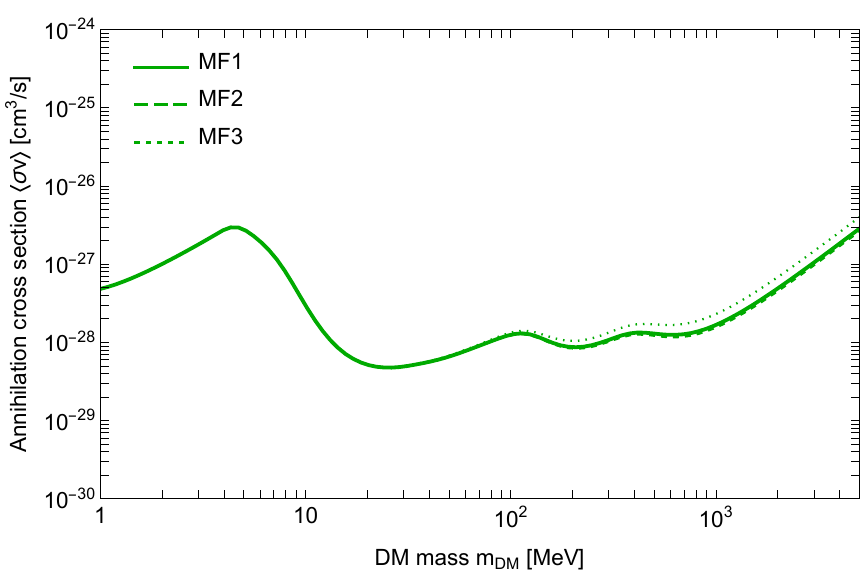}
	\end{subfigure}
	\caption{Illustration of the impact of astrophysical uncertainties on the limits on DM annihilating in $e^+e^-$. The illustrated sources of uncertainties are the choice of the DM profile (top left), the normalisation of the gas density (top right), the normalisation of the SL and IR components of the ambient photon density (bottom left), and the GMF configuration (bottom right).}
	 \label{fig:uncertainties}
	\bigskip
	\bigskip
	\begin{subfigure}[c]{0.49\linewidth}
		\centering
		\includegraphics[width=\linewidth]{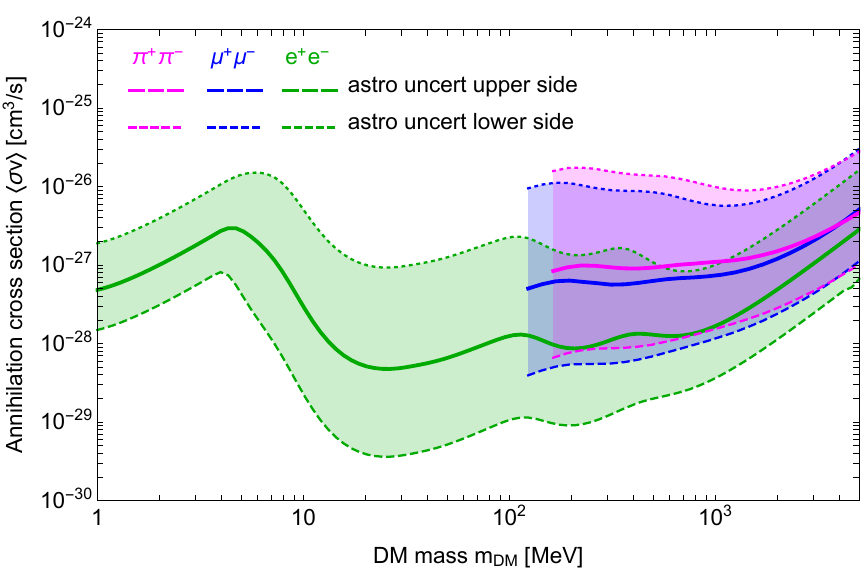}
	\end{subfigure}
	\hfill
	\begin{subfigure}[c]{0.49\linewidth}
		\centering
		\includegraphics[width=\linewidth]{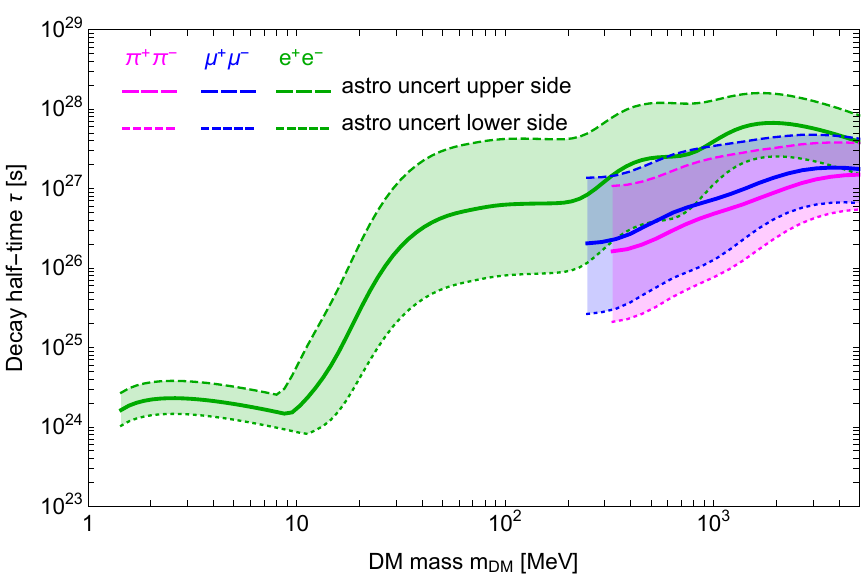}
	\end{subfigure}
	\caption{Illustration of the impact of total astrophysical uncertainties on DM annihilation (left panel) and decay (right panel) constraints.}
	 \label{fig:Combuncertainties}
\end{figure}

In Figure~\ref{fig:uncertainties} we show the impact of astrophysical uncertainties on the {\sc Xmm-Newton} constraints on DM annihilating in $e^+e^-$, by following the same strategy as in~\cite{Cirelli:2020bpc}. We vary the DM profile, the gas density in the Galaxy (which affects the energy loss of DM-produced $e^\pm$ through ICS), the radiation field density (impacting the energy losses through ICS but also the ICS radiative power directly) and the GMF (affecting energy loss through synchrotron radiation). More precisely: we adopt a Burkert profile (Equation~\ref{eq:Burkert}) and a cNFW one (Equation~\ref{eq:gNFW} with $\gamma = 1.26$), we vary the normalisation of the gas and radiation field density by a factor of $2$ above and below their central values, and we adopt the different configurations of the GMF discussed in~\cite{Cirelli:2010xx} (MF1, MF2 and MF3). We then compute the upper and lower envelopes of the $X$-ray fluxes from these combined variations, and we derive the corresponding bounds, resulting in the uncertainty bands of Figure~\ref{fig:Combuncertainties}. Note that the constraints can (generously) vary within two orders of magnitude.

For annihilating DM, the dominant uncertainty comes from the choice of the DM density profile, due to the most constraining ROI pointing near the GC. We recall Figure~\ref{fig:DMprofiles} which shows the large difference in the DM energy density between profiles when approaching the GC. For decaying DM, this uncertainty decreases since the DM density has a single power contribution (compared to squared for annihilating DM), and therefore is comparable to the uncertainties due to variations of the normalisation of gas and SL+IR photon densities. The choice of the GMF configuration does not impact much the results, since it only affects the energy loss of DM-produced $e^\pm$ through synchrotron radiation, which are subdominant in our scenario.

\section{Comparison with related work}
\label{sec:comparison}

Using a compilation of $X$-ray and soft $\gamma$-ray data from {\sc Heao-1}, {\sc Integral}, {\sc Comptel}, {\sc Egret} and {\sc Fermi}, Essig et al.~\cite{Essig:2013goa} have derived bounds on the $e^+e^-$ channel, shown as a dot-dashed line in Figure~\ref{fig:FinalResultsAnn}. This work does not include the ICS emission: indeed it leads to bounds that are comparable to ours in the small range where ICS is not relevant ($m_\textrm{DM} \lesssim 10$ MeV) and are instead much weaker than ours at any larger mass where the ICS is the leading contribution to our limits. Using low-energy measurements by {\sc Voyager 1} of the $e^\pm$ CR flux outside of the heliosphere, Boudaud et al.~\cite{Boudaud:2016mos} have derived constraints on the $e^+e^-$ and $\mu^+\mu^-$ channel, shown as dashed lines in Figure~\ref{fig:FinalResultsAnn}. We report the bounds of their propagation model $B$, characterised by weak reacceleration. Their constraints are stronger than ours only in a small mass interval around $10$ MeV, for the $e^+e^-$ case. They are always weaker for the $\mu^+\mu^-$ case. Using the impact on the CMB anisotropies of the $e^+e^-$ injection by DM annihilation events, Slatyer~\cite{Slatyer:2015jla} and Lopez-Honorez et al.~\cite{Lopez-Honorez:2013cua} derived the constraints represented by the dotted lines in Figure~\ref{fig:FinalResultsAnn}. Our {\sc Xmm-Newton} bounds reach deeper than these CMB constraints, in the portion of large mass where the ICS effect is important ($m_\textrm{DM} \gtrsim 180$ MeV and $m_{\rm DM} \gtrsim 400$ MeV, respectively for the $e^+e^-$ and $\mu^+\mu^-$, $\pi^+\pi^-$ channels). The CMB constraints are still more stringent elsewhere.
However, as discussed in~\cite{Cirelli:2020bpc}, the CMB constraints hold under the assumption that DM annihilations are $s$-wave. If the DM annihilation is $p$-wave, \emph{i.e.}\ $\langle \sigma v \rangle \propto v^2$, they weaken considerably. Our constraints are instead essentially insensitive to these difference~\cite{Cirelli:2020bpc}, which implies that, for the $p$-wave scenario, our limits represent the most stringent bounds for $m_\textrm{DM} \gtrsim 15$ MeV. Finally, Wadekar and Wang~\cite{Wadekar:2021qae} derived bounds from the requirement that $\textrm{DM DM} \to e^+e^-$ annihilations do not overheat the gas in the Leo T dSph. This constraint is represented by a dotted line in Figure~\ref{fig:FinalResultsAnn} and it is more stringent than ours for $m_\textrm{DM} \lesssim 20$ MeV. However, this bound would relax significantly if DM annihilations were $p$-wave (see~\cite{Wadekar:2021qae} for details), similarly to the CMB constraints.

For the case of decaying DM, the existing constraints in the literature are shown in Figure~\ref{fig:FinalResultsDec}. The diffuse $\gamma$-ray constraints of Essig et al.~\cite{Essig:2013goa} are shown as a dot-dashed line, while the CMB and the Leo T dSph gas heating constraints of~\cite{Liu:2016cnk} and~\cite{Wadekar:2021qae} respectively are shown as a dotted curve. 
The {\sc Voyager 1} constraints~\cite{Boudaud:2016mos} are dashed. Recently Calore et al.~\cite{Calore:2022pks} have considered the $\textrm{DM} \to e^+e^-$ channel (as well as the direct decaying channel $\textrm{DM}\to \gamma \gamma$, which is not of interest for us) and has used {\sc Integral/Spi} diffuse data, their bounds are displayed as a thin dot-dashed line. The constraints derived in this work (thick lines) are the most stringent limits for decaying DM for $m_\textrm{DM} \gtrsim 50$ MeV. For large masses, we improve upon the existing bounds by up to three orders of magnitude. Besides the $\mu^+\mu^-$ {\sc Voyager 1} ones, we are not aware of other existing constraints for the $\mu^+\mu^-$ and $\pi^+\pi^-$ channels in this mass interval.

\section{Summary}

In this chapter we have focused on light, sub-GeV DM ID, following up on the exploratory analysis performed in~\cite{Cirelli:2020bpc}. DM in this mass range ($1$ MeV to $\sim5$ GeV) is notoriously difficult to probe with indirect searches, given the scarcity of $100$ keV $-$ $100$ MeV range experiments which could probe its soft $\gamma$-ray prompt emission. We have therefore concentrated on its secondary emission, which produces $X$-rays via ICS of DM-produced $e^\pm$ over the Galactic ambient light. We have used data from the {\sc NuStar}, {\sc Suzaku}, {\sc Integral} and {\sc Xmm-Newton} satellites, in a number of different fields of observation in the Galaxy (see Figure~\ref{fig:galaxymap}). We have compared these measurements to the predicted flux from annihilating or decaying DM, considering the three relevant channels $\textrm{DM (DM)} \to e^+e^-$, $\mu^+\mu^-$ and $\pi^+\pi^-$. 

We find that the constraints imposed by the {\sc Xmm-Newton} whole-sky survey greatly improve upon the existing limits. For decaying DM, they are the most stringent to date, for $m_\textrm{DM} \gtrsim 50$ MeV, improving upon the existing bounds up to three orders of magnitude (see Figure~\ref{fig:FinalResultsDec}). For annihilating DM, our limits are the most constraining to date for $m_\textrm{DM} \gtrsim 180 \ \textrm{MeV}$;  for smaller masses, they are competitive with diffuse $\gamma$-ray constraints and $e^\pm$ constraints from {\sc Voyager 1}, but the CMB $s$-wave bounds are still more stringent (see Figure~\ref{fig:FinalResultsAnn}). The sizeable astrophysical uncertainties related to the Galactic DM distribution and the Galactic environment can affect these results and make them tighter or looser by up to one order of magnitude in each direction (see Figures~\ref{fig:uncertainties} and \ref{fig:Combuncertainties}). 

We remind the reader that the limits shown in this chapter were derived upon very conservative assumption on the propagation of DM-produced $e^\pm$ in the Galactic environment. In the next chapter, we investigate the impact on these results when adopting a realistic propagation setup, motivated by recent fits to fluxes and flux ratios of different CR species measured by {\sc Ams-02}~\cite{Luque:2021nxb,delaTorreLuque:2022vhm}.

%% file: Chapters/chap4.tex

\lettrine[lines=3, nindent=1pt]{A}{s} we saw in the previous chapter, annihilating or decaying sub-GeV DM can inject low-energy $e^\pm$, which can leave imprints in the diffuse $e^\pm$ flux that we detect at Earth, as well as the $X$- and soft $\gamma$-ray Galactic diffuse emission due to ICS on ambient photons. For instance, in recent years the {\sc Voyager 1} probe has measured the local flux of $e^\pm$ outside of the heliosphere~\cite{Stone:2013zlg} at energies below tens of MeV. These measurements have the key advantage that they are not significantly affected by solar screening~\cite{Potgieter:2013pdj}, which greatly reduces the flux of low-energy charged CRs at Earth. Additionally, we consider the production of secondary $X$-rays, in particular due to ICS of DM-produced $e^\pm$ on Galactic ambient radiation. The main goal of this chapter is to perform a more realistic computation of the DM-induced $X$- and $\gamma$-ray signals by considering the full-fledged propagation of DM-produced $e^\pm$ in the MW, in particular including momentum space diffusion (reacceleration). In turn, we reevaluate the constraints from {\sc Xmm-Newton} derived in Chapter~\ref{chap:subGeV} and assess the impact of $e^\pm$ reacceleration on the constraints. In the same manner, we also update the limits obtained in~\cite{Boudaud:2016jvj} using {\sc Voyager 1} $e^\pm$ data. For this study we use a fully numerical approach that does not need approximations and utilises current state-of-the-art propagation setups. In Section~\ref{sec:eepropag} we outline the propagation methodology and the computation of the DM-produced $e^\pm$ flux as well as secondary photon emissions. In Section~\ref{sec:results2} we discuss about the derivation of our limits on annihilating and decaying DM, and the impact of different uncertainty sources on the results. Finally, in Section~\ref{sec:comparison2} we compare these results with other relevant constraints on sub-GeV DM.

\newpage

\section{Electron-positron propagation and secondary radiations}
\label{sec:eepropag}

Once again we analyse each DM annihilation and decay channel ($\textrm{DM (DM)} \to e^+e^-$, $\mu^+\mu^-$ and $\pi^+\pi^-$) separately even though, in principle, DM could annihilate or decay to a mix of modes in specific models. In the case of the $X$-ray signals generated by the $e^\pm$ injected by DM, we consider that the total photon flux is composed of prompt emissions (FSR and Rad) and the secondary emissions from ICS. The first is straightforward and computed in the same manner as in Section~\ref{subsec:prompt}. For the second, we follow the same idea as in Section~\ref{subsec:secondary}, but we adopt instead a realistic propagation setup which we detail in the next section.

\subsection{Propagation equation and parameters}

As a first step, in order to compute the distribution and energy spectrum of DM-produced $e^\pm$, we use a customised version of the \verb|DRAGON2| code~\cite{Evoli:2016xgn, Evoli:2017vim}, which is publicly available at~\cite{DRAGON2Git}. \verb|DRAGON2| itself is a dedicated CR propagation code designed to self-consistently solve the diffusion-convection-loss equation described by Equation~\ref{eq:transporteq}, in order to obtain the density of CR species $i$ per momentum unit $f_i(p,\vec{x})$. Its source term, described by Equation~\ref{eq:source}, remains the same as in Chapter~\ref{chap:subGeV} since the injection of DM-produced $e^\pm$ comes from the same annihilation or decay channels.

In the \verb|DRAGON2| code, we use a spatial grid with a resolution of $150$ pc and a $e^\pm$ energy grid ranging from $50$ keV to $10$ GeV, with a $5\%$ resolution. We have tested that our results do not change appreciably for lower energy or spatial resolution. All relevant sources of energy losses (ionisation, Coulomb interactions, bremsstrahlung, ICS and synchrotron losses) are included and solved numerically without using approximations. We use the Galactic gas distribution model implemented by the \verb|GALPROP| group~\cite{Moskalenko:2001ya,Ackermann:2012pya}\footnote{A recent update on the gas model was published in~\cite{Porter_2022}, which should not differ very significantly from the one used in previous versions.}, with a gas composition assumed to be a mixture of $90\%$ hydrogen and $10\%$ helium. Then, we parametrise the energy dependence of the diffusion coefficient $D$ as a broken power-law with a break at rigidity $R_b=312$ GV~\cite{Genolini:2017dfb}
\begin{equation}
	D (R) = D_0 \beta^{\eta}\frac{\left(R/R_0 \right)^{\delta}}{\left[1 + \left(R/R_b\right)^{\Delta \delta / s}\right]^s}\;,
	\label{eq:diff_eq}
\end{equation}
and assume for simplicity that this coefficient is uniform everywhere in the Galaxy. We also incorporate $e^\pm$ reacceleration by Alfv\'enic turbulences in the Galaxy, which is parametrised in our evaluations by the Alfv\'en speed $v_A$ and is directly linked to the momentum diffusion coefficient $D_{pp}$ in Equation~\ref{eq:momdiffcoef}. We emphasise that the diffusive motion of CRs grants an unavoidable level of reacceleration on the particles, that can be thought as the exchange of energy of charged particles with plasma waves. Both~\cite{Boudaud:2016mos,Cirelli:2023tnx} were the first sub-GeV DM constraints from {\sc Xmm-Newton} and {\sc Voyager 1} data respectively, however they did not study how these signals are affected by reacceleration in detail. However, as we show in Section~\ref{sec:e+e-}, this may cause dramatic changes in the sub-GeV $e^\pm$ signals that we study here.

The propagation parameters used are given in Table~\ref{tab:params}, which have been obtained from a combined fit to {\sc Ams-02} data~\cite{AMS:2013fma,AMS:2018tbl} for B, Be and Li ratios to C and O, using the analysis reported in~\cite{DeLaTorreLuque:2021nxb,DeLaTorreLuque:2021ddh}. While the parameters $R_b$, $\Delta\delta$ and $s$ are only important for the regime above $312$ GeV, we fit the parameters $D_0$, $\eta$, $\delta$, $L$ and $v_A$ in this analysis, leaving $R_0$ fixed to $4$~GV, since it is just the rigidity where the diffusion coefficient is normalised. We discuss the impact on the results of the uncertainties involved in the diffusion model employed in Section~\ref{sec:results2}. 

Current CR analyses used to characterise the propagation processes are mostly restricted to the availability of data on secondary CRs (mainly B, Be and Li), for which the {\sc Ams-02} collaboration~\cite{AMS:2018tbl} has measured data only from a few hundred MeV. This implies that our knowledge on the transport of charged CRs in the Galaxy below hundreds of MeV is still very limited and there is no robust estimation of the diffusion coefficient below $\simeq 100$ MeV since different assumptions of the diffusion setup are able to reproduce the current local data with relative accuracy~\cite{Weinrich:2020cmw,Silver:2024ero}. In fact, the recent data from {\sc Ams-02} on B, Be and Li has revealed a statistically significant change on the energy dependence of the diffusion coefficient at sub-GeV energies~\cite{Weinrich:2020cmw, DeLaTorreLuque:2021nxb} which could be explained by the damping of Alfv\'enic waves~\cite{2006ApJ...642..902P, Reichherzer:2019dmb, Fornieri:2020wrr}, and different regimes of turbulence can also appear at lower energies. Here, we include the change on the trend of diffusion at sub-GeV energies, which can result in sizeable differences on the flux of $e^\pm$ produced by sub-GeV DM. In particular, we employ a diffusion coefficient that includes a factor $\beta^{-0.75}$ that implies a rise in diffusion at sub-GeV energies, see Equation~\ref{eq:diff_eq} and Table~\ref{tab:params}. 

\begin{table}[t!]
    \centering
    \begin{tabular}{|c|c|c|}
    	\hline
	\textbf{Halo height} 			 &  $L$ 		   & $8.00^{+2.35}_{-1.96}~\rm{kpc}$\\
	\textbf{Norm. of diffusion coeff.} & $D_{0}$  	   & $1.02^{+0.12}_{-0.10}\times10^{29}~\rm{cm}^{2}\rm{s}^{-1}$ \\
	\textbf{Norm. rigidity}    		 &  $R_{0}$ 	   & $4~\rm{GV}$ \\
	\textbf{Diffusion spectral index}  &  $\delta$	   & $0.49\pm 0.01$  \\
	\textbf{Velocity index} 		 &  $\eta$		   & $-0.75^{+0.06}_{-0.07}$\\
	\textbf{Alfv\'en velocity} 		 &  $v_{A}$	   & $13.40^{+0.96}_{-1.02}~\rm{km/s}$\\
	\textbf{Convection velocity}	 &  $v_c$		   & $0~\rm{km/s}$\\
	\textbf{Break rigidity} 		 &  $R_{b}$	   & $312 \pm 31~\rm{GV}$ \\
	\textbf{Index break} 			 &  $\Delta\delta$ & $0.20 \pm 0.03$ \\
	\textbf{Smoothing param.} 	 &  $s$		   & $0.04 \pm 0.0015$\\
	\hline
    \end{tabular}
    \caption{Main propagation parameters used in our analysis with the uncertainties related to their determination. Uncertainties in the parameters $R_b$, $\Delta\delta$ and $s$ come from \cite{Genolini:2017dfb}.}
    \label{tab:params}
\end{table}

\subsection{Electron-positron flux prediction}
\label{sec:e+e-}

\begin{figure}[t]
    \centering
    \begin{subfigure}[c]{0.49\linewidth}
        \centering
        \includegraphics[width=\linewidth,trim= 0.5cm 0.5cm 0.5cm 0]{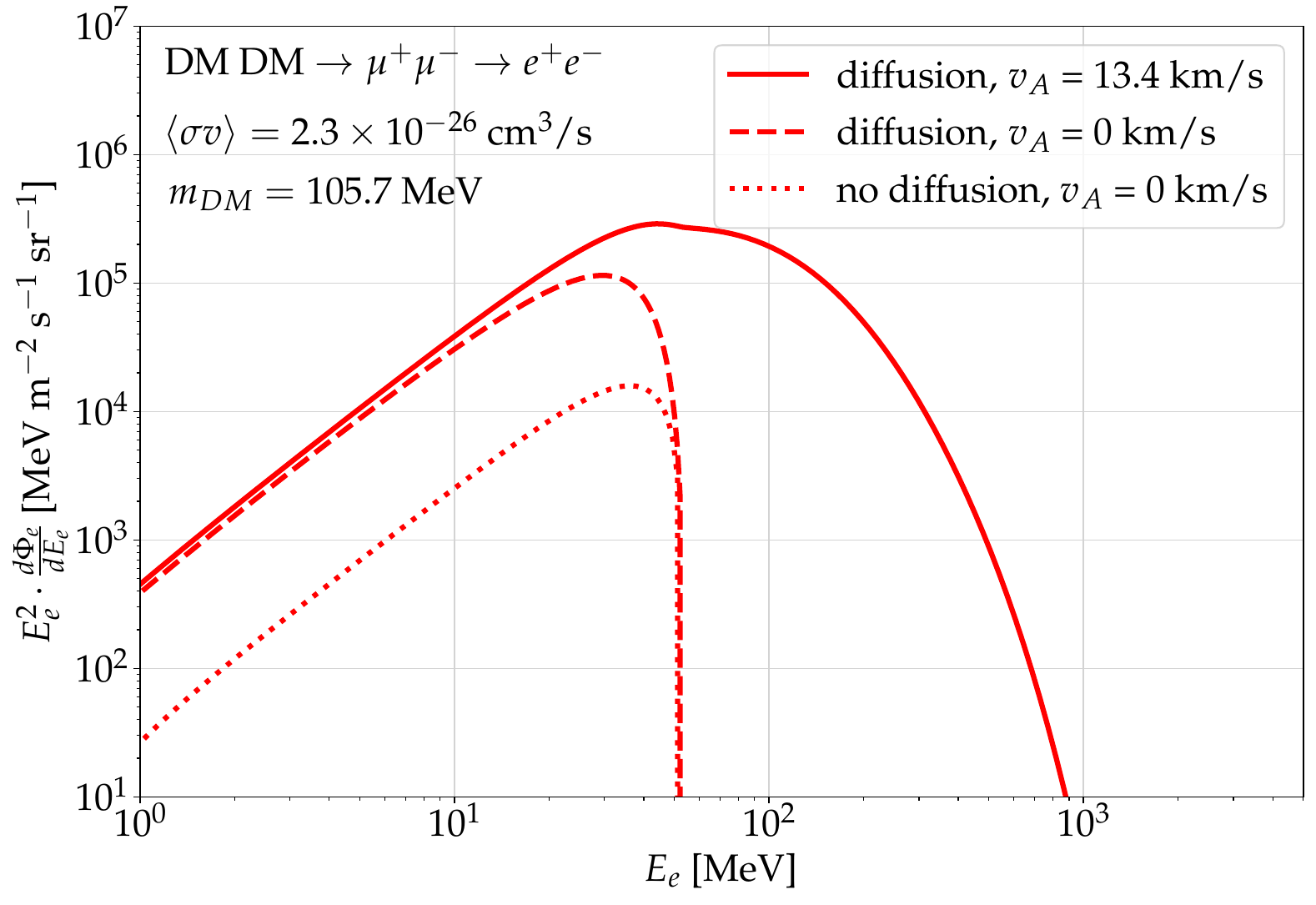}
    \end{subfigure}
    \hfill
    \begin{subfigure}[c]{0.49\linewidth}
        \centering
        \includegraphics[width=\linewidth,trim= 0.5cm 0.5cm 0.5cm 0]{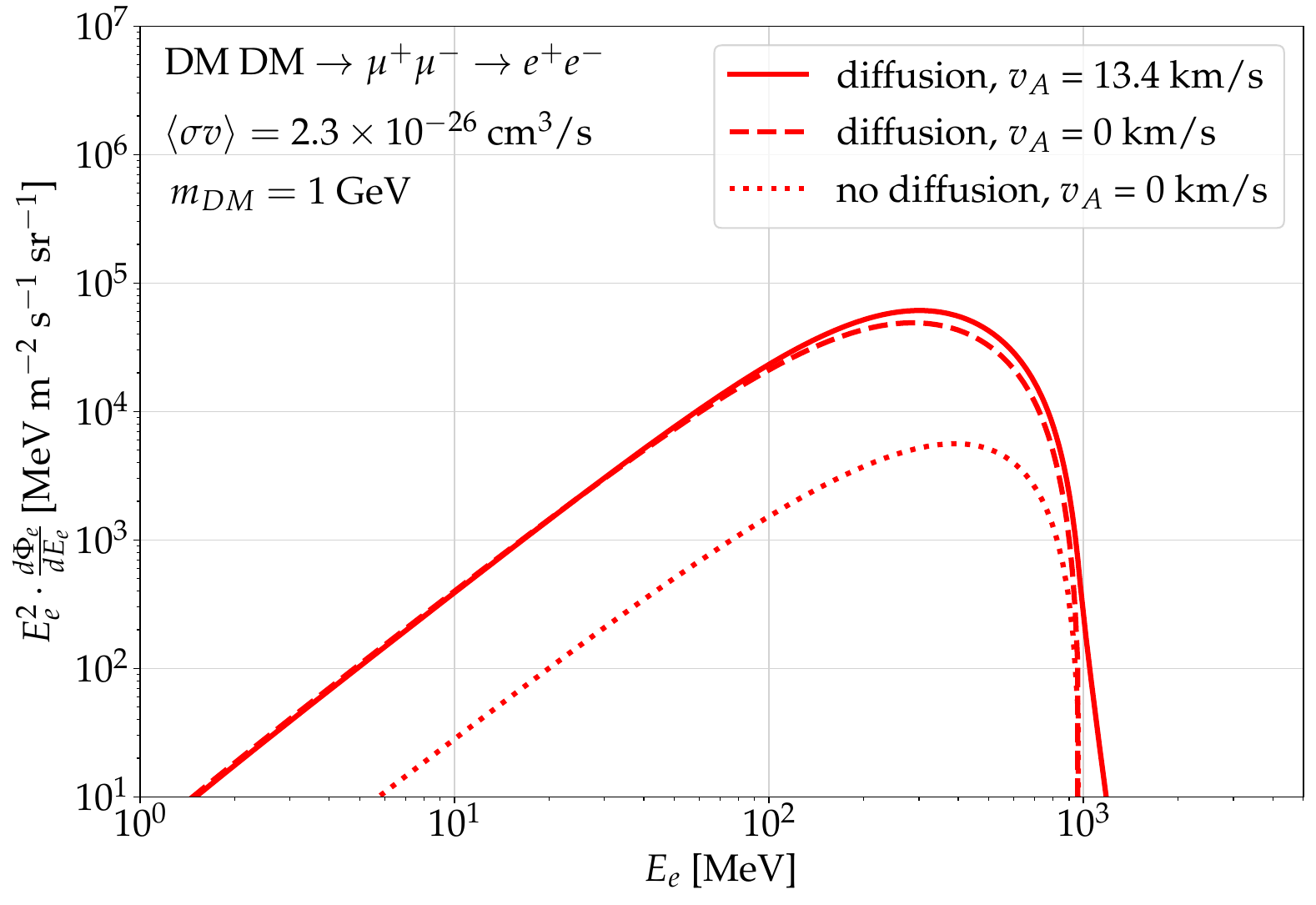}
    \end{subfigure}
    \caption{Comparison of the predicted $e^\pm$ fluxes at Earth from DM annihilation to $\mu^+\mu^-$ for $m_\textrm{DM} = m_\mu = 105.7$ MeV (left panel) and $m_\textrm{DM} = 1$ GeV (right panel) showing the impact of spatial diffusion and reacceleration. The solid lines describe the scenario including $e^\pm$ diffusion and reacceleration, the dashed lines the predicted signal with no reacceleration included and the dotted line represents the case where no propagation of $e^\pm$ is considered (only energy losses).}
    \label{fig:diff+vA}
\end{figure}

To emphasise the importance of including all the relevant effects associated with the propagation of $e^\pm$ in our results, we show in Figure~\ref{fig:diff+vA} the local flux of DM-produced $e^\pm$ from DM annihilating to $\mu^+\mu^-$ with a thermally averaged annihilation cross section of $\langle \sigma v\rangle= 2.3\times 10^{-26}$ cm$^3$\,s$^{-1}$, and for different propagation setups. In the left panel, we fix $m_\textrm{DM}$ to $m_\mu = 105.7$ MeV (to produce muons at rest) while in the right panel we fix it to $1$ GeV. Here we are comparing a setup (labelled `no diffusion') where we switch off any particle diffusion effects, only including the energy losses of DM-produced $e^\pm$ in the ISM. Then, we consider a more realistic situation where we enable the diffusion of particles, with the parameters in Table~\ref{tab:params}, except without reacceleration ($v_A = 0$). Finally, we show the results for the full propagation setup. We see in the left panel of Figure~\ref{fig:diff+vA} that reacceleration promotes low-energy $e^\pm$ to energies well above the input energy from the DM annihilation $\sqrt{s} = 2m_\textrm{DM}$. This causes huge changes in the predicted signal. The right panel of Figure~\ref{fig:diff+vA} shows that, when $\sqrt{s}$ is higher and so is the initial energy of DM-produced $e^\pm$, there is no significant broadening of the spectrum due to reacceleration since much more energy is needed to promote particles above the GeV scale. 

\begin{figure}[t]
    \centering
    \begin{subfigure}[c]{0.49\linewidth}
        \centering
        \includegraphics[width=\linewidth,trim= 0.5cm 0.5cm 0.5cm 0]{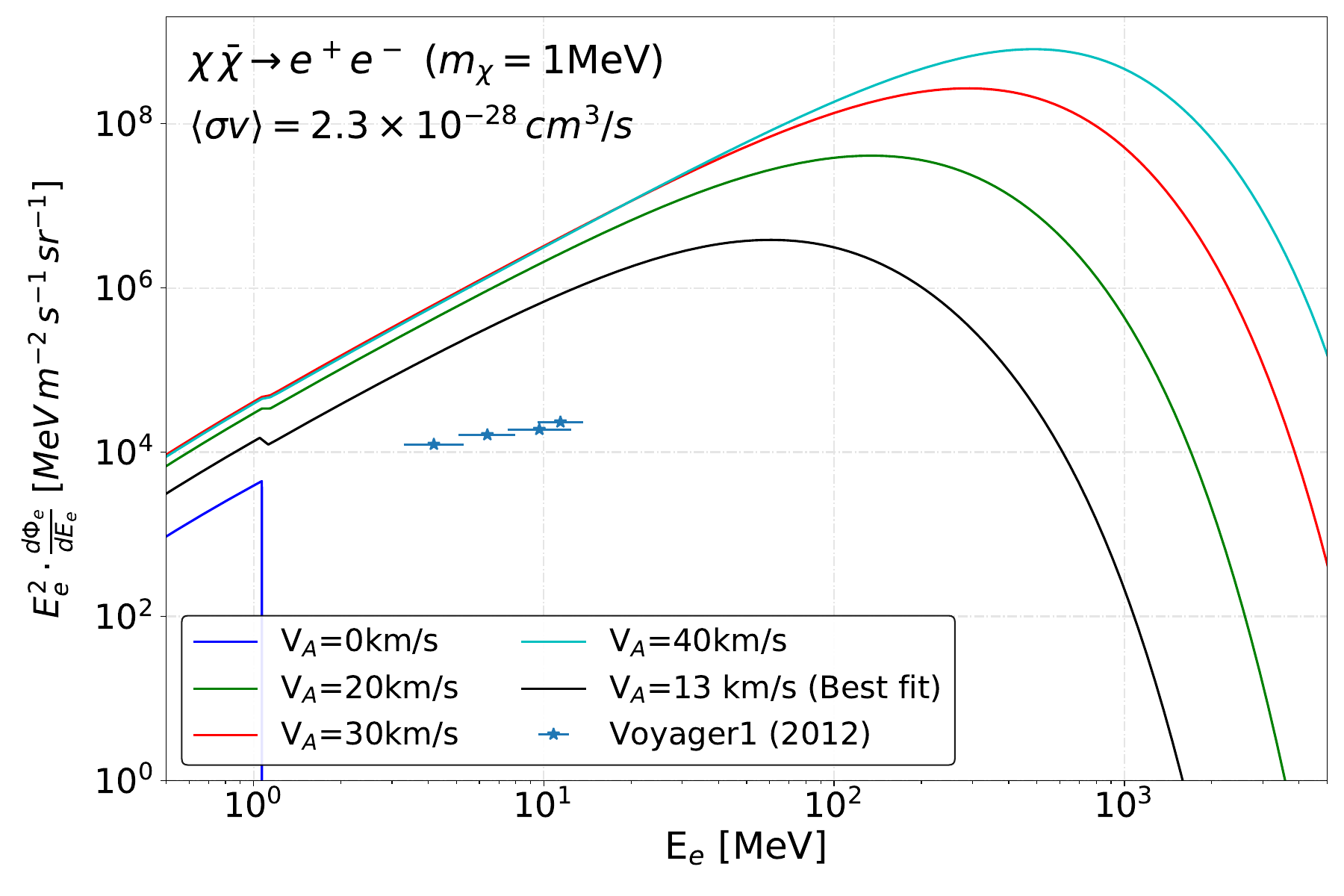}
    \end{subfigure}
    \hfill
    \begin{subfigure}[c]{0.49\linewidth}
        \centering
        \includegraphics[width=\linewidth,trim= 0.5cm 0.5cm 0.5cm 0]{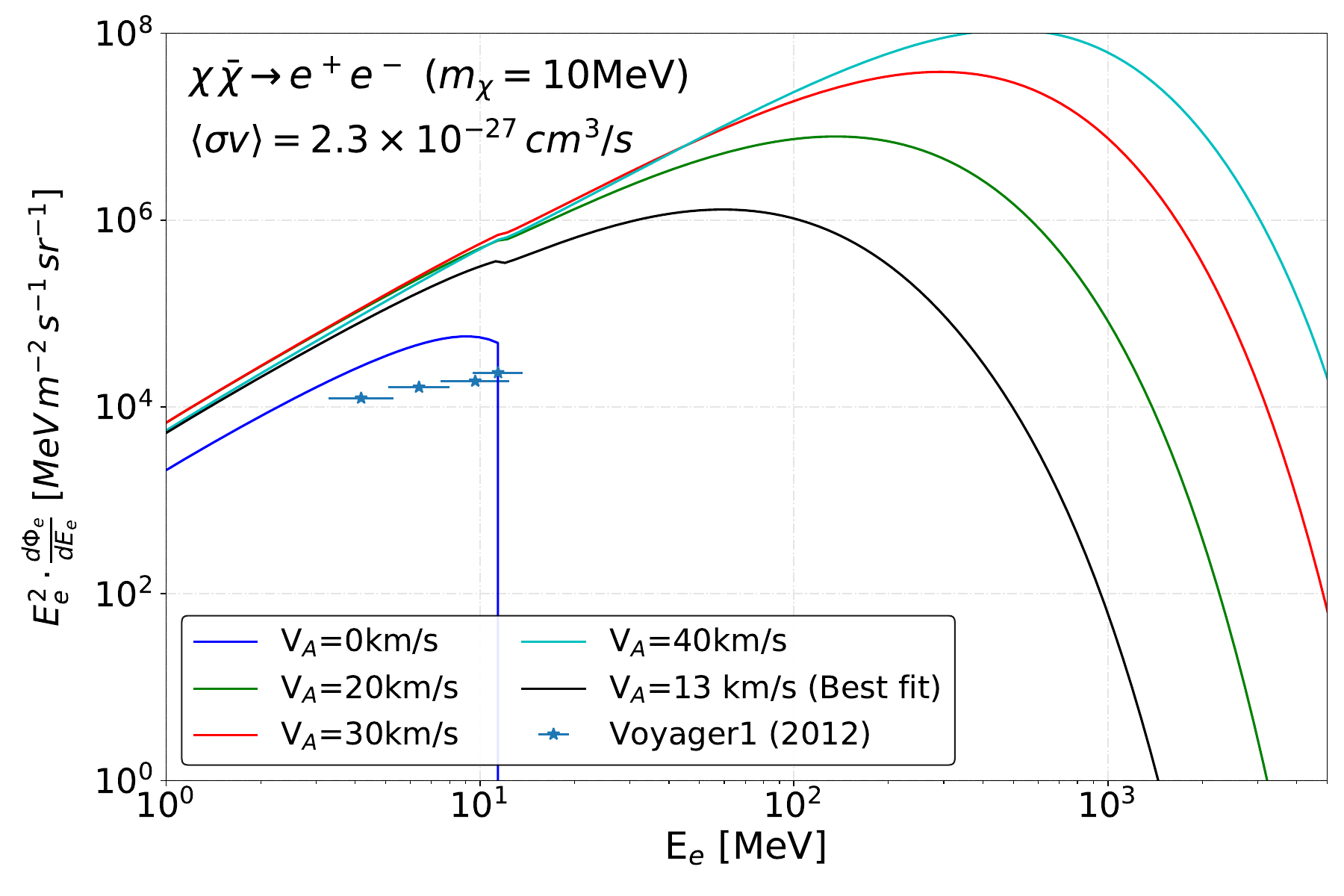}
    \end{subfigure}
    \hfill
    \begin{subfigure}[c]{0.49\linewidth}
        \centering
        \includegraphics[width=\linewidth,trim= 0.5cm 0.5cm 0.5cm 0]{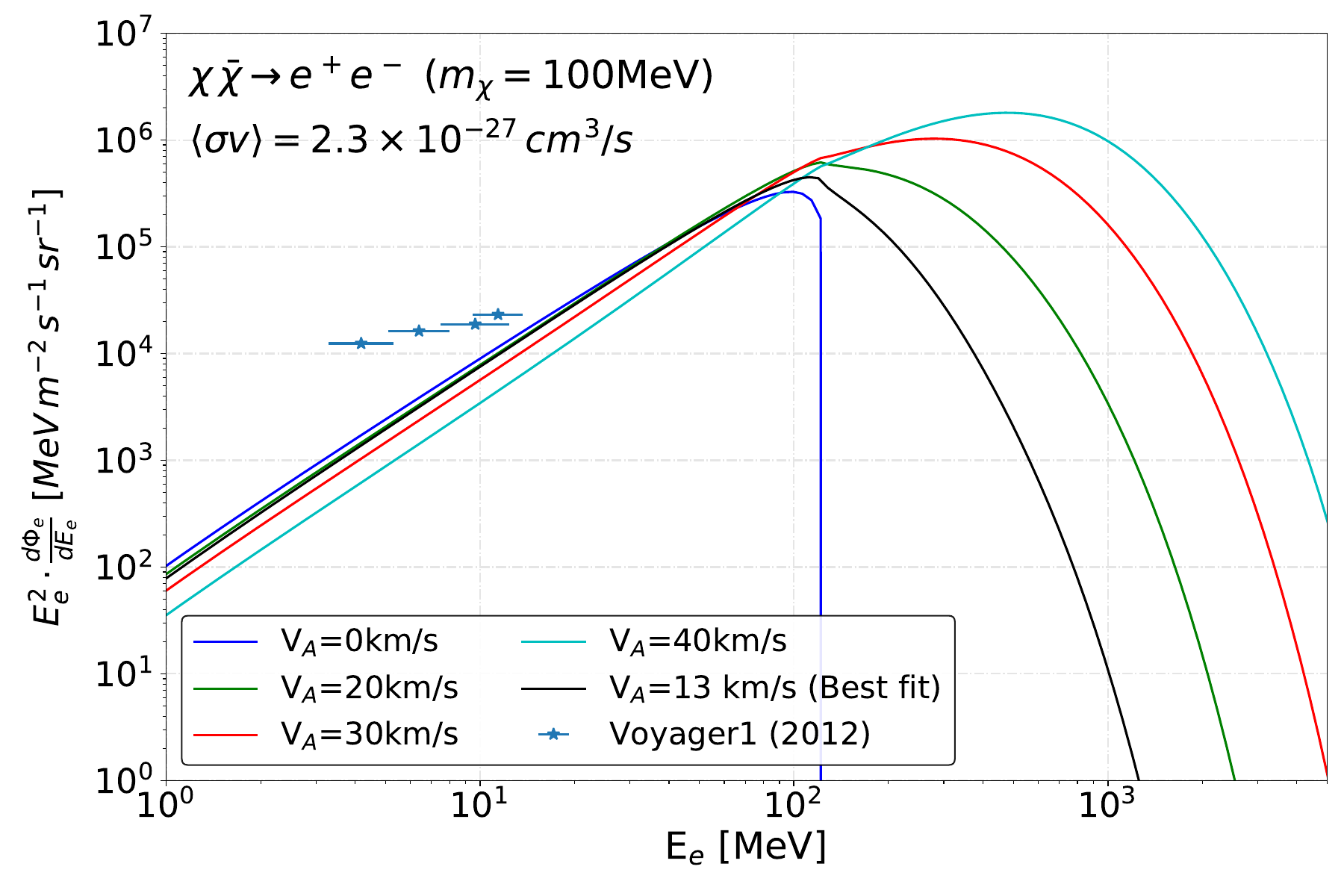}
    \end{subfigure}
    \hfill
    \begin{subfigure}[c]{0.49\linewidth}
        \centering
        \includegraphics[width=\linewidth,trim= 0.5cm 0.5cm 0.5cm 0]{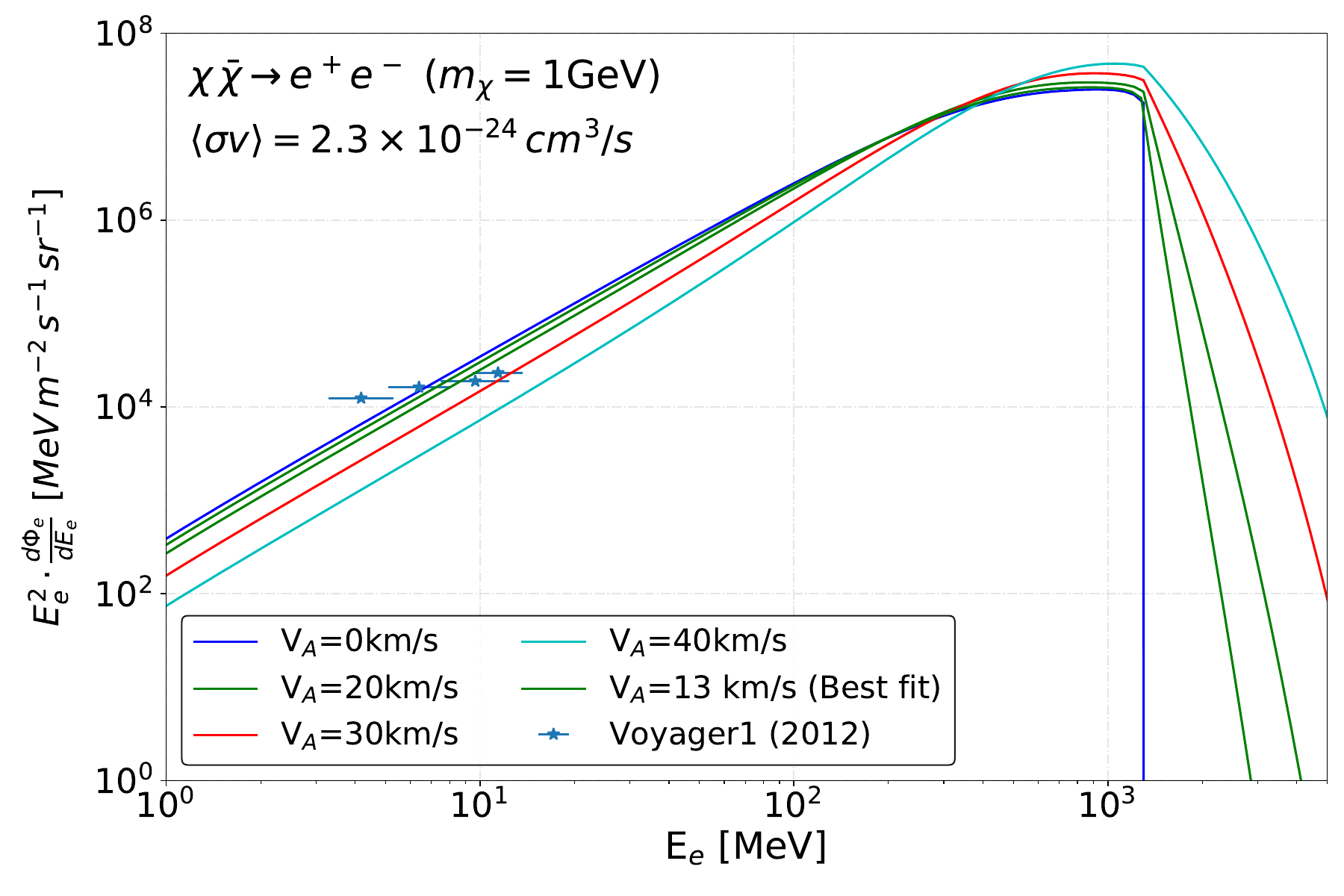}
    \end{subfigure}
    \caption{Comparison of the predicted $e^\pm$ flux from DM annihilating into $e^+e^-$ with {\sc Voyager 1} data. We consider different values of $m_\textrm{DM}$: $1$ MeV (top left panel), $10$ MeV (top right panel) and $100$ MeV (bottom left) and $1$ GeV (bottom right panel). We show the {\sc Voyager 1} data points in blue and the predicted DM-produced $e^\pm$ flux for various reacceleration scenarios: $v_a=0$ (blue), $13$ (black), $20$ (green), $30$ (red) and $40$ (cyan) km/s.}
    \label{fig:Va_ee_Scan_Voy}
\end{figure}

Given the importance of reacceleration in the constraints on low-mass DM particles, we show, in Figure~\ref{fig:Va_ee_Scan_Voy}, the local $e^\pm$ spectrum produced by annihilating DM into $e^+ e^-$, for DM masses from $1$ MeV (top left panel) to $1$ GeV (bottom right panel) and different values of the Alfv\'en velocity, ranging from $v_A=0$ km/s to $v_A=40$ km/s. In each case, the predicted best-fit value of $v_A$ is indicated as a black line, and constitutes our fiducial prediction. We do not consider values greater than $v_A=40$ km/s, following the conclusions of recent detailed CR analyses~\cite{DeLaTorreLuque:2021nxb,Weinrich:2020cmw} and because a larger value of $v_A$ would imply that the CR acquire more energy via interactions with interstellar turbulence than from their injection in supernova remnants~\cite{2006ApJ...642..902P}. For DM masses below the energy range of {\sc Voyager 1} data~\cite{Stone:2013zlg}, even a low level of reacceleration ($v_A\gtrsim 5$~km/s) is enough to move the peak of the DM-produced $e^\pm$ flux beyond the energy range probed by {\sc Voyager 1}, thus strengthening DM constraints.

\subsection{Secondary emissions of photons}

Once the distribution in the Galaxy of the propagated DM-produced $e^\pm$ is obtained, we make use of the \verb|HERMES| code~\cite{Dundovic:2021ryb} to integrate along the l.o.s.\ the CR emissions obtained with \verb|DRAGON2|, using detailed ambient SL and IR models~\cite{Vernetto:2016alq} to get high-resolution sky maps of the diffuse $X$- and soft $\gamma$-ray emission at the relevant energies. With this setup, we compute the ICS emission from DM-produced $e^\pm$ interacting with the ambient photons. Following~\cite{Vernetto:2016alq}, we estimate that the uncertainties coming from the SL and IR model used must be lower than $30\%$.

\begin{figure}[t]
    \centering
    \begin{subfigure}[c]{0.49\linewidth}
        \centering
        \includegraphics[width=\linewidth,height=0.22\textheight,trim= 0.5cm 0 1cm 2cm]{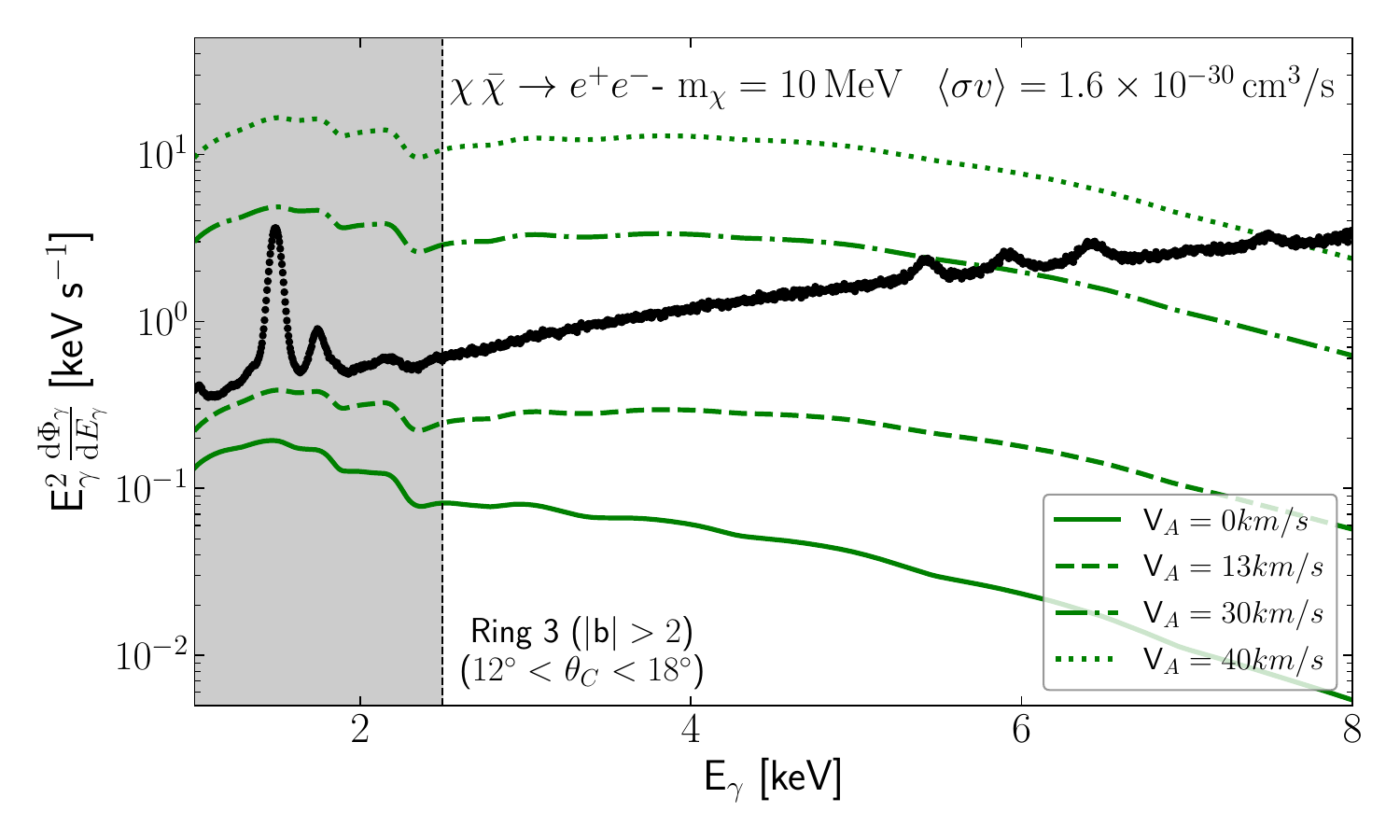}
    \end{subfigure}
    \hfill
    \begin{subfigure}[c]{0.49\linewidth}
        \centering
        \includegraphics[width=\linewidth,height=0.22\textheight,trim= 0.5cm 0 1cm 2cm]{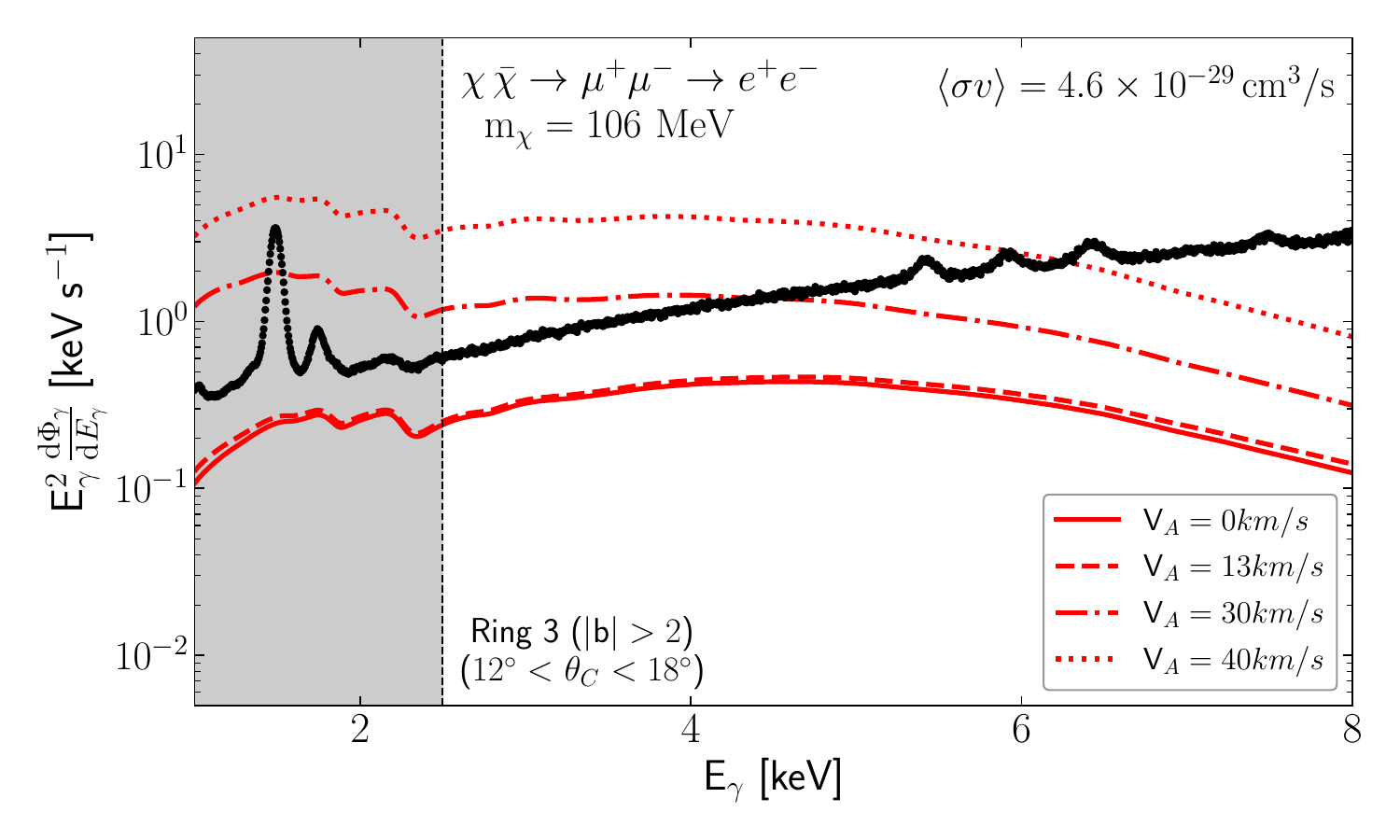}
    \end{subfigure}
    \caption{Comparison of {\sc Xmm-Newton/Mos} data in the third ring ($12^\circ < \theta < 18^\circ$) with the predicted DM-induced $X$-ray signal in that region, for DM with $m_\textrm{DM}=10$ MeV annihilating in $e^+e^-$ (left panel) and with $m_\textrm{DM} \simeq m_\mu$ annihilating in $\mu^+ \mu^-$ (right panel) for different levels of reacceleration: $v_A$ values of $0$ (solid), $13$ (dashed), $30$ (dot-dashed) and $40$ km/s (dotted).}
     \label{fig:XMM_VA}
\end{figure}

Given the high impact of reacceleration and propagation in the DM-produced $e^\pm$ signals, the associated secondary radiations will be similarly affected. Here we focus on the effect that the diffusion setup has on the constraints obtained from {\sc Xmm-Newton/Mos} data~\cite{Foster:2021ngm}, since we demonstrated in Chapter~\ref{chap:subGeV} that this dataset leads to the most constraining DM limits among all the current astrophysical experiments sensitive to the keV$-$MeV energy range. We illustrate the predicted $X$-ray signals in the third ring ($12^\circ < \theta < 18^\circ$; the most constraining one) for different values of $v_A$ in Figure~\ref{fig:XMM_VA}, in the case of DM with $m_\textrm{DM}= 10$ MeV annihilating into $e^+e^-$ (left panel) and with $m_\textrm{DM} \simeq m_\mu$ annihilating in $\mu^+ \mu^-$. Since we consider only the range of $2.5-8$ keV to obtain our constraints, due to the background noise in the detector at lower and higher energies respectively, we shade the energy region below $2.5$ keV in these plots.

As one can see, for $m_\textrm{DM} \lesssim 100$ MeV, the $X$-ray signals change very significantly for different $v_A$ values, while, for $m_\textrm{DM} \gtrsim 100$ MeV, the difference between the predicted signals using no reacceleration becomes similar to the ones predicted from our best-fit setup. More extreme $v_A$ values still lead to sizeable differences up to $m_\textrm{DM} \simeq 1$ GeV. This means that our limits including reacceleration will dramatically strengthen our limits for low DM masses in the case of DM annihilating or decaying into $e^+e^-$, whereas for the other channels the impact will be mild. Once again, bremsstrahlung emissions have a negligible contribution to these signals across the whole DM mass range, due to the masking of the GP in the dataset we use.

\section{Results and discussion}
\label{sec:results2}

In order to derive the limits on $\langle\sigma v\rangle$ and $\Gamma$, we use the same test statistic as in Chapter~\ref{chap:subGeV} (Equation~\ref{eq:chi2}) for each DM mass and DM annihilation or decay channel, which we perform to over all of the rings in the {\sc Xmm-Newton/Mos} dataset. We also perform this test to compare the local flux of DM-produced $e^\pm$ with measurements from {\sc Voyager 1} in order to derive bounds complementary to the {\sc Xmm-Newton} ones. We insist that the bounds are conservative, both for the case of {\sc Xmm-Newton} and {\sc Voyager 1}, since the several backgrounds that are expected, such as astrophysical emissions or extragalactic background light, are not included in the analysis. This insures a certain level of robustness in our limits, due to the uncertainties involved in the prediction of such backgrounds.

\begin{figure}[t]
    \centering
    \begin{subfigure}[c]{0.49\linewidth}
        \centering
        \includegraphics[width=\linewidth,trim= 0.5cm 0 1.5cm 2cm]{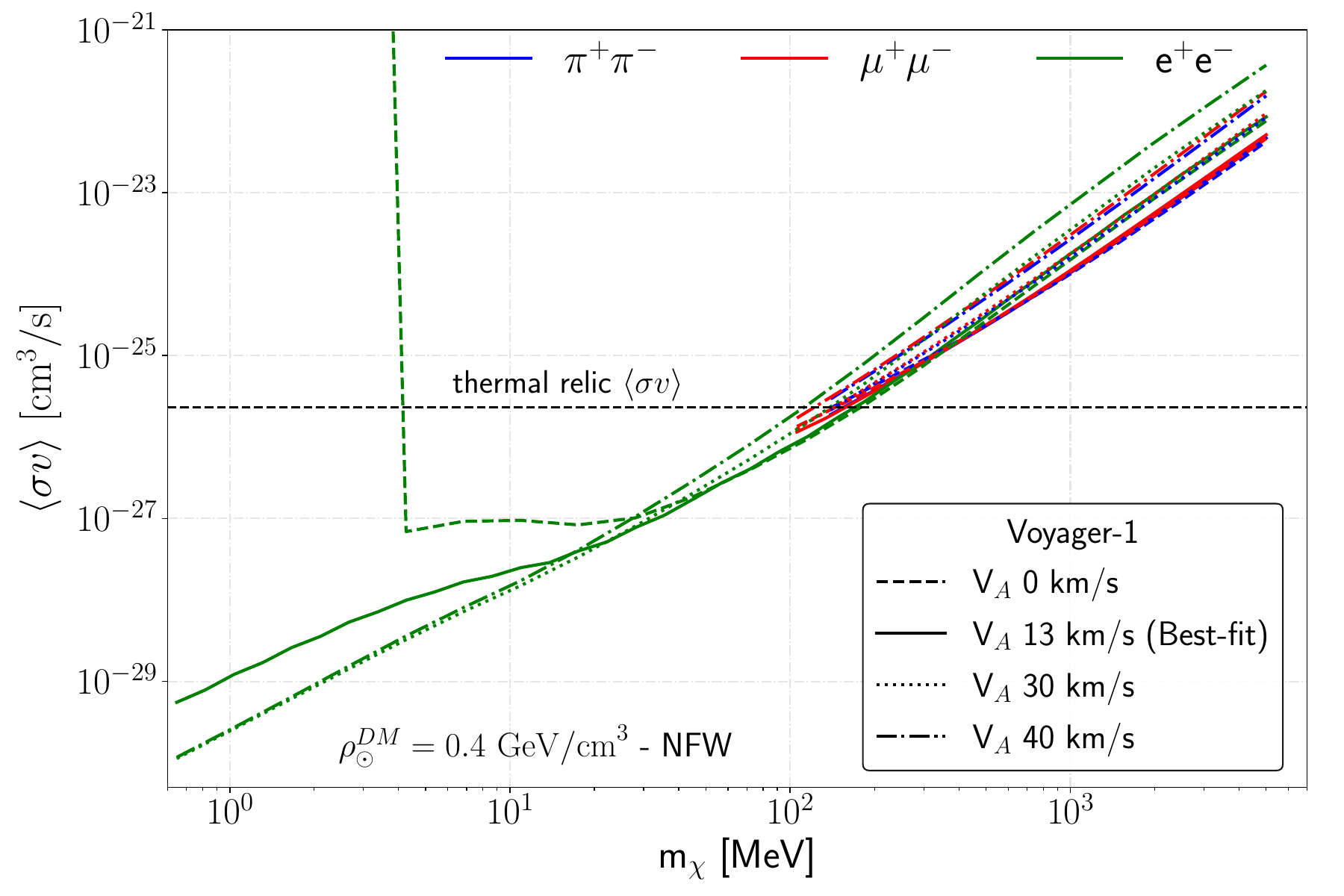}
    \end{subfigure}
    \hfill
    \begin{subfigure}[c]{0.49\linewidth}
        \centering
        \includegraphics[width=\linewidth,trim= 0.5cm 0 1.5cm 2cm]{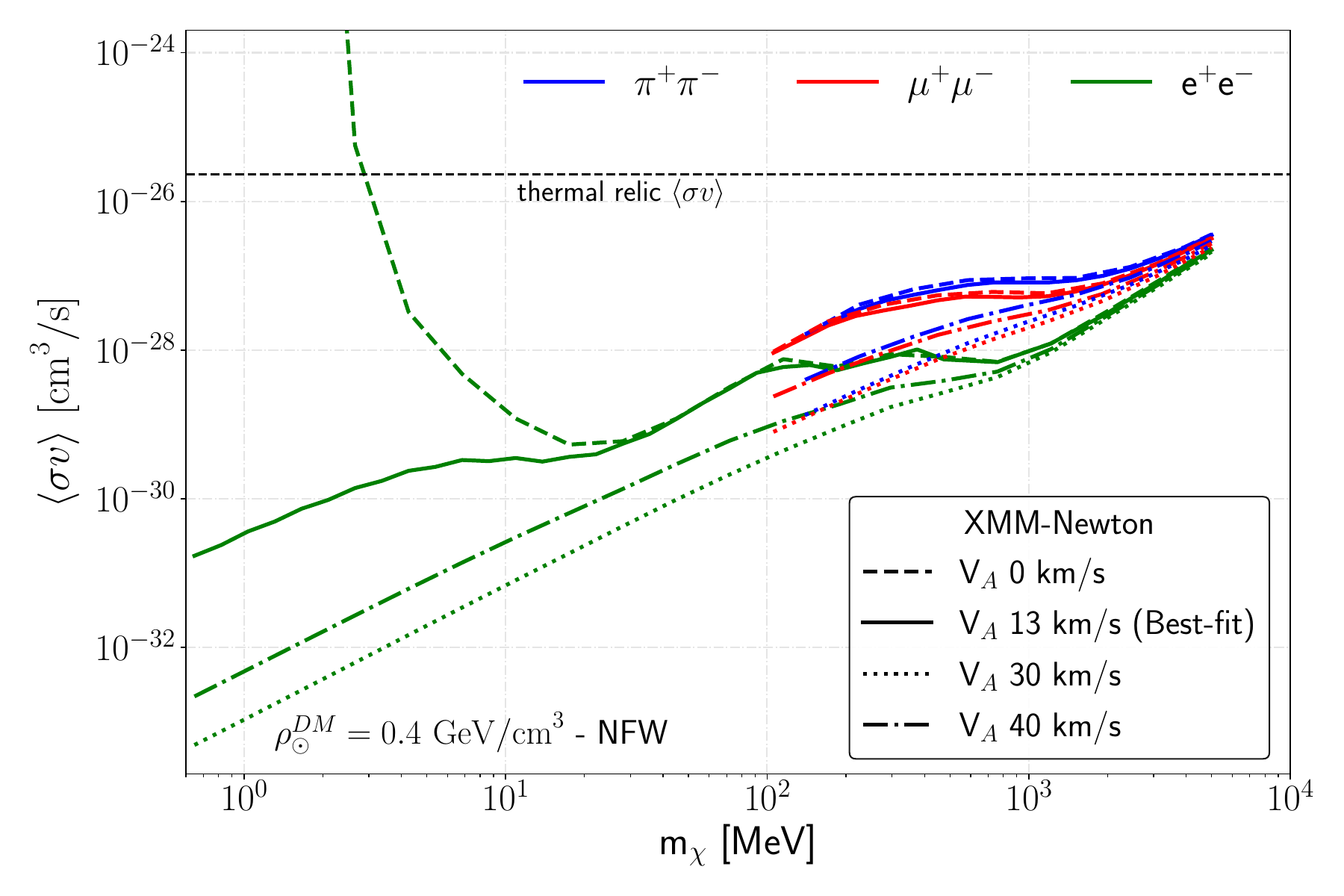}
    \end{subfigure}
    \caption{Limits on the DM annihilation cross section $\langle \sigma v \rangle$ derived using {\sc Voyager 1} data (left panel) and from the combined analysis of all rings from the {\sc Mos} dataset (right panel), for different levels of reacceleration, and for the $\pi^+ \pi^-$ (blue), $\mu^+ \mu^-$ (red) and $e^+ e^-$ (green) channels. We consider Alfv\'en velocities $v_A$ of $0$ (dashed), $13$ (\emph{i.e.}\ our best-fit value; solid), $30$ (dot-dashed) and $40$ (dotted) km/s. To facilitate the comparison of the results for different levels of reacceleration, we do not include prompt emission in the right panel.}
     \label{fig:lims_VA}
\end{figure}

To emphasise the importance of using a more realistic model of $e^\pm$ diffusion, including their reacceleration, we show in Figure~\ref{fig:lims_VA} a comparison of the bounds derived from {\sc Voyager 1} data (left panel) and {\sc Xmm-Newton} (right panel) for different values of $v_A$, from $0$ km/s (no reacceleration) to $40$ km/s, leaving the remaining propagation parameters untouched. Bounds on the cross section in the $\pi^+ \pi^-$, $\mu^+ \mu^-$ and $e^+ e^-$ final states are indicated as blue, red and green lines, respectively. We remark that, in the case of no reacceleration, we cannot constrain $\langle\sigma v \rangle$ for DM lighter than a few MeV using {\sc Voyager 1}, given that the DM-produced $e^\pm$ spectra lie below the {\sc Voyager 1} data (see the top left panel of Figure~\ref{fig:Va_ee_Scan_Voy}). The same occurs for {\sc Xmm-Newton}, since the energy of DM-produced $e^\pm$ is too low to up-scatter the ambient photons to the energies in the range considered for {\sc Xmm-Newton}. To clearly show what the effect of reacceleration on these limits is, we do not include the prompt emissions of photons in the right panel, which can only lead to a difference in the limits for the case of no reacceleration and for DM lighter than $10$ MeV. As we previously said, the effect of reacceleration is huge towards low DM masses, increasing the flux of DM-produced $e^\pm$, in turn the photon flux from ICS and consequently strengthening the limits.

\begin{figure}[t]
    \centering
    \begin{subfigure}[c]{0.49\linewidth}
        \centering
        \includegraphics[width=\linewidth,trim= 0.5cm 0 2.5cm 2cm]{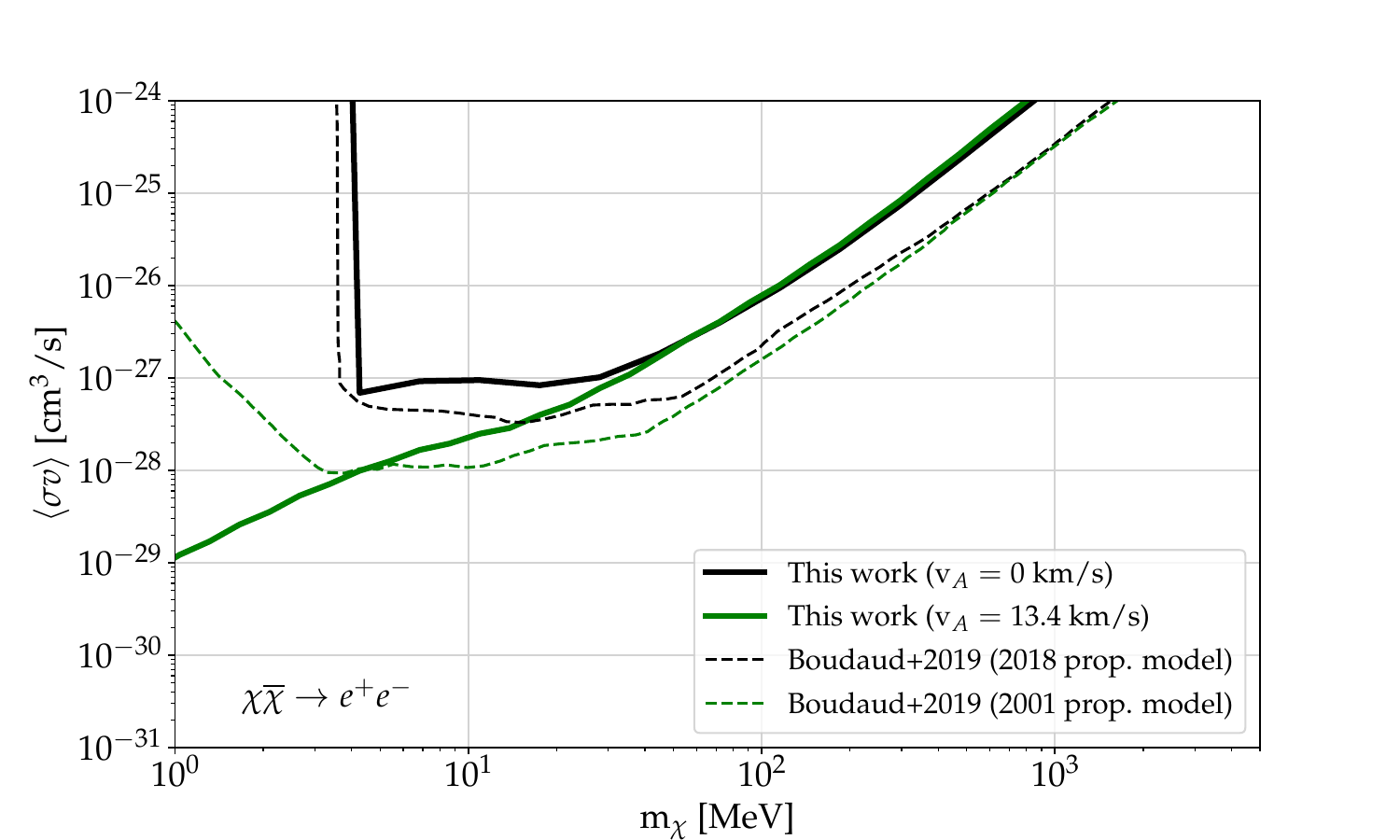}
    \end{subfigure}
    \hfill
    \begin{subfigure}[c]{0.49\linewidth}
        \centering
        \includegraphics[width=\linewidth,trim= 0.5cm 0 2.5cm 2cm]{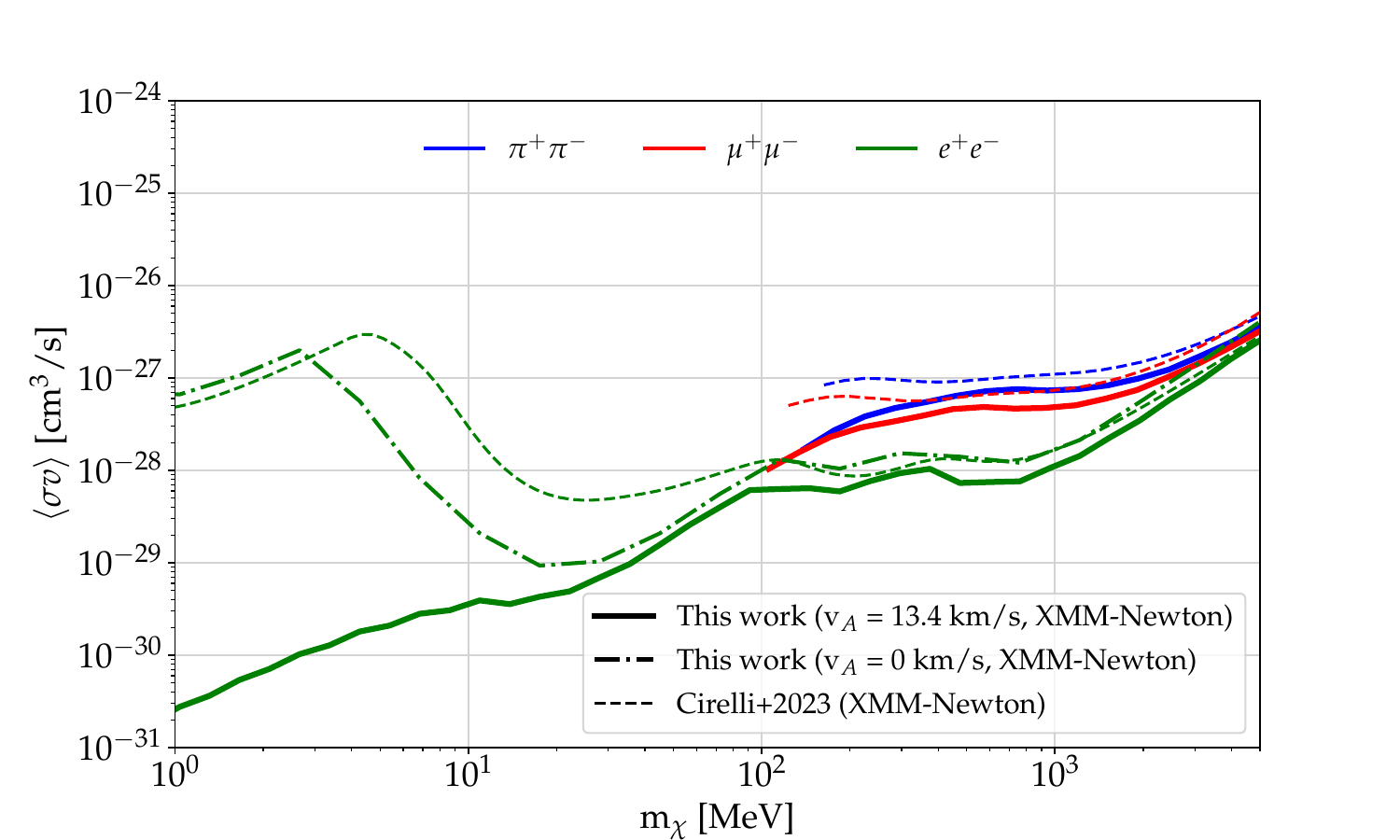}
    \end{subfigure}
    \hfill
    \begin{subfigure}[c]{0.49\linewidth}
        \centering
        \includegraphics[width=\linewidth,trim= 0.5cm 0 3cm 1cm,clip]{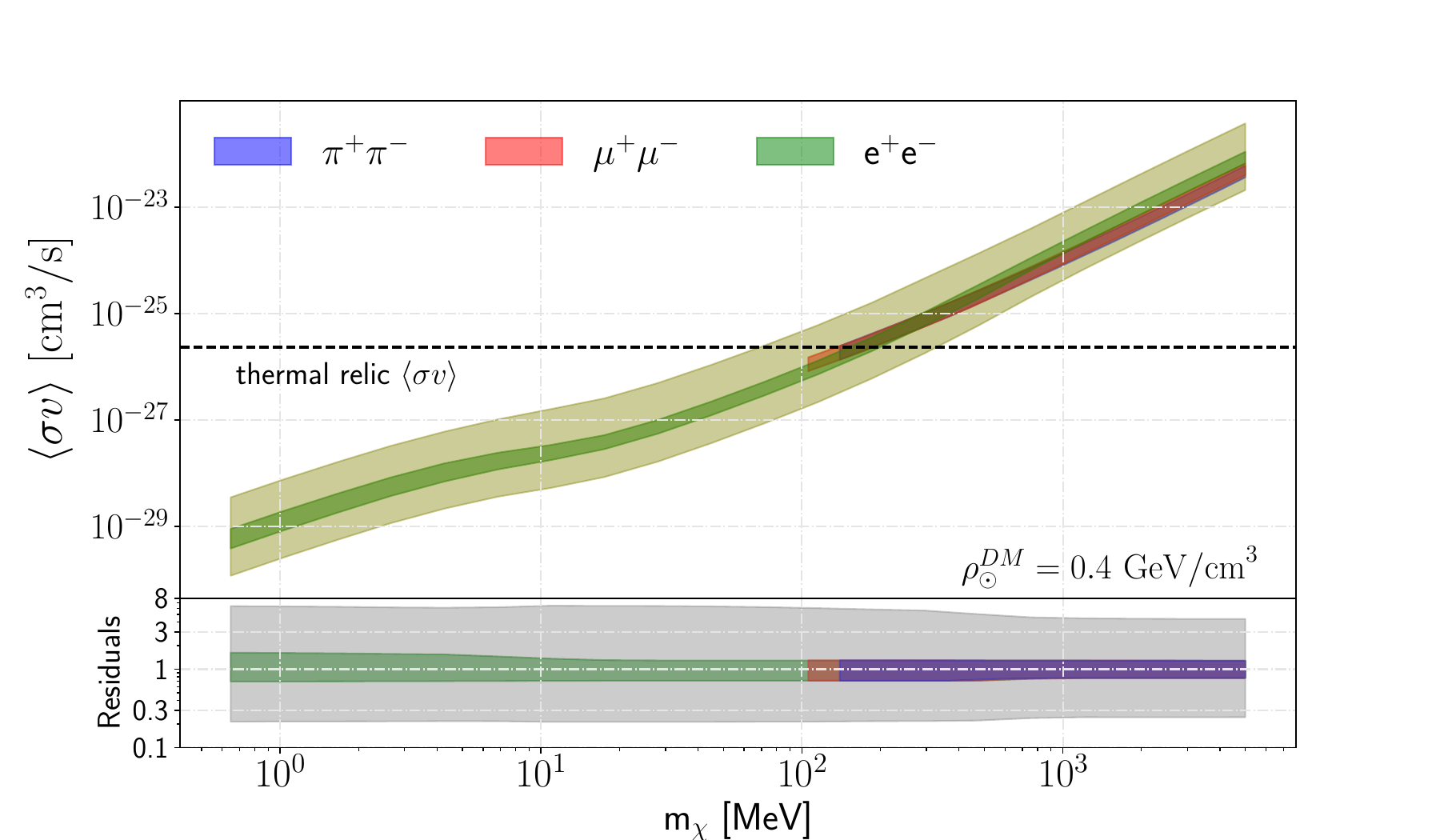}
    \end{subfigure}
    \hfill
    \begin{subfigure}[c]{0.49\linewidth}
        \centering
        \includegraphics[width=\linewidth,trim= 0.5cm 0 3cm 1cm,clip]{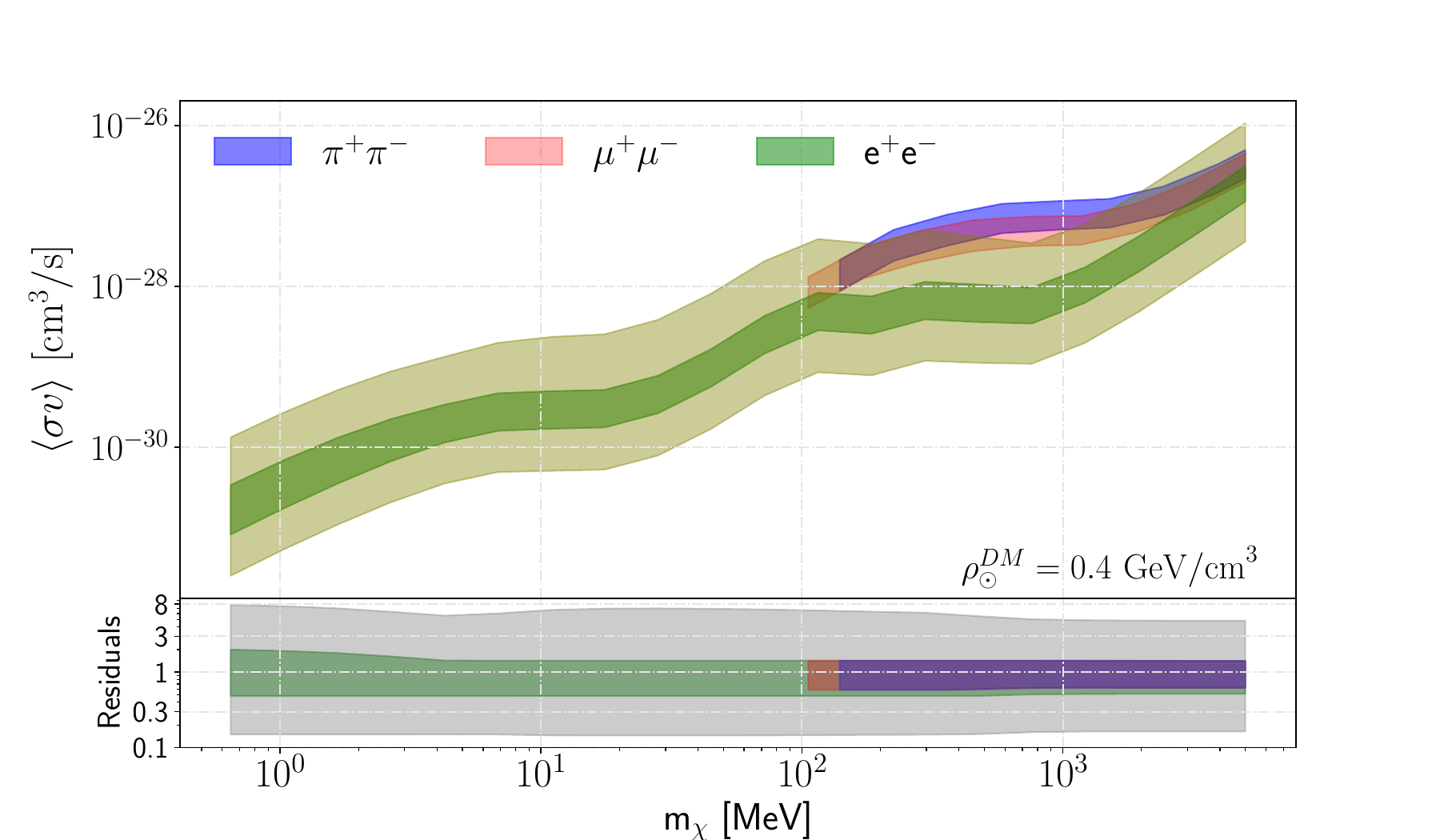}
    \end{subfigure}
    \caption{The top panels show the limits on annihilating DM from {\sc Voyager 1}  and from {\sc Xmm-Newton}, compared with previous constraints using the same datasets. In the top left panel we show our {\sc Voyager 1} bounds for two different $v_A$ values: $0$ (solid black line) and the best-fit value $13.4$ km/s (solid green line). The bounds are plotted alongside with previous {\sc Voyager~1} bounds where two propagation models were used: one without reacceleration from~\cite{Reinert:2017aga} (dashed black line) and one with reacceleration from~\cite{Maurin:2001sj, Donato:2003xg} (dashed green line). In the top right panel we show our {\sc Xmm-Newton} bounds for $v_A = 0$ (dot-dashed line) and $13.4$ km/s (solid lines) with the ones in Chapter~\ref{chap:subGeV} (dashed lines) for the $\pi^+\pi^-$ (blue line), $\mu^+ \mu^-$ (red line) and $e^+ e^-$ (green line) channels respectively. Bottom panels show the estimated uncertainties in our predictions, as discussed in the main text.}
     \label{fig:new+oldbounds}
\end{figure}

Then, we discuss the difference between our updated bounds with the previous ones that used the same datasets, illustrated by the top panels of Figure~\ref{fig:new+oldbounds}. The authors of~\cite{Boudaud:2016mos} published the first constraints on sub-GeV DM using {\sc Voyager 1}. However, their evaluation of $e^\pm$ propagation differs significantly from the one we use here, which considers the most recent CR data. They use the semi-analytical code {\sc USINE}~\cite{Maurin:2018rmm}, which employs an effective treatment to account for the energy losses, called the `pinching method'~\cite{Boudaud:2016jvj}. In addition, we also insist on the effect of reacceleration in our constraints, which was also treated differently in {\sc USINE}. More importantly, the authors of~\cite{Boudaud:2016mos} used a parametrisation obtained from a study in 2001~\cite{Maurin:2001sj}, when the precision of data was more limited and the systematic uncertainties (\emph{e.g.}, on spallation cross sections) was less under control, meaning that their diffusion coefficient differs significantly from our updated one. Specifically, they use a parametrisation with fixed $\eta = 1$ (\emph{i.e.}\ $D\propto \beta$, see Equation~\ref{eq:diff_eq}), while the one obtained in analyses of recent {\sc Ams-02} data when including this parameter in the fit is negative~\cite{DeLaTorreLuque:2021nxb, Weinrich:2020cmw} and, in our case, $\eta = -0.75$, implying much more confinement at the relevant sub-GeV energies considered here. This means that, in the sub-GeV regime, our diffusion coefficient is quite different, which can lead to important differences in our predicted signals and constraints. Moreover, the limits from~\cite{Boudaud:2016jvj} are obtained with a halo height value of $L\simeq15$~kpc, which explains why they are a factor of a few stronger than ours above a few tens of MeV.

In the top left panel of Figure~\ref{fig:new+oldbounds}, we are also including the limit derived by the same authors of~\cite{Boudaud:2016jvj} using a more updated diffusion coefficient (from 2018, coming from~\cite{Reinert:2017aga}), and no reacceleration. We have taken these bounds from a 2019 talk by M. Boudaud~\cite{BoudaudTalk} which we refer to as `Boudaud+2019' with the corresponding propagation models in Figure~\ref{fig:new+oldbounds} shown in dashed black and dashed green respectively. These comparisons evidence the fact that, given the uncertainties in the evaluation of the diffusion coefficient at low energies and the halo height, different propagation setups can differ significantly, and therefore, it would be crucial moving forward to update the constraints we derive here with updated propagation setups as more CR data becomes available at low energies and uncertainties are further reduced. We then compare our results with those obtained in Chapter~\ref{chap:subGeV}, which we refer to as `Cirelli+2023', shown as a dashed green line in the top right panel of Figure~\ref{fig:new+oldbounds}. One can also expect differences from one diffusion setup to another. In particular, in Chapter~\ref{chap:subGeV} we implemented a minimalistic propagation setup, only including the injection of DM-produced $e^\pm$ and their energy losses through the ISM. In this case, the resultant bounds only differ slightly for DM masses above the some tens of MeV due to the use of older ambient SL and IR photon maps in Chapter~\ref{chap:subGeV}, which tend to have a less significant optical component. Remarkably, we are able to extend the constraining power to DM lighter than $100$ MeV thanks to the incorporation of reacceleration in our scheme. Reacceleration dramatically increases the energy of low-mass DM-produced $e^\pm$. The energy of the associated photons from ICS thus increases and therefore reach the energy range of the {\sc Xmm-Newton} data. We can also point out that, for $m_\textrm{DM} \lesssim 10$ MeV, the limit from FSR is subdominant by orders of magnitude compared to ICS from reaccelerated $e^\pm$.

In the bottom left panel of Figure~\ref{fig:new+oldbounds}, we show the uncertainties related to our derived limits for {\sc Voyager 1}, obtained by varying the propagation parameters in Table~\ref{tab:params} up to their $1\sigma$ uncertainties, but also the value of the DM density at Earth (taken to be $\rho_{\odot}=0.420^{+0.011}_{-0.009} \pm0.025$~GeV/cm$^3$ from~\cite{Pato:2015dua}) and include a conservative $10\%$ factor to account for uncertainties in the gas distribution and energy losses. In the case of the limit from {\sc Xmm-Newton}, in the bottom right of Figure~\ref{fig:new+oldbounds}, we include also a quite conservative $30\%$ factor that accounts for uncertainties in the SL and IR components of the ambient photons, as reported in~\cite{Vernetto:2016alq}. In addition, we have estimated the uncertainties by considering two extremal DM density distributions, similarly to Chapter~\ref{chap:subGeV}: a Burkert profile (Equation~\ref{eq:Burkert}) which leads to significantly weaker constraints, and a cNFW profile (Equation~\ref{eq:gNFW} with $\gamma = 1.26$) that leads to stronger constraints. In Figure~\ref{fig:new+oldbounds} we display in olive, for the $e^+e^-$ channel, the uncertainty band representing the impact of these two extremal profiles on the limits, on top of the other astrophysical uncertainties, represented in green. Given that these uncertainties are expected to be roughly the same for the other channels, we indicate them as a grey band in the residual panels.

\begin{figure}[p]
	\centering
	\begin{subfigure}[c]{0.9\linewidth}
		\centering
		\includegraphics[width=\linewidth]{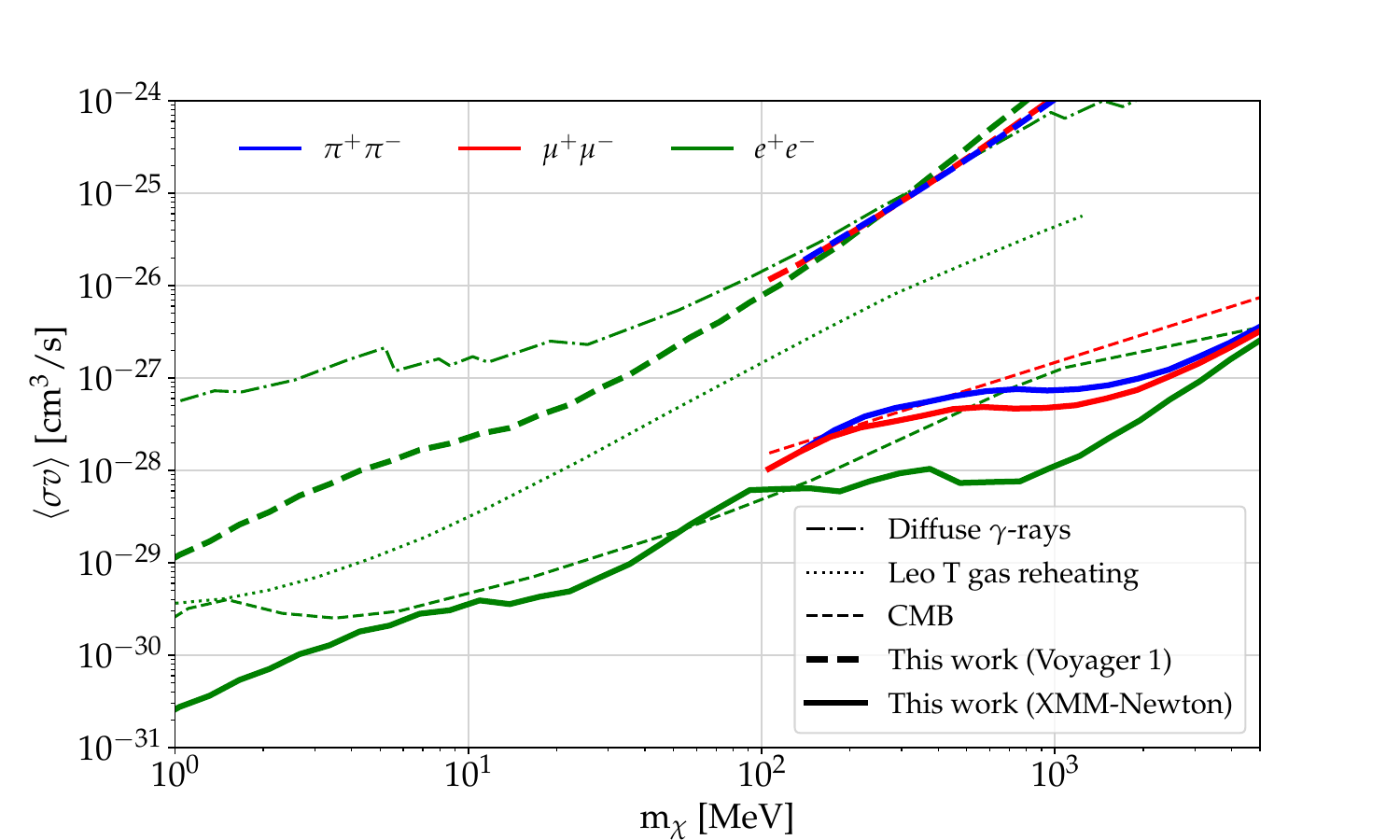} 
	\end{subfigure}
	\caption{Comparison of the bounds on annihilating DM derived with the best-fit propagation parameters (thick dashed for {\sc Voyager 1} and solid lines for {\sc Xmm-Newton} respectively) with other existing constraints. We show the diffuse $\gamma$-ray bound from~\cite{Essig:2013goa} (dot dashed line), the CMB bounds from~\cite{Slatyer:2015jla} and~\cite{Lopez-Honorez:2013cua} (dashed lines) and the bounds from gas reheating in the Leo T dwarf galaxy from~\cite{Wadekar:2021qae} (dotted line). Once again we show the channels $\pi^+\pi^-$ (blue), $\mu^+\mu^-$ (red) and $e^+e^-$ (green) channels, respectively.}
	\label{fig:bounds+litterature_ann} 
	\begin{subfigure}[c]{0.9\linewidth}
		\centering
		\includegraphics[width=\linewidth]{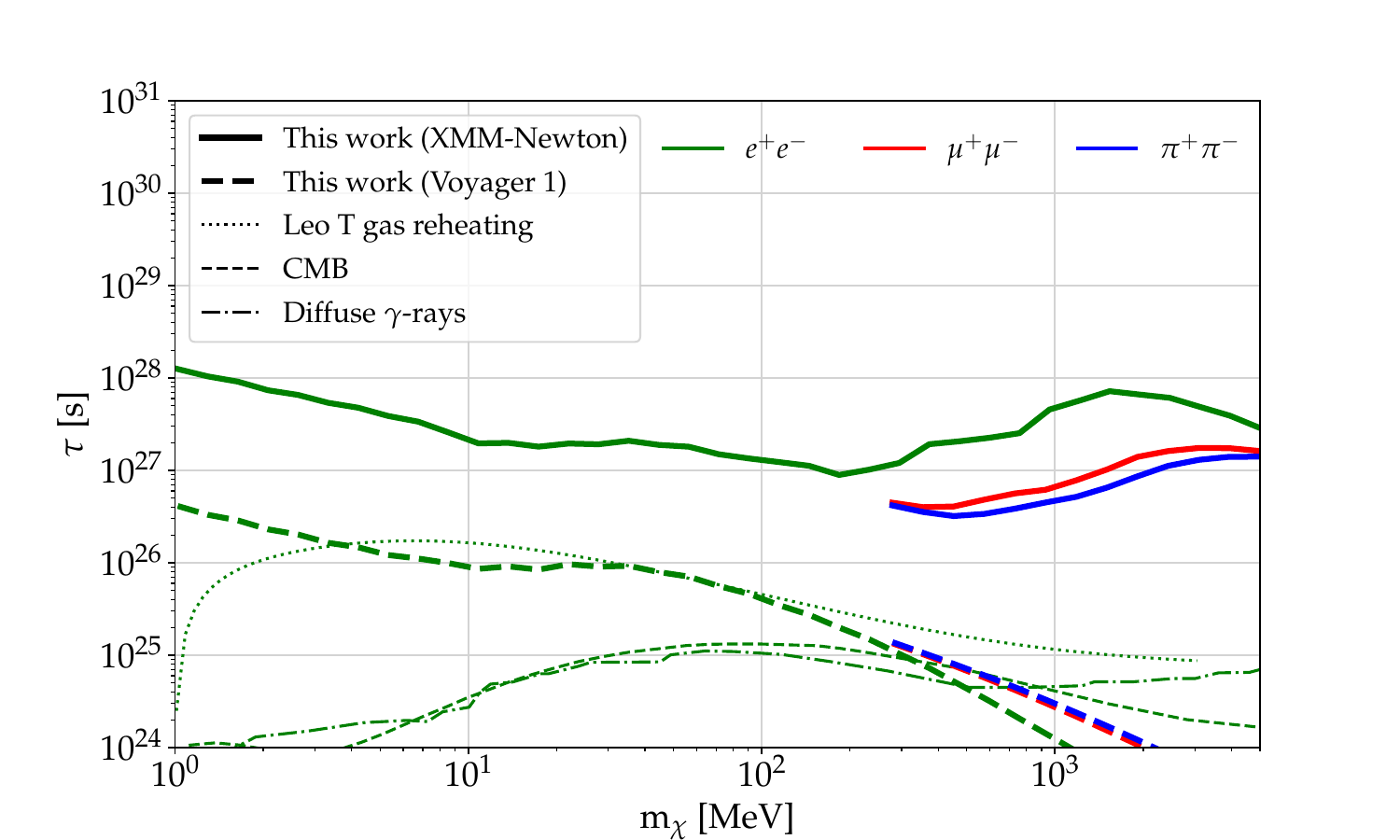} 
	\end{subfigure}
	\caption{Comparison of the bounds on annihilating DM derived with the best-fit propagation parameters (thick dashed for {\sc Voyager 1} and solid lines for {\sc Xmm-Newton} respectively) with other existing constraints. The constraints and the references are the same as in Figure~\ref{fig:bounds+litterature_ann}, except that the CMB ones are derived in Liu et al.~\cite{Liu:2016cnk}.}
	\label{fig:bounds+litterature_dec}
\end{figure}

\section{Comparison to other work}
\label{sec:comparison2}

Once again we compare our fiducial bounds on annihilating and decaying DM obtained with other ones from the literature. We refer the reader to the discussion in Section~\ref{sec:comparison} for more information about the latter. In summary, using our model of $e^\pm$ propagation, gas and ambient photon maps and a NFW profile, the bound we derive on DM annihilating into $e^+ e^-$ and $\mu^+ \mu^-$, using {\sc Xmm-Newton} is more stringent than the CMB one across almost the entire considered DM mass range, as shown in Figure~\ref{fig:bounds+litterature_ann}. However we emphasise that the CMB bounds are more robust, \emph{i.e.}\ they depend on fewer parameters and have smaller uncertainties. For decaying sub-GeV DM (shown in Figure~\ref{fig:bounds+litterature_dec}), the CMB bounds remain strong but not among the strongest in the literature (unlike the annihilating DM case). This is due to the fact DM clusters at redshifts $z \lesssim 100$, which greatly enhances the DM annihilation rate and therefore the $e^\pm$ injection, whereas the decay rate remains constant. Hence our bounds on decaying DM from {\sc Xmm-Newton} is orders of magnitude more stringent than the CMB ones. In our study we limited ourselves to $s$-wave DM annihilations. In~\cite{Boudaud:2018oya} the authors set bounds on the $p$-wave annihilation cross section using the same {\sc Voyager 1} dataset and CR propagation scheme as in~\cite{Boudaud:2016mos}. One can show that their bounds can vary by less than an order of magnitude if $p$-wave DM is assumed. If the same analysis as~\cite{Boudaud:2018oya} was done in our case, we would expect the same variation in our bounds in the case of $p$-wave DM since we also studied DM annihilations in the MW at the present time, so the DM velocity dispersion would also be the same. We therefore conclude that if DM annihilations were $p$-wave, our constraints would remain stronger than the CMB and Leo T ones.

\section{Summary}

In this chapter, we improved the annihilating and decaying DM constraints from {\sc Xmm-Newton} (see the top right panel of Figure~\ref{fig:new+oldbounds}) by including a realistic propagation setup for $e^\pm$, whose parameters are obtained from detailed analyses of the most recent CR data. This improvement is mainly due to the reacceleration of low-energy DM-produced $e^\pm$ from their interaction with the turbulent component of the GMFs, which drastically increases their energy.

As shown in Figures~\ref{fig:bounds+litterature_ann} and \ref{fig:bounds+litterature_dec}, our fiducial limits exclude thermally averaged cross sections down to $10^{-31}$ cm$^3$\,s$^{-1}$ and decay lifetimes up to $10^{28}\,\textrm{s}$, both for $m_\textrm{DM} \simeq 1$ MeV. These yield the strongest astrophysical constraints for this mass range of DM and surpasses cosmological bounds across a wide range of DM masses as well. However, our {\sc Xmm-Newton} limits suffer from the uncertainties on the propagation parameters, DM profile, Galactic gas and ambient photons densities. This is less the case for our {\sc Voyager 1} bounds, although they are significantly weaker than the {\sc Xmm-Newton} ones. The uncertainties on the bounds are shown in the bottom panels of Figure~\ref{fig:new+oldbounds}.

%% file: Chapters/chap5.tex

\lettrine[lines=3, nindent=1pt]{I}{n} this final chapter, we explore another credible candidate for DM: PBHs. To this day, PBHs with a mass between $\sim 5\times10^{17}$ and $\sim 10^{22}$ g remain totally unconstrained, meaning that, in this mass range, there is a possibility that they can constitute the entirety of DM~\cite{Carr:2020gox}. On the lower end of this range, constraints on $f_\textrm{PBH}$ come from PBH evaporation, which consists of a source of indirect signals from PBHs, in the same manner as decaying DM. In the following, we improve these limits by considering three different diffuse probes of PBH evaporation: i) the flux of PBH-evaporated $e^\pm$, ii) the flux of secondary $X$-rays from the propagation of the $e^\pm$ in the ISM and iii) the flux of $511$ keV photons coming from the annihilation of PBH-produced $e^+$ with free $e^-$ in the ISM. For the first two probes, we employ the same computation scheme as in Chapter~\ref{chap:prop}, \emph{i.e.}\ dealing with the propagation of PBH-produced $e^\pm$ using the numerical code \verb|DRAGON2| with the same propagation setup, and computing the secondary emission of $X$-rays from ICS of these $e^\pm$ on Galactic ambient photons using \verb|HERMES|. This chapter is organised as follows: in Section~\ref{sec:injection} we cover the injection of $e^\pm$ from PBH evaporation, in Section~\ref{sec:diffuse} we explore the computation of the flux from the three aforementioned diffuse probes and finally in Section~\ref{sec:result+discussion} we discuss our main results and compare them with other ones from the literature.

\newpage

\section{Electron-positron injection from primordial BHs}
\label{sec:injection}

Recalling the second part of Section~\ref{subsec:particleprod}, Schwarzschild BHs evaporate over time by emitting particles at a rate described by Equation~\ref{eq:evap}. In this chapter, we however generalise this equation to the case of Kerr (electrically neutral, spinning) BHs, which are characterised by their reduced spin parameter $a^\star \equiv J/(GM^2)$, with $J$ being its angular momentum and $G$ the Newtonian gravitational constant. The temperature of a Kerr BH is written~\cite{Arbey:2019jmj}
\begin{equation}
    \label{eq:BHtemp}
    T = \frac{1}{4\pi GM} \frac{\sqrt{1-{a^\star}^2}}{1+\sqrt{1-{a^\star}^2}}\;,
\end{equation}
retrieving $T = 1/(8\pi GM)$ for Schwarzschild BHs (\emph{i.e.}\ $a^\star = 0$). Equation~\ref{eq:BHtemp} also imposes that $a^\star < 1$, since if $a^\star =1$ then $T=0$, which is forbidden by thermodynamics.
Then, the emission spectrum of primary particles $i$ is given by
\begin{equation}
    \frac{d^2N_i}{dtdE_i} = \frac{1}{2\pi} \sum_\text{d.o.f.} \frac{\Gamma_i(E_i,M,a^\star)}{e^{E'_i/T}\pm 1}\;,
    \label{eq:prim_spec_BH}
\end{equation}
where
\begin{equation}
	E'_i \equiv E_i - m\Omega \quad \textrm{and} \quad \Omega \equiv \frac{a^\star}{2GM\left(1+\sqrt{1-{a^\star}^2}\right)}\;,
\end{equation}
are the total energy of the emitted particle and the BH horizon rotational energy, respectively, and $m = \{-l, ..., l\}$ is the projection on the BH axis of the particle angular momentum $l$. The $\pm$ signs depend on the spin of the particles radiated: $+$ for fermions and $-$ for bosons. Once again, the $\Gamma_i$ are the grey body factors which now depend on $a^\star$. In order to compute the spectra of primary particles, we use the numerical code \verb|BlackHawk|.

Then, evaporated particles can hadronise, decay or emit soft radiations. \verb|BlackHawk| also has the possibility of dealing with such processes by including tables from the particle physics codes \verb|PYTHIA|, \verb|Herwig| and \verb|Hazma|. However, their domains of validity differ, \verb|PYTHIA| and \verb|Herwig| handle processes with particle energies above $10$ GeV very well, whereas \verb|Hazma| excels below the QCD scale ($\sim 250$ MeV). Since we are interested in the production of sub-GeV $e^\pm$, we only use \verb|Hazma| to treat secondary processes, and limit ourselves to its domain of validity, which corresponds to BHs with $M \gtrsim 10^{14.5}$ g. An upper limit on the BH mass can also be defined when the evaporation into $e^\pm$ is not possible anymore (\emph{i.e.}\ $T \ll m_e$) corresponding to $M \lesssim 10^{17.5}$ g. In this BH mass range, $e^\pm$, neutrinos and photons are emitted through evaporation, and to some extent (for lower BH masses) muons and pions. \verb|Hazma| handles the decay of the two latter, as well as FSR from all charged particles.

Although it is believed that PBHs cannot acquire a substantial spin from their production process~\cite{Harada:2020pzb} unless formed in the matter-dominated Universe~\cite{Harada:2017fjm}, it has been argued that they can do so by repeatedly merging with other BHs. Moreover, BHs can also acquire a spin due to the accretion of gas surrounding them. The maximum spin value a BH can obtain through this process is $a_\textrm{lim}^\star \approx 0.998$~\cite{1974ApJ...191..507T}, known as the \textit{Thorne limit}, and can slightly vary depending on the considered accretion model~\cite{Sadowski:2011ka}. A similar limit can be derived for BH mergers~\cite{Kesden:2009ds}. Nevertheless, PBHs formed during the matter-dominated Universe could evade these limits, providing a smoking gun signature of their existence. In our study, we decide to remain agnostic on the PBH production process and consider two extreme benchmarks to quantify the impact of the choice of the spin distribution on our results. We therefore examine the case where all PBHs are Schwarzschild BHs, and another one where they are all near-extremal\footnote{`Extremal' denotes the case where $a^\star=1$, which is forbidden by thermodynamics. Near-extremal is then associated to the class of Kerr BHs for which $a^\star \rightarrow 1$.} Kerr BHs with a spin of $a^\star = 0.9999$.

\begin{figure}[t]
    \centering
    \begin{subfigure}[c]{0.49\linewidth}
        \centering
        \includegraphics[width=\linewidth,trim= 0.5cm 0 2.5cm 1.5cm]{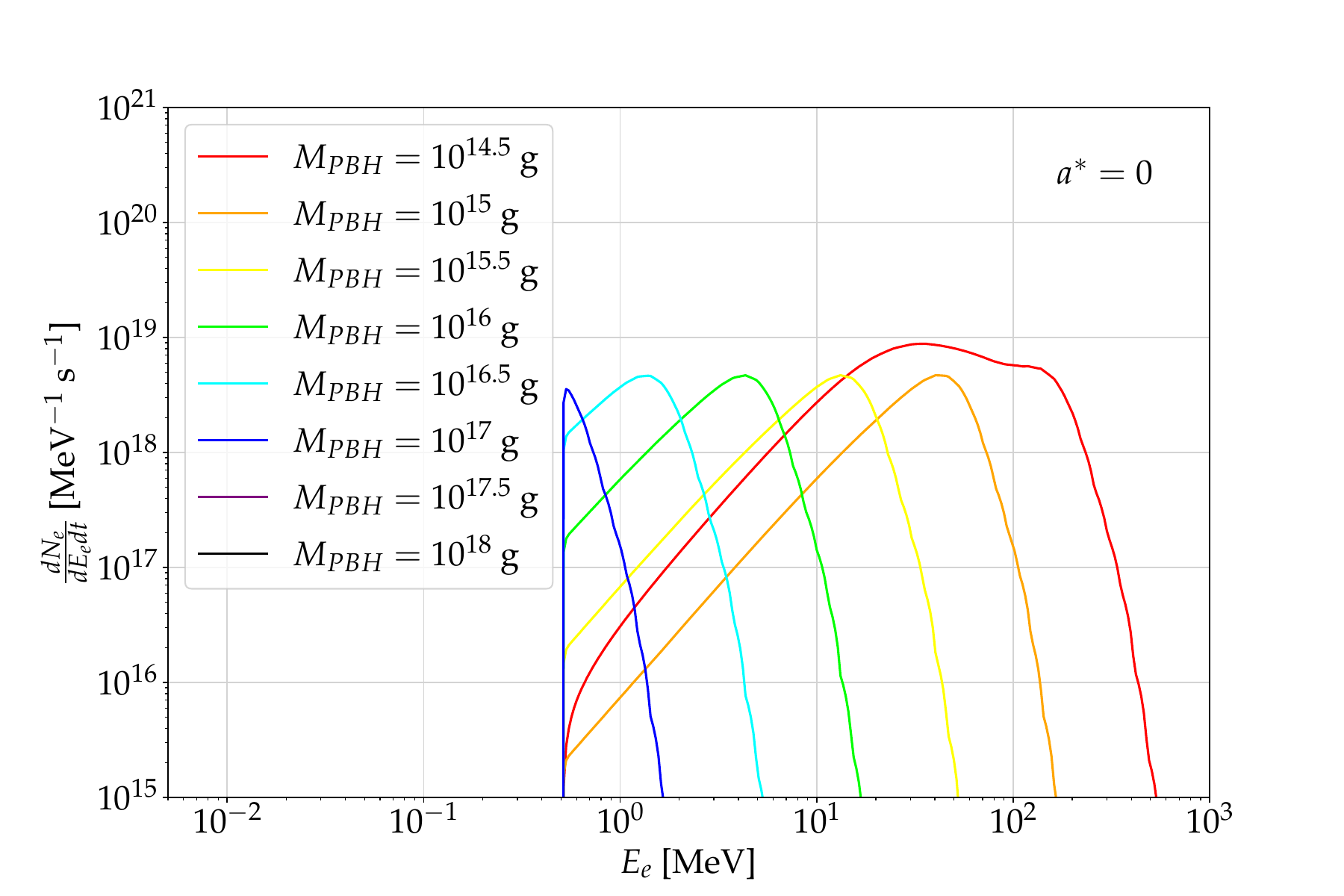}
    \end{subfigure}
    \hfill
    \begin{subfigure}[c]{0.49\linewidth}
        \centering
        \includegraphics[width=\linewidth,trim= 0.5cm 0 2.5cm 1.5cm]{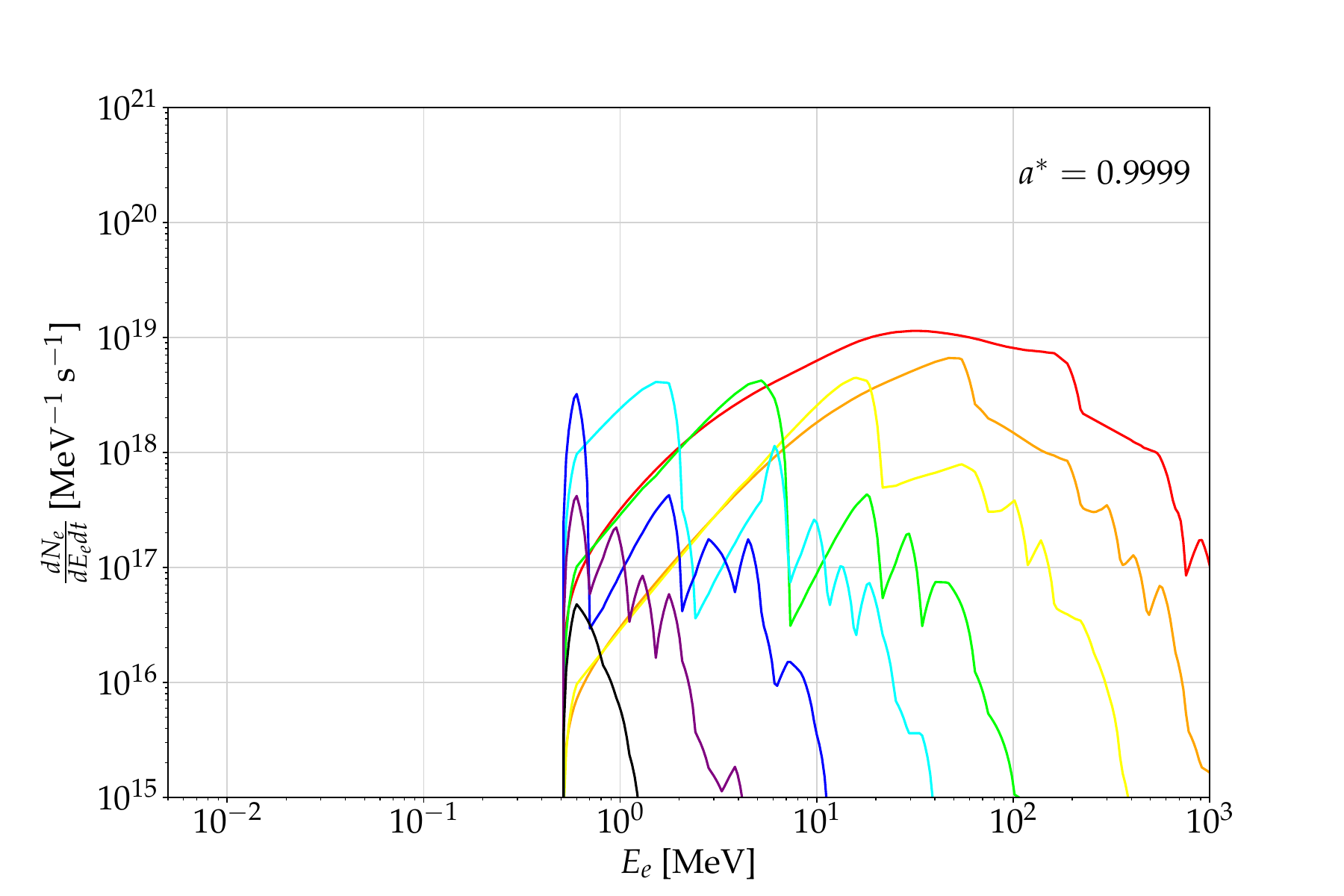}
    \end{subfigure}
    \begin{subfigure}[c]{0.49\linewidth}
        \centering
        \includegraphics[width=\linewidth,trim= 0.5cm 0 2.5cm 1.5cm,clip]{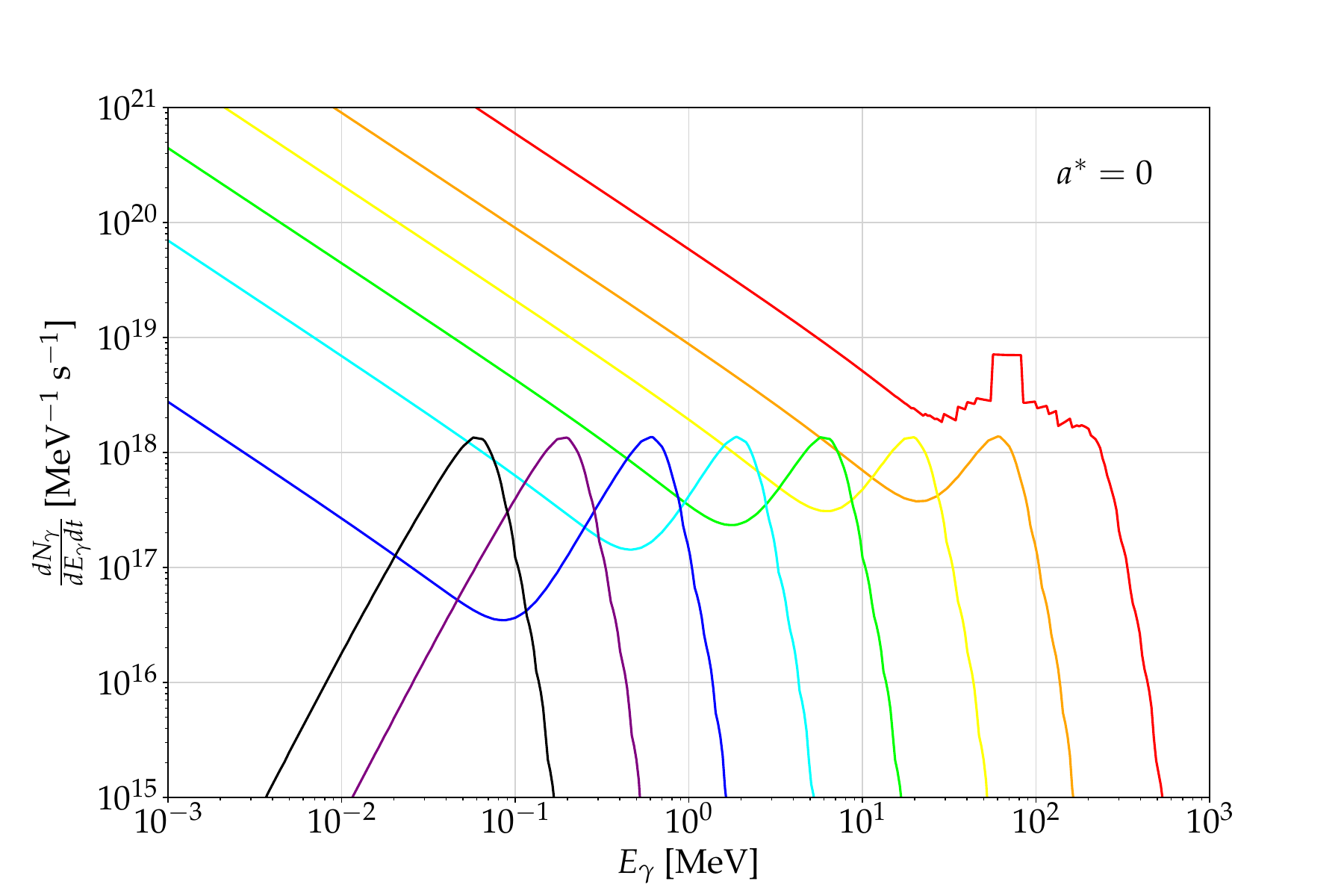}
    \end{subfigure}
    \hfill
    \begin{subfigure}[c]{0.49\linewidth}
        \centering
        \includegraphics[width=\linewidth,trim= 0.5cm 0 2.5cm 1.5cm,clip]{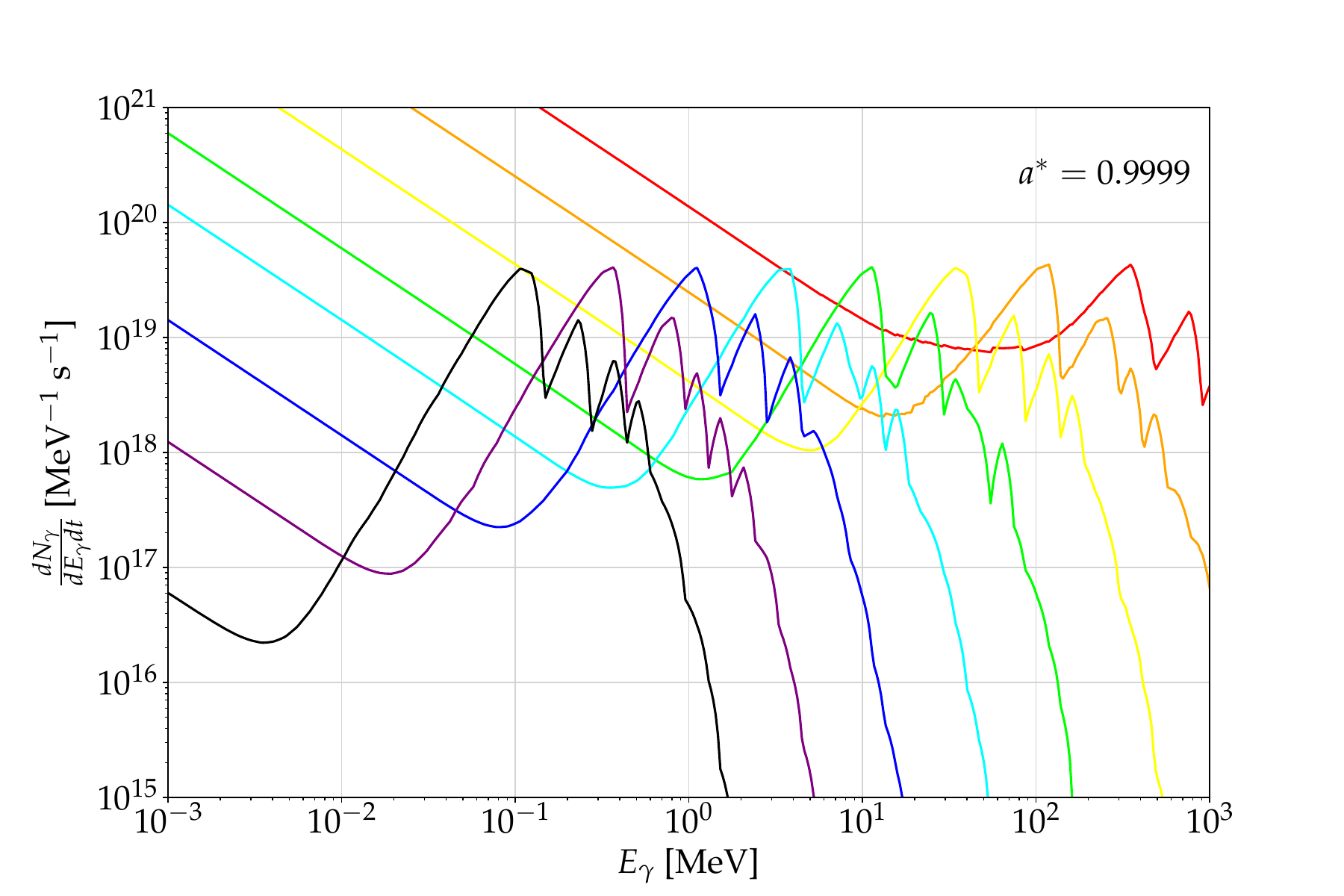}
    \end{subfigure}
    \cprotect\caption{Spectra of secondary $e^\pm$ (top panels) and photons (bottom panels) from the evaporation of a single BH with spin of $a^\star = 0$ (left panels) and $a^\star = 0.9999$ (right panels), for the following BH masses:  $M = 10^{14.5}$ g (red), $10^{15}$ g (orange), $10^{15.5}$ g (yellow), $10^{16}$ g (lime), $10^{16.5}$ g (cyan), $10^{17}$ g (blue), $10^{17.5}$ g (purple) and $10^{18}$ g (black). Output of \verb|BlackHawk+Hazma|.}
    \label{fig:evapspec}
\end{figure}

In Figure~\ref{fig:evapspec} we show the total spectra of $e^\pm$ and photons from the evaporation of a single PBH for a range of masses and for $a^\star = 0$ ($0.9999$) in the left (right) panels. The spectrum of $e^\pm$ solely comes from their emission from the PBH for $M_\textrm{PBH} \gtrsim 10^{14}$ g. However, for lower masses, muons and charged pions start to be produced and their decay into $e^\pm$ contribute to the low-energy bump in the $e^\pm$ spectrum for $M_\textrm{PBH} = 10^{14.5}$ g. The photon spectra show a similar behavior (despite the inappropriate numerical sampling in $E_\gamma$) due to the production of neutral pions that decay into photons. Also, in the photon spectra and for masses where $e^\pm$ start to be produced efficiently, the low-energy slope corresponds to FSR, for which $dN/dE \propto 1/E$. Finally, we can witness that Kerr PBHs emit more particles at higher energies than Schwarzschild ones, due to the transfer of angular momentum from the BH to the emitted particles.

\begin{figure}[t]
    \centering
    \includegraphics[width=0.8\textwidth]{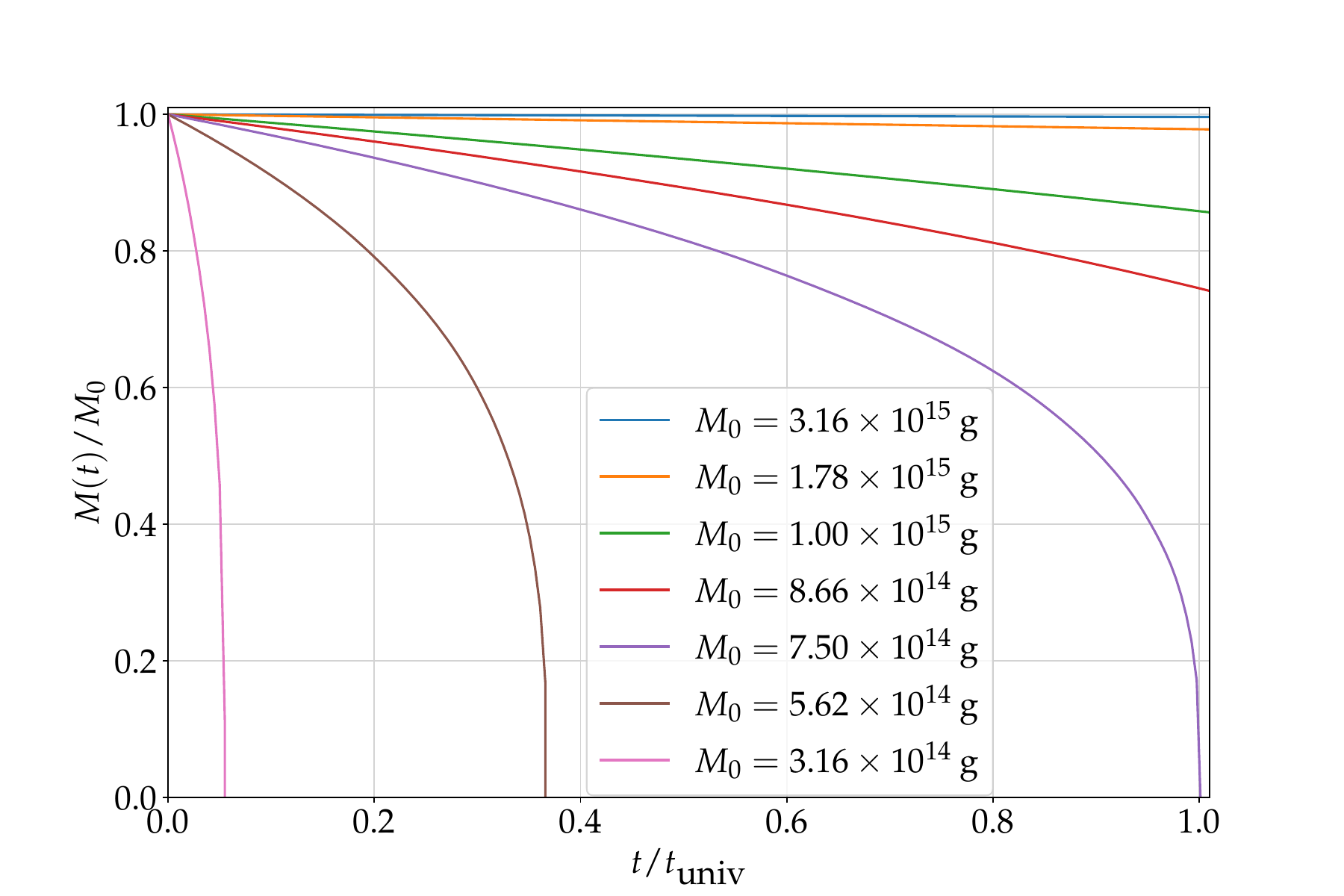}
    \caption{Evolution of the mass $M$ of Schwarzschild BHs for different initial masses $M_0$ at $t=0$. The $x$-axis represents the time in terms of fractions of the age of the Universe and $y$-axis the BHs mass in terms of fractions of its initial mass.}
    \label{fig:PBHevol}
\end{figure}

To compute the flux of propagated $e^\pm$ injected by PBH evaporation, we once again have to solve Equation~\ref{eq:transporteq}, in the same manner as described in Chapter~\ref{chap:prop} however with the source term of Equation~\ref{eq:sourcePBH}. We rewrite it here for the convenience of the reader, including the integration bounds
\begin{equation}
   Q_e(E_e,\vec{x}) = f_\textrm{PBH}\rho_\textrm{DM}(\vec{x}) \int_{M_\textrm{min}}^\infty \frac{dM}{M}\frac{dN_\textrm{PBH}}{dM} \frac{d^2N_e}{dtdE_e}\;,
\end{equation}
where $M_\textrm{min} \approx 7.5 \times 10^{14}$ g corresponds to the minimal mass of PBHs today. As shown in Figure~\ref{fig:PBHevol}, PBHs formed in the early Universe with a mass below $M_\textrm{min}$ should have all evaporated by now, whereas PBHs with an initial mass of $10^{15}$ g have experienced their mass decreasing by $\mathcal{O}(20\%)$. Thus, we follow the approximation where all PBHs with $M_\textrm{PBH} \leq M_\textrm{min}$ do not exist today, and the remaining ones have the same mass distribution $dN_\textrm{PBH}/dM$ as in their formation in the early Universe.

We consider the so-called \emph{log-normal} PBHs mass distribution
\begin{equation}
    \label{eq:lognorm}
    \frac{dN_\textrm{PBH}}{dM} = \frac{1}{\sqrt{2\pi}\sigma M}\exp\left(-\frac{\log^2(M/M_\textrm{PBH})}{2\sigma^2}\right)\;,
\end{equation}
where $M_\textrm{PBH}$ is the peak PBH mass, and $\sigma$ is the width of the distribution. This mass function is relevant when assuming the formation of PBHs from a large enhancement in the inflationary power spectrum~\cite{Dolgov:1992pu,Carr:2017jsz}. We explore values of $\sigma$ varying from $0$ to $2$, the case $\sigma \rightarrow 0$ corresponding to a \emph{monochromatic} mass distribution ($dN_\textrm{PBH}/dM=\delta(M-M_\textrm{PBH})$).

\section{Diffuse emissions}
\label{sec:diffuse}

As previously mentioned, we are interested in three distinct probes of PBHs. First, the PBH-produced $e^\pm$ emission to which we can compare with {\sc Voyager 1} data. Second, the emission of secondary $X$-rays, from ICS on PBH-produced $e^\pm$ on Galactic ambient light, where we use {\sc Xmm-Newton/Mos} data to set limits. Lastly, the emission of $511$ keV photons, due to the annihilation of PBH-produced $e^+$ with $e^-$ in the ISM, or the decays of positronium (Ps; \emph{i.e.}\ a bound state of $e^+e^-$) that we can compare with $511$ keV line measurements by {\sc Integral/Spi}. To compute the flux of the first two probes, we employ essentially the same framework as in Chapter~\ref{chap:prop}, recalled in Sections~\ref{subsec:PBHee} and \ref{subsec:Xrays}, whereas we detail the computation of the flux of secondary $511$ keV photons in Section~\ref{subsec:511keV}, developed in~\cite{Keith:2021guq, DelaTorreLuque:2023cef}.

\subsection{Diffuse electron-positron emission}
\label{subsec:PBHee}

\begin{figure}[t]
    \centering
    \begin{subfigure}[c]{0.49\linewidth}
        \centering
        \includegraphics[width=\linewidth,trim= 0.5cm 0 1cm 0]{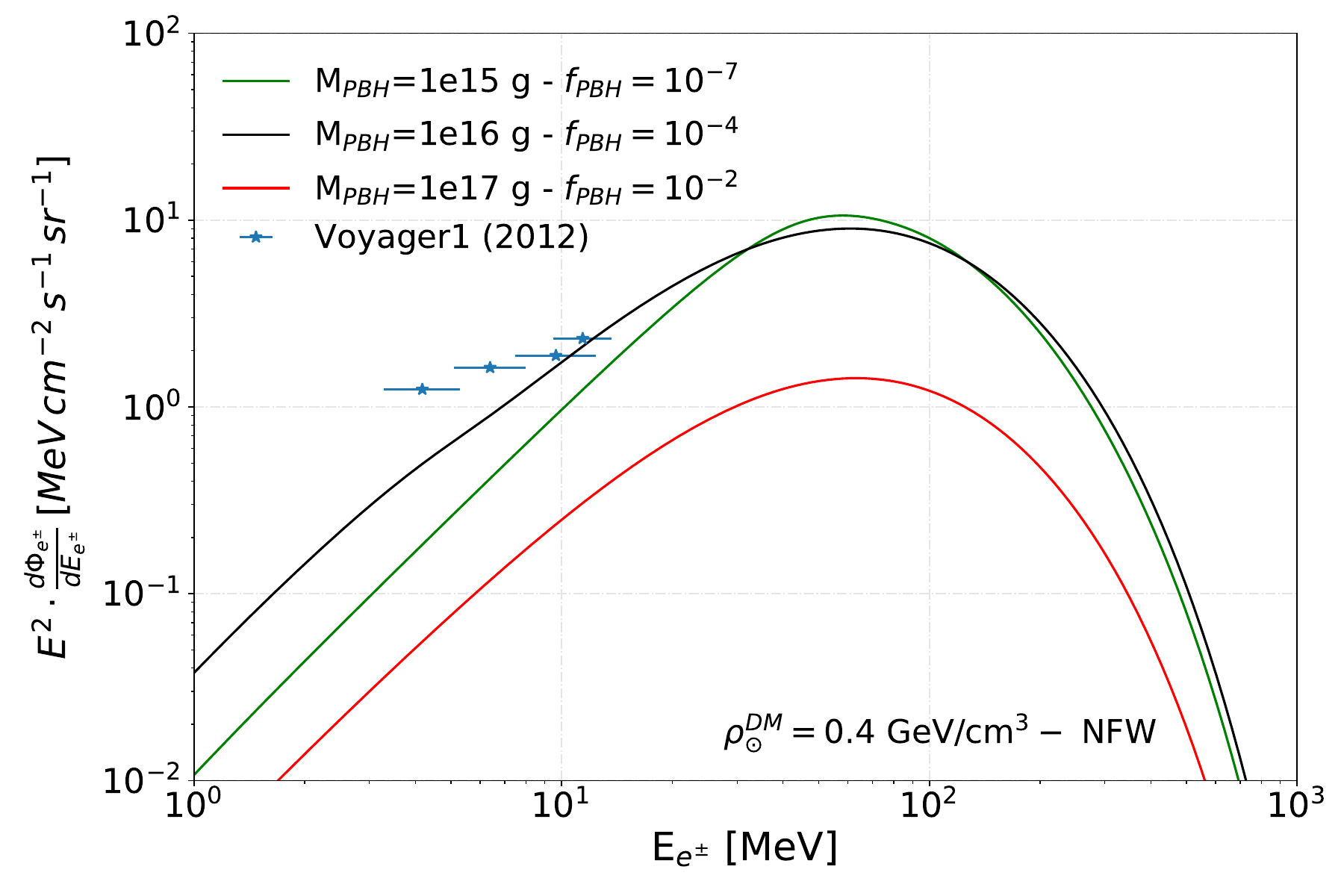}
    \end{subfigure}
    \hfill
    \begin{subfigure}[c]{0.49\linewidth}
        \centering
        \includegraphics[width=\linewidth,trim= 0.5cm 0 1cm 0]{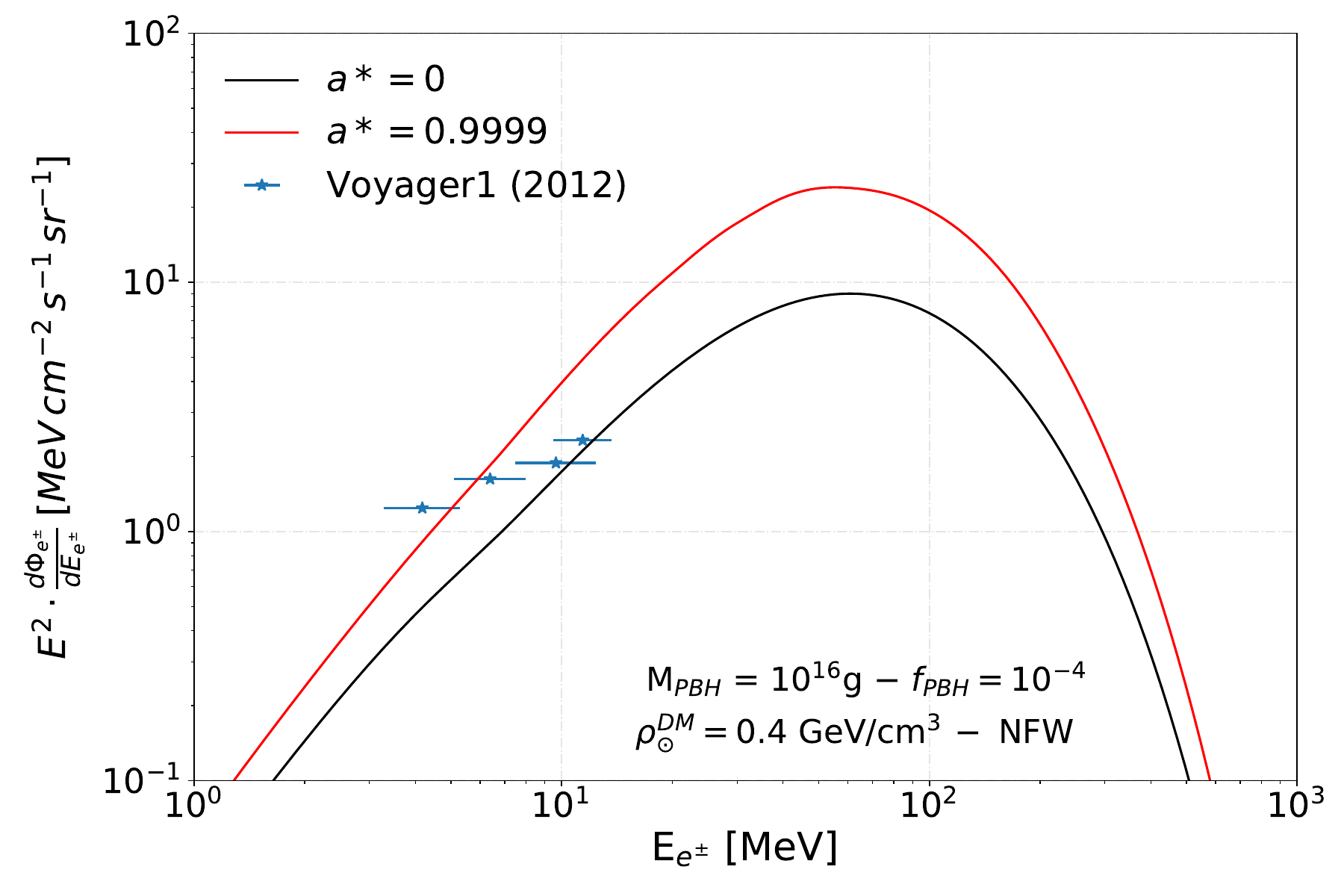}
    \end{subfigure}
    \begin{subfigure}[c]{0.49\linewidth}
        \centering
        \includegraphics[width=\linewidth,trim= 0.5cm 0 1cm 0]{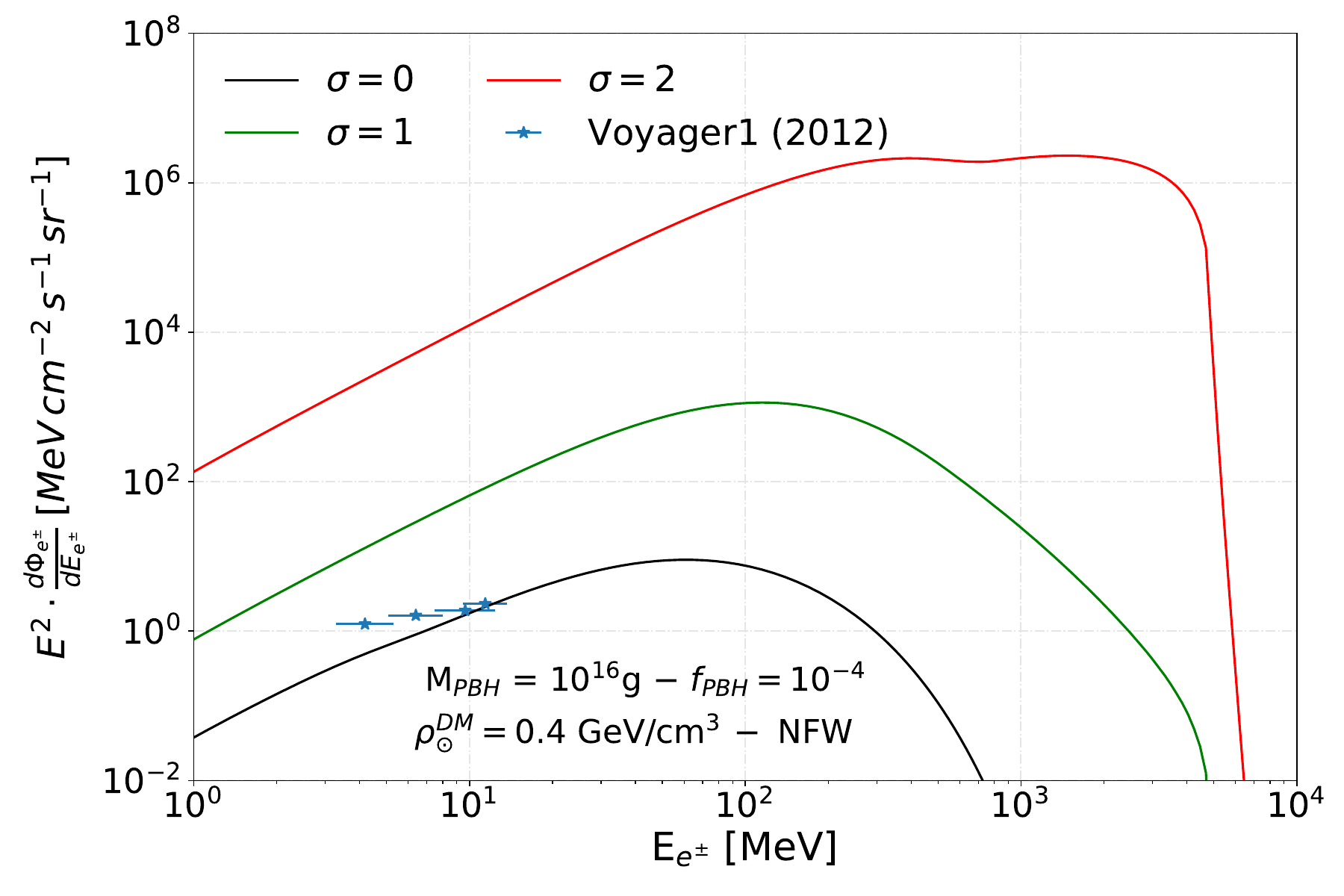}
    \end{subfigure}
    \hfill
    \begin{subfigure}[c]{0.49\linewidth}
        \centering
        \includegraphics[width=\linewidth,trim= 0.5cm 0 1cm 0]{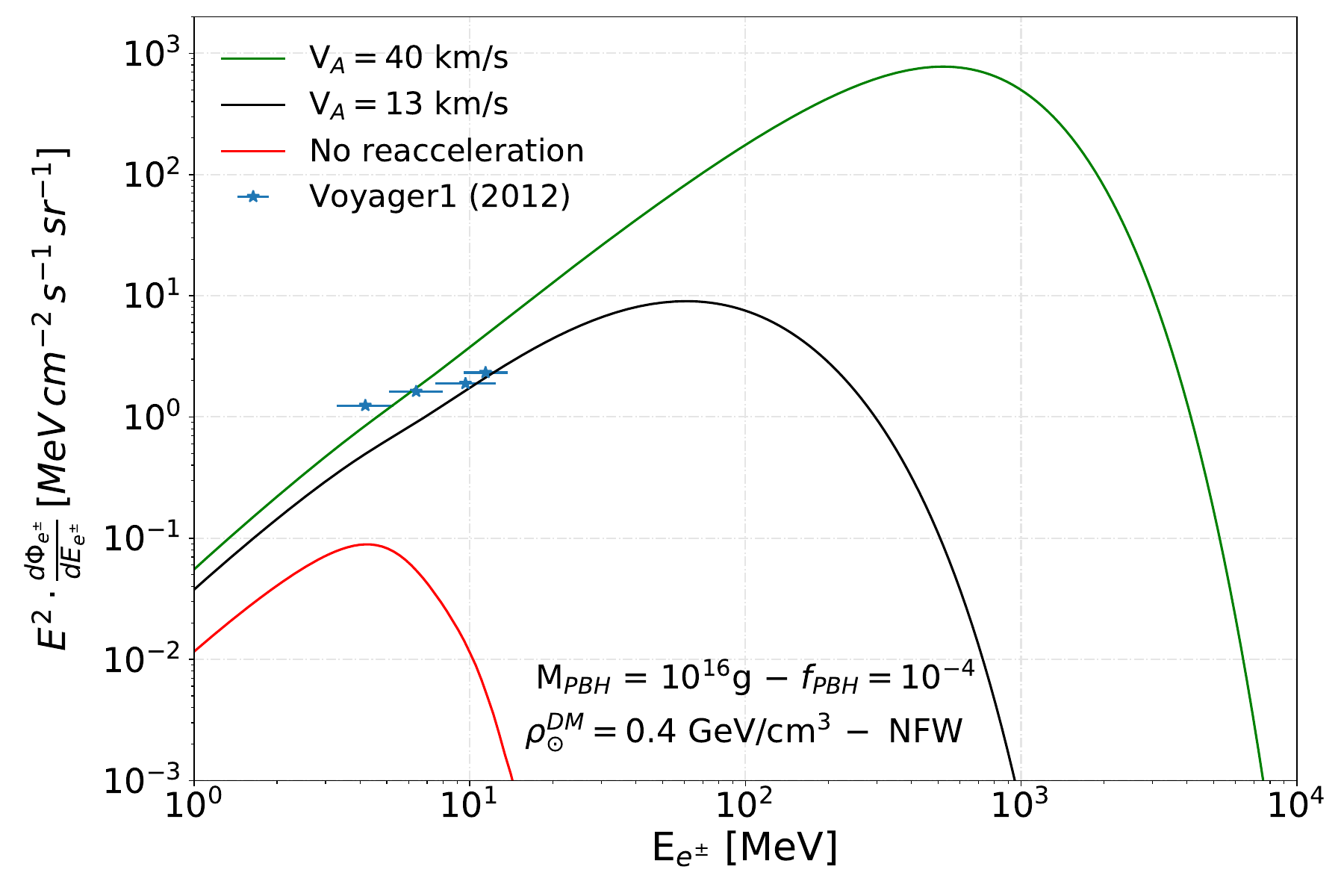}
    \end{subfigure}
    \caption{Local $e^{\pm}$ spectrum generated from PBH evaporation under different assumptions. Top left: Comparison of the spectrum for Schwarzschild PBH of different masses, assuming their mass distribution to be monochromatic. Top right: Comparison of the expected local $e^{\pm}$ spectrum from Schwarzschild ($a^\star=0$) and near-extremal Kerr ($a^\star=0.9999$) PBHs, assuming they are monochromatic in mass. Bottom left: Comparison of the predicted local $e^{\pm}$ spectrum for Schwarzschild PBHs assuming a log-normal mass distribution with different values of the width $\sigma$. Bottom right: Comparison of the local $e^{\pm}$ spectrum for monochromatic Schwarzschild PBHs, for different levels of reacceleration.}
     \label{fig:VoyagervsVa}
\end{figure}

In order to deal with the propagation of $e^\pm$ injected by PBH evaporation, to then compute the flux of diffuse $e^\pm$, $X$-rays and contribution to the $511$ keV line, we have to solve the diffusion-convection-loss equation (Equation~\ref{eq:transporteq}) in the same manner as in Chapter~\ref{chap:prop}. We use the numerical solver \verb|DRAGON2| with the same propagation setup described in Section~\ref{sec:eepropag}, since the PBH-produced $e^\pm$, in the PBH mass range we consider, have the same energies as in the case of annihilating or decaying sub-GeV DM. However, given that systematic uncertainties are difficult to assess and different CR analyses can find slightly different results~\cite{Strong:1998pw, 1987ICRC....2..218O, 1994ApJ...431..705S}, we consider other extreme propagation scenarios that allow us to evaluate the impact of these uncertainties in the predicted $e^{\pm}$ spectra and the bounds on the fraction of DM consisting of PBHs. We first repeat our calculations for `realistic' variations of the propagation parameters found in our analysis, which consist of taking the values that maximise the difference in flux from the benchmark case at $3\sigma$. Similarly to the case of DM decay, the parameters with a greater effect on the diffuse spectra produced from PBHs are the Alfv\'en velocity $v_A$, which controls the level of diffuse reacceleration~\cite{DelaTorreLuque:2023nhh, DelaTorreLuque:2023olp}, and the height $L$ of the halo, which dictates the volume where CRs are confined and where PBHs produce particles that can reach us. In this way, to obtain a realistic uncertainty band in our predictions, we use a conservative setup where $L = 4$ kpc and $v_A=7$ km/s, which produces a lower flux of $e^\pm$ and therefore more conservative limits. In turn, the more aggressive setup is meant to increase the flux of $e^\pm$ from PBHs, and uses values of $L = 12$ kpc and $v_A=20$ km/s. As a point of reference, we recall that our benchmark values are $L = 8$~kpc and $v_A=13.4$~km/s. We tested an even more `general' and extreme variation of propagation setup, that ensures that the flux of particles must be between the two extremes: the `optimistic' case, where $L = 16$ kpc and $v_A=40$ km/s are much higher than the typical values\footnote{We consider that $v_A=40$~km/s is the maximum realistic value for $v_A$, as discussed in Section~\ref{sec:e+e-}.}. Then, the most `pessimistic' case will be the one with no reacceleration ($v_A=0$) and $L = 3$ kpc. These two cases are unlikely, given the fact that the propagation of CRs implies energy exchange with plasma waves and therefore non-zero reacceleration, and values below $L = 3$ kpc seem to be strongly disfavoured from CR analyses and other existing constraints~\cite{DeLaTorreLuque:2021yfq, Evoli:2019iih, Weinrich:2020ftb, delaTorreLuque:2022vhm}. As an example, we show in the bottom right panel of Figure~\ref{fig:VoyagervsVa} the dramatic effect of reacceleration in the $e^\pm$ spectrum at Earth from a $M_\textrm{PBH}=10^{16}$ g PBH, for different values of $v_A$, comparing our benchmark scenario with the aforementioned optimistic and pessimistic ones.

We will obtain constraints on $f_\textrm{PBH}$ using {\sc Voyager 1} measurements of the local $e^\pm$ flux. A comparison of the $e^\pm$ flux measured by {\sc Voyager 1} and the predicted local $e^\pm$ spectrum for monochromatic PBHs with different masses, for our benchmark propagation setup and NFW DM profile, is shown in the top left panel of Figure~\ref{fig:VoyagervsVa}. In addition, we also illustrate the spectra predicted assuming a log-normal PBH mass distribution with $\sigma=1$ (green line) and $\sigma=2$ (red line) in the bottom left panel. This allows one to see how the $\sigma$ parameter affects our predictions, given that realistically one must expect a non-zero $\sigma$. It can be seen that the higher $\sigma$ the higher is the expected flux and the higher is the energy reached by the PBH-produced $e^\pm$. The main reason for this is that the contribution from lower mass PBHs is very important and dominates the spectra of these particles. Then, in the top right panel, we compare the spectra produced from PBHs with different values of the spin parameter $a^\star$. In particular, we show the cases of Schwarzschild ($a^\star=0$) and near-extremal Kerr ($a^\star=0.9999$) PBHs, as well as for the intermediate case $a^\star=0.99$. As one can see from the figure, spinning PBHs always lead to a higher flux although not changing its spectral shape appreciably, in agreement with what was found in~\cite{Arbey:2020yzj, Arbey:2021mbl}.

\subsection{Diffuse \texorpdfstring{$X$}{X}-ray emission}
\label{subsec:Xrays}

\begin{figure}[t]
    \centering
    \begin{subfigure}[c]{0.49\linewidth}
        \centering
        \includegraphics[width=\linewidth,trim= 0.5cm 0 3cm 0.5cm]{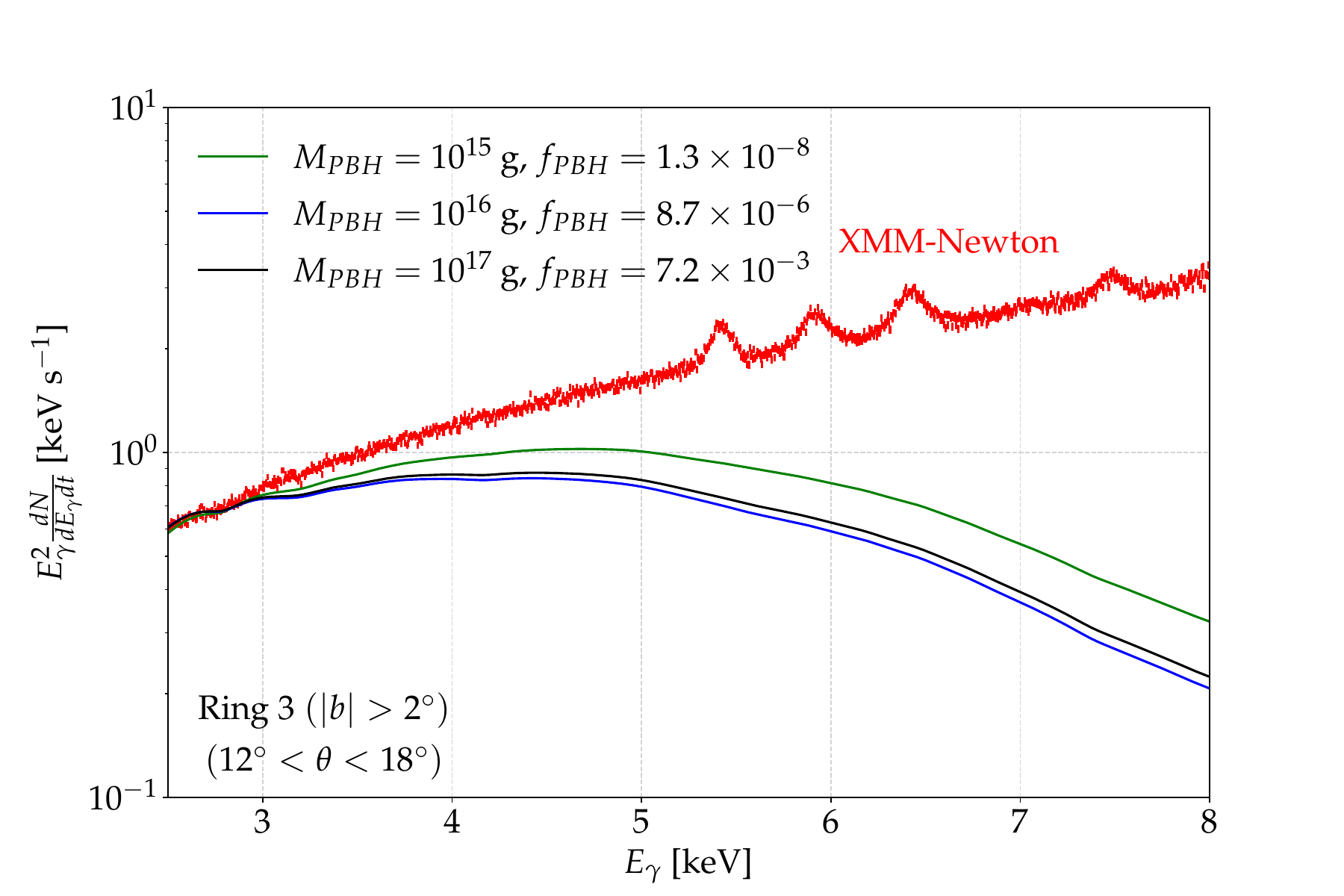}
    \end{subfigure}
    \hfill
    \begin{subfigure}[c]{0.49\linewidth}
        \centering
        \includegraphics[width=\linewidth,trim= 0.5cm 0 3cm 0.5cm]{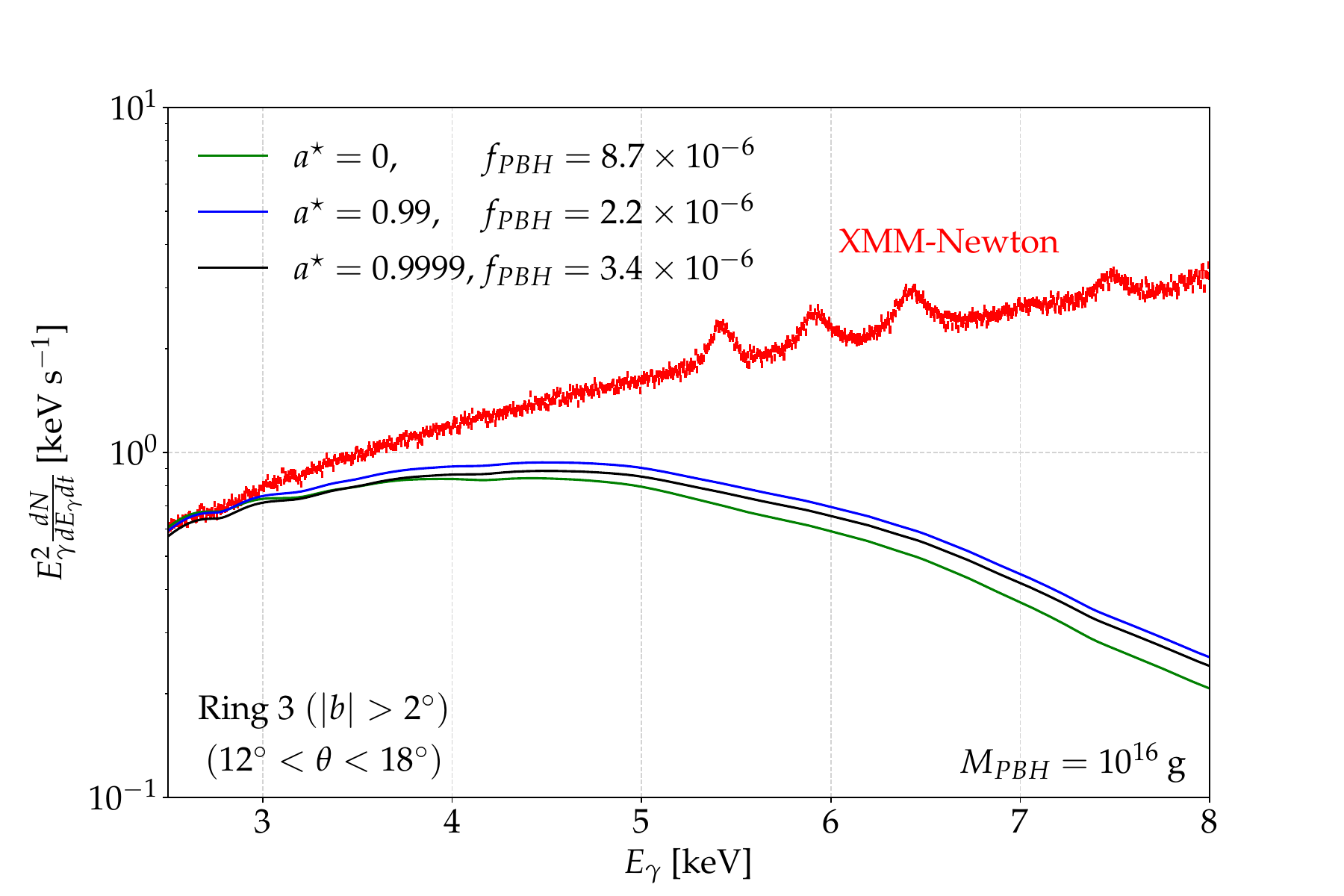}
    \end{subfigure}
    \begin{subfigure}[c]{0.49\linewidth}
        \centering
        \includegraphics[width=\linewidth,trim= 0.5cm 0 3cm 0.5cm]{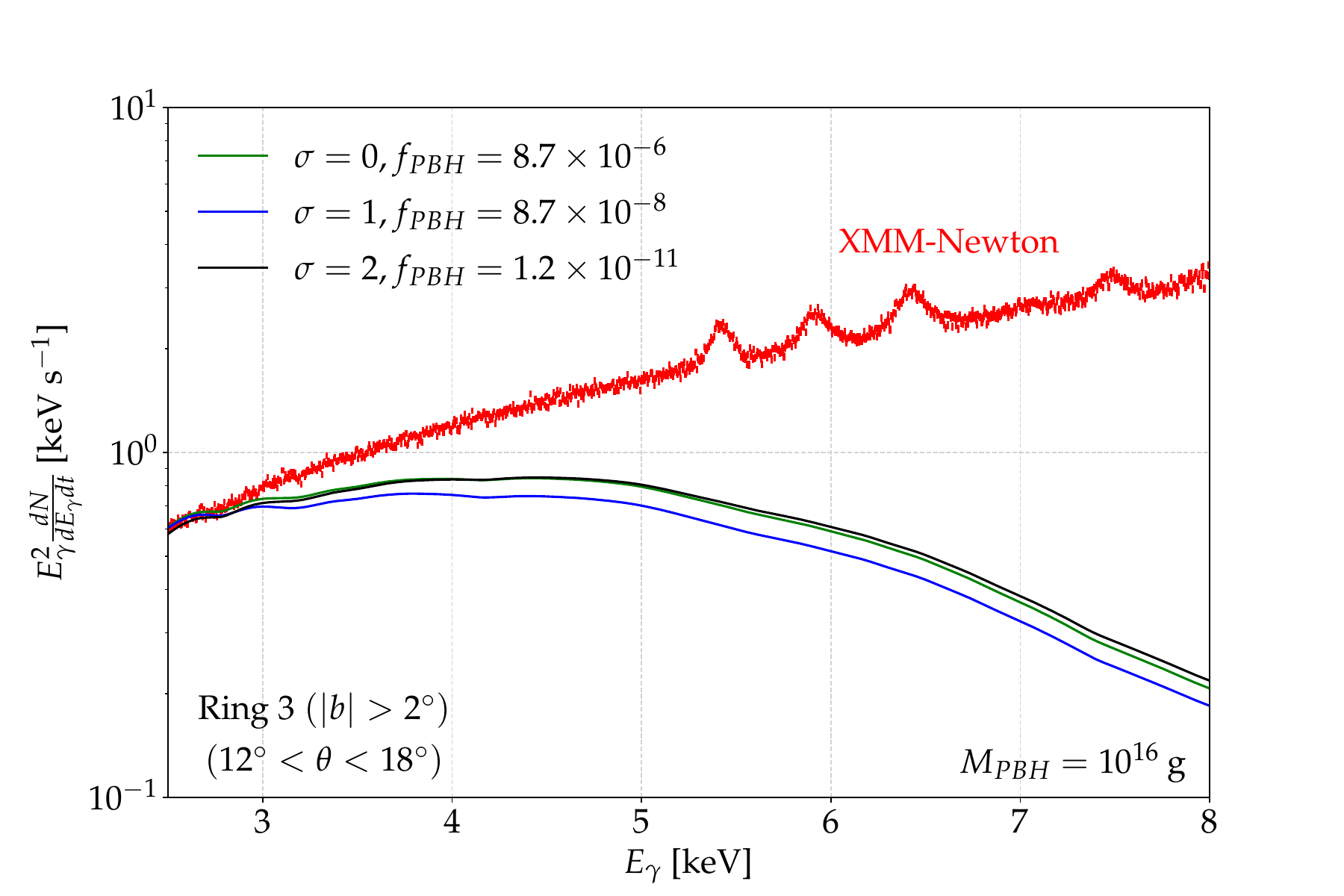}
    \end{subfigure}
    \caption{Comparison of the predicted PBH-induced $X$-ray emission with diffuse $X$-ray data from {\sc Xmm-Newton} in a region close to the GC. We show the prediction is shown for different values of $M_\textrm{PBH}$ when the PBH mass distribution is monochromatic (top left panel), of $a^\star$ (top right panel) and $\sigma$ when the distribution is log-normal (lower panel).}
     \label{fig:XMMComp}
\end{figure}

During their propagation, the population of $e^\pm$ injected in the Galaxy produces different secondary radiations that can be used to track their distribution and density. Similarly to Chapters~\ref{chap:subGeV} and \ref{chap:prop}, the dominant source of secondary emissions in the $X$-ray range is due to ICS of injected $e^\pm$ on Galactic ambient photons.

To calculate the diffuse $X$-ray emissions generated from the diffuse distribution of $e^\pm$ in the Galaxy that we obtain with \verb|DRAGON2|, we employ once more the \verb|HERMES| code. The total $X$-ray flux also includes photons directly emitted during PBH evaporation, as well as FSR produced by evaporated charged particles. It turns out that this component is sub-dominant compared to the $X$-ray flux emitted during the transport of evaporated $e^\pm$. We compute $2\sigma$ bounds from the diffuse Galactic $X$-ray emission observed by {\sc Xmm-Newton} in the $2.5-8$~keV band, as done in Chapters~\ref{chap:subGeV} and \ref{chap:prop}, to which we refer the reader for more details. In Figure~\ref{fig:XMMComp}, we compare the $X$-ray diffuse emission expected from PBH evaporation. In the top left panel, we show the case of monochromatic PBHs with masses of $10^{15}$ g, $10^{16}$ g and $10^{17}$ g, for our benchmark propagation setup and NFW DM profile. In the top right panel, we compare the emission expected from a Schwarzschild PBH ($a^\star=0$) and Kerr one with $a^\star=0.99$ and a near-extremal one with $a^\star=0.9999$, all for $M_\textrm{PBH}=10^{16}$~g. Note that the value of $f_\textrm{PBH}$ is different for every case, as indicated in the legend. In the bottom panel, we show results for Schwarzschild PBHs distributed log-normally with a mean mass of $10^{16}$ g and different values of $\sigma$, ranging from $\sigma=0$ (monochromatic case) to the wider $\sigma=2$. The conclusions for the impact of these parameters in the $X$-ray Galactic diffuse emission are similar to those found for the local $e^\pm$ flux in Figure~\ref{fig:VoyagervsVa}.

\subsection{511 keV line}
\label{subsec:511keV}

\begin{figure}[t]
    \centering
    \begin{subfigure}[c]{0.49\linewidth}
        \centering
        \includegraphics[width=\linewidth,trim= 0.5cm 0 0.5cm 0]{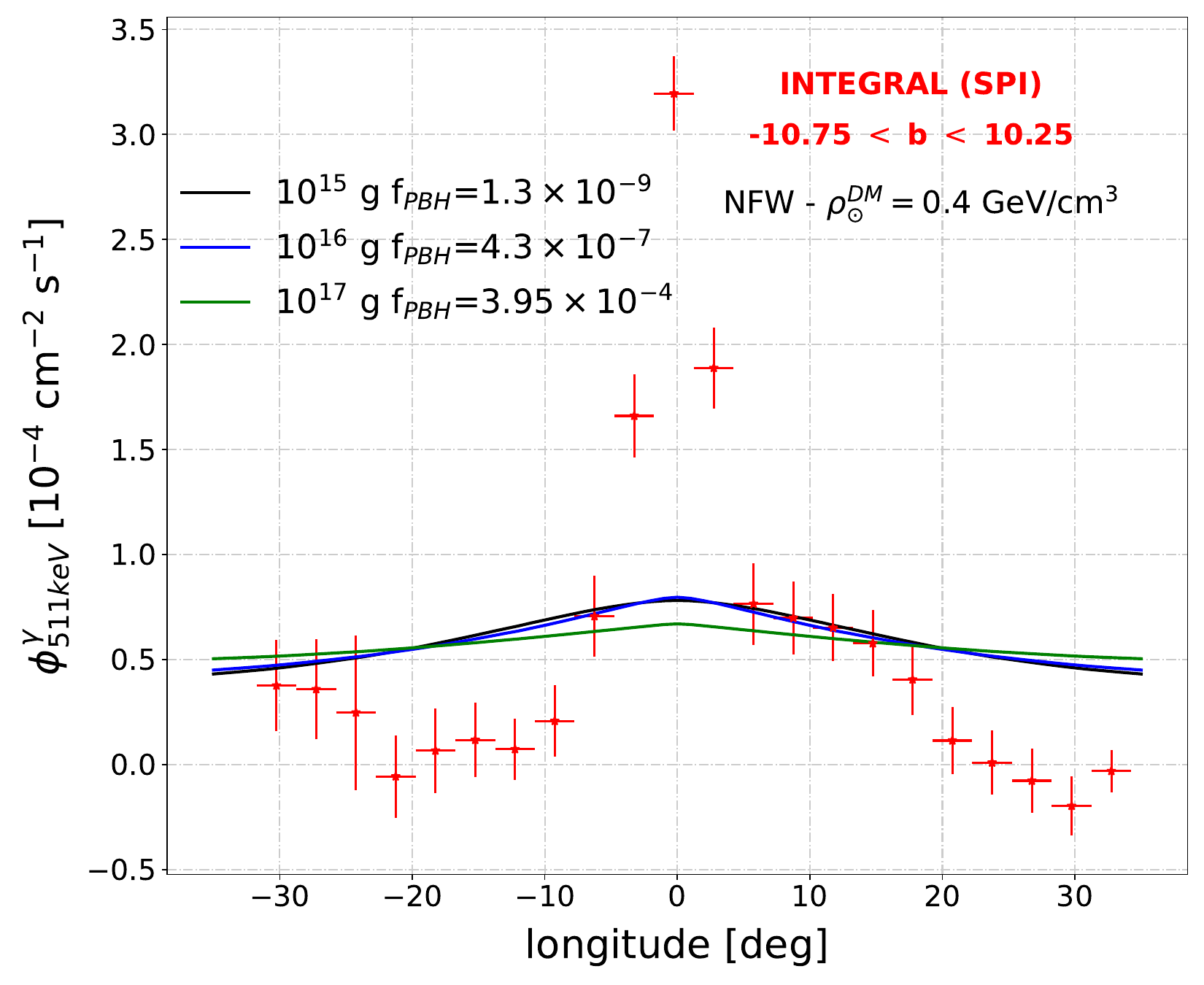}
    \end{subfigure}
    \hfill
    \begin{subfigure}[c]{0.49\linewidth}
        \centering
        \includegraphics[width=\linewidth,trim= 0.5cm 0 0.5cm 0]{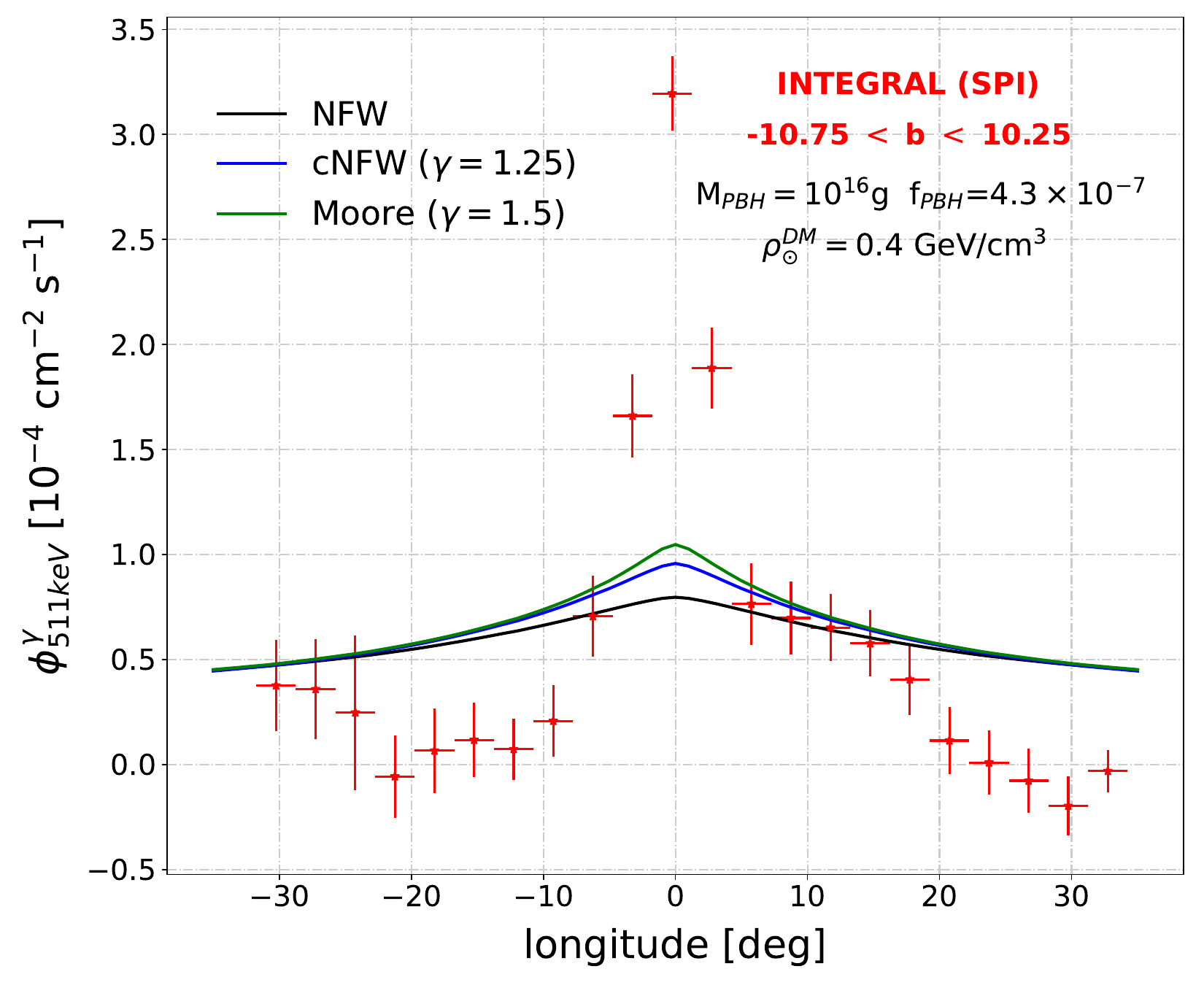}
    \end{subfigure}
    \caption{Comparison of the expected longitude profile of the $511$ keV emission from PBH evaporation with {\sc Integral/Spi} data. The left panel shows a comparison of the signals expected from different PBH masses, assuming again Schwarzschild and monochromatically distributed PBHs,  while the right panel shows a comparison of the predicted profile for different DM distributions, for the case of Schwarzschild PBH monochromatically distributed with $M_\text{PBH}=10^{16}$ g.}
    \label{fig:511_profs}
\end{figure}

PBH-induced $e^+$ are expected either to annihilate with $e^-$ in the ISM, producing photons with an energy of $511$ keV, or, if they lose enough energy during their propagation, form Ps. Ps exists in four states, depending on the spin states of the $e^+$ and $e^-$: one para-Ps state with antiparallel spins that decays into two $511$ keV photons\footnote{Para-Ps can theoretically decay into any even number of photons, although the probability decreases quickly with the number of photons: para-Ps decaying into four photons has a branching ratio of $\simeq 10^{-6}$.}, and three ortho-Ps states with parallel spins that decay into three photons\footnote{Ortho-Ps could also decay into any odd number of photons, but the branching ratios of the decay into more than three photons is negligible.}. Therefore the number of emitted $511$~keV photon per $e^+$ is $2[(1-f_\textrm{Ps}) + f_\textrm{Ps}/4]$, where $f_\textrm{Ps}$ is the Ps formation probability. This probability essentially scales with the density of free $e^-$ at a given location at the Galaxy. In the GC, the value of $f_\textrm{Ps}$ is measured to be $f^\textrm{GC}_\textrm{Ps} = 0.967\pm0.022$~\cite{Jean:2005af}, and we consider that the distribution of free $e^-$ follows the \verb|NE2001| model~\cite{Cordes:2002wz,Cordes:2003ik}, with the correction proposed in~\cite{Gaensler:2008ec}, consisting of: i) a thick disk whose scaling follows $\exp(-\lvert z \rvert/H_1)$, with $H_1 = 1.83$~kpc, corresponding to the warm ionised medium, and ii) a thin disk associated to low-altitude HII regions, whose scaling follows $\exp(-\lvert z \rvert/H_2)$, with $H_2 = 0.14$~kpc. In summary, we consider that the number of emitted $511$ keV photon per $e^+$, at any position in the Galaxy, is
\begin{equation}
	N_\textrm{511 keV}(\vec{x}) = 2\left[(1-f^\textrm{GC}_\textrm{Ps}) + \frac{f^\textrm{GC}_\textrm{Ps}}{4}\right] \left(e^{-\lvert z \rvert/H_1}+ e^{-\lvert z \rvert/H_2}\right)\;.
\end{equation}
Then, we can write the differential flux of the $511$ keV line from a given direction
\begin{equation}
	\frac{d\Phi_\gamma^{511\,\textrm{keV}}}{d\Omega} =  \frac{1}{4\pi}\int_\textrm{l.o.s.} ds\, N_\textrm{511 keV}(\vec{x})\,\Phi_e(\vec{x})\;,
\end{equation}
where $\Phi_e = \int dE_e\,d\Phi_e/dE_e$, whose integrand is defined in Equation~\ref{eq:CRflux}.

In this work, we obtain constraints from the longitude profile of the $511$ keV line emission measured by {\sc Integral/Spi}, following the procedure described in~\cite{DelaTorreLuque:2023cef}. We have tested that our results are compatible with previous evaluations applied to other exotic sources of $e^+$~\cite{Calore:2021klc, Calore:2021lih, DelaTorreLuque:2023huu, DelaTorreLuque:2023nhh, Carenza:2023old}.

In the left panel of Figure~\ref{fig:511_profs}, we show the predicted longitude profile of the $511$ keV emission for PBHs of masses between $10^{15}$ and $10^{17}$ g compared to {\sc Integral/Spi} data~\cite{Siegert:2015knp}, assuming a NFW profile and with $f_\textrm{PBH}$ specified in the legend for each case. It can be seen that the most constraining data points are those obtained at high longitudes. Given that these points are also those expected to be more affected by systematic uncertainties (mainly background noise and the need of templates to extract measurements) and the limited statistics of the measurements, the bounds that we derive are conservatively calculated by multiplying by a factor of $2$, as a proxy for the effect of systematic uncertainties, as done in~\cite{DelaTorreLuque:2023nhh, DelaTorreLuque:2024zsr}. We show in the right panel of Figure~\ref{fig:511_profs} a comparison of the predicted $511$ keV line profile with the NFW DM distribution with other popular DM profiles, namely a Moore profile (Equation~\ref{eq:gNFW} with $\gamma=1.5$)~\cite{Moore:1999nt}, a cNFW profile (Equation~\ref{eq:gNFW} with $\gamma=1.25$), for a monochromatic $10^{16}$ g mass PBH. We expect cored DM distributions (such as Burkert or Isothermal, respectively Equations~\ref{eq:Burkert} and \ref{eq:Isothermal}) to lead to a flatter $511$ keV profile. As one can see, the predicted profiles are very similar at high longitudes and only change significantly around the central longitudes. Therefore, the uncertainties in the derived constraints from the DM distribution are very small. Similarly, adopting $a^\star \neq 0$ or $\sigma\neq0$ has no significant consequences on the shape of the profile and essentially changes only the  normalisation of the signal.

\section{Results and comparison with other work}
\label{sec:result+discussion}

In this section, we discuss the limits on PBHs we derived in this work, for our benchmark scenario. In addition to displaying the limits using the three probes (diffuse $e^\pm$, $X$-rays and $511$ keV line), we also show the impact on these limits when assuming different PBH mass and spin distributions, and propagation setups. Here, we set $2\sigma$ bounds on $f_\textrm{PBH}$ and $M_\textrm{PBH}$ by applying the same criterion as in Equation~\ref{eq:chi2}, rewritten here in the case of PBHs
\begin{equation}
    \sum_i \left( \frac{\textrm{Max}\left[\phi_{\textrm{PBH}, i}(f_\textrm{PBH}M_\textrm{PBH}) - \phi_{i}, 0 \right]}{\sigma_i}\right)^2 =4 \;, 
\end{equation} 
where $i$ denotes the data point, $\phi_\textrm{PBH}$ is the predicted flux induced by PBH evaporation, $\phi_{i}$ is the measured flux and $\sigma_i$ the associated standard deviation of the measurements.

\begin{figure}[t]
    \centering
    \includegraphics[width=0.8\linewidth]{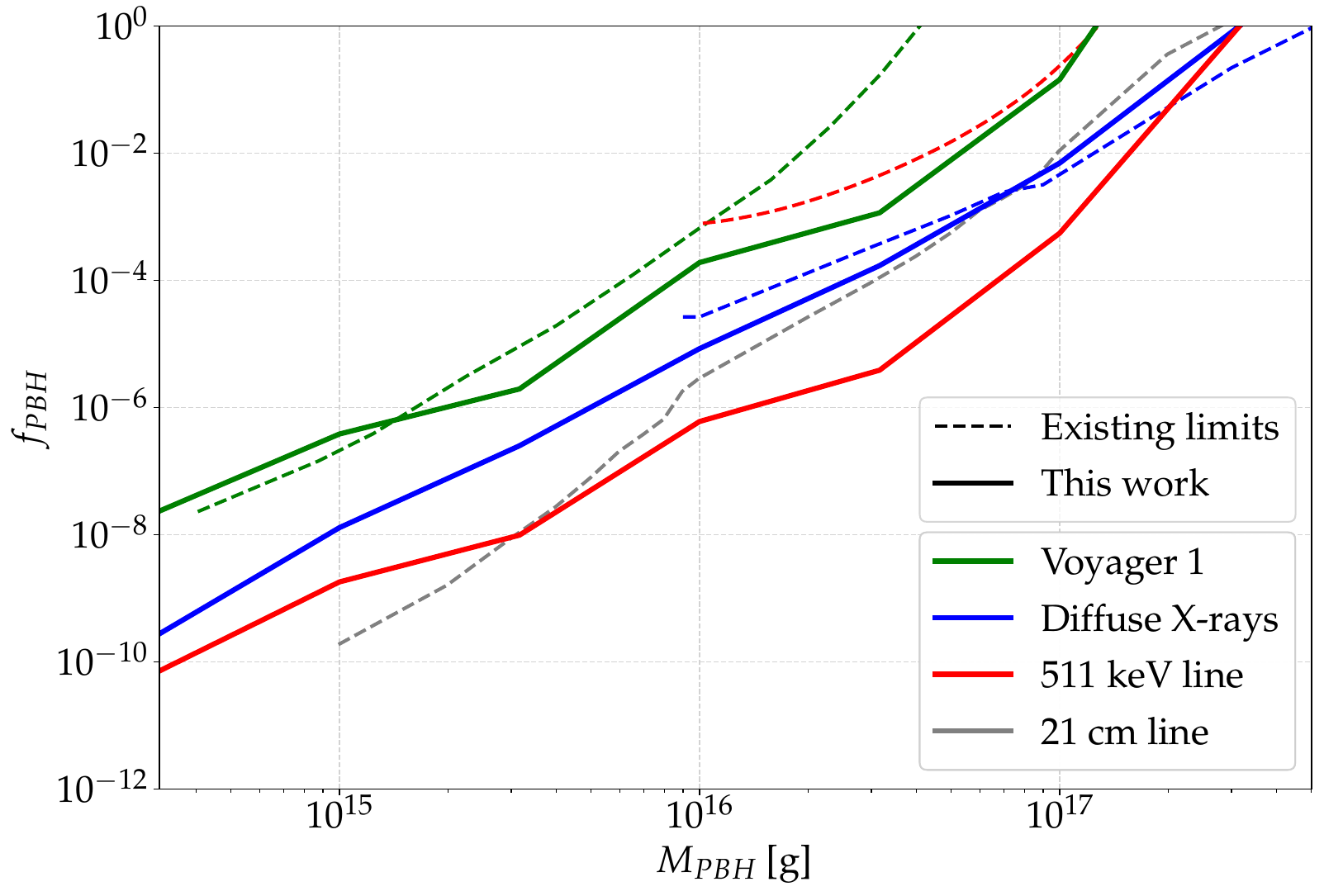}
    \cprotect\caption{Comparison of our limits on $f_\textrm{PBH}$ with other existing ones. The color of the lines represent the different probes used to set the constraints: green for the $e^\pm$ measurements from {\sc Voyager 1}, blue for $X$-ray diffuse observations from {\sc Xmm-Newton}, red for the $511$~keV excess reported by {\sc Integral} and grey for the $21$ cm line measurements from {\sc Edges}. The two different line styles correspond to either the bounds derived either in this work (solid) or in the literature~\cite{Boudaud:2018hqb,Laha:2019ssq,Tan:2024nbx,Mittal:2021egv} (dashed) reported in \verb|PBHbounds|~\cite{kavanagh_2019}.}
    \label{fig:moneyplot}
\end{figure}

In Figure~\ref{fig:moneyplot} we show our benchmark limits on Schwarzschild PBHs, assuming a monochromatic mass distribution and a NFW DM profile, and compare them to existing ones. The solid lines represent the bounds derived in this work, while the dashed lines represent some of the most stringent limits on $f_\textrm{PBH}$ reported in \verb|PBHbounds|~\cite{kavanagh_2019} across the $10^{15}-5\times10^{17}$ g PBH mass range.

We show the {\sc Voyager 1} limits in green on Figure~\ref{fig:moneyplot}, where the dashed line corresponds to the limit reported in~\cite{Boudaud:2018hqb} without background subtraction. The authors used a propagation model with strong reacceleration named `Model $B$'. Our {\sc Voyager 1} limit is comparable to the existing one for $M_\textrm{PBH} \lesssim 10^{16}$ g and becomes more stringent for higher PBH masses. The effect is likely due to the differences in how reacceleration is implemented in the \verb|DRAGON2| code with respect to the semi-analytical code \verb|USINE| used in their work, where reacceleration only takes place in a thin disk, instead of adopting uniform reacceleration across the whole Galaxy, that is important given that CR particles spend most of their time in the Galactic halo while propagating. In addition, to model energy losses, which are key for MeV particles, \verb|USINE| needs to make use of the pinching method~\cite{Boudaud:2016jvj}.

\begin{figure}[t]
    \centering
    \begin{subfigure}[c]{0.46\linewidth}
        \centering
        \includegraphics[width=\linewidth,trim= 0 0 1.5cm 0.5cm]{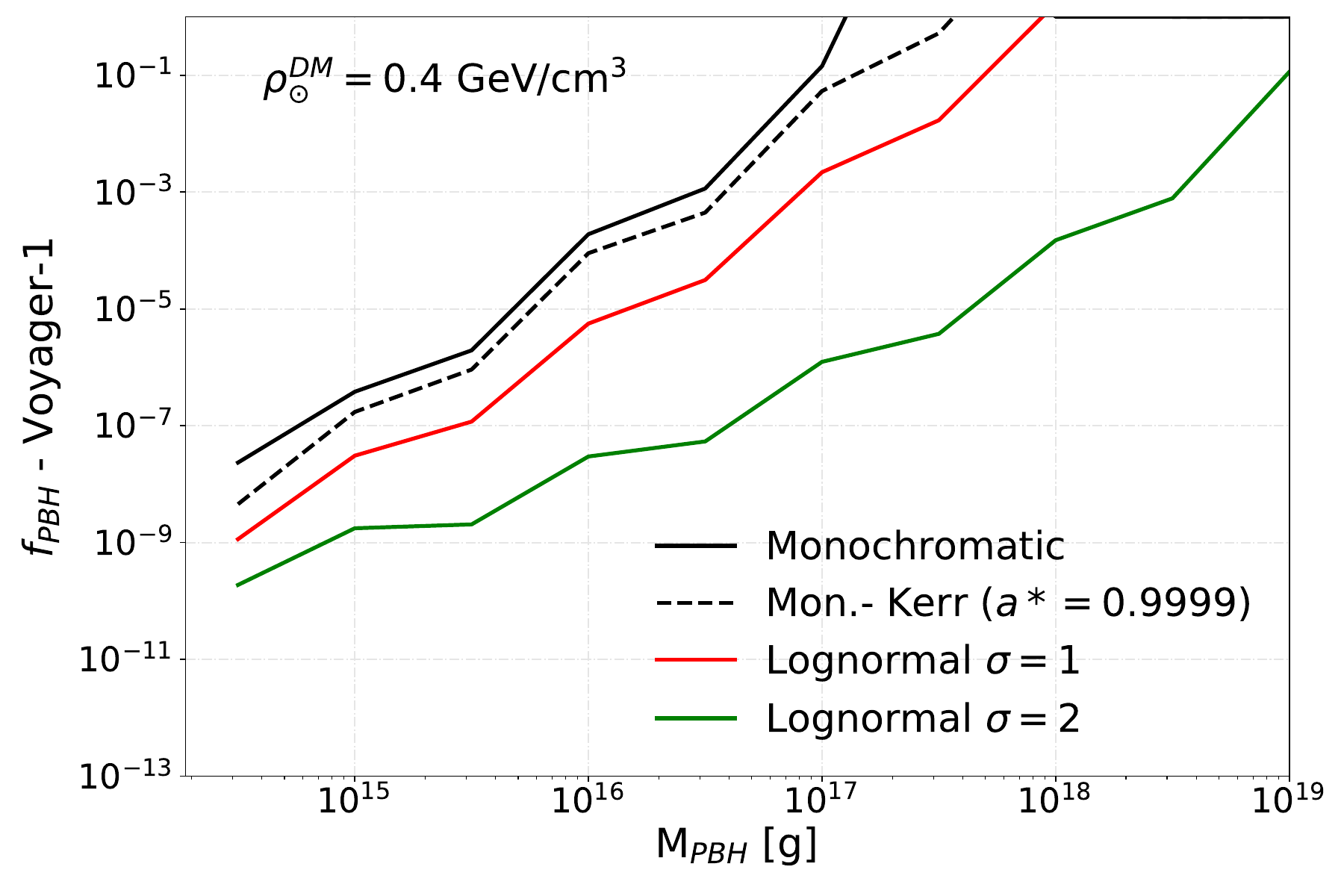}
    \end{subfigure}
    \hfill
    \begin{subfigure}[c]{0.53\linewidth}
        \centering
        \vspace{-0.5cm}
        \includegraphics[width=\linewidth,trim= 0 0 1.5cm 0.5cm]{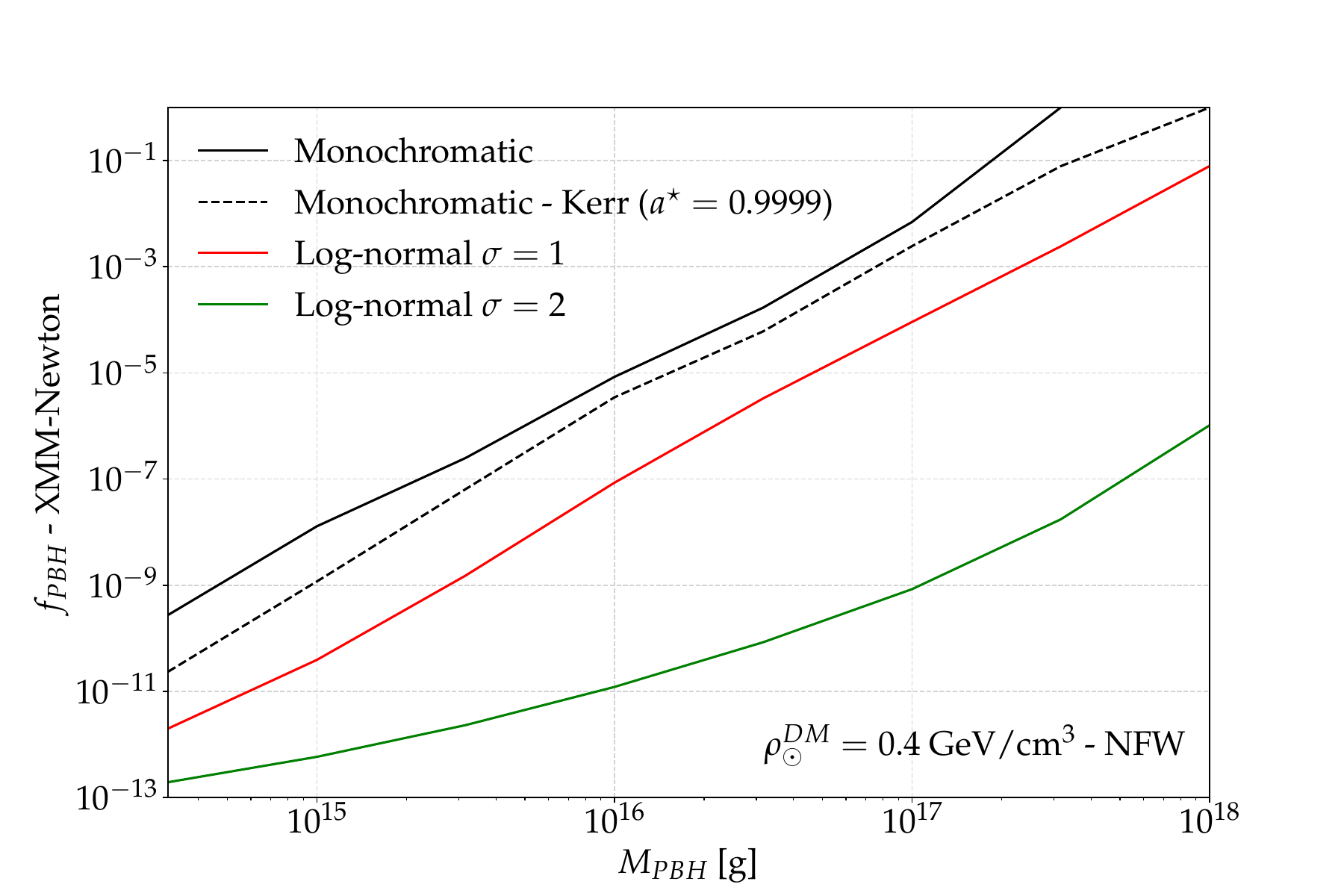}
    \end{subfigure}\vspace{0.2cm}
    \begin{subfigure}[c]{0.46\linewidth}
        \centering
        \includegraphics[width=\linewidth,trim= 0 0 1.5cm 0.5cm]{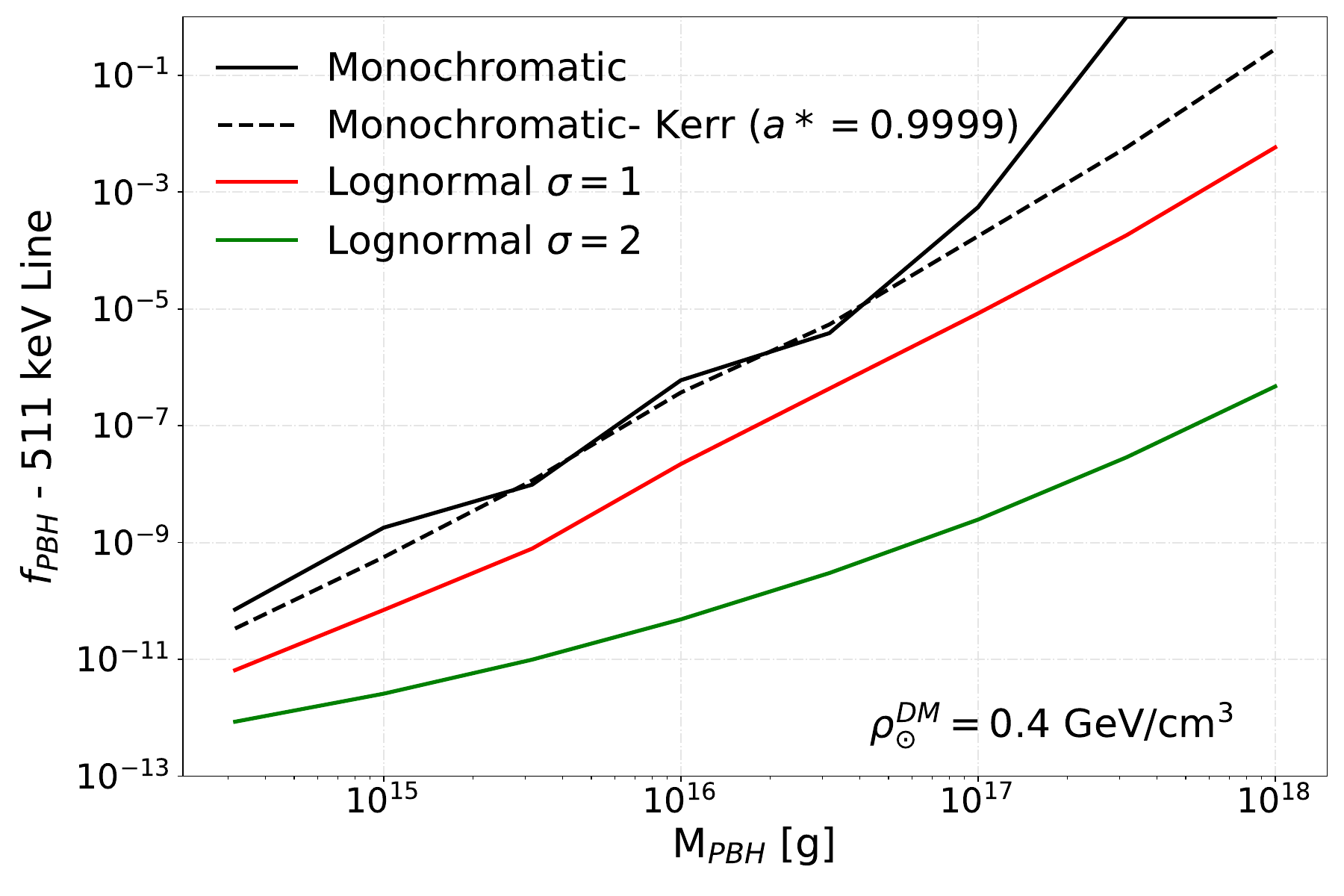}
    \end{subfigure}
    \caption{Limits on Schwarzschild PBHs we derive using {\sc Voyager 1} (top left panel), {\sc Xmm-Newton} (top right panel) and the $511$~keV line reported by {\sc Integral} (bottom panel) for mass distributions with $\sigma = 0$ (monochromatic, solid black line), 1 (red line) and 2 (green line), as well as for monochromatic near-extremal Kerr PBHs (dashed line).}
     \label{fig:SomeLimits}
\end{figure}

The limits from diffuse $X$-ray emissions are shown in blue in Figure~\ref{fig:moneyplot}, where the dashed line is the limit set in~\cite{Tan:2024nbx}. The authors have computed the flux of prompt $X$-ray emissions from the evaporation of extragalactic PBHs and compared it to the isotropic cosmic $X$-ray background measurements, without considering the secondary ICS emissions, to set a limit on $f_\textrm{PBH}$. Remarkably, the low-energy part of the $X$-ray measurements are those most constraining. Therefore, $X$-ray diffuse measurements at lower energies are expected to improve these limits significantly. However, $X$-ray emission starts to be severely absorbed by the ISM gas, which can make it more difficult to improve these constraints using lower energy data.

Our $511$ keV bound, which we weaken by a factor of $2$ to account for systematic uncertainties in the data (as mentioned when discussing the calculation of the $511$ keV line in Section~\ref{subsec:511keV}), is shown as a red solid line in Figure~\ref{fig:moneyplot}, where we compare with the limit reported in~\cite{Laha:2019ssq} (red dashed line). They used the rate of $e^+$ injection needed to explain the total $511$ keV flux from {\sc Integral} in the bulge. Given that the high-longitude measurements of the $511$ keV line emission are the most constraining measurements, the use of the longitudinal profile leads to more stringent results compared to using the bulge emission~\cite{DelaTorreLuque:2023cef}. In addition, the authors include only the emission from a NFW DM profile within the inner $3$~kpc from the GC and did not model $e^+$ propagation. As a result, our $511$ keV bound appears to be more stringent than the one of~\cite{Laha:2019ssq}. It has been shown in~\cite{DelaTorreLuque:2024zsr} that using in-flight $e^+$ annihilation emission can improve the $511$ keV limits on sub-GeV DM, given that measurements of the diffuse $\gamma$-ray emission above a few MeV have a reduced systematic uncertainty and more reliable background models can be used.

Finally, we report the bound derived by requiring that the amount of heating of the intergalactic medium from PBH evaporation is constrained by $21$ cm observations by the {\sc Edges} experiment~\cite{Mittal:2021egv}, which we show as a dotted gray line in Figure~\ref{fig:moneyplot}. All in all, our limit from the longitude profile of the $511$ keV line is competitive with the {\sc Edges} limit below $M_\textrm{PBH} \simeq 10^{16}$~g and becomes the most stringent limit to date for PBH masses between $3\times10^{15}$ and $2\times 10^{17}$ g, for the fiducial astrophysical scenario.

In Figure~\ref{fig:moneyplot}, we assumed PBHs to be Schwarzschild ones with a monochromatic mass distribution, representing the most conservative case. However, if their mass distribution were instead log-normal, as in Equation~\eqref{eq:lognorm}, there would be a low-mass PBH population that contributes to most of the flux of evaporated $e^\pm$ and photons, leading to a strengthening of the limits. Actually, for increasing values of the width of the distribution $\sigma$, the low-mass population increases and therefore the limits become more and more stringent. Alternatively, if PBHs were Kerr BHs, they would produce more particles at high energies, ending up with a strengthening of the limits as well. Figure~\ref{fig:SomeLimits} illustrates the impact of the choice of mass and spin distributions on the limits on $f_\textrm{PBH}$. We also note that in the case of a cuspier DM distribution than the NFW, the {\sc Xmm-Newton} constraints could beat those from {\sc Edges} or the $511$ keV line. In turn, for a cored one, the limits will be significantly weaker.

\begin{figure}[t]
    \centering
    \begin{subfigure}[c]{0.465\linewidth}
        \centering
        \includegraphics[width=\linewidth,trim= 0 0 1.5cm 0.5cm]{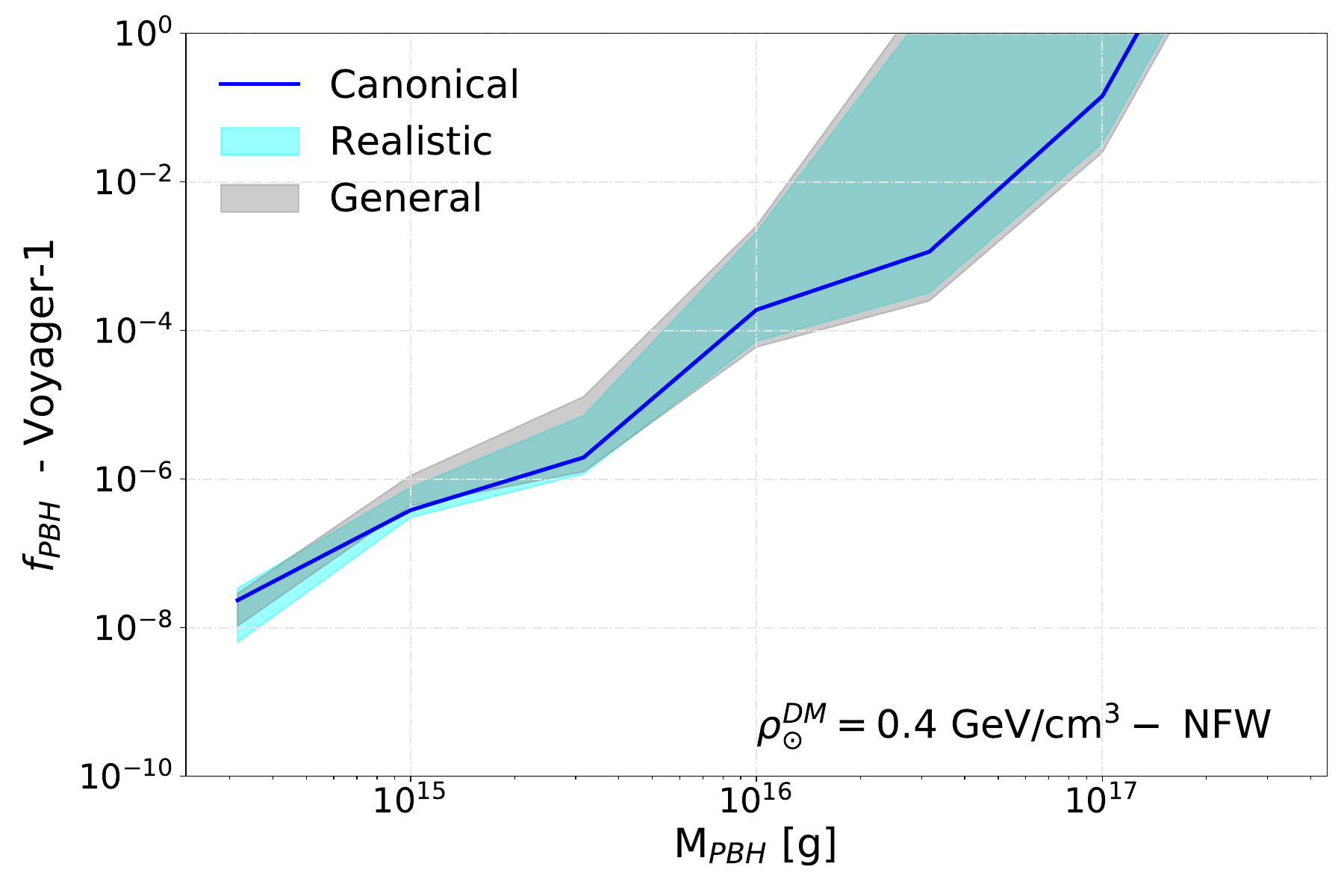}
    \end{subfigure}
    \hfill
    \begin{subfigure}[c]{0.515\linewidth}
        \centering
        \vspace{-0.5cm}
        \includegraphics[width=\linewidth,trim= 0 0 1.5cm 0.5cm]{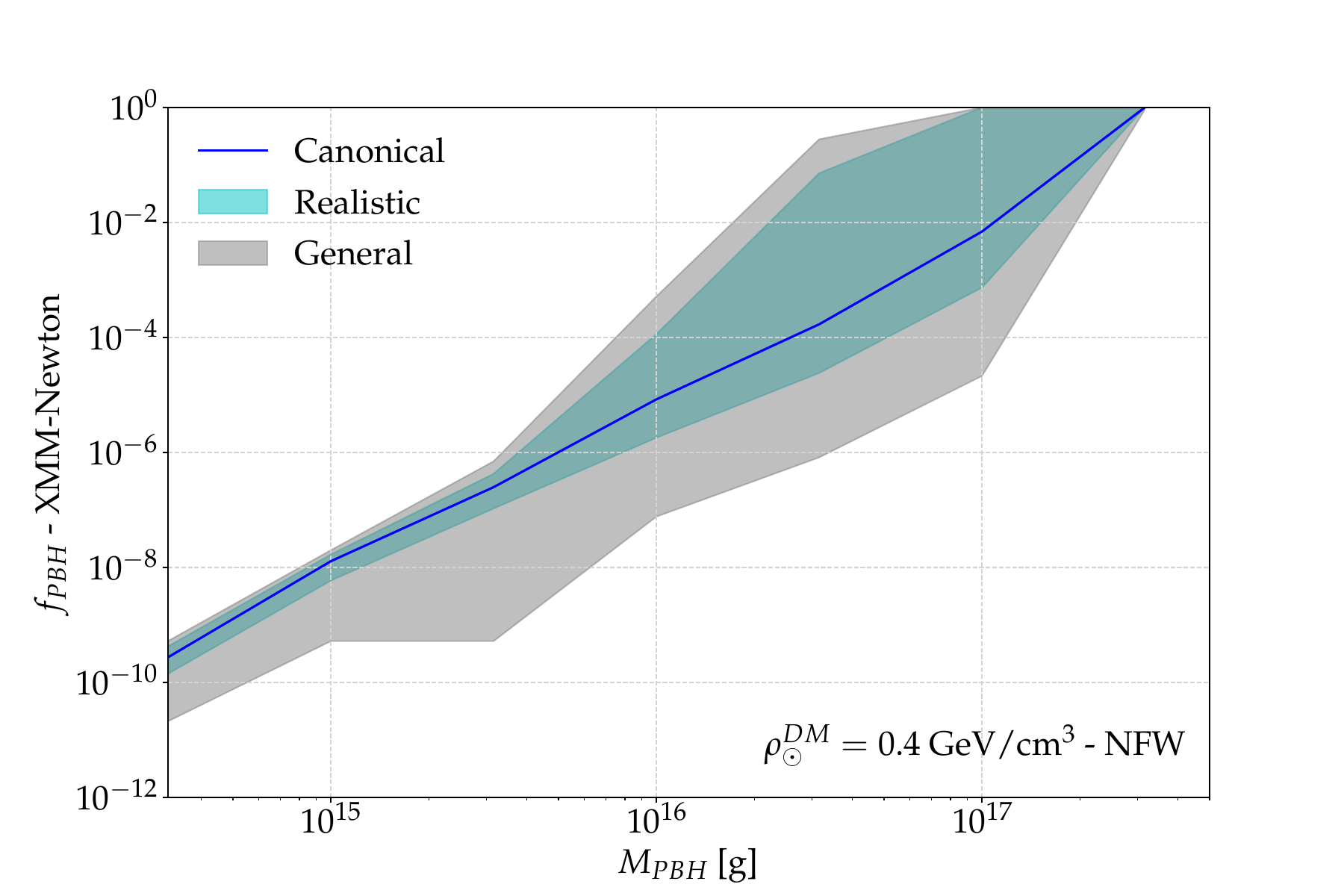}
    \end{subfigure}\vspace{0.2cm}
    \begin{subfigure}[c]{0.465\linewidth}
        \centering
        \includegraphics[width=\linewidth,trim= 0 0 1.5cm 0.5cm]{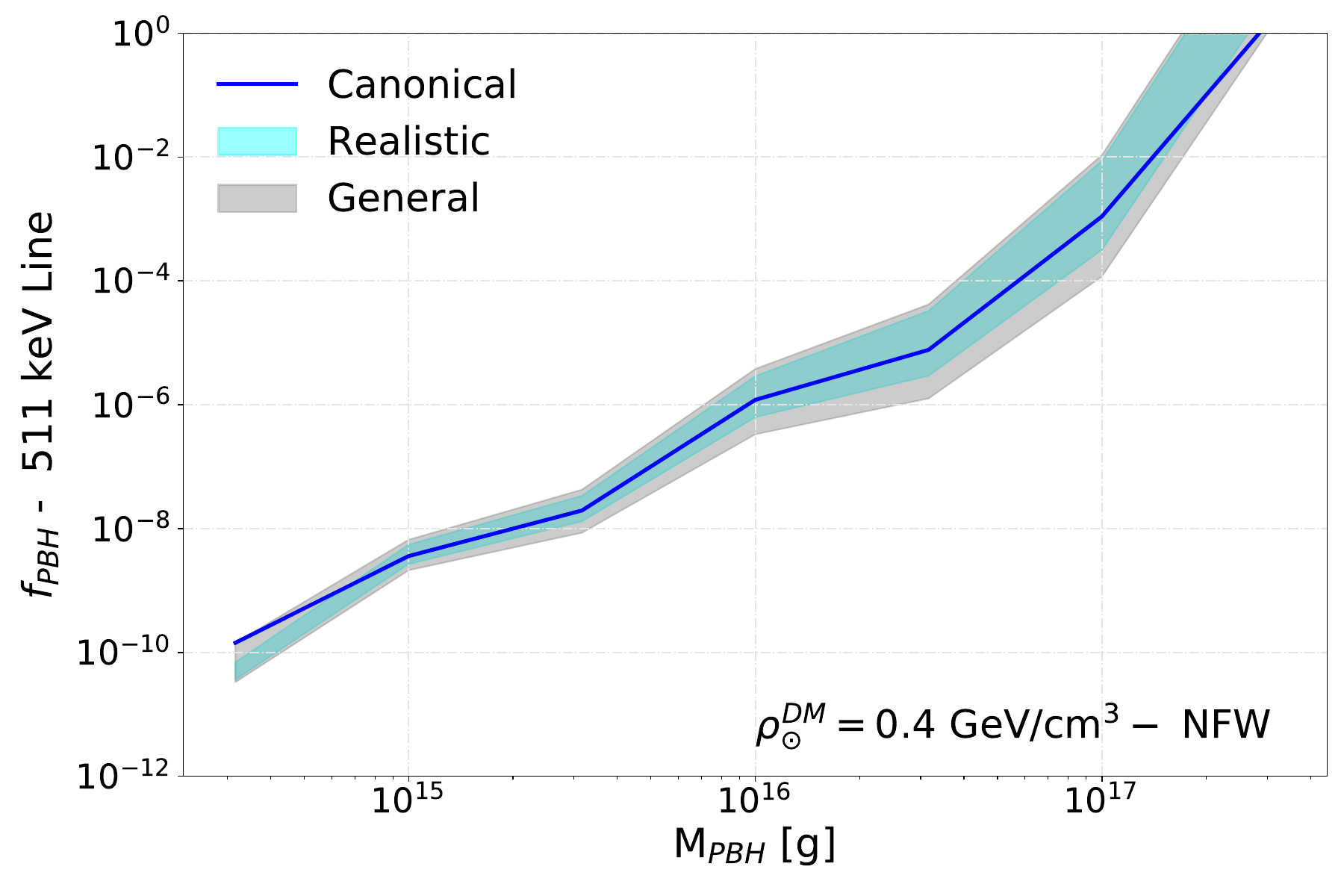}
    \end{subfigure}
    \caption{Uncertainties in the limits we derive using {\sc Voyager 1} (top left panel), {\sc Xmm-Newton} (top right panel) and the $511$~keV line reported by {\sc Integral} (bottom panel). The solid blue lines correspond to the limits using our fiducial propagation model. The blue bands show how our limits are impacted when varying the Alfv\'en speed $v_A$ and the halo height $L$ within their $3\sigma$ uncertainty. The grey bands correspond to a more conservative scenario where we vary $v_A$ between $0$ and $40$ km/s and $L$ between $3$ and $16$ kpc.}
     \label{fig:UncertsLimits}
\end{figure}

Finally, in Figure~\ref{fig:UncertsLimits} we report the uncertainties on the limits we derive in this work, showing the impact of the choice of the propagation model, by using the propagation scenarios explained in Section~\ref{subsec:PBHee}. The blue bands (labeled `realistic') correspond to the variation of $v_A$ and $L$ up to their 3$\sigma$ uncertainties. For the lower side of the band we use $v_A = 20$ km/s and $L = 12$ kpc, while for the upper side we use $v_A = 7$ km/s and $L = 4$ kpc. Then the gray bands in Figure~\ref{fig:UncertsLimits} (labeled `general') represent more conservative uncertainties, where for the lower side we adopt $v_A = 40$ km/s and $L = 16$ kpc, and for the upper side $v_A = 0$ km/s and $L = 3$ kpc. In general, these variations may affect our limits by up to an order of magnitude or more. In the case of the $511$ keV line, we observe that the uncertainty bands are in general smaller. This is due to the fact that the morphology of the predicted $511$ keV line does not change significantly for different values of reacceleration (at high longitudes, where the main constraints come from, reacceleration does not appreciably change the emission, which is spatially very flat in any case). In the case of the diffuse $e^\pm$ and $X$-ray constraints, we observe that the uncertainty band typically broadens at higher PBH masses. The reason is that heavy PBHs inject lower energy $e^{\pm}$, which reacceleration affects much more. For example, in the no reacceleration case the emission from PBHs of mass higher than a few times $10^{16}$ g lies below the {\sc Voyager 1} data points (therefore, no constraint can be set). Additionally, we note that uncertainties from the choice of DM distribution will have very little effect in the constraints from {\sc Voyager 1} and the $511$ keV line, while the limits from {\sc Xmm-Newton} can be significantly affected, given that the most constraining $X$-ray data is coming from the inner regions of the Galaxy, which is where our predictions are more affected by uncertainties in the DM distribution.

\section{Summary}

In this chapter, we derived the limits on the fraction of DM consisting of PBHs using three different yet complementary indirect probes of their evaporation: PBH-produced $e^{\pm}$, secondary $X$-rays from ICS of the latter on Galactic ambient light, and the $511$ keV line emissions from the annihilation of PBH-produced $e^+$ in the ISM, or decay of Ps states. These limits are illustrated in Figure~\ref{fig:moneyplot}, assuming all PBHs to be Schwarschild ones with a monochromatic mass distribution, as well as the same $e^\pm$ propagation setup as in Chapter~\ref{chap:prop}. The most compelling result comes from the $511$ keV line emission from the Galactic disk, where our analysis, assuming a NFW DM profile, yields the most stringent limits to date for PBH masses between $3\times 10^{15}$ and $2\times 10^{17}$ g. We additionally remind the reader that our limits are conservative, derived without including astrophysical background modeling. We note, however, that the secondary $X$-ray constraints are stronger than those from the $511$ keV line for optimistic diffusion parameters, as shown in Figure~\ref{fig:UncertsLimits}, and also in the case of a cuspier DM distribution. Uncertainties on our limits, depicted in our analysis, underscore the sensitivity of our results to the choice of propagation model parameters, such as the Alfv\'en speed and halo height. 

Moreover, considering alternative mass distributions, such as a log-normal distribution, or the inclusion of Kerr BHs, would lead to even more stringent limits, as shown in Figure~\ref{fig:SomeLimits}, due to the increased flux of evaporated particles and photons. We note that PBHs can constitute a significant fraction of the DM only in the gap between $\sim 10^{18}-10^{21}$ g in the monochromatic mass case.

%% file: Chapters/conclusion.tex

For almost a century since its first postulation, the nature of DM has been a mystery to us, even with the ongoing efforts of the world's most brilliant minds. Although this statement is quite pessimistic for the future, the community has over time managed to continuously constrain the properties of DM, as well as developed increasingly sensitive experiments in attempt to detect it. Among the numerous detection techniques employed in DM searches, we focused in this thesis on the possibility that DM can produce indirect signals in the form of a variety of particle messengers. In particular, we focused on the possibility that DM consists of sub-GeV particles or PBHs that can emit low-energy $e^\pm$, and $X$-rays through the propagation of the former ones in the ISM. The non-detection of these products enabled us to set some of the strongest constraints to date on these candidates.

\medskip

In Chapter~\ref{chap:subGeV}, we have focused on sub-GeV DM, which is challenging to probe due to solar screening and the lack of $\gamma$-ray observatories for energies around one MeV. The secondary emission of $X$-rays, mainly through the ICS of DM-produced $e^\pm$ on ambient photons in the Galaxy, was already proven to be a way to circumvent these issue, in addition to being a powerful method to set stringent constraints on sub-GeV DM. Here, we have done an analysis using data from the {\sc NuStar}, {\sc Suzaku}, {\sc Integral} and {\sc Xmm-Newton} satellites in different ROIs, and computing the secondary $X$-rays flux predictions, by adopting a minimalistic CR propagation setup.

As a result, we find that the constraints imposed by {\sc Xmm-Newton} data greatly improve upon the previous limits on annihilating DM. For this case, our bounds are the strongest to date from $m_\textrm{DM} \gtrsim 180$ MeV, excluding DM annihilation cross sections down to $\sim 10^{-28}$~cm$^3$\,s$^{-1}$ for the $e^+e^-$ channel and $m_\textrm{DM} \simeq 1$ GeV, as shown in Figure~\ref{fig:FinalResultsAnn}. On the other hand, for decaying DM, our limits are the most stringent in the literature for $m_\textrm{DM} \gtrsim 100$ MeV, excluding DM lifetimes up to $\sim 10^{28}$~s, for $m_\textrm{DM} \simeq 1$ GeV as well, as shown in Figure~\ref{fig:FinalResultsDec}. We also have evaluated the impact of the different sources of uncertainties (choice of the DM profile, gas, ambient photon densities and GMF configuration) on our limits, which can strengthen or weaken them up to one order of magnitude (see Figures~\ref{fig:uncertainties} and \ref{fig:Combuncertainties}).

\medskip

In Chapter~\ref{chap:prop}, we improved the annihilating and decaying sub-GeV constraints imposed by {\sc Xmm-Newton} derived in Chapter~\ref{chap:subGeV} by adopting a more realistic CR propagation scheme, which is dealt with the numerical code \verb|DRAGON2|. In particular, the inclusion of the stochastic reacceleration of DM-produced $e^\pm$ due to their interaction with the turbulent component of the GMF significantly improved the limits for $m_\textrm{DM} \lesssim 20$ MeV. As this process is more efficient for low energy $e^\pm$, the energy of the latter is increased up to the point where their ICS on ambient photons produce $X$-rays that are in the energy range probed by {\sc Xmm-Newton}, thus strengthening the constraints for lower DM masses.

We showed in Figures~\ref{fig:bounds+litterature_ann} and \ref{fig:bounds+litterature_dec} the limits we derive in the fiducial astrophysical scenario (that includes the DM profile, gas and ambient photon densities and also the propagation parameters). These bounds are the strongest ones to date in almost the entire considered DM mass range ($1$ MeV $-$ $5$ GeV), excluding DM annihilation cross sections down to $10^{-31}$~cm$^3$\,s$^{-1}$ and DM lifetimes up to $10^{28}$ s, both for $m_\textrm{DM} \simeq 1$ MeV. Once again, these bounds suffer from uncertainty sources. For complementarity purpose, we also derived bounds using the predicted local flux of DM-produced $e^\pm$ which we compare to measurements from {\sc Voyager 1}. This results into weaker bounds than the {\sc Xmm-Newton} ones, although they are more robust since the uncertainties associated to the different astrophysical ingredients at the vicinity of Earth are lower. We show the impact of astrophysical uncertainties on our bounds in the bottom panels of Figure~\ref{fig:new+oldbounds}.

\medskip

In Chapter~\ref{chap:PBH}, we have focused this time on PBHs as a DM candidate. We thus have derived limits on the fraction of DM consisting of PBHs using three complementary probes of their evaporation. Similarly to Chapter~\ref{chap:prop}, we computed the bounds using the predicted local flux of PBH-produced $e^\pm$ and the flux of secondary $X$-ray emissions to compare them with {\sc Voyager 1} and {\sc Xmm-Newton} data respectively. What changed from Chapter~\ref{chap:prop} was the injection of $e^\pm$, which comes from PBH evaporation this time. We used the numerical code \verb|BlackHawk| in order to compute it. In addition to that, we also evaluated the $511$ keV line emissions from both the annihilation of PBH-produced $e^+$ in the ISM and the decay of para-Ps formed by PBH-produced $e^+$ and free $e^-$ in the ionised ISM. To set the limit, we then used the longitudinal profiles of the $511$ keV line measured by {\sc Integral}.

The limits we derived using the three probes are shown in Figure~\ref{fig:moneyplot}, where we assumed that all PBHs are Schwarzschild ones and have the same mass, as well as the fiducial astrophysical scenario. The $511$ keV limit is the most stringent out of the three, and is strongest one to date in the $3\times 10^{15} - 2\times 10^{17}$ g PBH mass range. In Figure~\ref{fig:UncertsLimits} we also show how robust each of the bound is, finding that the $511$ keV one is also the most robust among the three. However, for a more optimistic astrophysical scenario, the {\sc Xmm-Newton} bound could be stronger than the $511$ keV one.

Finally, in Figure~\ref{fig:SomeLimits} we evaluated the impact of different spin and mass distributions on the limits. In particular, Kerr PBHs are expected to evaporate more particles at higher energies, which strengthen our bounds. Although it is unlikely that a PBH can acquire a near-extremal spin, they could achieve $a^\star \simeq 0.7$ through successive mergers. Moreover, if the PBH masses are distributed log-normally, which is a more realistic distribution than the monochromatic one, increasing values of the width of the distribution $\sigma$ results in a strengthening of the constraints as well, due to the increasing populating of lighter PBHs (whose evaporation rate is drastically higher than heavier ones).

\medskip

We conclude by exploring some possible future developments in the research conducted in this thesis. First, all of our bounds were derived without any assumptions on the astrophysical background. These result therefore in conservative constraints which could be improved by incorporating a realistic model of the astrophysical $X$-ray background. Second, our investigation on sub-GeV DM was done under the assumption of $s$-wave annihilations. Therefore a possible development would be to assume $p$-wave annihilations instead, in order to compute limits on the $p$-wave annihilation cross section. Third, we derived model-independent bounds on sub-GeV DM, but it would be interesting to see where these bounds lie when considering different BSM models that predict a sub-GeV DM candidate. This would be straightforward as long as we have the expressions of the branching ratios for the different DM annihilation or decay channels, the annihilation cross section or decay rate in terms of the couplings between DM and SM particles.

Finally, data from the recent $X$-ray telescope {\sc eRosita} have started to be released. These would be a great opportunity to improve our limits on sub-GeV DM at lower masses, or on PBHs at higher masses, as {\sc eRosita} can probe $X$-rays down to $0.2$ keV (instead of $2.5$~keV in the {\sc Xmm-Newton} data we used). In the further future, we can expect upcoming $\gamma$-ray observatories such as {\sc Amego} and {\sc e-Astrogam} to fill the MeV gap, and thus open new prospects for DM ID in this energy range.

\medskip

Although the true nature of DM remains an enigma, we can place our confidence in the vibrant and dedicated scientific community that tirelessly pushes the boundaries of both theoretical and experimental developments. The future is bright, and with patience and unwavering dedication, we will inch ever closer to unraveling the mysteries of DM.

%% file: Chapters/appendixA.tex

\pagestyle{fancy}

\lettrine[lines=3, nindent=1pt]{I}{n} this appendix we introduce some trigonometry tools to integrate the emissions of CR and radiation over a squared $b\times\ell$ or an annulus ROI. For that we formulate a generalised coordinates system that describes the position of a point on the ridge of a cone of aperture~$\theta$ pointing towards a position $(b,\ell)$ in the sky. This enables us to compute the expressions of cylindrical and radial coordinates in terms of the generalised ones, as well as the differential solid angle for each of the two ROI types. We then write the maximum value of the l.o.s.\ coordinate in the case where we integrate the CR and radiation emissions over a ROI encapsulated in the spatial boundary of the Galaxy, here a cylinder of radius $R_\textrm{max}$ and height $L$.

\section{Generalised coordinate system}
\label{sec:gencoords}

We formulate a coordinate system that enable us to perform integrations over $b\times\ell$ regions and annuli in the MW, starting by introducing some of the useful systems that describe the position of a target point in the Galaxy, which are illustrated in Figure~\ref{fig:coords}:
\begin{itemize}
	\item \textbf{Radial coordinate $r$}: We use this coordinate to parametrise the DM density profile, which is considered to be spherically symmetric. $r$ simply corresponds to the distance of the target from the GC.
	\item \textbf{Cylindrical coordinates $(R,z)$:} This Galactocentric system is useful to describe the distribution of baryonic matter and radiation fields which are contained in the Galactic disk. $R$ is the projection of $r$ on the GP, and $z$ is the height of the target from the GP.
	\item \textbf{Galactic coordinates $(s,b,\ell)$:} This is the heliocentric system we use to perform integrations over ROIs, where $s$ is the coordinate that runs along the l.o.s, and $(b,\ell)$ are the Galactic latitude and longitude, respectively.
\end{itemize}

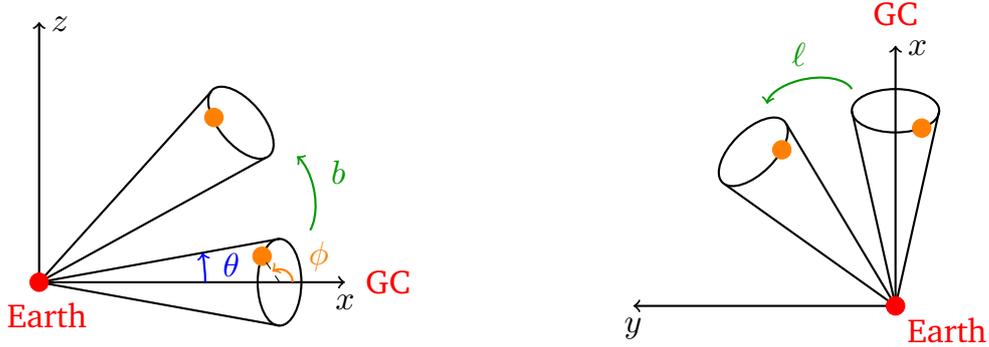
\begin{figure}
    \begin{subfigure}[c]{0.49\linewidth}
        \centering
        \begin{tikzpicture}[scale=1.15, every node/.style={transform shape}]
        
            \draw[thick,->] (0,0,0) -- (3.5,0,0) node[anchor=north]{$x$};
            \draw[thick,->] (0,0,0) -- (0,3,0) node[anchor=north west,yshift=0.2cm]{$z$};
            
            
            \draw[thick] (3,0,0) arc (0:360:0.25 and 0.5);
            
            \draw[thick] (0,0,0) --  (2.75,-0.5,0);
            \draw[thick] (0,0,0) --  (2.75,0.5,0);
            
            \draw[thick] (2.5,2,0) {[rotate=40] arc (0:360:0.25 and 0.5)};
    
            \draw[thick] (0,0,0) --  (1.97,2.2,0);
            \draw[thick] (0,0,0) --  (2.6,1.44,0);
            
            \draw[thick, ->, black!40!green] (3.1,0.6,0) {[rotate=10] arc (-60:60:0.25 and 0.5)} node[anchor=west,xshift=0.1cm,yshift=-0.2cm,color=black!40!green]{$b$};
            
            \filldraw [red] (0,0) circle (3pt) node[below right,xshift=-0.5cm,yshift=-0.1cm] {Earth} coordinate (E);        
            \draw [red] (3.6,0) node[right] {GC} coordinate (GC);
            \draw (2.75,0.5,0) node[below]{} coordinate (ap);
            
            \pic[draw, ->, thick, angle radius=1.9cm, blue]{angle=GC--E--ap} node[midway,xshift=2.2cm,yshift=0.2cm,blue]{$\theta$};
            
            \filldraw [orange] (2.55,0.3) circle (3pt) coordinate (T);      
            \draw (2.75,0) node[below]{} coordinate (PT);
            \draw (3,0) node[below]{} coordinate (YT);
            \draw[dashed] (PT) -- (T);
            
            \filldraw [orange] (2,1.9) circle (3pt);    
            
            \pic[draw, ->, thick, angle radius=0.15cm,orange]{angle=YT--PT--T} node[midway,xshift=3.2cm,yshift=0.3cm,orange]{$\phi$};
                        
        \end{tikzpicture}
    \end{subfigure}
    \hfill
    \begin{subfigure}[c]{0.49\linewidth}
        \centering
        \begin{tikzpicture}[scale=1.15, every node/.style={transform shape}]
        
            \draw[thick,->] (0,0,0) -- (-3,0,0) node[anchor=north]{$y$};
            \draw[thick,->] (0,0,0) -- (0,3,0) node[anchor=north west,yshift=0.2cm]{$x$};
            
            
            \draw[thick] (0.5,2.25,0) arc (0:360:0.5 and 0.25);
            
            \draw[thick] (0,0,0) --  (0.5,2.25,0);
            \draw[thick] (0,0,0) --  (-0.5,2.25,0);
            
            \draw[thick] (-1.45,1.6,0) {[rotate=-45] arc (0:360:0.25 and 0.5)};
    
            \draw[thick] (0,0,0) --  (-1.25,2.1,0);
            \draw[thick] (0,0,0) --  (-1.95,1.4,0);
            
            \draw[thick, ->, black!40!green] (-0.5,2.5,0) {[rotate=10] arc (10:170:0.5 and 0.25)} node[anchor=east,yshift=0.4cm,xshift=0.6cm,color=black!40!green]{$\ell$};
            
            \filldraw [red] (0,0) circle (3pt) node[below right] {Earth};
            \draw [red] (0,3.1) node[above] {GC};
            
            \filldraw [orange] (0.3,2.05) circle (3pt);
            \filldraw [orange] (-1.3,1.8) circle (3pt);

        \end{tikzpicture}
    \end{subfigure}
    \caption{Illustration of the rotations used to obtain the generalised coordinate system. Starting with a cone pointing towards the GC, we rotate it in the $xz$-plane by an angle $b$ (left panel), then in the $xy$-plane by an angle $\ell$ (right panel).}
    \label{fig:3Drotations}
\end{figure}

To generalise these coordinate systems, we consider a ROI in the form of a cone of aperture $\theta$ centered towards the GC, as pictured in the left panel of Figure~\ref{fig:3Drotations}. The position of a target point on the ridge of the cone's base is parametrised by the angle $\phi$ so that its cartesian coordinates are
\begin{equation}
	\left\{
	\begin{array}{l}
		x = s\cos\theta \\
		y = s\sin\theta\cos\phi \\
		z = s\sin\theta\sin\phi
	\end{array}
	\right.\;.
\end{equation}
We then proceed by rotating the cone by an angle $b$ in the $xz$-plane, as shown in the left panel of Figure~\ref{fig:3Drotations}, and write the new coordinates of the target
\begin{equation}
	x'+iz' = e^{ib}(x+iz) \implies 
	\left\{
	\begin{array}{l}
		x' = x\cos b - z\sin b \\
		y' = y \\
		z' = x\sin b + z\cos b
	\end{array}
	\right.\;.
\end{equation}
Finally, we rotate the system by an angle $\ell$ in the $xy$-plane, as shown in the right panel of Figure~\ref{fig:3Drotations}. The new coordinates are then
\begin{equation}
	x''+iy'' = e^{i\ell}(x'+iy') \implies 
	\left\{
	\begin{array}{l}
		x'' = x'\cos\ell - y'\sin\ell \\
		y'' = x'\sin\ell + y'\cos\ell \\
		z'' = z'
	\end{array}
	\right.
\end{equation}
\begin{equation}
	\label{eq:gencoords}
	\implies 
	\left\{
	\begin{array}{l}
		x'' = s[\cos b\cos\ell\cos\theta - (\sin\ell\cos\phi+\sin b\cos\ell\sin\phi)\sin\theta] \\
		y'' = s[\cos b\sin\ell\cos\theta + (\cos\ell\cos\phi-\sin b\sin\ell\sin\phi)\sin\theta] \\
		z'' = s(\sin b\cos\theta + \cos b\sin\theta\sin\phi)
	\end{array}
	\right.\;.
\end{equation}
We are now able to know the coordinates of a point on the ridge of a cone's base that points towards the position $(b,\ell)$ in the sky. This system can be used to both integrate over $b\times\ell$ regions (by setting $\theta$ to $0$) and annuli in the sky (by fixing $b$ and $\ell$). From there, we can express $R$ and $r$ in terms of the generalised coordinates by recalling that $R = \sqrt{(x-r_\odot)^2+y^2}$ and $r = \sqrt{R^2+z^2}$:
\begin{gather}
	\label{eq:R}
	\begin{split}
		R = \big[r_\odot^2 + s^2 &(\cos^2 b\cos^2\theta + \sin^2\theta \cos^2\phi + \sin^2 b\cos^2 \ell \sin^2\theta\sin^2\phi \\ 
		& + \sin^2b\sin^2\ell\sin^2\theta\sin^2\phi - 2 \cos b\sin b\cos\theta\sin\theta\sin\phi) \\
		+ 2s&r_\odot  (\sin b\cos\ell\sin\theta\sin\phi - \cos b\cos\ell\cos\theta + \sin\ell\sin\theta\cos\phi)\big]^{1/2}\;,
	\end{split} \\
	\label{eq:r}
	r = \left[r_\odot^2+s^2+2sr_\odot(\sin b\cos\ell\sin\theta\sin\phi-\cos b\cos\ell\cos\theta+\sin\ell\sin\theta\cos\phi)\right]^{1/2}\;.
\end{gather}

In a given coordinate system, the differential solid angle $d\Omega$ spanning from $\hat{\theta}$ to $\hat{\theta}+d\hat{\theta}$ and $\hat{\phi}$ to $\hat{\phi}+d\hat{\phi}$ on a spherical surface is expressed by
\begin{equation}
	\label{eq:diffsolidangle}
	d\Omega = \frac{1}{s^2}\left\lvert\left\lvert \frac{d\vec{r}}{\partial\hat{\theta}} \times \frac{d\vec{r}}{\partial\hat{\phi}} \right\rvert\right\rvert d\hat{\theta}d\hat{\phi}\;,
\end{equation}
where $\vec{r}$ is the position vector of the target, whose components are in Equation~\ref{eq:gencoords} in the generalised coordinate system (in which $(\hat{\theta},\hat{\phi})=(b,\ell)$ or $(\theta,\phi)$ depending on the considered ROI type).

\section{Integration over a \texorpdfstring{$\protect\bxl$}{b x l} region}

In the case where the ROI is a $b\times\ell$ region, we set $\theta=0$ in Equations~\ref{eq:R} and \ref{eq:r}, giving
\begin{equation}
	\left\{
	\begin{array}{l}
		R = \sqrt{r_\odot^2+s^2\cos^2 b - 2sr_\odot\cos b\cos\ell} \\
		z = s\sin b \\
		r = \sqrt{r_\odot^2+s^2 - 2sr_\odot\cos b\cos\ell}
	\end{array}
	\right.\;.
\end{equation}

By combining Equations~\ref{eq:gencoords} and \ref{eq:diffsolidangle} when $(\hat{\theta},\hat{\phi})=(b,\ell)$ and $\theta=0$, we obtain the differential solid angle for this type of region, which is
\begin{equation}
	d\Omega = \cos b\,db\,d\ell\;,
\end{equation} 
and therefore the total solid angle is
\begin{equation}
	\Delta\Omega = \int_{b_\textrm{min}}^{b_\textrm{max}}  \int_{\ell_\textrm{min}}^{\ell_\textrm{max}}  \cos b\,db\,d\ell = (\ell_\textrm{max}-\ell_\textrm{min})(\sin b_\textrm{max}-\sin b_\textrm{min})\;,
\end{equation}
where $b_\textrm{min}$, $b_\textrm{max}$, $\ell_\textrm{min}$ and $\ell_\textrm{max}$ delimit the $b\times\ell$ region.

\section{Integration over an annulus region pointed at \texorpdfstring{$\protect\bl$}{(b,l)}}

In the case where the ROI is an annulus centered on a position $(b,\ell)$ in the sky, the $R$ and $r$ are simply given by~\ref{eq:R} and \ref{eq:r}, where $b$ and $\ell$ are now fixed. In particular, when $(b,\ell)=(0,0)$ (\emph{i.e.}\ an annulus centered on the GC), we obtain\footnote{We remind the reader that $\cos^2\theta+\sin^2\theta\cos^2\phi=1-\sin^2\theta\sin^2\phi$.}
\begin{equation}
	\left\{
	\begin{array}{l}
		R = \sqrt{r_\odot^2+s^2(\cos^2 \theta + \sin^2\theta\cos^2\phi) - 2sr_\odot\cos\theta} \\
		z = s\sin\theta\sin\phi \\
		r = \sqrt{r_\odot^2+s^2 - 2sr_\odot\cos\theta}
	\end{array}
	\right.\;.
\end{equation}

And we also compute the differential solid angle for this type of ROI, by inserting Equation~\ref{eq:gencoords} into Equation \ref{eq:diffsolidangle} when $(\hat{\theta},\hat{\phi})=(\theta,\phi)$, for any $(b,\ell)$:
\begin{equation}
	d\Omega = \sin \theta\,d\theta\,d\phi\;,
\end{equation} 
and therefore the total solid angle is
\begin{equation}
	\Delta\Omega = \int_{\theta_\textrm{min}}^{\theta_\textrm{max}}  \int_0^{2\pi}  \sin\theta\,d\theta\,d\phi = 2\pi(\cos\theta_\textrm{min}-\cos\theta_\textrm{max})\;,
\end{equation}
where $\theta_\textrm{min}$ and $\theta_\textrm{max}$ delimit the aperture of the annulus.

\section{Upper integration bound for the line of sight coordinate}

In this section, we compute the maximum value of the l.o.s.\ coordinate $s$ in the setup where the spatial boundary of the Galaxy is a cylinder of radius $R_\textrm{max}$ and height $L$, useful for Section~\ref{subsec:secondary}. $s$ is at its maximum $s_\textrm{max}$ whenever one of the two following conditions is satisfied: $R(s) = R_\textrm{max}$ or $\lvert z(s)\rvert = L$. If the first condition is reached first, we have
\begin{gather}
	\label{eq:cond1}
	R(s_\textrm{max}) = R_\textrm{max} \implies (x(s_\textrm{max})-r_\odot)^2+y^2(s_\textrm{max})-R_\textrm{max} = 0 \\
	\label{eq:cond1sol}
	\implies s_\textrm{max}^\pm = \frac{r_\odot f_x \pm r_\odot\sqrt{f_x^2+\left(f_x^2+f_y^2\right)\left[\left(\frac{R_\textrm{max}}{r_\odot}\right)^2-1\right]}}{f_x^2+f_y^2}\;,
\end{gather}
where $f_i= i/s$ encapsulates the angular dependency of the coordinate $i$. Actually, the physical solution of Equation~\ref{eq:cond1} is when $+$ is taken in Equation~\ref{eq:cond1sol}, since the second term in the square root is always positive. We therefore have $s_\textrm{max} = s_\textrm{max}^+$. This solution corresponds to the maximum of $s$ for a cylinder of infinite height and finite radius $R_\textrm{max}$. On the other hand, if the second condition is satisfied first, we obtain
\begin{equation}
	\label{eq:cond2sol}
	\lvert z(s_\textrm{max}) \rvert = L \implies s_\textrm{max} = \frac{L}{\lvert f_z \rvert}\;,
\end{equation}
which corresponds to the maximum of $s$ for a cylinder of infinite radius and finite height $L$. To combine the two solutions, we simply take the minimum value between the two, giving
\begin{equation}
	s_\textrm{max} = \min\left\{\frac{r_\odot f_x + r_\odot\sqrt{f_x^2+\left(f_x^2+f_y^2\right)\left[\left(\frac{R_\textrm{max}}{r_\odot}\right)^2-1\right]}}{f_x^2+f_y^2}, \frac{L}{\lvert f_z \rvert}\right\}\;.
\end{equation}
In particular, in Galactic coordinates $(b,\ell)$, by taking $\theta = 0$ in Equation~\ref{eq:gencoords}, we obtain
\begin{equation}
	s_\textrm{max} = \min\left\{\frac{r_\odot}{\cos b}\left[\cos\ell + \sqrt{\left(\frac{R_\textrm{max}}{r_\odot}\right)^2-\sin^2\ell}\right], \frac{L}{\lvert\sin b\rvert}\right\}\;,
\end{equation}
and in spherical coordinates $(\theta,\phi)$, by taking $(b,\ell) = (0,0)$ in Equation~\ref{eq:gencoords}
\begin{equation}
	s_\textrm{max} = \min\left\{\frac{r_\odot}{1-\sin^2\theta\sin^2\phi}\left[\cos\theta + \sqrt{\cos^2\theta+\left(1-\sin^2\theta\sin^2\phi\right)\left[\left(\frac{R_\textrm{max}}{r_\odot}\right)^2-1\right]}\right], \frac{L}{\lvert\sin\theta\sin\phi\rvert}\right\}\;.
\end{equation}

%% file: Chapters/appendixB.tex

\begin{figure}[t]
    \centering
    \begin{subfigure}[c]{0.49\linewidth}
        \centering
        \includegraphics[width=\linewidth]{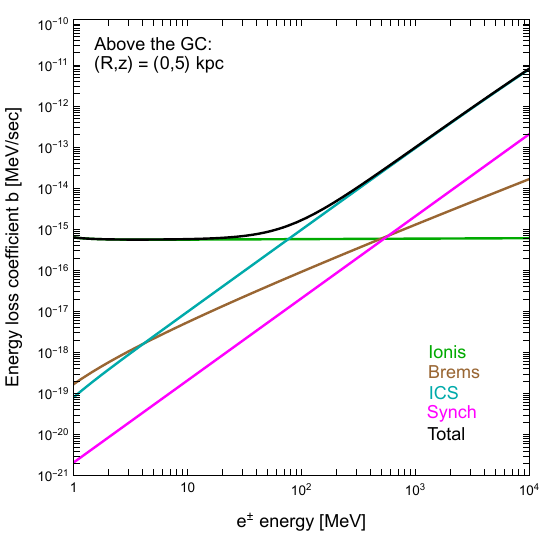}
    \end{subfigure}
    \hfill
    \begin{subfigure}[c]{0.49\linewidth}
        \centering
        \includegraphics[width=\linewidth]{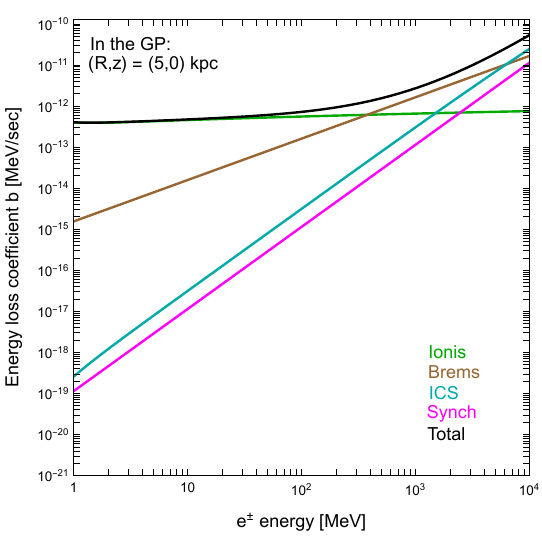}
    \end{subfigure}
    \caption{$e^\pm$ energy loss rates through ionisation and Coulomb interactions (green), bremsstrahlung (brown), ICS (cyan) and synchrotron radiation (magenta) above the GC (left panel) and in the GP (right panel). The total energy loss rates are shown in black.}
    \label{fig:elosses}
\end{figure}

\lettrine[lines=3, nindent=1pt]{I}{n} this appendix we provide the expression of the energy-loss function $b(E_e, \vec{x}) \equiv -\dot{E}_e$ for $e^\pm$ propagating through the Galaxy. They can suffer energy losses due to their scattering on radiation fields (ICS), interactions with the GMF (synchrotron), when passing by nuclei in the ISM (bremsstrahlung), or by ionising gas in the ISM and through Coulomb scattering on free $e^-$ in ionised gas. We essentially use the same formalism as in~\cite{Schlickeiser:2002pg,Buch:2015iya}. We also show in Figure~\ref{fig:elosses} these energy loss functions at two positions in the MW and in the $1-10^4$ MeV energy range, which was the relevant in research conducted in this thesis.

\section{Inverse Compton scattering}
\label{apxsec:ICS}

In the Thomson limit (which was used throughout the thesis), the $e^\pm$ energy losses through ICS are encoded in
\begin{equation}
	b_\textrm{ICS}(E_e,\vec{x}) = \frac{4c\sigma_\textrm{T}}{3m_e^2}E_e^2\int_0^\infty dE_\gamma\, n_\gamma(E_\gamma, \vec{x})\;,
\end{equation}
where $\sigma_\textrm{T} \equiv 8\pi r_e^2/3 = 6.65\times10^{-25}$ cm$^2$ is the Thomson scattering cross section (where $r_e$ is the $e^\pm$ classical radius), and $n_\gamma(E_\gamma,\vec{x})$ is the ambient photon number density at a position $\vec{x}$ in the Galaxy. The integral in this expression represents the total energy density of these photons.

\section{Synchrotron emission}
\label{apxsec:syn}

The energy losses of $e^\pm$ through the synchrotron radiation are written
\begin{equation}
	b_\textrm{syn}(E_e,\vec{x}) = \frac{4c\sigma_\textrm{T}}{3m_e^2}E_e^2\frac{B^2(\vec{x})}{8\pi}\;,
\end{equation}
where $B(\vec{x})$ is the intensity of the GMF at the position $\vec{x}$, summing its continuous and turbulent components. This expression is analogous to the ICS energy-loss coefficient, since $B^2/8\pi$ is the total energy density of the GMF.

\section{Bremsstrahlung}
\label{apxsec:brems}

The energy losses through bremsstrahlung are described by
\begin{equation}
	b_\textrm{brems}(E_e,\vec{x}) = c\sum_i n_i(\vec{x}) \int_0^{E_e}dE_\gamma\, E_\gamma \frac{d\sigma_i}{dE_\gamma}(E_\gamma,E_e)\;,
\end{equation}
where $n_i(\vec{x})$ is the number density of the gas species $i$ at a position $\vec{x}$ in the Galaxy and $d\sigma_i/dE_\gamma$ is the differential bremsstrahlung cross sections, which are written~\cite{Bethe:1934za}
\begin{equation}
	\frac{d\sigma_i}{dE_\gamma}(E_\gamma,E_e) = \frac{3\alpha\sigma_\textrm{T}}{8\pi E_\gamma}\left\{\left[1+\left(1-\frac{E_\gamma}{E_e}\right)^2\right]\phi^i_1(E_e,E_\gamma)-\frac{2}{3}\left(1-\frac{E_\gamma}{E_e}\right)\phi^i_2(E_e,E_\gamma)\right\}\;,
\end{equation}
where $\phi_{1,2}^i$ are scattering functions whose expressions are quite cumbersome to formulate, and out of the scope of the thesis. We however consider two limiting regimes. When the impinging $e^\pm$ is highly relativistic ($\gamma_e \gtrsim 10^3$), the electronic cloud around the atoms in the gas screens their nucleus, effectively decreasing the bremsstrahlung emission. The scattering functions are essentially constant in this \emph{strong shielding} regime. Their value for HI and HeI are given in Table~2 in~\cite{Blumenthal:1970gc} and rewritten here for the convenience of the reader:
\begin{gather}
	\phi_{1,\textrm{ss}}^\textrm{HI} = 45.79\;, \quad \phi_{2,\textrm{ss}}^\textrm{HI} = 44.46\;, \\
	\phi_{1,\textrm{ss}}^\textrm{HeI} = 134.60\;, \quad \phi_{2,\textrm{ss}}^\textrm{HeI} = 131.40\;,
\end{gather}
and the energy losses are then given by
\begin{equation}
	b_\textrm{brems}^\textrm{neut,ss}(E_e,\vec{x}) = \frac{3c\alpha\sigma_\textrm{T}}{8\pi}E_e^2 \sum_i n_i (\vec{x}) \left(\frac{4}{3}\phi_{1,\textrm{ss}}^i-\frac{1}{3}\phi_{2,\textrm{ss}}^i \right)\;.
\end{equation}
Otherwise, for less relativistic incoming $e^\pm$ ($\gamma_e \lesssim 10^2$), electronic screening becomes inefficient. This is referred to as the \emph{weak shielding} regime. It essentially corresponds to the case where the gas is ionised. In this regime, the scattering functions are written
\begin{equation}
	\phi^i_{1,\textrm{ion+ws}}(E_e,E_\gamma) = \phi^i_{2,\textrm{ion+ws}}(E_e,E_\gamma) = 4Z_i(Z_i+1)\left\{\log\left[\frac{2E_e}{m_ec^2}\left(\frac{E_e-E_\gamma}{E_\gamma}\right)\right]-\frac{1}{2}\right\}\,,
\end{equation}
where $Z_i$ is the number of protons in a nucleus of gas species $i$.
When the gas is ionised, these scattering functions are valid for the whole impinging $e^\pm$ energy range, since there is no more electronic screening. The energy loss are written, in the case the gas is ionised or neutral and weakly shielded
\begin{equation}
	b_\textrm{brems}^\textrm{ion+ws}(E_e,\vec{x}) = \frac{3c\alpha\sigma_\textrm{T}}{2\pi}E_e^2 \left[\log\left(\frac{2E_e}{m_e}\right) -\frac{1}{3}\right]\sum_i n_i (\vec{x}) Z_i(Z_i+1)\;.
\end{equation}
In order to study the intermediate regime between the strong- and weak-shielding ones (\emph{i.e.}\ $10^2 \lesssim \gamma_e \lesssim 10^3$), we can perform an interpolation.

\section{Coulomb scattering and ionisation}
\label{apxsec:coul+ion}

The energy loss rate of $e^\pm$ through their scattering on free $e^-$ in an ionised plasma is
\begin{equation}
	b_\textrm{Coul}(E_e,\vec{x}) = \frac{3c\sigma_\textrm{T}}{4}n_e(\vec{x})\left[\log\left(\frac{E_e}{m_e}\right)+2\log\left(\frac{m_e}{E_p(\vec{x})}\right)\right]\;,
\end{equation}
where $n_e(\vec{x})$ is the number density of free $e^-$ at a position $\vec{x}$ in the MW and $E_p \equiv \sqrt{4\pi n_e r_e^3}m_e/\alpha$ is the minimum excitation energy of the plasma. $e^\pm$ can also lose energy from ionising elements in a neutral gas. The energy loss function is written in this case
\begin{equation}
	b_\textrm{ion}(E_e,\vec{x}) = \frac{9c\sigma_\textrm{T}}{4}\sum_i n_i(\vec{x})Z_i \left[\log\left(\frac{E_e}{m_e}\right)+\frac{2}{3}\log\left(\frac{m_e}{\Delta E_i}\right)\right]\;,
\end{equation}
where $\Delta E_i$ is the average excitation energy of the gas species $i$, which is $15.0$ eV for HI and $41.5$~eV for HeI.